\documentclass[11pt]{paper}
\usepackage[latin1]{inputenc}
\usepackage{amsfonts,amsmath,amsthm,amssymb,mathrsfs}
\usepackage{fullpage,fancyhdr,hyperref}
\usepackage{enumitem,graphicx,booktabs,blkarray,subcaption}
\usepackage[dvipsnames]{xcolor}
\usepackage{esint}
\usepackage{float}
\usepackage{faktor}
\usepackage{cancel}
\usepackage{comment}
\usepackage[top=2cm,bottom=2cm,left=1.9cm,right=1.9cm,marginparwidth=1.75cm]{geometry}
\usepackage{overpic}
\usepackage{etoolbox}
\apptocmd{\thebibliography}{
    \setlength{\itemsep}{-1pt}
    \setlength{\parsep}{0pt}
}{}{}

\def\J{\mathcal{J}}
\def\L{\mathcal{L}}

\DeclareMathOperator{\Ai}{Ai}
\DeclareMathOperator{\arccosh}{arccosh}
\DeclareMathOperator{\arcsinh}{arcsinh}
\DeclareMathOperator{\arctanh}{arctanh}
\DeclareMathOperator{\arccoth}{arccoth}
\DeclareMathOperator*{\argmax}{arg\,max}
\DeclareMathOperator*{\argmin}{arg\,min}
\DeclareMathOperator{\dist}{dist}
\DeclareMathOperator*{\sgn}{sgn}
\DeclareMathOperator{\Cl}{Cl}

\renewcommand\Re{\operatorname{Re}}
\renewcommand\Im{\operatorname{Im}}

\makeatletter
\newcommand{\oref}[3][2ex]{
    \mathrel{\mathop{#3}\limits^{\vbox to#1{\kern-2\ex@
\hbox{$\scriptstyle\text{#2}$}\vss}}}}

\long\def\tmpcomment#1\endtmpcomment{}
\long\def\backupcomment#1\endbackupcomment{}

\newtheorem{theorem}{Theorem}
\newtheorem{corollary}{Corollary}

\newtheorem{proposition}{Proposition}
\newtheorem{lemma}{Lemma}
\newtheorem*{cuprinciple*}{Conformal-Uniformal Principle}

\theoremstyle{remark}
\newtheorem{remark}{Remark}

\theoremstyle{definition}
\newtheorem{definition}{Definition}

\title{Quantum walks and conformal mappings}

\author{
    \begin{minipage}[t]{0.25\textwidth}
        Fedor Kuyanov \\[5pt]
        {\small \textnormal{\textbf{University of Oxford}}}
    \end{minipage}
    \begin{minipage}[t]{0.25\textwidth}
        Michael Drmota \\[5pt]
        {\small \textnormal{\textbf{TU Wien}}}
    \end{minipage}
    \begin{minipage}[t]{0.25\textwidth}
        Mikhail Skopenkov \\[5pt]
        {\small \textnormal{\textbf{KAUST}}}
    \end{minipage}
    \begin{minipage}[t]{0.25\textwidth}
        Alexey Ustinov \\[5pt]
        {\small \textnormal{\textbf{HSE University}}}
    \end{minipage} \\ \vspace{-10pt}
}

\begin{document}

\maketitle

\begin{abstract}
    We study the most elementary discrete model of an electron moving along a line, known as Feynman checkers, quantum walk, or the six-vertex model with special complex weights, and provide a uniform asymptotic formula for the wave function in the whole $xt$-plane. In addition to the well-known appearance of the Airy function that describes the transition between the oscillating and exponentially decaying regimes, we observe a second transition governed by the gamma function when the electron speed approaches the speed of light. We prove two general theorems producing uniform asymptotic formulae involving the Airy and gamma functions, using conformal mappings associated with the underlying integral representations. They are applicable in several other situations as well.


\end{abstract}

\section{Introduction}

\begin{figure}[htb]
    \centering
    \begin{tabular}{cccccc}
        \multicolumn{2}{c}{
            \begin{overpic}[width=0.3\textwidth]{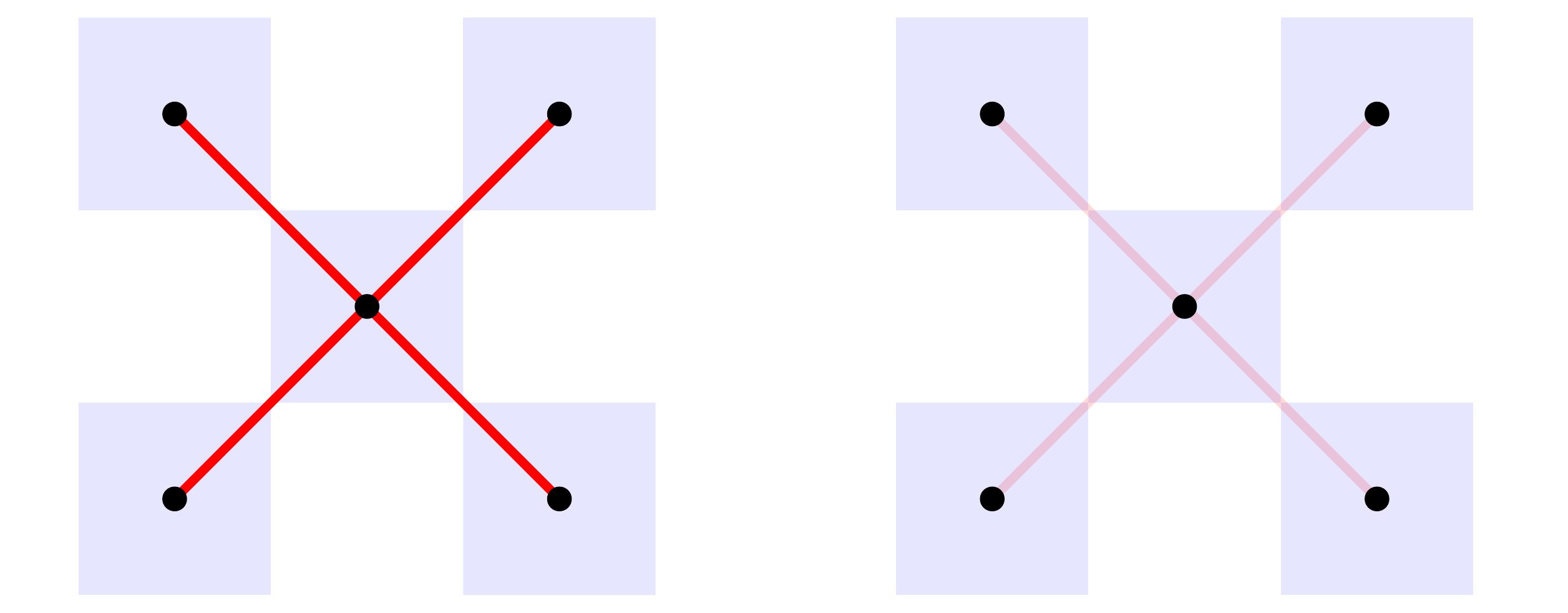}
                \footnotesize{
                    \put(4.5,0){$\underbrace{\phantom{\hspace{0.27\textwidth}}}$}
                }
            \end{overpic}
        } &
        \multicolumn{2}{c}{
            \begin{overpic}[width=0.3\textwidth]{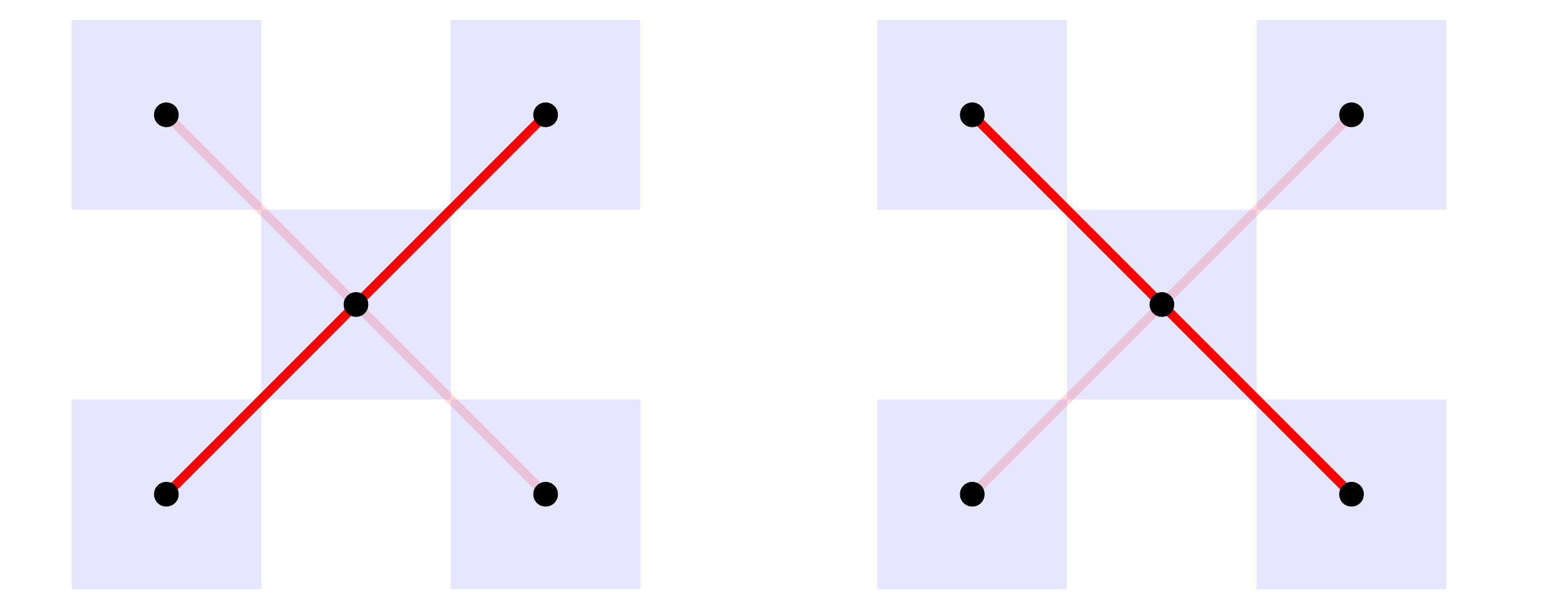}
                \footnotesize{
                    \put(4,0){$\underbrace{\phantom{\hspace{0.27\textwidth}}}$}
                }
            \end{overpic}
        } &
        \multicolumn{2}{c}{
            \begin{overpic}[width=0.3\textwidth]{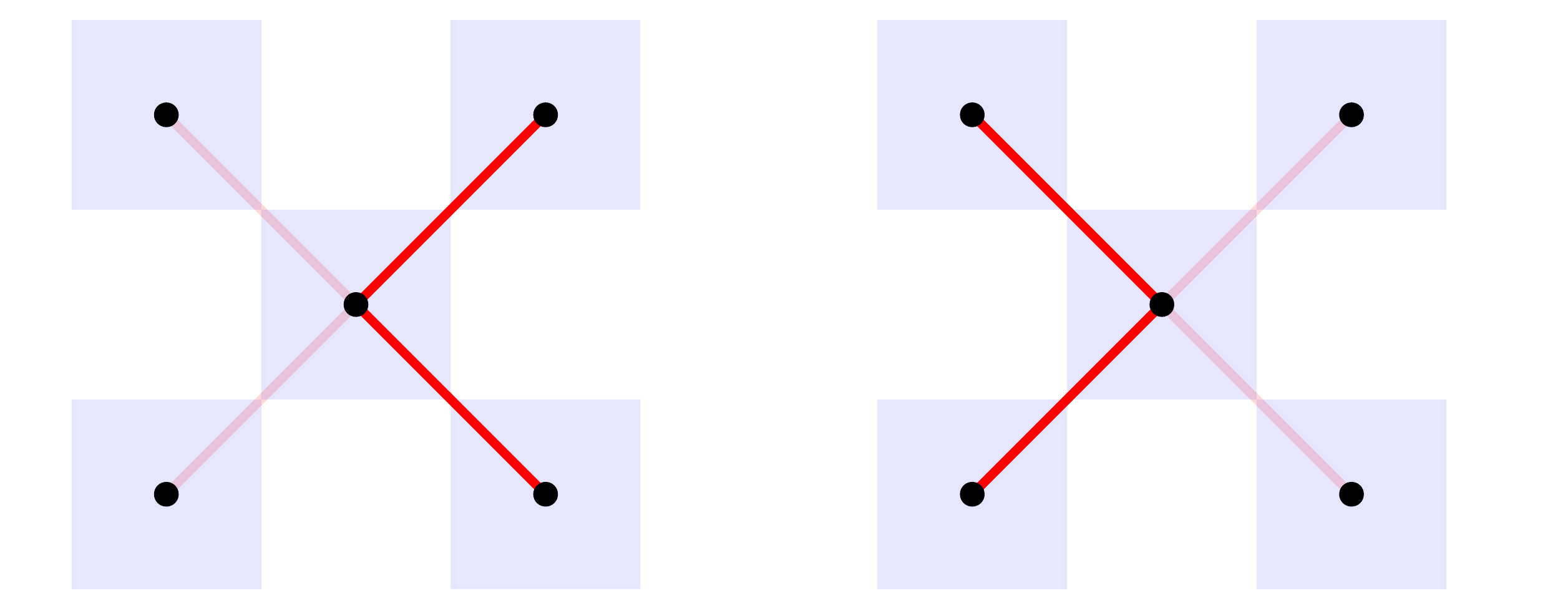}
                \footnotesize{
                    \put(4,0){$\underbrace{\phantom{\hspace{0.27\textwidth}}}$}
                }
            \end{overpic}
        }
        \vspace{4pt} \\
        \multicolumn{2}{c}{$1$} &
        \multicolumn{2}{c}{$1 / \sqrt{2}$} &
        \multicolumn{2}{c}{$-i / \sqrt{2}$} \\
        \vspace{0pt}
    \end{tabular}
    \begin{overpic}[width=0.1\linewidth]{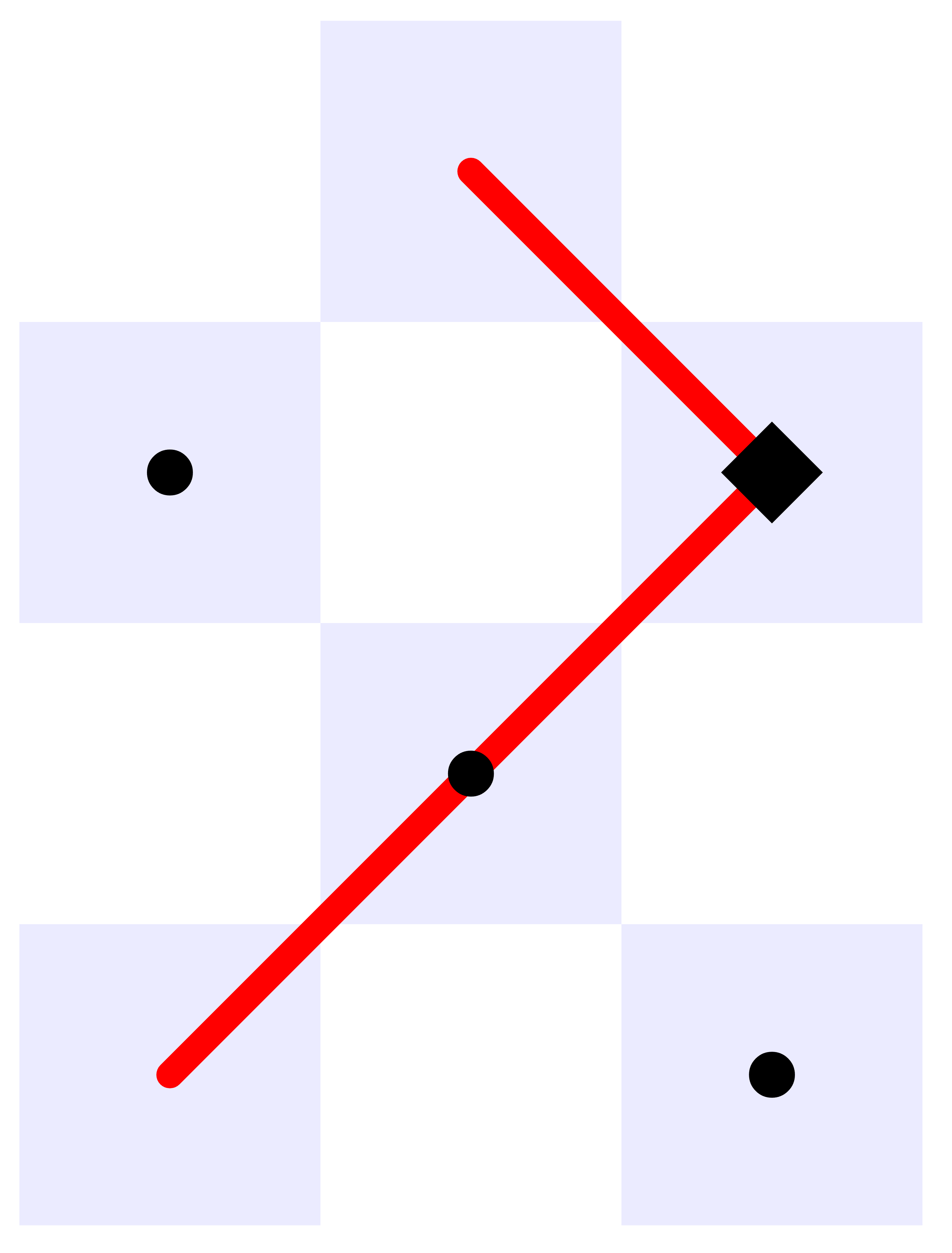}
        \put(-3,0){\footnotesize{$(0, 0)$}}
        \put(24,94){\footnotesize{$(x, t)$}}
        \put(35,45){\footnotesize{$s$}}
        \put(6,15.0){\rotatebox{-45}{$\blacktriangle$}}
        \put(30.5,94.0){\rotatebox{-135}{$\blacktriangledown$}}
    \end{overpic}
    \qquad
    \begin{overpic}[width=0.1\linewidth]{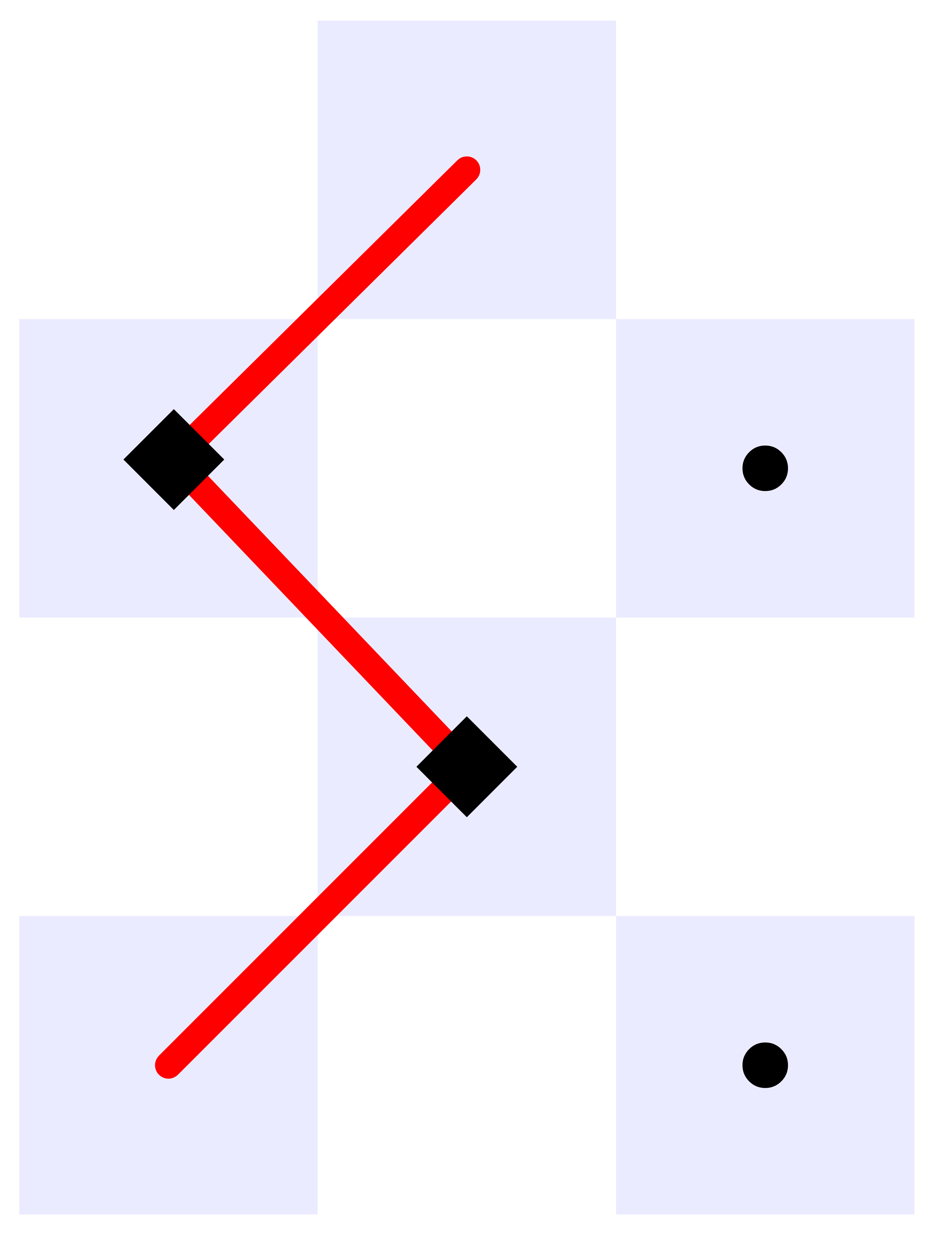}
        \put(-3,0){\footnotesize{$(0, 0)$}}
        \put(24,94){\footnotesize{$(x, t)$}}
        \put(35,45){\footnotesize{$s'$}}
        \put(6,15.0){\rotatebox{-45}{$\blacktriangle$}}
        \put(30.5,88.7){\rotatebox{-45}{$\blacktriangle$}}
    \end{overpic}
    \qquad
    \includegraphics[width=0.5\textwidth]{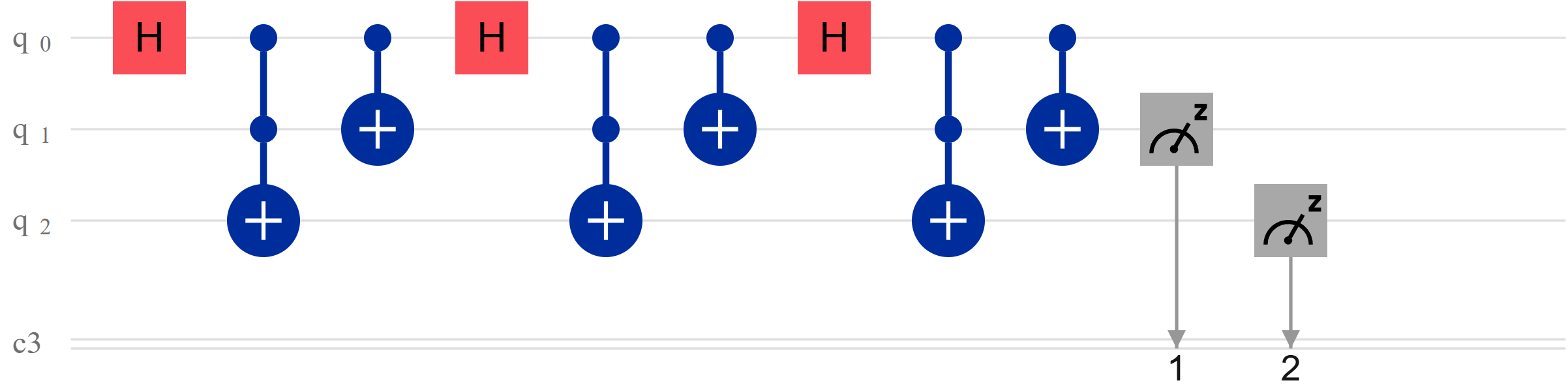} \\
    \vspace{1em}
    \caption{A quantum walk (namely, the Hadamard walk) viewed as: (top) a six-vertex model with the complex weights shown under the configurations;
    (bottom-left) a sum over paths; (bottom-right) a quantum algorithm in the quantum-circuit language. See Sec.~\ref{sec-checkers-notions} and~\cite[Figure~6]{SkoUst-22} for details.}
    \label{fig-checkers-model}
\end{figure}

\begin{figure}[htb]
    \centering
    \begin{subfigure}[c]{0.4\textwidth}
        \includegraphics[width=\textwidth]{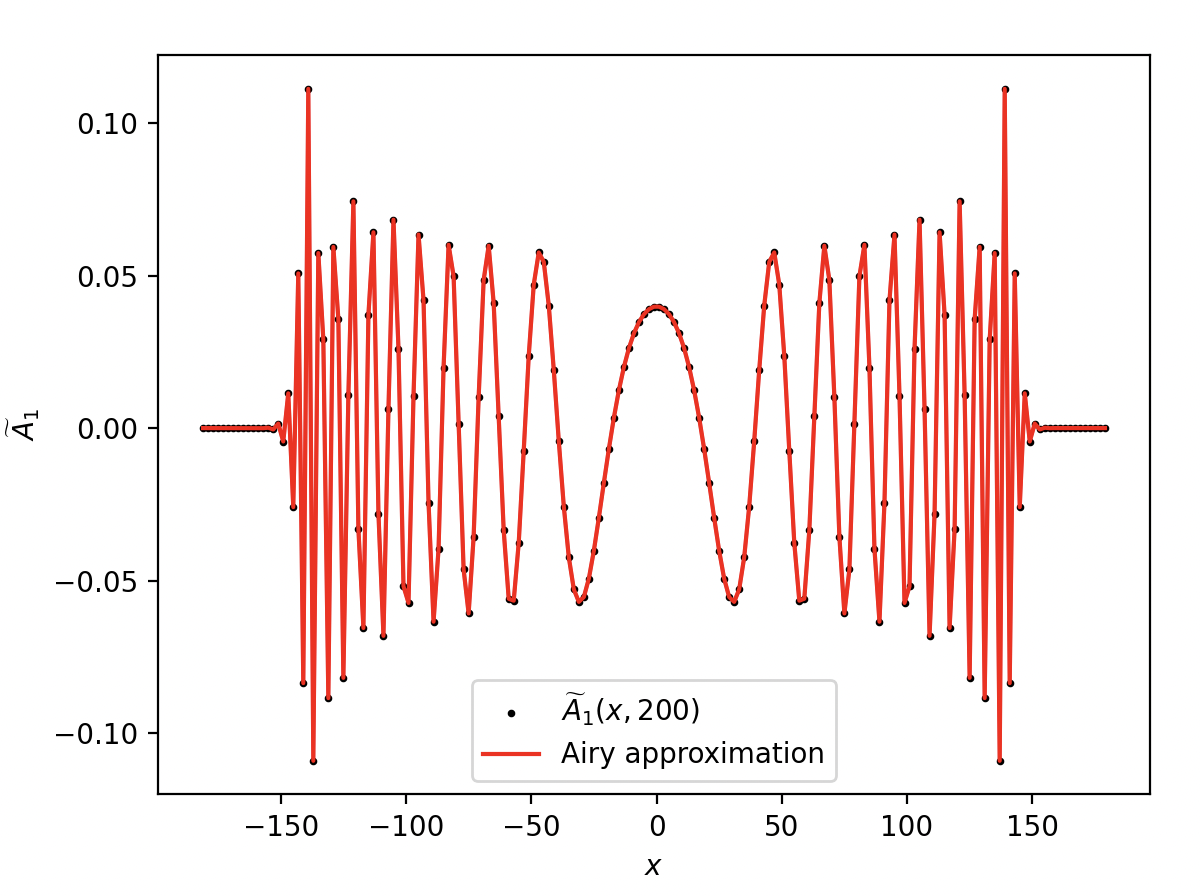}
        \caption{real part between the peaks}
        \label{fig-checkers-a1-between}
    \end{subfigure}
    \begin{subfigure}[c]{0.4\textwidth}
        \includegraphics[width=\textwidth]{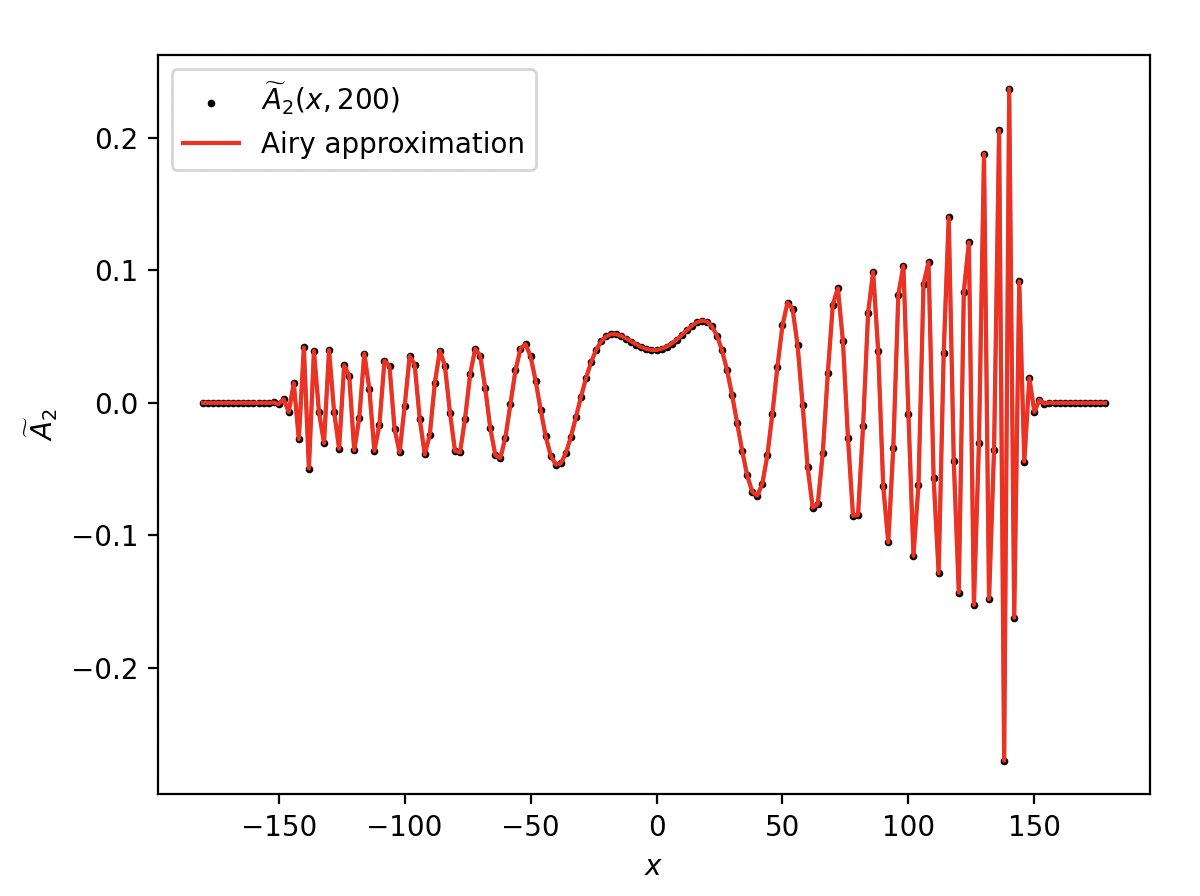}
        \caption{imaginary part between the peaks}
    \end{subfigure}
    \begin{subfigure}[c]{0.4\textwidth}
        \includegraphics[width=\textwidth]{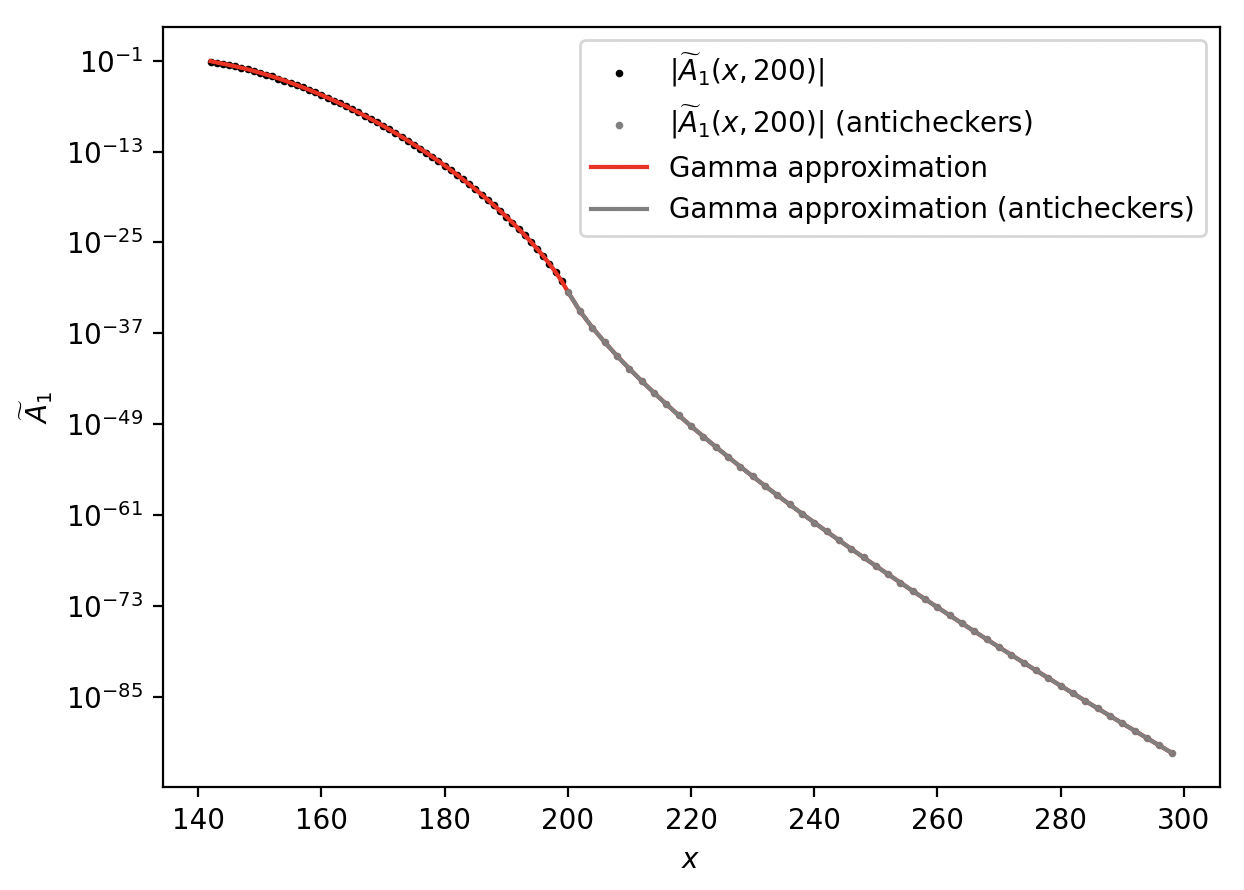}
        \caption{real part outside the peaks}
    \end{subfigure}
    \begin{subfigure}[c]{0.4\textwidth}
        \includegraphics[width=\textwidth]{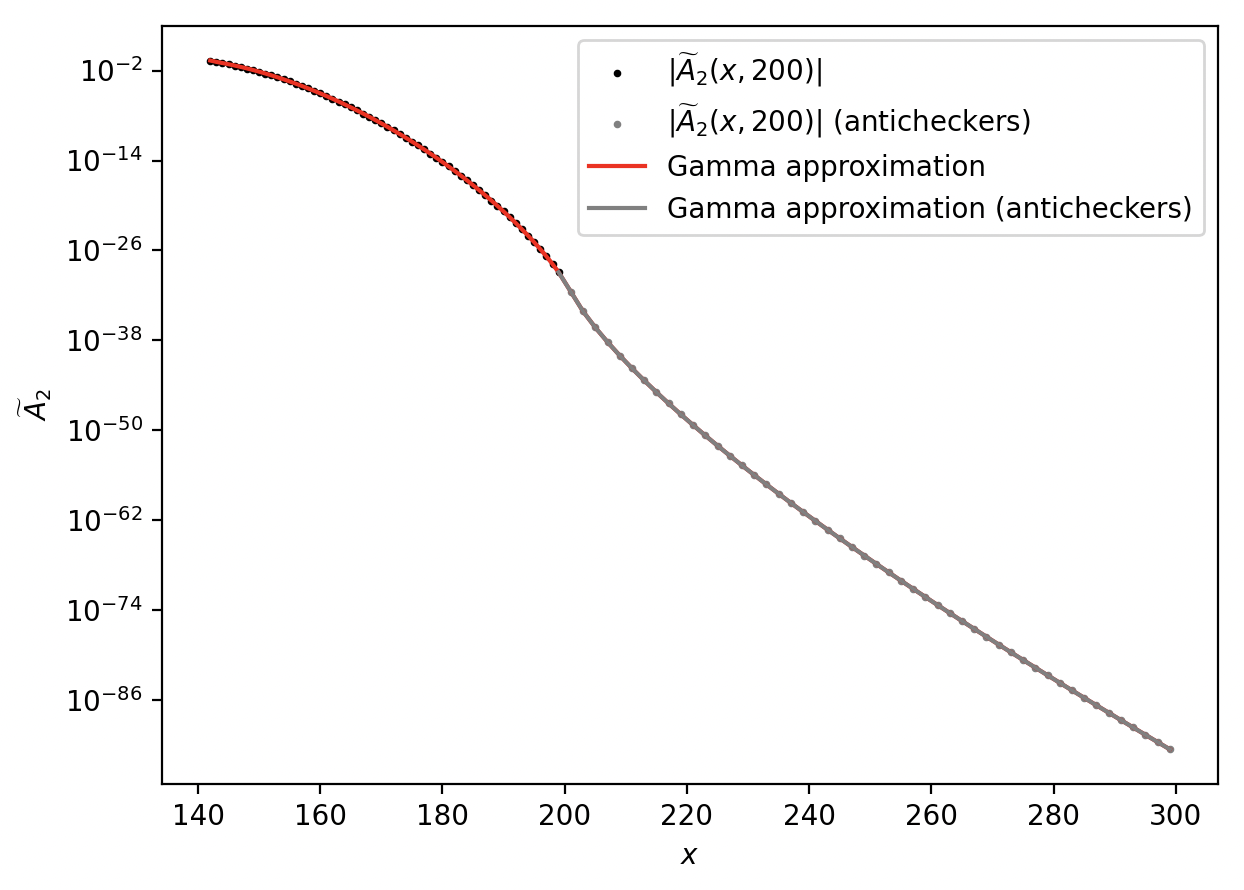}
        \caption{imaginary part outside the peaks}
    \end{subfigure}
    \caption{Real and imaginary parts of the wave function of a Hadamard walk at the fixed time $t = 200$ and their uniform approximations from Theorems~\ref{th-airy-checkers}~and~\ref{th-gamma-checkers}. Gray color depicts the values for Feynman \emph{anti}checkers, a generalization involving the creation and annihilation of particle-antiparticle pairs.}
    \label{fig-checkers-plot}
\end{figure}

One of the main open problems in mathematics is a rigorous construction of quantum field theory. For instance, the case of $4$-dimensional Yang--Mills theory with a mass gap is a Millennium Prize Problem. A standard approach is \emph{constructive field theory}, which often constructs a continuum theory as a limit of discrete theories~\cite{Glimm-Jaffe-12}. This involves two limits: the \emph{continuum limit}, as the discretization step tends to zero, and the \emph{infinite-volume/large-time limit}, as the system size/evolution time tends to infinity. In this work, we find uniform large-time asymptotic formulae for the most elementary discrete model of an electron moving along a line, known as \emph{Feynman checkers}, \emph{quantum walks}, or \emph{six-vertex model} with special complex weights (Figs.~\ref{fig-checkers-model} and~\ref{fig-checkers-plot}). See an elementary introduction~\cite{Ozhegov-25} and surveys~\cite{Kempe-09,Konno-20,SkoUst-22,Venegas-Andraca-12}.

This model is a promising direction in the development of constructive field theory because it involves \emph{real} time, whereas in most other theories, time is \emph{imaginary} \cite{Rothe-12,Zee-10}. However, computations in models with real time lead to oscillatory integrals rather than exponentially damped ones. For such integrals, estimates are obstructed by the \emph{sign problem} of huge cancellation during integration. Getting asymptotic formulae that are \emph{uniform} in the involved parameters (and therefore most useful) is even more challenging.

Well-developed tools for such estimates are \emph{stationary-phase}, \emph{saddle-point}, and \emph{steepest-descent} methods~\cite{Huxley-96}. Their natural limitation is the collision of critical points and/or singularities as parameters vary. The latter cases can still be analyzed using the \emph{semiclassical approximation}, such as \emph{Wentzel--Kramers--Brillouin (Liouville--Green) method} \cite{Maslov-01}. Those methods provide the main term in the asymptotic formula (and even the whole asymptotic series, which need not converge). Collision of critical points typically leads to the \emph{Airy phenomenon}, i.e., the Airy function in the main term \cite{Banderier-random-01,Banderier-12,Olver-97}. Collision of a critical point with a singularity can lead to other special functions, in particular the gamma function \cite{Bleistein-66,Borodin-Olshansky-03}. The latter case can be called the \emph{gamma phenomenon}. Despite numerous illuminating examples and theories~\cite{Maslov-87,NIST-10,Olver-97,Pemantle-24}, we are not aware of a single general theorem providing a \emph{nontrivial uniform bound on the approximation error throughout the entire parameter domain} in these collision cases (before this work). Here, a nontrivial bound means that the error term is ``uniformly asymptotically smaller'' than a ``typical'' main term.

\begin{figure}[htb]
    \centering
    \begin{subfigure}[c]{0.49\textwidth}
        \hspace{-0.3cm}
        \includegraphics[height=0.89\textwidth]{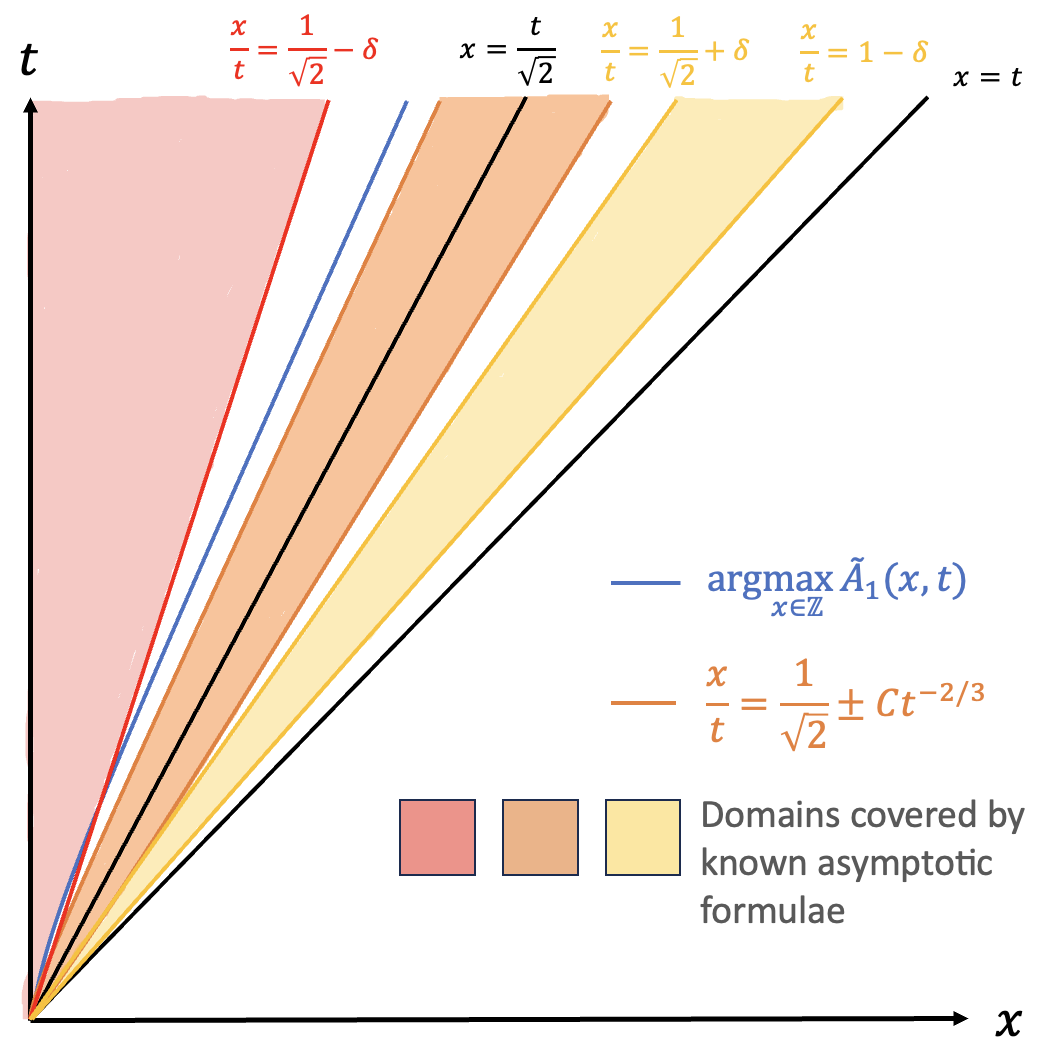}
        \caption{known results}
        \label{fig-intro-domains-before}
    \end{subfigure}
    \begin{subfigure}[c]{0.49\textwidth}
        \hspace{-0.7cm}
        \includegraphics[height=0.89\textwidth]{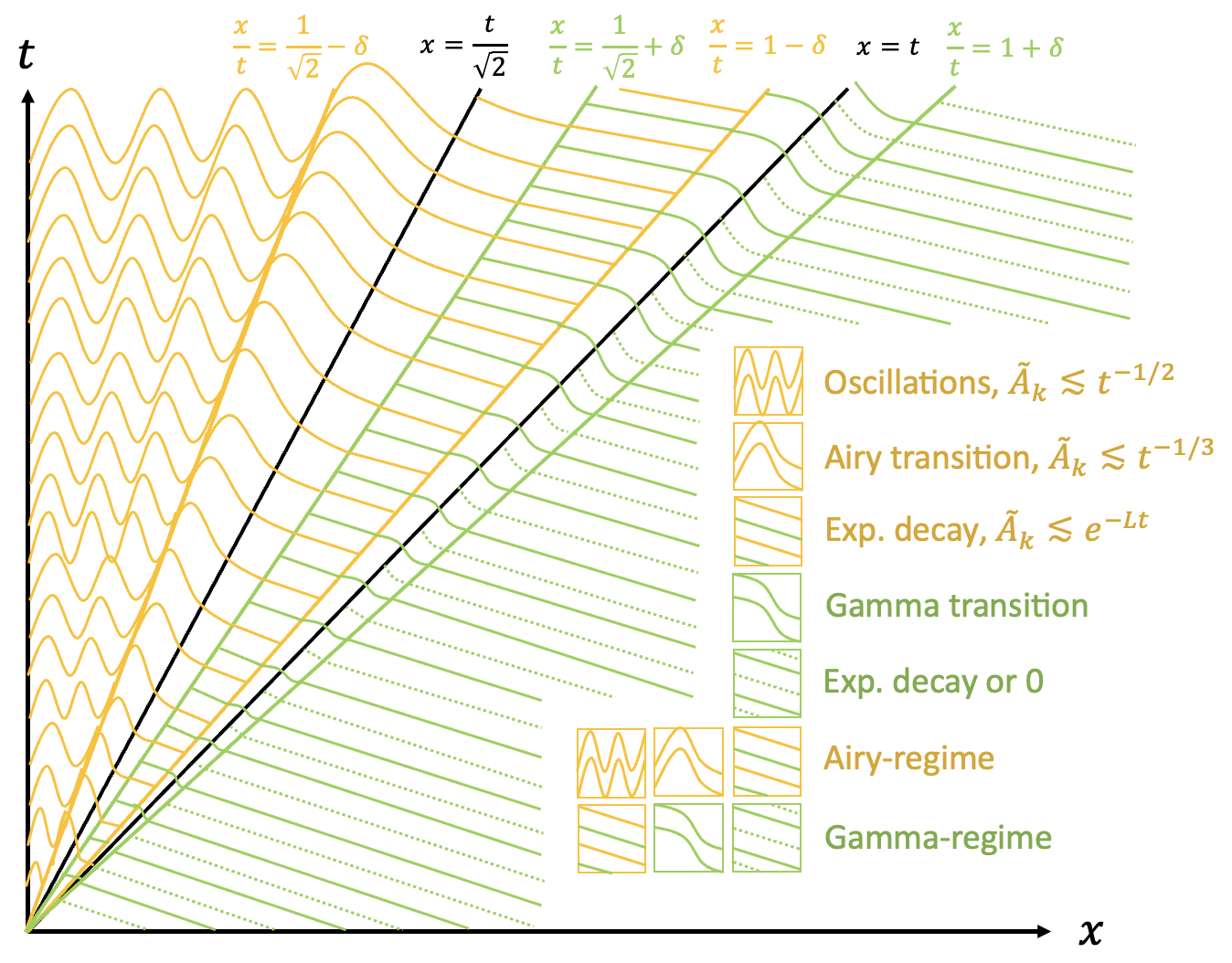}
        \caption{our results}
        \label{fig-intro-domains-after}
    \end{subfigure}
    \caption{(a): Domains where large-time asymptotic formulae for the wave function of a one-dimensional quantum walk (Hadamard walk) were known before with mathematical proofs. The point of maximum is near the boundary of those domains; thus, the known results are insufficient to estimate the maximum. (b): Domains, in which the asymptotic formulae for new Theorems~\ref{th-airy-checkers} and~\ref{th-gamma-checkers} hold. Together, they cover the whole plane. Different regimes and transitions between them are observed as specified in Tab.~\ref{tab-checkers-asymp}. For $x / t > 1$, the wave function vanishes for a quantum walk, but not for its modification involving the creation and annihilation of particle-antiparticle pairs.}
    \label{fig-intro-domains}
\end{figure}

For quantum walks, the large-time behavior of the wave function has been studied extensively, as summarized in Fig.~\ref{fig-intro-domains}; see also Tab.~\ref{tab-checkers-asymp} in Sec~\ref{sec-checkers-results}. R.~Feynman~\cite[Problem~2.6]{Feynman-Hibbs-65} initiated this research by posing a similar problem related to the particular case of small average velocity $x / t$. A.~Ambainis et al.~\cite{Ambainis-etal-01} demonstrated that the wave function has two peaks when the average velocity is roughly $\pm 1 / \sqrt{2}$ (blue curve in Fig.~\ref{fig-intro-domains-before}), oscillates between the peaks (red region), and exponentially decays outside (yellow region). See also Fig.~\ref{fig-checkers-plot}. Carteret et al.~\cite{Carteret-etal-03} showed that the transition between the regimes is described by the Airy function, and computed the main term of the asymptotic formula away from the light cone $x / t = \pm 1$. N.~Konno~\cite{Konno-20} and G.~Grimmett, S.~Janson, and P.F.~Scudo \cite{Grimmett-etal-04} found the weak large-time limit, which ``averages'' oscillations; cf.~simpler proofs by A.~Gottlieb et al.~\cite{Gottlieb-et-al-05} and two of the authors~\cite{SkoUst-22}.

First rigorous error estimates were proved by T.~Sunada and T.~Tate \cite[Propositions~2.2, 3.1, and~4.1]{Sunada-Tate-12} in the regions shown in red, orange, and yellow in Fig.~\ref{fig-intro-domains-before}. They were not yet uniform, leaving open the question of the precise asymptotic behavior of the peaks, known as \emph{dispersive estimates}~\cite{Maeda-etal-18, Maeda-etal-22, Mazutani-11}. A \emph{uniform} error bound between the peaks, excluding (respectively, including) their vicinity, was obtained by two of the authors \cite[Theorem~2]{SkoUst-22} (respectively, by P.~Zakorko \cite[Theorem~1]{Zakorko}, private communication), and led to the first rigorous solution of the original Feynman's problem \cite[Corollary~5]{SkoUst-22}. A.~Borodin (private communication) posed the problem of extending these results to the region outside the peaks. Here, the main difficulty is describing the \emph{transition} between different regimes (white regions), when we are not too close and not too far from the peaks, and especially, when we are close to the light cone~$x / t = \pm 1$.

\subsection{Contributions}

For the first time, we provide an asymptotic formula for the wave function of the one-dimensional quantum walk with a uniform error bound in the whole $xt$-plane. Namely, we cover the plane by two regions, where the formula involves Airy and gamma functions, respectively (see Theorems~\ref{th-airy-checkers} and~\ref{th-gamma-checkers} and Fig.~\ref{fig-intro-domains-after}). So, quantum walks exhibit both the common Airy phenomenon (near so-called turning points) and a less-known \emph{gamma phenomenon} (near the light cone). The asymptotic formula extends to the modification of the quantum walk, allowing the creation and annihilation of particle-antiparticle pairs \cite{SkoUst-24}.

We prove the first general theorems producing such uniform asymptotic formulae from certain conformal mappings (see Theorems~\ref{th-airy} and~\ref{th-gamma}). This way we demonstrate that the gamma phenomenon is universal and bears deep connections to the well-known Airy phenomenon and conformal geometry (see Tab.~\ref{tab-airy-gamma}).

As an illustrative application of our uniform estimates, we find asymptotic formulae for the height and position of the peak of the wave function in the one-dimensional quantum walk (see Proposition~\ref{prop-maximum}).

\begin{table}[h]
    \caption{Comparison of the Airy and gamma phenomena. The latter is characterized by a factor of $1 / \Gamma(x)$ in the asymptotic formula. For positive integer $x$, the factor decays exponentially, and for negative integer $x$, it vanishes due to the poles of the gamma function. Refer to Sec.~\ref{sec-main-idea} for the intuition regarding the Lagrangian, its normal form, and resulting asymptotic formulae.}
    \label{tab-airy-gamma}
    \centering
    \begin{tabular}{ll}
        \toprule
        \multicolumn{1}{c}{\textbf{Airy phenomenon (Theorem~\ref{th-airy})}} & \multicolumn{1}{c}{\textbf{Gamma phenomenon (Theorem~\ref{th-gamma})}} \\
        \midrule
        Airy function in the asymptotic formula & Gamma function in the asymptotic formula \\
        Transition from oscillations to exponential decay & Transition from exponential decay to identical zero \\
        Two zeros of the Lagrangian's derivative collide & A zero and a pole of the Lagrangian's derivative collide \\
        Lagrangian's normal form is $F(z) = z^3 / 3 - \theta z$ & Lagrangian's normal form is $F(z) = v\log z - z + \Delta$ \\
        Its derivative is $F'(z) = (z - \sqrt \theta)(z + \sqrt \theta)$ & Its derivative is $F'(z) = (v - z) / z$
        \\[5pt]
        \multicolumn{2}{c}{\textbf{Common connections to conformal geometry}} \\[5pt]
        \multicolumn{2}{c}{Lagrangian's derivative maps the fundamental domain conformally to a half-plane (Fig.~\ref{fig-airy-w'-inner}/Fig.~\ref{fig-gamma-p'-inner})} \\
        \multicolumn{2}{c}{Lagrangian maps the fundamental domain conformally to a standard domain (Fig.~\ref{fig-airy-L-F}/Fig.~\ref{fig-gamma-L-F})} \\
        \bottomrule
    \end{tabular}
\end{table}

\subsection{Main idea}
\label{sec-main-idea}

The main new idea of the paper is the following informal principle.

\begin{cuprinciple*}
    Conformal-geometric interpretations lead to uniform asymptotics.
\end{cuprinciple*}

Let us illustrate this idea on a toy example, which will later extend to more complicated scenarios.

\paragraph{Conformal interpretation of Einstein's formula}
Einstein's formula $E = mc^2$ is an instance of the relation $E = \sqrt{m^2c^4 + p^2c^2}$ between the energy $E$ and the momentum $p$ of a relativistic particle. In the natural units, where $c = \hbar = 1$, which we use in what follows, it takes the form
$$
E(p) = \sqrt{p^2 + m^2}.
$$
The derivative
\begin{equation}
    \label{eq-einstein-E'}
    E'(p) = \frac{p}{\sqrt{p^2 + m^2}}
\end{equation}
is interpreted as the particle velocity as a function of the momentum (by Hamilton's equations).

\begin{figure}
    \centering
    \begin{subfigure}[c]{0.44\textwidth}
        \includegraphics[width=\textwidth]{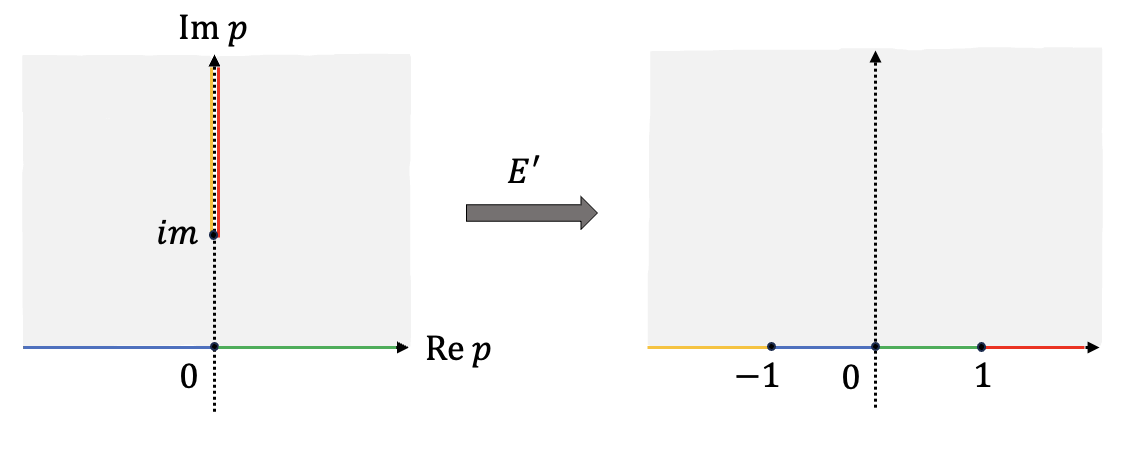}
        \caption{in Einstein's formula; see~\eqref{eq-einstein-E'}}
        \label{fig-intro-einstein-E'}
    \end{subfigure}
    \hspace{0.02\textwidth}
    \begin{subfigure}[c]{0.5\textwidth}
        \includegraphics[width=\textwidth]{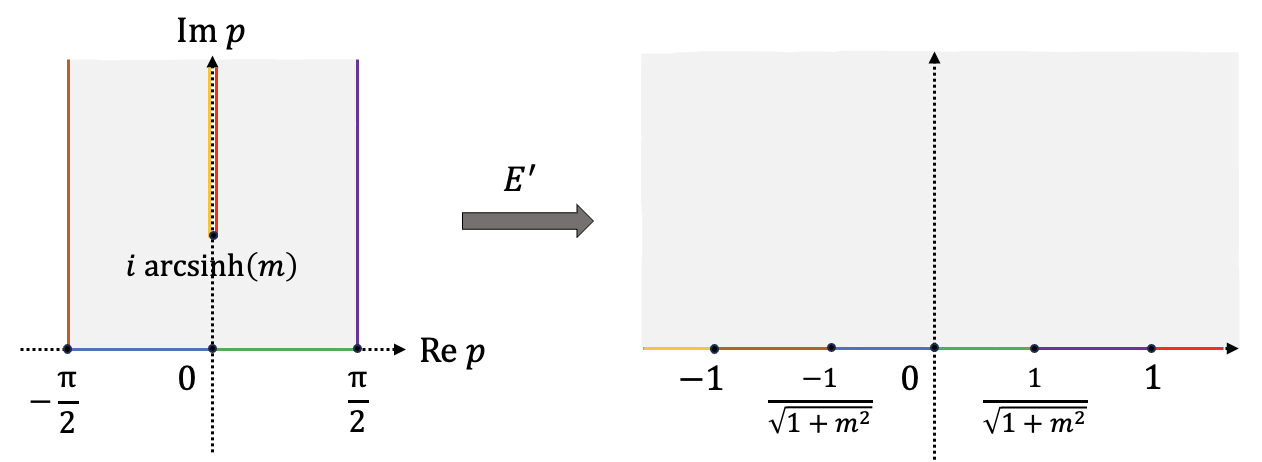}
        \caption{in quantum walks; see~\eqref{eq-checkers-E'}}
        \label{fig-intro-checkers-E'}
    \end{subfigure}
    \caption{The conformal map $E'(p)$. Boundary intervals and their images are shown in the same colour.}
    \label{fig-intro-E'}
\end{figure}

We now allow complex values of $p$ and extend the function $E'(p)$ to the upper half-plane $\Im p > 0$. Choosing the principal branch of the square root forces a slit from $im$ to $+i\infty$, resulting in the natural domain of definition of the function. The key observation is that $E'(p)$ is a conformal map (see Fig.~\ref{fig-intro-einstein-E'}):
\begin{itemize}
    \item[$\rhd$] \emph{the derivative $E'(p)$ is a 1--1 conformal map of its natural domain of definition to the upper half-plane}.
\end{itemize}

This remains true for a non-relativistic particle, for which the energy is expressed in terms of the momentum as $E(p) = {p^2}/{2m}$: the natural domain of definition is the upper half-plane, and $E'(p) = p / m$.

\paragraph{Conformal interpretation of the dispersion relation of a quantum walk}
The same structure persists in discrete models. Consider the relation between the energy $E$ and the momentum $p$ of a quantum walk \cite{SkoUst-22}:
$$
E(p) = \arccos\left(\frac{\cos p}{\sqrt{1 + m^2}}\right).
$$
The derivative
\begin{equation}
    \label{eq-checkers-E'}
    E'(p) = \frac{\sin p}{\sqrt{\sin^2 p + m^2}}
\end{equation}
is a discrete analog of \eqref{eq-einstein-E'}, still interpreted as the particle velocity as a function of the momentum.

Again, we extend the function $E'(p)$ to the complex plane. The function has a lot of symmetries, so that the values can be reconstructed from the ones in the so-called \emph{fundamental domain}, given by $\Im p > 0$, $|\Re p| < \pi / 2$. Choosing the principal branch of the square root forces a slit from $i\arcsinh m$ to $+i\infty$, resulting in the natural domain of definition of the function. Again (see Fig.~\ref{fig-intro-checkers-E'}):
\begin{itemize}
    \item[$\rhd$] \emph{the derivative $E'(p)$ is a 1--1 conformal map of its natural domain of definition to the upper half-plane},
\end{itemize}
showing the phenomenon is not tied to a specific model.

Surprisingly, this property of $E'(p)$ and the boundary behavior of $E(p)$ are the only essential assumptions of our Theorems~\ref{th-airy} and~\ref{th-gamma} (up to cosmetic changes) sufficient to prove the Airy and gamma phenomena.

\paragraph{Interpretation of imaginary momenta}
The conformal interpretation is clearer in quantum mechanics. By Schr\"odinger's equation, a particle with momentum $p$ and energy $E(p) = \sqrt{p^2 + m^2}$ has the wave function
\begin{equation}
    \label{eq-particle-wf}
    \psi(x, t) = e^{ipx-iE(p)t} = e^{it(pv-E(p))},
\end{equation}
where $v = x / t$. This expression makes perfect sense for complex $p$ with $\Im p > 0$ as the solution with the initial data $\psi(x, 0)$ exponentially decaying as $x \to +\infty$. The value $E(p)$ then describes the time dependence of the solution, and the exponent in \eqref{eq-particle-wf} is the quantity controlling the asymptotics.

\begin{figure}
    \centering
    \begin{subfigure}[c]{0.47\textwidth}
        \includegraphics[width=\textwidth]{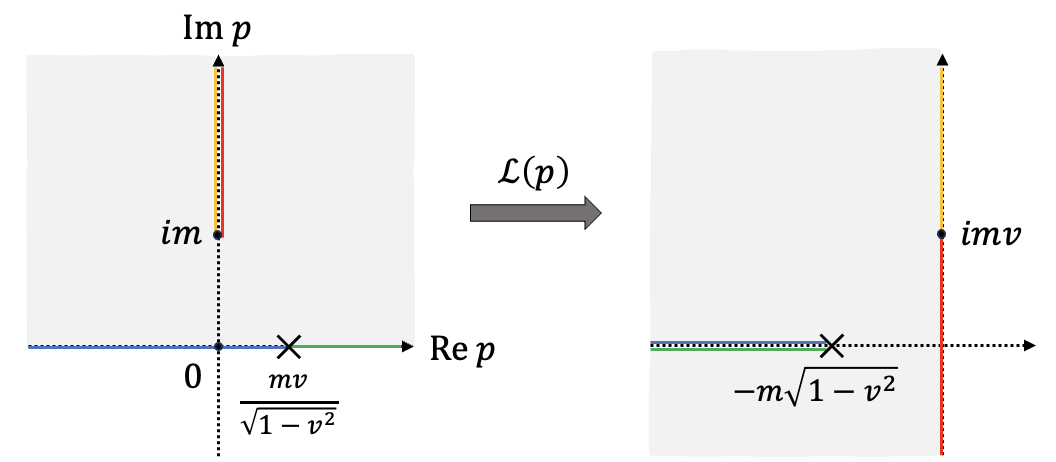}
        \caption{$\L(p) = pv - \sqrt{p^2 + m^2}$}
        \label{fig-intro-einstein-L}
    \end{subfigure}
    \hspace{0.02\textwidth}
    \begin{subfigure}[c]{0.47\textwidth}
        \includegraphics[width=\textwidth]{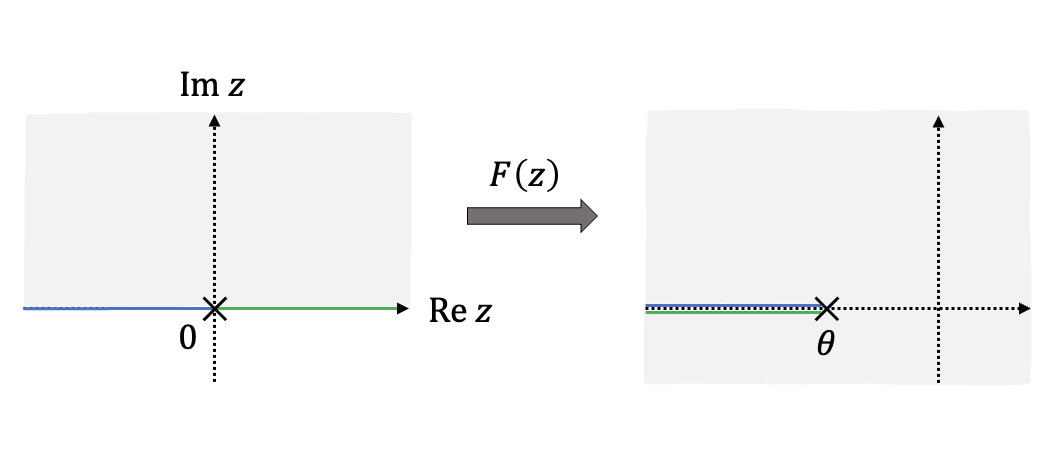}
        \caption{$F(z) = -z^2 / 2 + \theta$}
        \label{fig-intro-einstein-F}
    \end{subfigure}
    \caption{The relativistic phase-space Lagrangian $\L(p)$ and its normal form $F(z)$ as conformal mappings. Crosses show their critical points and values, which are the familiar expressions for momentum and the Lagrangian in terms of velocity. We choose $\theta = -m\sqrt{1 - v^2}$ to align the crosses in the image.}
    \label{fig-intro-einstein-L-F}
\end{figure}

\paragraph{Conformal interpretation of the relativistic Lagrangian}
The exponent in \eqref{eq-particle-wf} is known as
the \emph{phase-space Lagrangian}
$$
\L(p) = pv - E(p) = pv - \sqrt{p^2 + m^2},
$$
where the velocity $v \in (-1; 1)$ is an independent parameter. We extend the function $\L(p)$ to the natural domain of definition (the same as for $E'(p)$) while keeping $v \in (-1; 1)$. Since $\L'(p) = v - E'(p)$, we immediately conclude that
\begin{itemize}
    \item[$\rhd$] \emph{the derivative $\L'(p)$ is a 1--1 conformal map of its natural domain of definition to the lower half-plane}.
\end{itemize}
The phase-space Lagrangian itself also has a conformal interpretation, this time not universal (see Fig.~\ref{fig-intro-einstein-L}):
\begin{itemize}
    \item[$\rhd$] \emph{the relativistic phase-space Lagrangian $\L(p)$ is a 1--1 conformal map of its natural domain of definition to the left half-plane $\Re z < 0$ with the slit from $-m\sqrt{1 - v^2}$ to $-\infty$.}
\end{itemize}
This is easily verified by finding the image of the boundary and applying the argument principle.

Despite their simplicity, these conformal interpretations have probably not been observed before.

\paragraph{Normal form of the Lagrangian}
We wish to model the behavior of the conformal mapping $\L(p)$ using a simpler conformal mapping $F(z)$ with the same image of the real line. The simplest such \emph{normal form} is
$$
F(z) = -\frac{z^2}{2} + \theta, \qquad \text{where } \theta = -m\sqrt{1 - v^2}.
$$
It takes the upper half-plane to the complex plane with the slit from $\theta$ to $-\infty$ (see Fig.~\ref{fig-intro-einstein-F}).

\paragraph{Conformal change of variable}
A general solution of Schr\"odinger's equation has the form
\begin{equation}\label{eq-general-solution}
    \psi(x, t) = \int_{-\infty}^{+\infty} a(p) \, e^{it\L(p)} \, dp
\end{equation}
with some $a(p)$ decaying at infinity fast enough. To estimate the integral, perform the change of variable
$$
z(p) = F^{-1}(\L(p)),
$$
so $e^{it\L(p)}$ is replaced with $e^{itF(z)}$. The resulting integral is estimated easily, leading to an asymptotic formula for $\psi(x, t)$ that is uniform in the velocity $v = x / t$, if $v$ is bounded away from the speed of light.

\paragraph{Boundary behavior vs local analysis of singularities}
The proofs of our main results follow the same plan, but the Lagrangian and its normal form are more complicated (see Tab.~\ref{tab-airy-gamma} and Sec.~\ref{sec-airy-substitution}), and we need to shift the integration contour in our complex domain. Thus, we need the domain to be sufficiently large while keeping the Lagrangian a 1--1 conformal map. It is well-known \cite[Chap.~VII, Sec.~4, paragraph after Theorem~1]{Wong} that \emph{uniform} error bounds require a \emph{global} 1--1 change of variable rather than a local one.

Here, our key new insight is that \emph{boundary} behavior of conformal mappings enforces global bijectivity. Using symmetries and the above conformal interpretations further simplifies the analysis, leading to conditions that are easy to state and verify. The boundary behavior of the Lagrangian and the interior behavior of its derivative become the main limitations of our method, replacing the usual limitations on the collision of critical points and singularities.

\subsection{Organization of the paper}

In Sec.~\ref{sec-statements}, we state the main results: we define the quantum walks in Sec.~\ref{sec-checkers-notions}, provide uniform asymptotic formulae for their wave functions in Sec.~\ref{sec-checkers-results}, and introduce notation in Sec.~\ref{sec-notation}. The proofs are given in a more general setting in Secs.~\ref{sec-airy}--\ref{sec-gamma}, which follow closely parallel lines. These proofs differ little in complexity from those tailored to the quantum walk case, while clarifying the geometric mechanism. Throughout, we prefer geometric arguments to ad hoc computations (e.g., with the argument principle).

In Sec.~\ref{sec-airy}, we study the Airy phenomenon: we state a general theorem in Sec.~\ref{sec-airy-general}, prove it in Secs.~\ref{sec-airy-substitution}--\ref{sec-airy-proof}, and apply to several particular examples including quantum walks in Secs.~\ref{sec-airy-bessel}--\ref{sec-airy-checkers}.

In Sec.~\ref{sec-gamma}, we study the gamma phenomenon: we state a general theorem in Sec.~\ref{sec-gamma-general}, prove it in Secs.~\ref{sec-gamma-substitution}--\ref{sec-gamma-proof}, and apply to several particular examples including quantum walks in Secs.~\ref{sec-gamma-beta}--\ref{sec-gamma-checkers}.

In Sec.~\ref{sec-peak}, we apply our main results to studying the peak of the wave function of a quantum walk, and in Sec.~\ref{sec-future}, we discuss the directions of future work.

Some technical lemmas used in the above proofs are postponed until Appendices~\ref{sec-boundary-neighbourhoods}--\ref{sec-asymptotic-estimates}.

\section{Statements}
\label{sec-statements}

Let us define quantum walks, state our main results, and introduce notation and conventions.

\subsection{Main notions}
\label{sec-checkers-notions}

\begin{definition}
    \label{def-checkers}
    (See Fig.~\ref{fig-checkers-model}, bottom-left)
    Fix $m > 0$ called the \emph{particle mass}. A \emph{checker path} $s$ is a finite sequence of points of $\mathbb{Z}^2$ such that the vector from each point (except the last one) to the next one equals either $(1, 1)$ or $(-1, 1)$. Denote by $\mathrm{turns}(s)$ the number of points in $s$ (not the first and not the last one) such that the vectors from the point to the next and to the previous ones are orthogonal. To the path $s$, assign the complex number
    $$
    a(s) := (1 + m^2)^{1 - n/2} \, i(-im)^{\mathrm{turns}(s)},
    $$
    where $n$ is the total number of points in $s$. For each $(x, t) \in \mathbb{Z}^2$, where $t > 0$, denote by
    \begin{equation*}
        a(x, t, m) := \sum_s a(s)
    \end{equation*}
    the sum over all checker paths $s$ from $(0, 0)$ to $(x, t)$ containing $(1, 1)$. An empty sum is set to be zero by definition. The function $a(x, t, m)$ is called the \emph{wave function (for the quantum walk)}. Denote
    \begin{align*}
        \widetilde A_1(x, t, m) &= \Re a(x, t + 1, m), & & \text{for $x + t$ odd and $t \ge 0$}, \\
        \widetilde A_2(x, t, m) &= \Im a(x + 1, t + 1, m), & & \text{for $x + t$ even and $t \ge 0$};
    \end{align*}
    this shift of the coordinates results in particularly elegant expressions~\eqref{eq-A1-int-osc}, \eqref{eq-A2-int-osc}, and~\eqref{eq-symmetries} below.
\end{definition}

One interprets $|a(x, t, m)|^2$ as the probability to find an electron of mass $m$ in the interval of length $1$ around the point $x$ at the time $t > 0$, if the electron was emitted from the origin at the time $0$.

This elementary definition (due to Feynman) is equivalent to the one in the quantum-information literature \cite{Ambainis-etal-01}: $\Re a(x, t, m)$ and $\Im a(x, t, m)$ are, up to sign, the left and right components of the wave function of a \emph{quantum walk} starting at $x=0$ in the right direction, with the coin operator $\left(
    \begin{smallmatrix}
        1 & m \\ m & -1
    \end{smallmatrix}
\right)/\sqrt{1+m^2}$.

The particular case $m = 1$ is known as the \emph{Hadamard walk} because the latter operator becomes the Hadamard operator. In this case, we omit the argument $m$ and write $a(x, t) := a(x, t, 1)$ etc.

\begin{figure}[htbp]
    \centering
    \begin{tabular}{cccccc}
        \includegraphics[width=0.10\textwidth]{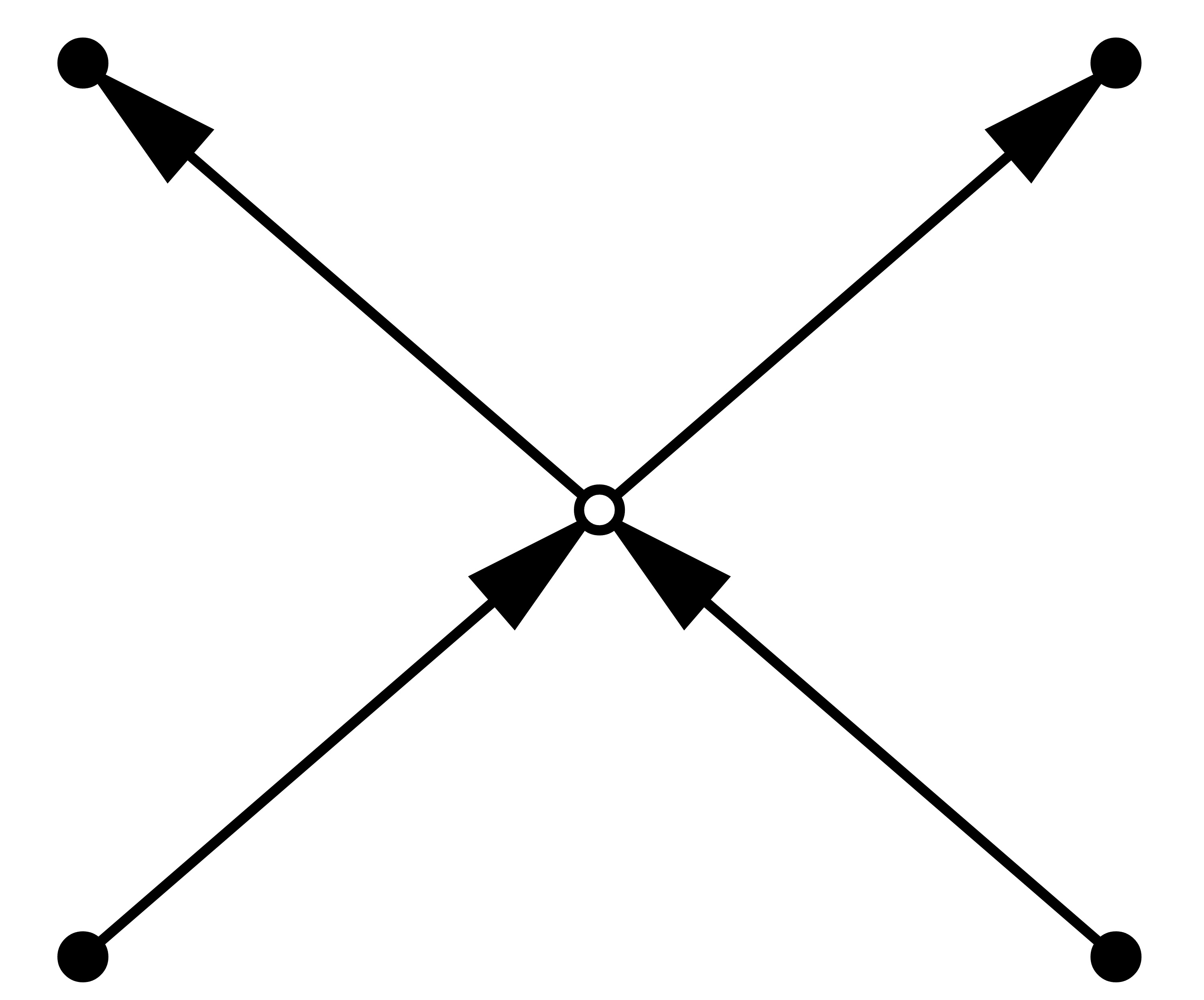} &
        \includegraphics[width=0.10\textwidth]{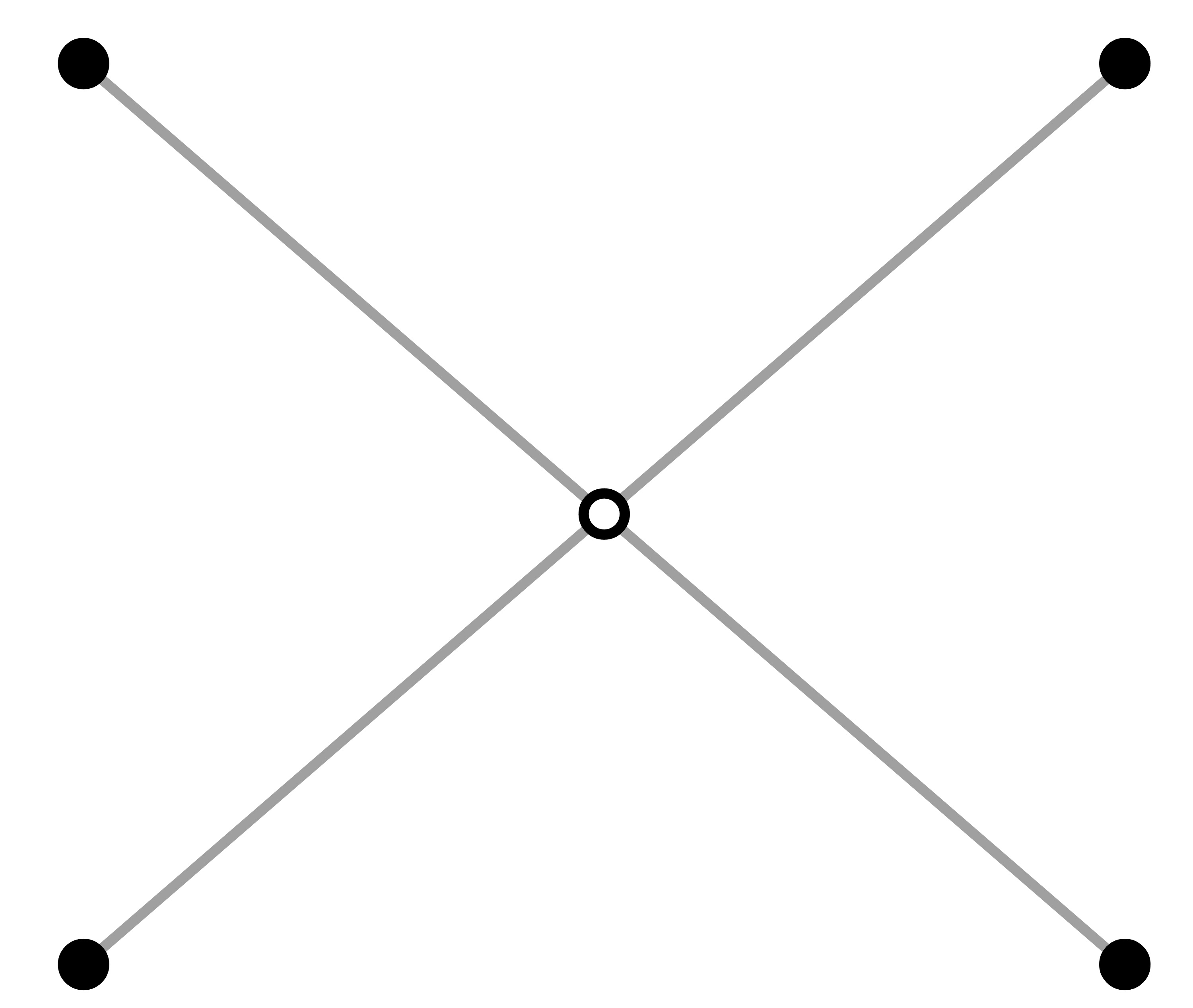} &
        \includegraphics[width=0.10\textwidth]{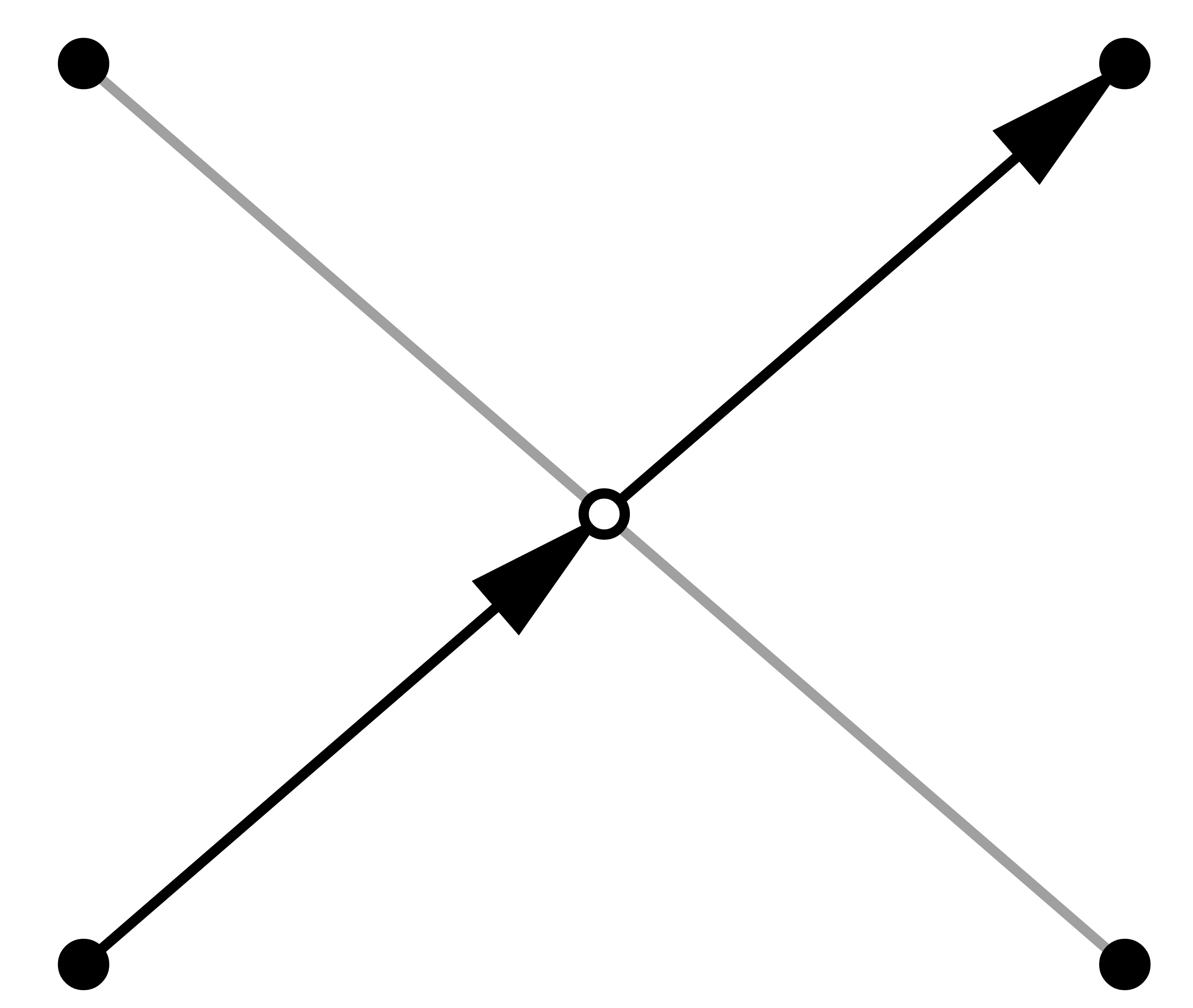} &
        \includegraphics[width=0.10\textwidth]{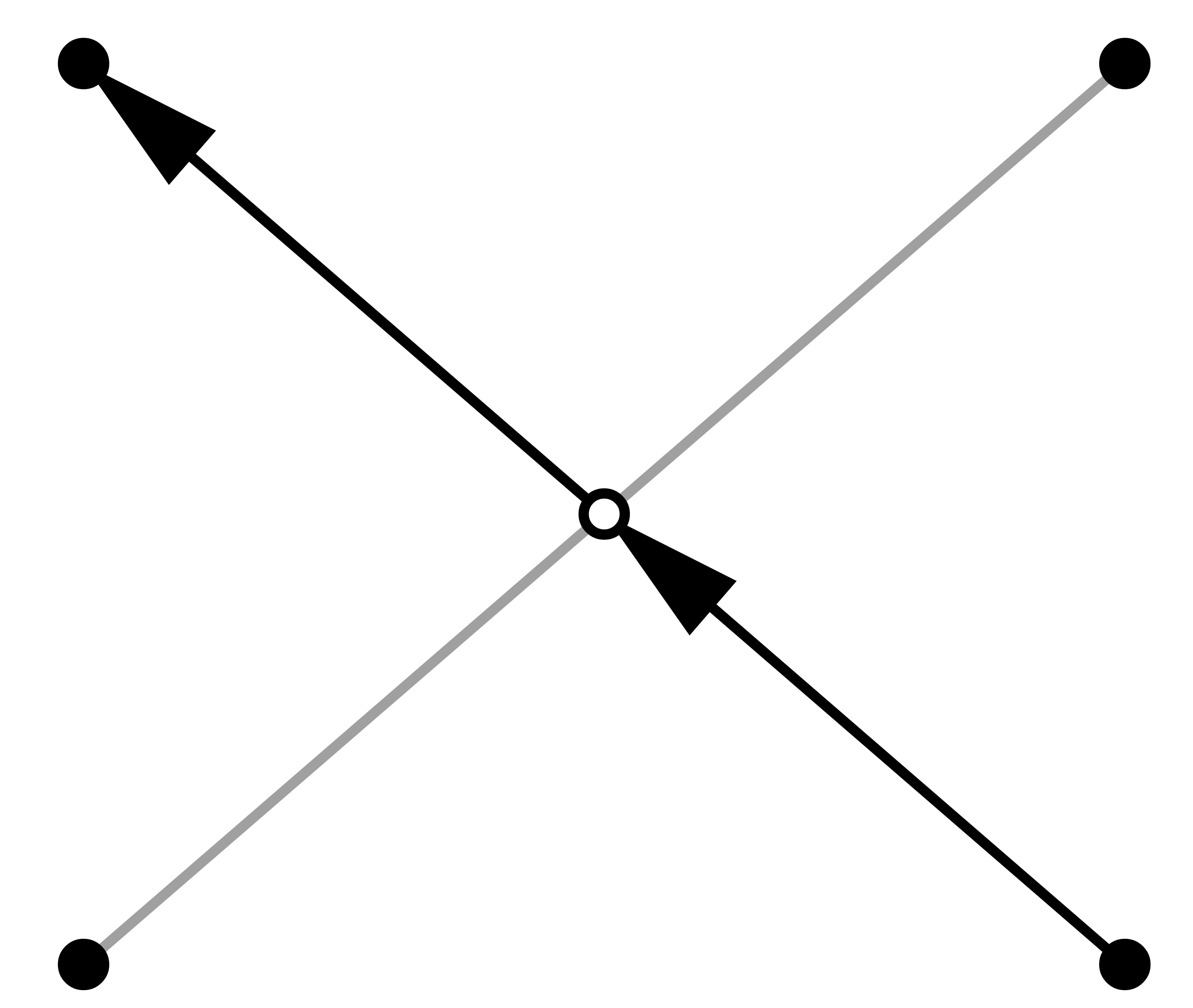} &
        \includegraphics[width=0.10\textwidth]{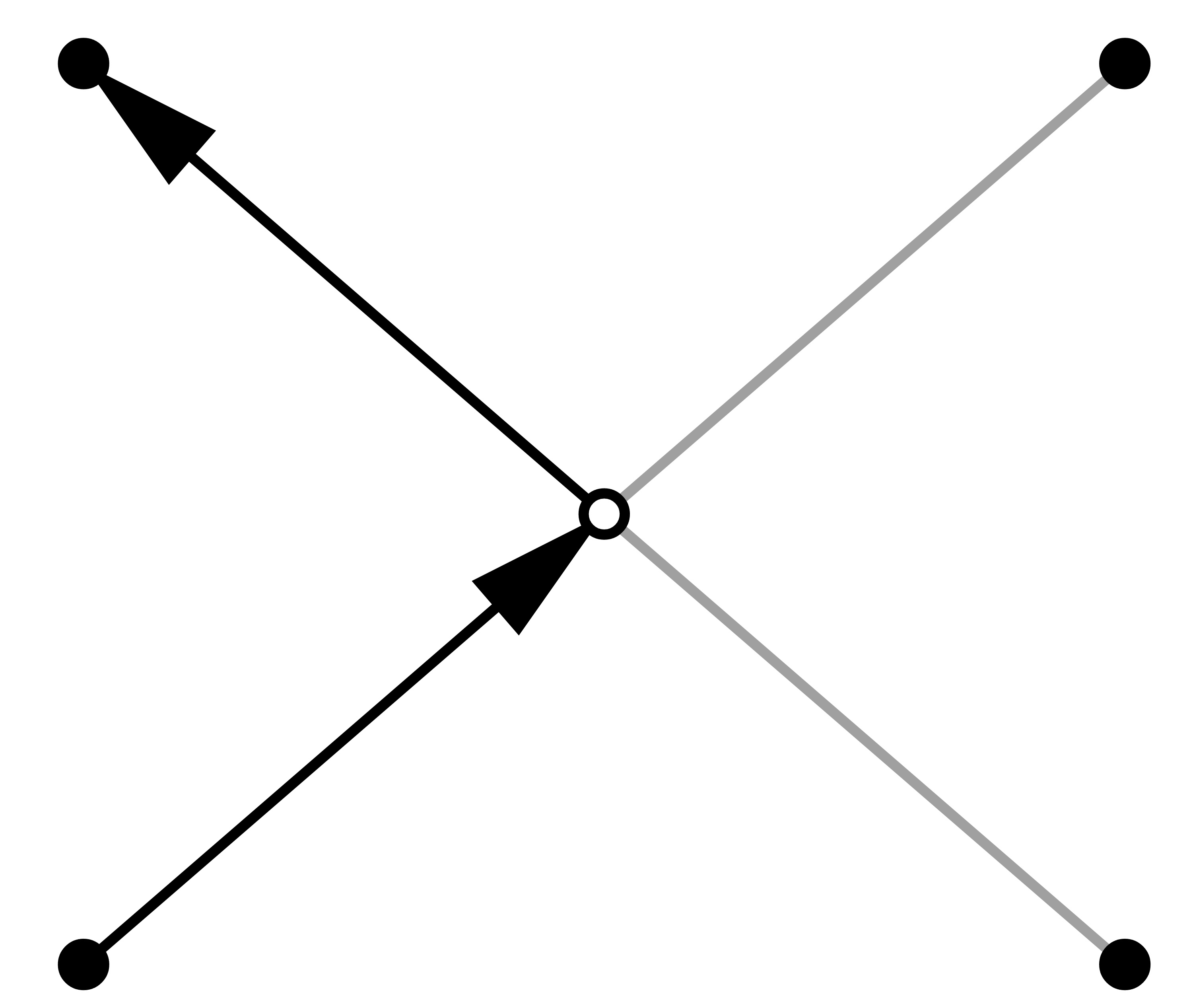} &
        \includegraphics[width=0.10\textwidth]{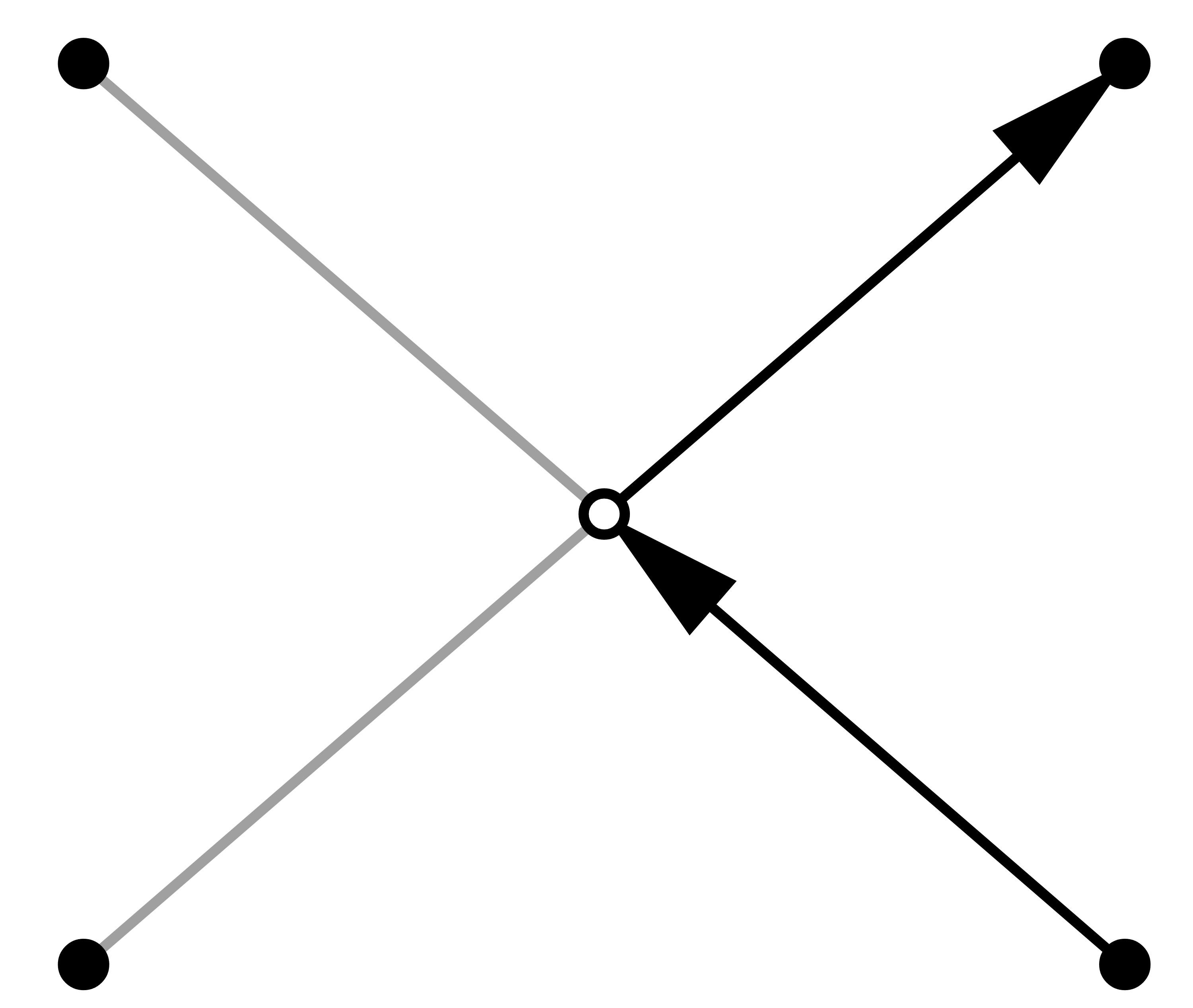} \\
        \multicolumn{2}{c}{$1$} &
        \multicolumn{2}{c}{$1 / \sqrt{1 + m^2}$} &
        \multicolumn{2}{c}{$-im / \sqrt{1 + m^2}$} \\[0.3cm]
        \includegraphics[width=0.10\textwidth]{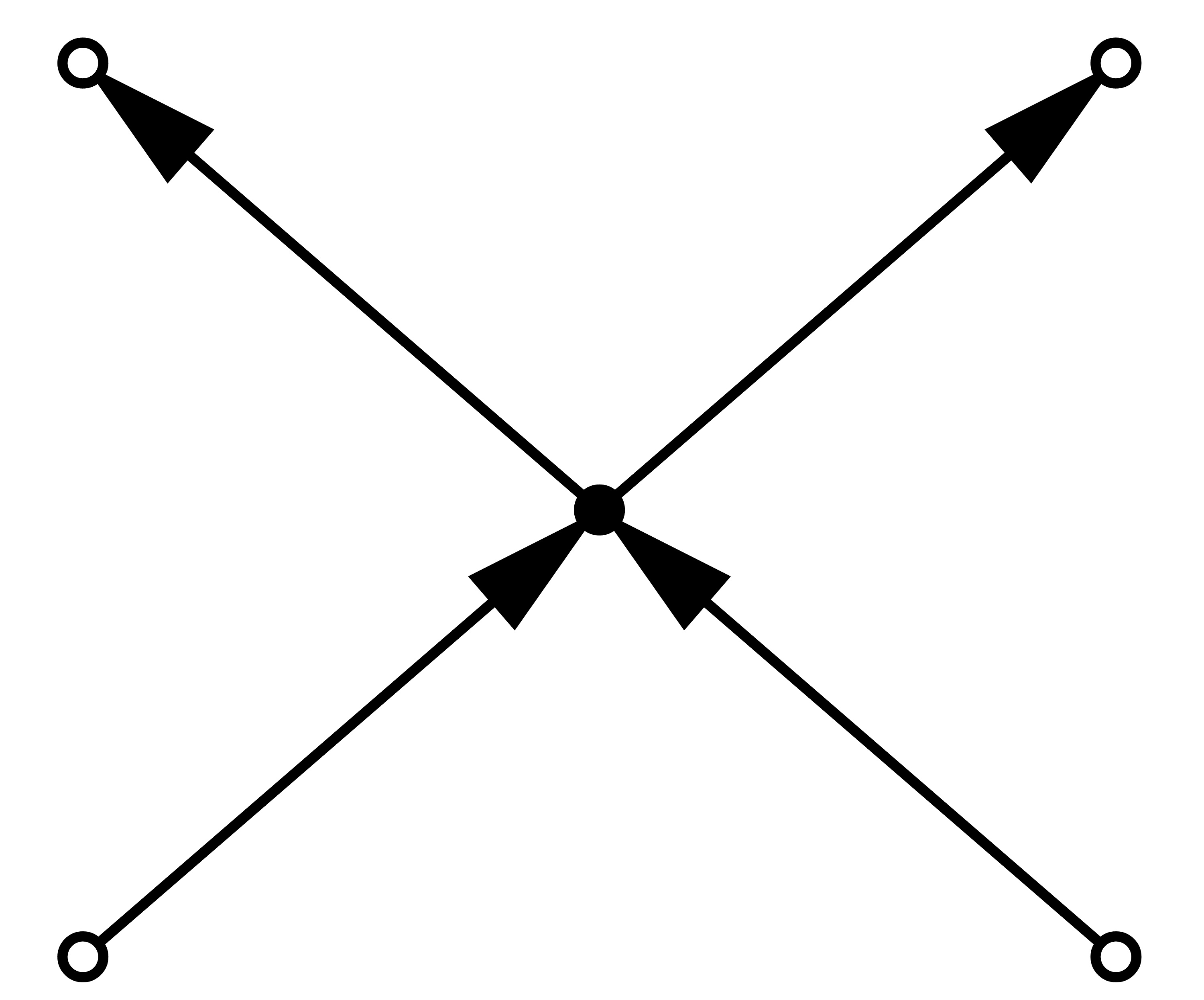} &
        \includegraphics[width=0.10\textwidth]{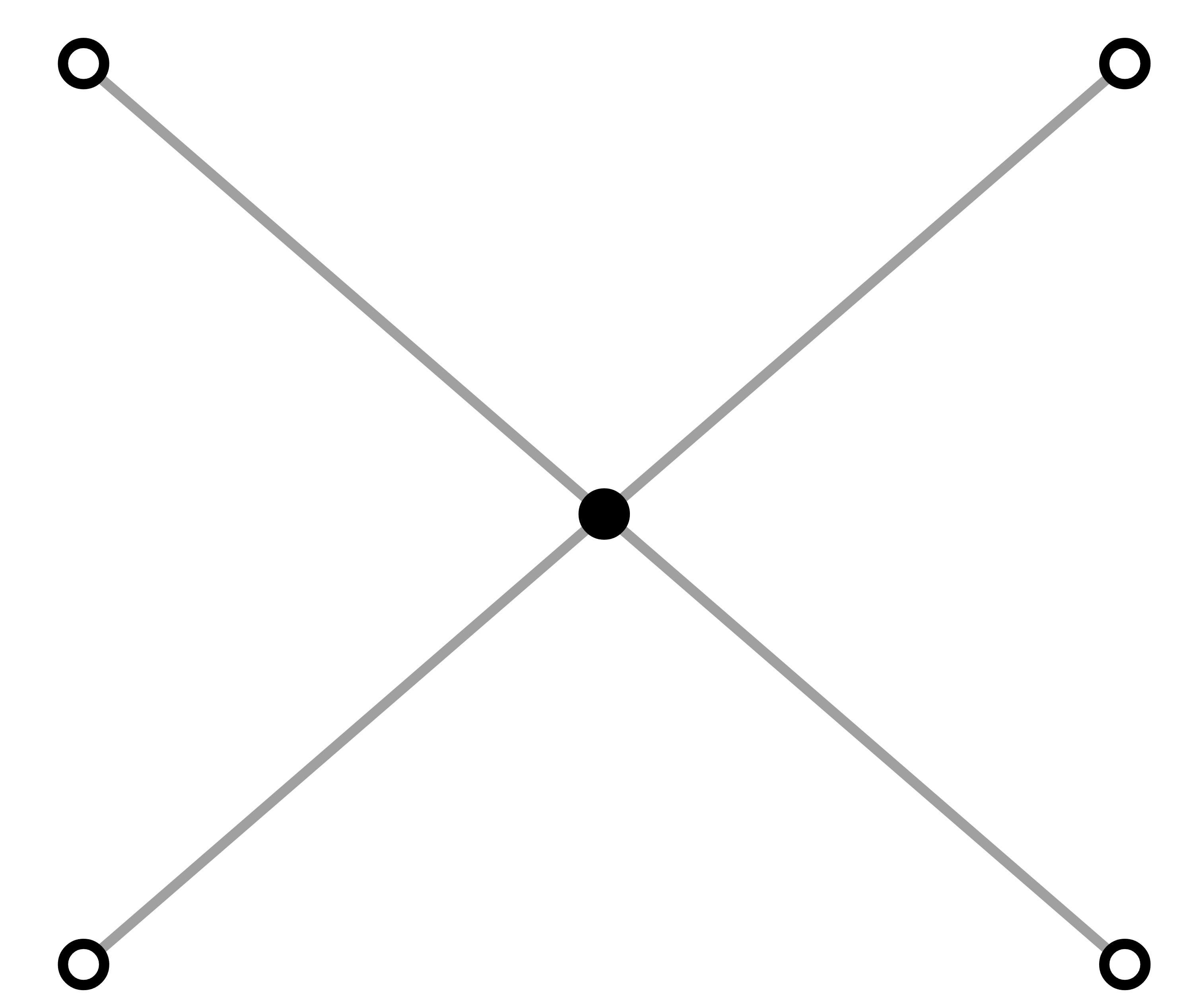} &
        \includegraphics[width=0.10\textwidth]{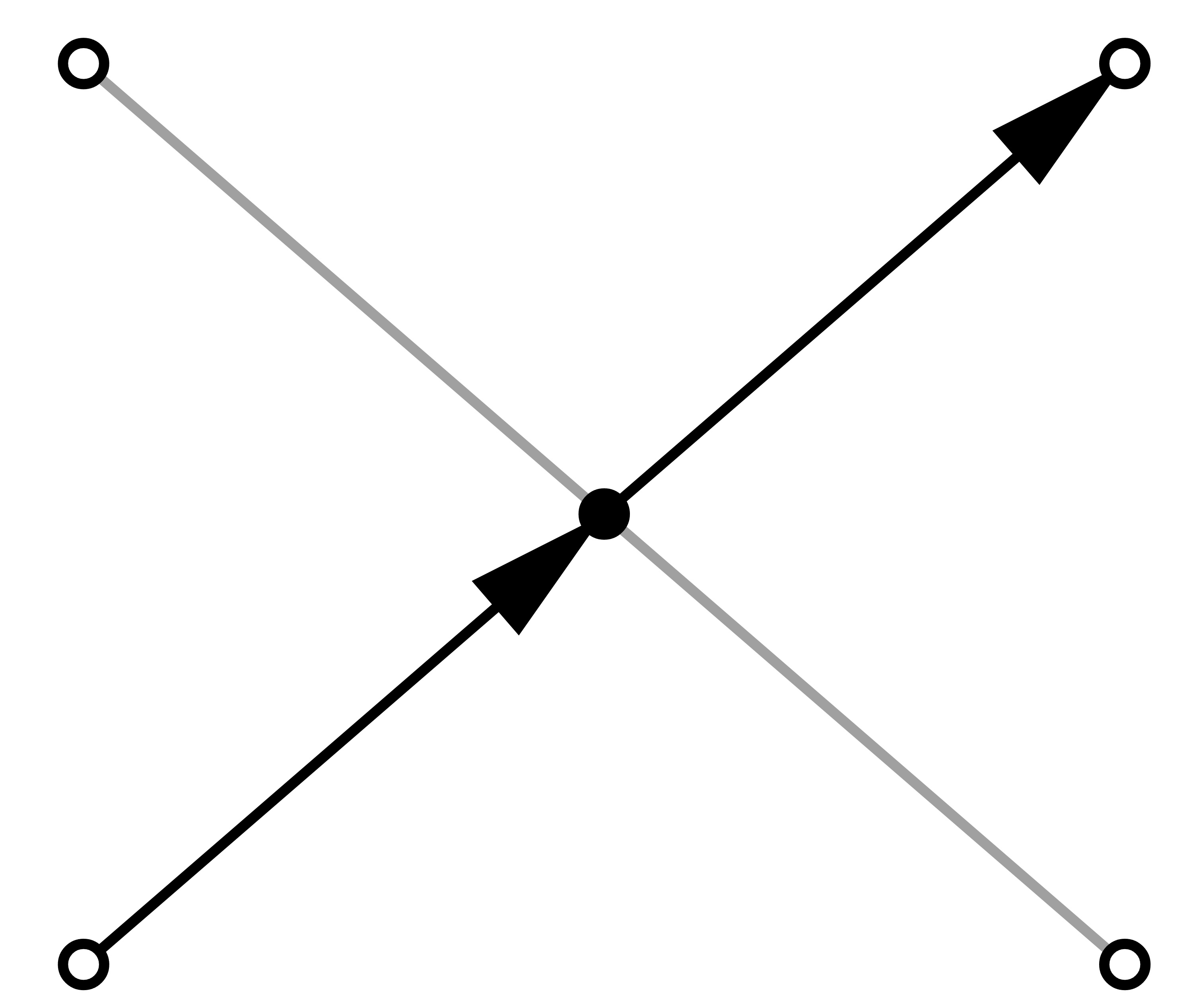} &
        \includegraphics[width=0.10\textwidth]{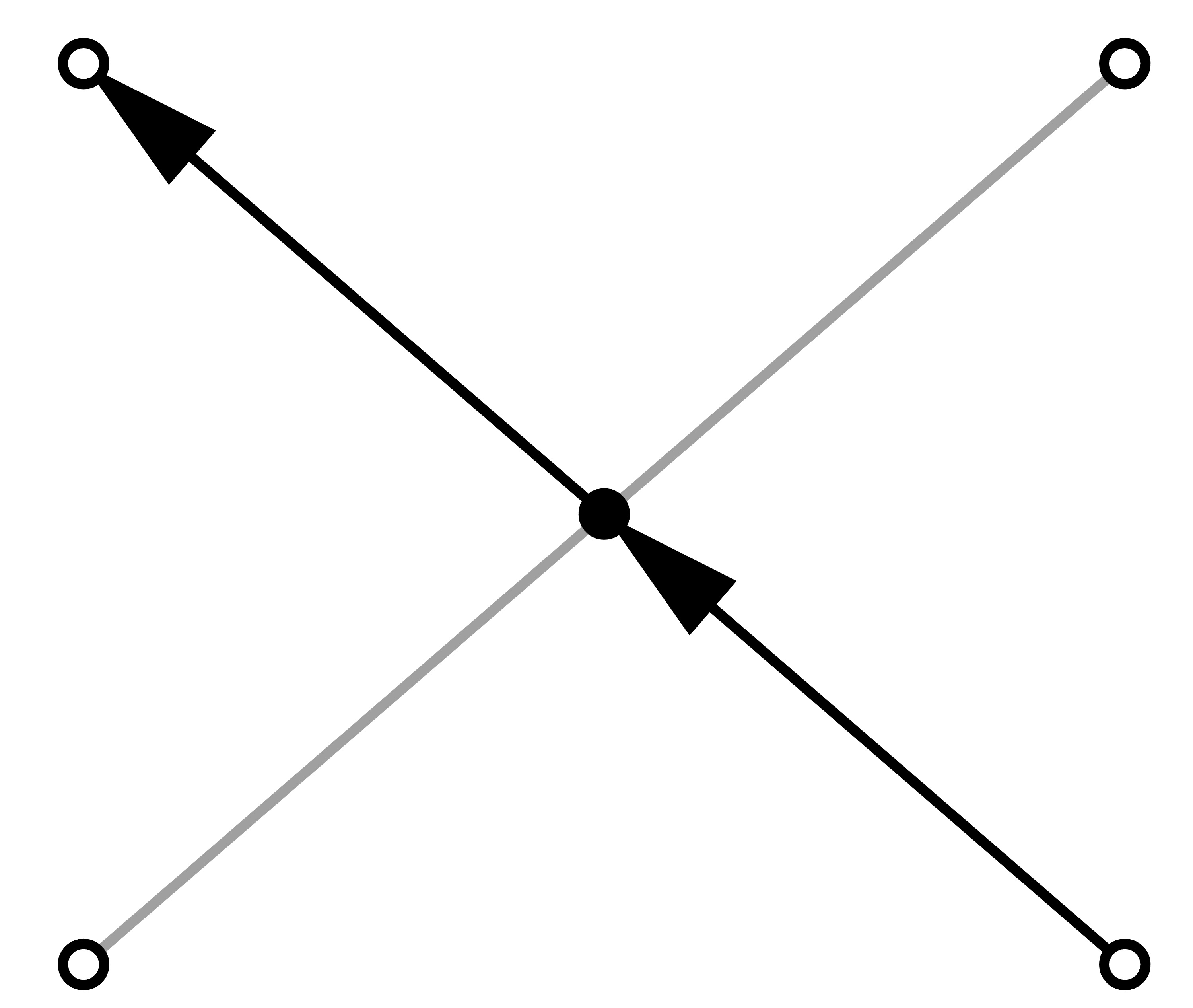} &
        \includegraphics[width=0.10\textwidth]{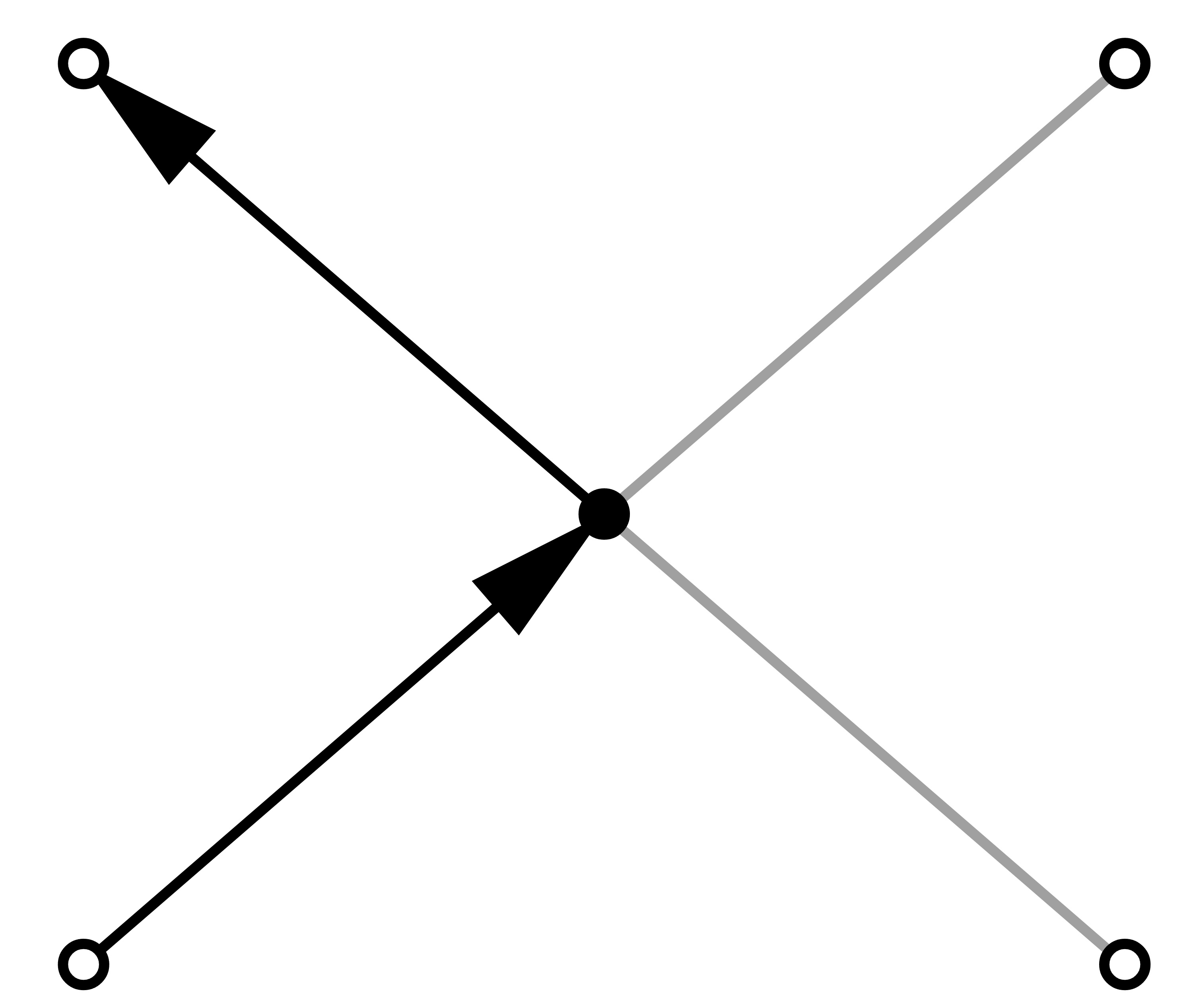} &
        \includegraphics[width=0.10\textwidth]{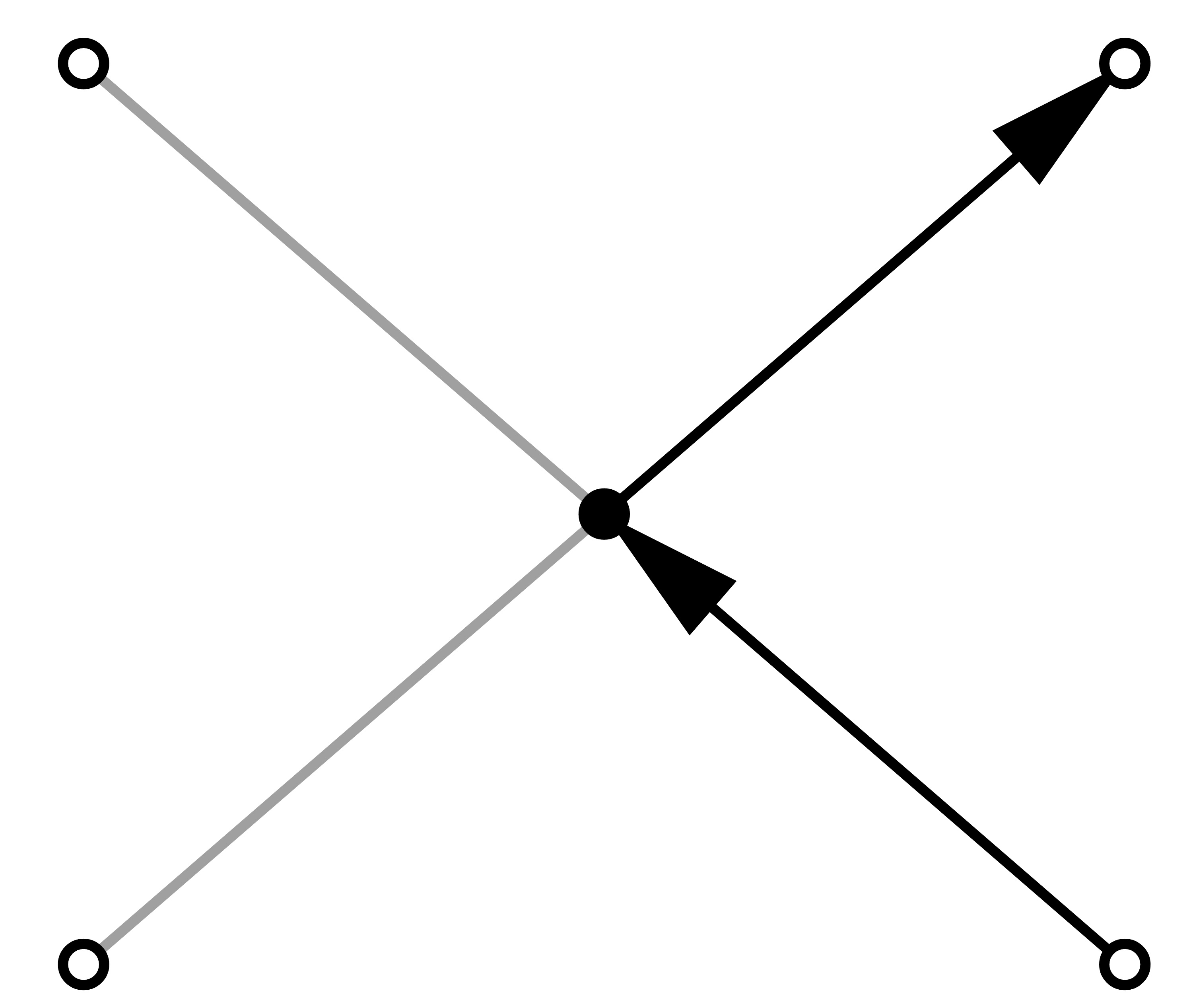} \\
        \multicolumn{2}{c}{$1$} &
        \multicolumn{2}{c}{$1 / \sqrt{1 - \varepsilon^2}$} &
        \multicolumn{2}{c}{$-\varepsilon / \sqrt{1 - \varepsilon^2}$}
    \end{tabular}
    \caption{Feynman (anti)checkers as a six-vertex model. Weights of integer (top) and half-integer (bottom) vertices depend on mass $m$ and small imaginary mass $\varepsilon$.
    Each configuration has also an overall sign defined globally in terms of a certain ``loop decomposition''. See~\cite[Definition~9]{SkoUst-24}.}
    \label{fig-checkers-6v}
\end{figure}

The model can be viewed as a \emph{six-vertex model} with certain complex weights. The number $a(s)$ can be alternatively defined as $i$ times the product of the vertex weights shown in the top row of Fig.~\ref{fig-checkers-6v}, over all lattice vertices. Here the thin gray segments are the ones not contained in the checker path $s$, and the leftmost vertex type cannot actually occur in the setup of Definition~\ref{def-checkers} but occurs in generalizations.

The key to getting asymptotic formulae is the following integral formulae, which can be viewed as an alternative definition of the functions $\widetilde A_1(x, t, m)$ and $\widetilde A_2(x, t, m)$ allowing us to extend them to $t \le 0$ and arbitrary parity of $x + t$. Such an extension is interpreted as allowing creation and annihilation of electron-positron pairs during motion and is therefore called \emph{Feynman anticheckers}. It is also a version of the six-vertex model; see Fig.~\ref{fig-checkers-6v}. The particular case when $t = 0$ and $x$ is even is equivalent to a \emph{massive random walk} \cite[Remark~3]{SkoUst-24}. We refer to \cite{SkoUst-24} for a detailed introduction to the resulting generalization.

\begin{definition}
    \label{def-checkers-int-osc}
    (See \cite[Definition~5]{SkoUst-22}, \cite[Proposition~1]{SkoUst-24})
    For each $m > 0$ and $(x, t) \in \mathbb{Z}^2$, set
    \begin{align}
        \label{eq-A1-int-osc}
        \widetilde A_1(x, t, m) &:= \pm \frac{im}{2\pi} \int_{-\pi}^{\pi} \frac{e^{ipx-iE(p)t} \, dp}{\sqrt{m^2 + \sin^2 p}} = \pm \frac{i^{x-t+1} \, m}{2\pi} \int_{-\pi}^{\pi} \frac{e^{ipx-i\omega(p)t} \, dp}{\sqrt{m^2 + \cos^2 p}}; \\
        \label{eq-A2-int-osc}
        \widetilde A_2(x, t, m) &:= \pm \frac{1}{2\pi}\int_{-\pi}^{\pi} \left(1 + \tfrac{\sin p}{\sqrt{m^2 + \sin^2 p}}\right) e^{ipx-iE(p)t} \, dp
        = \pm \frac{i^{x - t}}{2\pi}  \int_{-\pi}^{\pi} \left(1 + \tfrac{\cos p}{\sqrt{m^2 + \cos^2 p}}\right) e^{ipx - i\omega(p)t} \, dp,
    \end{align}
    where the minus sign in $\pm$ in the expression for $\widetilde A_k$ is taken for $t < 0$ and $x + t + k$ even,
    \begin{equation}
        \label{eq-w}
        E(p) := \arccos\left(\frac{\cos p}{\sqrt{1 + m^2}}\right) \qquad\text{and}\qquad \omega(p) := E\left(p + \frac{\pi}{2}\right) - \frac{\pi}{2} = \arcsin\left(\frac{\sin p}{\sqrt{1 + m^2}}\right).
    \end{equation}
\end{definition}

To derive the Airy phenomenon, the odd function $\omega(p)$ is more convenient than the even function $E(p)$.

To derive the gamma phenomenon, we need the following Hankel-type representation proved in Sec.~\ref{sec-gamma-checkers}:

\begin{proposition}
    \label{prop-checkers-int-damp}
    For each $m > 0$ and integer $x, t \ge 0$, we have
    \begin{align}
        \label{eq-A1-int-damp}
        \widetilde A_1(x, t, m) &= \frac{i^{x-t} \, m}{2\pi (1 + m^2)^{t/2}} \int_{-\infty}^{(0+)} \frac{e^{\omega(t+x+1)/2 - p(\omega)(t-x+1)/2} \, d\omega}{1 + m^2 - e^{\omega}}, \\
        \label{eq-A2-int-damp}
        \widetilde A_2(x, t, m) &= \frac{i^{x-t-1}}{2\pi (1 + m^2)^{t/2}} \cdot \frac{t + x}{t - x} \cdot \int_{-\infty}^{(0+)} e^{\omega(t+x)/2 - p(\omega)(t-x)/2} \, d\omega,
    \end{align}
    where $x \ne t$ in \eqref{eq-A2-int-damp} and
    \begin{align}
        \label{eq-p}
        p(\omega) := \log(e^{\omega} - 1) - \log(1 + m^2 - e^{\omega}).
    \end{align}
\end{proposition}

Hereafter, we take the principal branch $\log \colon \mathbb{C} \setminus (-\infty; 0] \to \{z: -\pi < \Im z < \pi\}$, and $\int_{-\infty}^{(0+)} f(\omega) \, d\omega$ denotes the integral over a contour $\gamma(s)$ that encircles the ray $(-\infty; 0]$ once counter-clockwise in its sufficiently small neighbourhood and satisfies $\Re \gamma(s) \to -\infty$ as $s \to \pm \infty$.

\subsection{Evidence of the gamma phenomenon}
\label{sec-gamma-evidence}

Although the gamma phenomenon for quantum walks has not been reported before, its presence is easy to spot. The well-known combinatorial formula for the wave function of the Hadamard walk \cite[Proposition~3]{SkoUst-22}
$$
a(x, t) = 2^{(1-t)/2} \sum_{r=0}^{(t-|x|)/2} (-1)^r \binom{(x + t - 2) / 2}{r} \left(\binom{(t - x - 2) / 2}{r} + i\binom{(t - x - 2) / 2}{r - 1}\right),
$$
where $|x|<t$ and $t-x$ is even, implies that for each fixed integer $n \ge 1$, we have
$$
a(t - 2n, t) = i \, 2^{(1-t)/2} \frac{(-1)^n \, t^{n}}{n!} \left(1 + O_n\left(\frac{1}{t}\right)\right),
$$
where the index in $O_n$ emphasizes that the constant understood in the big-Oh notation depends on $n$.

This observation shows that \emph{any uniform asymptotic formula for the wave function of a quantum walk in the whole $xt$-plane (with the error term uniformly dominated by the main term) must involve a gamma function} (or other special function reproducing the factorial for fixed $t - x$). In particular, the assumption $x / t \le 1 - \delta$ in the known rigorous uniform asymptotic formulae \cite{Sunada-Tate-12} and in Theorem~\ref{th-airy-checkers} below is essential.

\subsection{Main results}
\label{sec-checkers-results}

We provide uniform asymptotic formulae for $\widetilde A_1(x, t, m)$ and $\widetilde A_2(x, t, m)$ for fixed $m > 0$ and $t \to +\infty$ in terms of \emph{Airy} and \emph{gamma functions}
\begin{align*}
    \Ai(u) := \frac{1}{\pi} \int_0^{+\infty} \cos\left(up + \frac{p^3}{3}\right) \, dp \qquad\text{and}\qquad
    \Gamma(z) := \int_0^{+\infty} p^{z-1} e^{-p} \, dp.
\end{align*}
Here, the latter can be analytically continued to all real $z$ except for the non-positive integers. In the latter case, we set $1 / \Gamma(z) := 0$. Analogously, some other functions in equations~\eqref{eq-A1-airy-asymp} and~\eqref{eq-A1-gamma-asymp} below are extended by continuity as immediately clarified.

\begin{theorem}[Airy phenomenon]
    \label{th-airy-checkers}
    For each $m > 0$, $\delta \in (0; 1)$, and $(x, t) \in \mathbb{Z}^2$, satisfying $t > 0$ and
    \begin{align*}
        v := x / t \in \left[0; 1 - \delta\right],
    \end{align*}
    we have
    \begin{align}
        \label{eq-A1-airy-asymp}
        \widetilde A_1(x, t, m) &= \frac{i^{x-t+1} \sqrt{m}}{t^{1/3}} \left(\frac{4\theta(v, m)}{1 - (1 + m^2)v^2}\right)^{1/4} \Ai\left(-\theta(v, m)\,t^{2/3}\right)
        \begin{cases}
            +\,O_m\left(\tfrac{1}{t^{3/2}}\right), & \text{if } v \le \hat v, \\
            \times\left(1 + O_{m, \delta}\left(\tfrac{1}{t}\right)\right), & \text{if } v \ge \hat v,
        \end{cases} \\
        \nonumber
        \widetilde A_2(x, t, m) &= \frac{i^{x-t} \sqrt{m}}{t^{1/3}} \left(\frac{4\theta(v, m)}{1 - (1 + m^2)v^2}\right)^{1/4} \Ai\left(-\theta(v, m)\,t^{2/3}\right) \sqrt{\frac{1 + v}{1 - v}} \,\,
        \begin{cases}
            +\,O_m\left(\tfrac{1}{t^{3/2}}\right), & \text{if } v \le \hat v, \\
            \times\left(1 + O_{m, \delta}\left(\tfrac{1}{t}\right)\right), & \text{if } v \ge \hat v,
        \end{cases}
        \intertext{where the ratio $\frac{4\theta(v, m)}{1 - (1 + m^2)v^2}$ is continuously extended by $2^{4/3} m^{-2/3}$ at $v = \hat v := 1 / \sqrt{1 + m^2}$, and}
        \label{eq-theta-def}
        \theta(v, m) &:=
        \begin{cases}
            \left(\frac{3}{2} \arctan\frac{1}{m}\right)^{2/3}, & \text{if } v = 0, \\
            \left(\frac{3}{2}\left(-v\arctan\frac{\sqrt{1 - (1 + m^2)v^2}}{mv} + \arctan\frac{\sqrt{1 - (1 + m^2)v^2}}{m}\right)\right)^{2/3}, & \text{if } 0 < v \le \hat v, \\
            -\left(\frac{3}{2}\left(v\arctanh\frac{\sqrt{(1 + m^2)v^2 - 1}}{mv} - \arctanh\frac{\sqrt{(1 + m^2)v^2 - 1}}{m}\right)\right)^{2/3}, & \text{if } v \ge \hat v.
        \end{cases}
    \end{align}
\end{theorem}

\begin{theorem}[Gamma phenomenon]
    \label{th-gamma-checkers}
    For each $m, \delta > 0$ and each $(x, t) \in \mathbb{Z}^2$, satisfying $t > 0$ and
    \begin{align*}
        v := x / t \in \left[1 / \sqrt{1 + m^2} + \delta; +\infty\right),
    \end{align*}
    we have
    \begin{align}
        \label{eq-A1-gamma-asymp}
        \widetilde A_1(x, t, m) &= \frac{i^{x-t+1} \, \sqrt{m}}{t^{1/2}} \cdot \frac{\exp(-tL(v, m))}{((1 + m^2)v^2 - 1)^{1/4}} \cdot \frac{(|t - x| / (2e))^{(t-x)/2}}{\Gamma((t - x + 1) / 2)} \left(1 + O_{m, \delta}\left(\frac{1}{x + t}\right)\right), \\
        \nonumber
        \widetilde A_2(x, t, m) &= \frac{i^{x-t} \, \sqrt{m}}{t^{1/2}} \cdot \frac{\exp(-tL(v, m))}{((1 + m^2)v^2 - 1)^{1/4}} \cdot \frac{(|t - x| / (2e))^{(t-x)/2}}{\Gamma((t - x + 2) / 2)} \sqrt{\frac{t + x}{2}} \left(1 + O_{m, \delta}\left(\frac{1}{x + t}\right)\right),
        \intertext{where the factor $(|t - x| / (2e))^{(t-x)/2}$ is interpreted as $1$ for $x = t$, and}
        \label{eq-L-def}
        L(v, m) &=
        \begin{cases}
            v\arctanh\frac{\sqrt{(1 + m^2)v^2 - 1}}{mv} - \arctanh\frac{\sqrt{(1 + m^2)v^2 - 1}}{m}, & \text{if } v < 1, \\
            \frac{1}{2} \log(1 + m^2), & \text{if } v = 1, \\
            v\arccoth\frac{\sqrt{(1 + m^2)v^2 - 1}}{mv} - \arccoth\frac{\sqrt{(1 + m^2)v^2 - 1}}{m}, & \text{if } v > 1.
        \end{cases}
    \end{align}
\end{theorem}

Hereafter, the notation $f(x, t, m) = g(x, t, m) + O_{m, \delta}(h(x, t))$ means that there is a constant $C(m, \delta) > 0$ (depending on $m, \delta$ but \emph{not} on $x, t$) such that for each $x, t, m, \delta$ satisfying the assumptions of the theorems, we have $|f(x, t, m) - g(x, t, m)| \le C(m, \delta) \, h(x, t)$. Similarly, we write $f(x, t, m) = g(x, t, m) + \Omega_{m, \delta}(h(x, t))$ if $|f(x, t, m) - g(x, t, m)| \ge C(m, \delta) \, h(x, t)$. The big-Oh and big-Omega notations with other indices have a similar meaning, including the case when some of the indices are functions rather than numbers.

Theorems~\ref{th-airy-checkers} and~\ref{th-gamma-checkers} extend to $x < 0$ and $t < 0$ by symmetries \cite[Proposition~6]{SkoUst-24}:
\begin{equation}
    \label{eq-symmetries}
    \begin{gathered}
        \widetilde A_1(x, t, m) = \widetilde A_1(-x, t, m) = \widetilde A_1(x, -t, m) = \widetilde A_1(-x, -t, m), \\ (t - x) \, \widetilde A_2(x, t, m) = (t + x) \, \widetilde A_2(-x, t, m),
        \quad \widetilde A_2(x, t, m) = -\widetilde A_2(-x, -t, m) \quad\text{for }(x, t) \ne (0, 0).
    \end{gathered}
\end{equation}
Furthermore, Theorem~\ref{th-gamma-checkers} also extends to the case $t = 0$ as shown in the last row of Tab.~\ref{tab-checkers-asymp}. In this case, it is necessary to replace the variable $t$ in \eqref{eq-A1-gamma-asymp} with $x / v$ and take the limit as $v \to +\infty$. So, Theorems~\ref{th-airy-checkers} and~\ref{th-gamma-checkers} together cover the whole $xt$-plane.

The formulae for $\widetilde A_1(x, t, m)$ from both theorems simplify on certain intervals of $v$, as shown in Tab.~\ref{tab-checkers-asymp}. In particular, this reproduces the known results of \cite{Ambainis-etal-01,Sunada-Tate-12}, this time with uniform error bounds. For instance, on the interval $0 \le v\le \hat v - \delta$, we observe oscillations of amplitude of order $t^{-1/2}$, whereas in the vicinity of $v = \hat v$, the wave function is of order $t^{-1/3}$; see Fig.~\ref{fig-intro-domains-after}. On the common interval $\hat v + \delta < v < 1 - \delta$ of the two theorems, both expressions involving Airy and gamma functions simplify to the same exponential decay. Remarkably, they do \emph{not} simplify on most intervals corresponding to the white angles in Fig.~\ref{fig-intro-domains-before}. Eq.~\eqref{eq-A1-gamma-asymp} has also the equivalent form~\eqref{eq-A1-gamma-asymp-simp}. In row 3 of Tab.~\ref{tab-checkers-asymp}, we require $\delta<\frac{\hat v-\hat v^2}{2}$ to
guarantee $v\in [0;1-\delta]$.

The simplified asymptotic formulae for $\widetilde A_2(x, t, m)$ for $0\le x<t$ can easily be obtained from the ones for $\widetilde A_1(x, t, m)$ and the following corollary.

\begin{table}[hbt]
    \caption{Uniform asymptotic formulae for the wave function $\widetilde A_1(x, t, m)$ for different intervals of $v = x / t$.}
    \label{tab-checkers-asymp}
    \centering
    \begin{tabular}{llp{0.18\textwidth}}
        \toprule
        \textbf{Speed interval} & \textbf{Asymptotic formula for $\widetilde A_1(x, t, m)$} & \textbf{Reference} \\
        \midrule

        $v = 0$ & $\sqrt{\frac{2m}{\pi t}} \cdot {\small
            \begin{cases}
                \cos\left(t\arctan m - \frac{\pi}{4}\right), & t \text{ odd;}\\
                -i\sin\left(t\arctan m - \frac{\pi}{4}\right), & t \text{ even;}
            \end{cases}
        \,+ O_m\left(\tfrac{1}{t^{3/2}}\right)}$ & \vspace{-0.45cm}
        Corollary~\ref{cor-airy-checkers};

        cf.\,\cite[Theorem\,3]{Bogdanov-20} \\[0.2cm]

        $0 < v \le \hat v - \delta < \hat v$ & $\sqrt{\frac{2m}{\pi t}} \cdot \frac{i^{x-t+1} \cos\left(\frac{2}{3}t\theta(v, m)^{3/2} - \frac{\pi}{4}\right)}{(1 - (1 + m^2)v^2)^{1/4}} + O_{m, \delta}\left(\tfrac{1}{t^{3/2}}\right)$
        & \vspace{-0.45cm}
        Corollary~\ref{cor-airy-checkers}

        or\,\cite[Theorem\,3]{SkoUst-24} \\[0.15cm]

        $\frac{\delta}{t^{2/3}}<|v - \hat v|<\delta< \frac{\hat v-\hat v^2}{2}$
        & Eq.~\eqref{eq-A1-airy-asymp} & Theorem~\ref{th-airy-checkers} \\[0.2cm]

        $|v - \hat v| \le \frac{\delta}{t^{2/3}}$ & $\sqrt[3]{\frac{2m}{t}} \cdot i^{x-t+1} \Ai\left(\frac{v - \hat v}{\hat v} \sqrt[3]{\frac{2t^2}{m^2}}\right) + O_{m, \delta}\left(\frac{1}{t}\right)$ & Corollary\,\ref{cor-airy-checkers};\,cf.\,\cite{Sunada-Tate-12} \\[0.4cm]

        $\hat v + \delta < v < 1 - \delta < 1$ & $\sqrt{\frac{\vphantom{2}m}{2\pi t}} \cdot \frac{i^{x-t+1}\exp\left(-tL(v, m)\right)}{((1 + m^2)v^2 - 1)^{1/4}} \cdot \left(1 + O_{m, \delta}\left(\frac{1}{t}\right)\right)$ & Corollary\,\ref{cor-airy-checkers};\,cf.\,\cite{Sunada-Tate-12} \\[0.4cm]

        $\frac{\delta}{t}<|1 - v|<\delta<1-\hat v$
        & Eq.~\eqref{eq-A1-gamma-asymp} & Theorem~\ref{th-gamma-checkers} \\[0.3cm]

        $|1 - v| < \frac{\delta}{t}$ & $\frac{m^{t - x} t^{(t-x-1)/2}}{(1 + m^2)^{t/2}} \cdot \frac{i^{x-t+1}}{\Gamma\left((t-x+1)/2\right)} \cdot \left(1 + O_{m, \delta}\left(\frac{1}{t}\right)\right)$
        & \vspace{-0.45cm}
        Corollary~\ref{cor-gamma-checkers};

        cf.~Sec.~\ref{sec-gamma-evidence} \\[0.2cm]

        $v > 1 + \delta > 1$ & $\sqrt{\frac{2m}{\pi t}} \cdot \frac{i\exp\left(-tL(v, m)\right)}{((1 + m^2)v^2 - 1)^{1/4}} \cdot \left(1 + O_{m, \delta}\left(\frac{1}{x + t}\right)\right)$, \, {\small $x - t$ even} & Corollary~\ref{cor-gamma-checkers} \\[0.4cm]

        $t = 0$ (``$v = +\infty$'') & $\sqrt{\frac{2m}{\pi x}} \cdot \frac{i\exp(-x\arcsinh m)}{(1 + m^2)^{1/4}} \cdot \left(1 + O_m\left(\frac{1}{x}\right)\right)$, \, {\small $x > 0$ even}
        & \vspace{-0.25cm}
        Corollary~\ref{cor-gamma-checkers} \\

        \bottomrule
    \end{tabular}
\end{table}

\begin{corollary}
    \label{cor-A2-A1}
    For each $m > 0$ there is $C = C(m) > 0$ such that for each $(x, t) \in \mathbb{Z}^2$ satisfying $0 \le x < t$ and $t > C$, we have
    \begin{align}
        \label{eq-cor-A2-A1-Gamma}
        \widetilde A_2(x, t, m) = -i\widetilde A_1(x, t, m) \cdot \frac{\Gamma((t - x + 1) / 2)}{\Gamma((t - x + 2) / 2)} \sqrt{\frac{t + x}{2}}
        \begin{cases}
            +\,O_m\left(\tfrac{1}{t^{4/3}}\right), & \text{if } 0 \le x / t \le \hat v; \\
            \times\left(1 + O_m\left(\tfrac{1}{t}\right)\right), & \text{if } \hat v < x / t < 1,
        \end{cases}
    \end{align}
    where $\hat v = 1 / \sqrt{1 + m^2}$. In particular, if $x / t \le 1 - \delta$ for some $\delta > 0$, then this simplifies to
    \begin{align}
        \label{eq-cor-A2-A1-Airy}
        \widetilde A_2(x, t, m) = -i\widetilde A_1(x, t, m) \sqrt{\frac{t + x}{t - x}}
        \begin{cases}
            +\,O_m\left(\tfrac{1}{t^{3/2}}\right), & \text{if } 0 \le x / t \le \hat v; \\
            \times\left(1 + O_{m, \delta}\left(\tfrac{1}{t}\right)\right), & \text{if } \hat v < x / t \le 1 - \delta.
        \end{cases}
    \end{align}
\end{corollary}

Note that this formula is \emph{not} a relation between the right and left components of the wave function for a quantum walk; it rather connects both to the ones in Feynman \emph{anti}checkers.

The bound on $x / t$ in Theorem~\ref{th-airy-checkers} is sharp in the sense that the assumption $|x / t| \le 1 - \delta$ cannot be relaxed to $|x / t| < 1$: the constant $C(m, \delta)$ understood in $O_{m, \delta}$ blows up as $\delta \to 0$. Indeed, if this were not the case, then we would get a contradiction to Theorem~\ref{th-gamma-checkers} (and Sec.~\ref{sec-gamma-evidence}).

The bound $x / t > 1 / \sqrt{1 + m^2} + \delta$ in Theorem~\ref{th-gamma-checkers} is sharp in the sense that it cannot be relaxed to $x / t > 1 / \sqrt{1 + m^2}$; otherwise, we get a contradiction to Theorem~\ref{th-airy-checkers}.

Numerical experiments demonstrate that the error estimate order in Theorem~\ref{th-airy-checkers} is sharp (at least in the case $v \le \hat v$), in the sense that $O_m\left(t^{-3/2}\right)$ cannot be replaced with $O_m\left(t^{-\alpha}\right)$ for $\alpha > 3/2$.

We shall see that $L(v, m)$ and $\frac{2}{3}\theta(v, m)^{3/2}$ have the meaning of Lagrangian (Fig.~\ref{fig-intro-einstein-L-F} and Secs.~\ref{sec-airy-general} and~\ref{sec-gamma-general}).

\subsection{Notation and conventions}
\label{sec-notation}

Our general results below are most elegantly stated for real analytic functions. Recall that an analytic function $f(z)$ in a domain $D \subset \mathbb{C}$ symmetric with respect to the real axis is called \emph{real analytic} if for each $z \in D$ we have $f(\overline z) = \overline{f(z)}$. A function $f \colon [\alpha; \beta] \to \mathbb{R}$ on $[\alpha; \beta] \subset \mathbb{R}$ is \emph{real analytic} if it extends to a complex-valued real analytic function in a complex neighbourhood of $[\alpha; \beta]$. Recall that $\hat z \in D$ is a critical point of an analytic function $f(z) \colon D \to \mathbb{C}$ if $f'(\hat z) = 0$. It is \emph{non-degenerate} if $f''(\hat z) \neq 0$, and it has \emph{order two} if $f''(\hat z) = 0$ but $f'''(\hat z) \neq 0$. An analytic function $f(z)$ with no critical points is called a \emph{conformal mapping} (which may not be globally injective). A conformal mapping preserves angles between curves.

For uniform estimates, we deal with functions defined on families of domains. By a \emph{family of domains} we mean an open subset $D \subset V \times \mathbb{C}$, where $V = [\alpha; \beta] \subset \mathbb{R}$, such that $D_v := D \cap (\{v\} \times \mathbb{C})$ is connected and non-empty for each $v \in V$. A family of functions $f_v \colon D_v \to \mathbb{C}$ is called \emph{analytic} if it extends to an analytic function $f_v(z) = \tilde f(v, z)$ in a neighbourhood of $D$ in $\mathbb{C} \times \mathbb{C}$. An analytic family of functions $f_v(z)$ is called \emph{(even/odd) real analytic} if for each fixed $v \in V$ the function $f_v(z)$ is (even/odd) real analytic. We often omit the index $v$ when it is fixed. The same applies to families depending on several parameters.

By a fractional power $z^{p/q}$ and the logarithm $\log z$ we mean their \emph{principal} branches, that is, the branches in $\mathbb{C} \setminus (-\infty; 0]$ that are positive for $z > 1$. For $z \in (-\infty; 0)$, define them by the continuity from the upper half-plane. In particular, we assume that $\sqrt z = z^{1/2}$ and $\sqrt[3]{z} = z^{1/3}$ have a positive imaginary part for $z < 0$. Under this convention, the identity $z^{r+s} = z^r z^s$ holds. We also set $\arg z := \Im \log z \in (-\pi; \pi]$.

Denote by $H^+$ and $H^-$ the upper and lower half-planes of the complex plane, and by $Q_1, \dots, Q_4$ the four open quadrants. Denote by $[\alpha; \beta]$ the segment in the complex plane with the endpoints $\alpha, \beta \in \mathbb{C}$ (in particular, $[\alpha; \beta] = [\beta; \alpha]$). Define $[\alpha; \beta)$, $(\alpha; \beta]$, and $(\alpha; \beta)$ analogously, including the possibilities $\alpha, \beta = c \pm d\infty$ for $c, d \in \mathbb{C}$. The notation $\int_\alpha^\beta f(z) \, dz$ stands for the contour integral over the oriented interval $(\alpha; \beta)$, understood as an improper Riemann integral when $\alpha$ or $\beta$ is infinite.

For a function $f \colon (\beta; \beta \pm \delta) \to \mathbb{R}$, where $\delta > 0$, denote $f(\beta \pm) := \lim_{\varepsilon \to 0, \varepsilon > 0} f(\beta \pm \varepsilon)$. For $\beta = c + d\infty$ with $c, d \in \mathbb{C}$, set $f(\beta) := \lim_{t \to +\infty, \, t \in \mathbb{R}} f(c + dt)$. This makes sense for real-valued and complex-valued $f$, with the limit belonging to $\mathbb{R} \cup \{\pm \infty\}$ and $\mathbb{C} \cup \{\infty\}$, respectively. Write $f(\pm d\infty):=f(0\pm d\infty)$, in particular, $(\pm \infty)^{-1} = 0$. Order $\mathbb{R} \cup \{\pm \infty\}$ naturally. For a subset $S \subset \mathbb{C}$ and a function $f \colon D \to \mathbb{R}$, denote
\begin{equation}
    \label{eq-liminf}
    \liminf_{z \to S} f(z) := \lim_{\varepsilon \searrow 0} \left(\inf_{z \in D \colon \dist(z, S) < \varepsilon} f(z)\right),
\end{equation}
where $\inf_{z \in \varnothing} f(z) := +\infty$. Define $\limsup_{z \to S} f(z)$ analogously, with $\sup_{z \in \varnothing} f(z) := -\infty$. We write $f(z) \sim g(z)$ as $z \to z_0 \in \mathbb{C} \cup \{\infty\}$ if $f(z) / g(z) \to 1$. The notation $f \colon S_1, \dots, S_k \mapsto T_1, \dots, T_k$, where $S_i, T_i \subset \mathbb{C}$, stands for $f(S_i) = T_i$ for each $i = 1, \dots, k$. Also, denote by $f^{-1}(u) := \{z \in D \colon f(z) = u\}$ the full preimage of a point $u \in \mathbb{C}$.

For any $S \subset \mathbb{C}$, the notation $\partial S$ stands for the boundary of $S$, and $\Cl S$ stands for its closure, that is, $\Cl S := S \cup \partial S$. For a subset $T \subset \Cl S$, by a slight abuse of terminology, by a \emph{neighbourhood of $T$ in $S$} we mean the intersection $N \cap S$ for some open set $N \subset \mathbb{C}$ containing $T$. A \emph{family of neighbourhoods} is understood analogously.

\section{Airy phenomenon}
\label{sec-airy}

\subsection{General Airy phenomenon}
\label{sec-airy-general}

Let us switch to the general form of integrals \eqref{eq-A1-int-osc}--\eqref{eq-A2-int-osc}:
\begin{align*}
    \psi(x, t) := \frac{1}{2\pi} \int_{-\pi}^{\pi} a(p) \, e^{ipx - i\omega(p)t} \, dp,
\end{align*}
where $x \ge 0$, $t > 0$ are integers and $a(p)$, $\omega(p)$ are even and odd real analytic $2\pi$-periodic functions, respectively, satisfying certain conditions. Define the \emph{fundamental domain} as follows (see, e.g., Fig.~\ref{fig-airy-w'-inner}):
\begin{align}
    \label{def-U}
    U := \{z \in \mathbb{C} \colon 0 < \Re z < \pi, \Im z > 0\}.
\end{align}

\begin{theorem}
    \label{th-airy}
    Let $D \subset \mathbb{C}$ be a $2\pi$-periodic simply-connected domain symmetric about the coordinate axes and containing $\partial U$. Let $a(p)$ and $\omega(p)$ be even and odd real analytic $2\pi$-periodic functions in $D$, such that $\omega'$ is a bijective conformal mapping of $D \cap U$ to $H^-$. Then
    \begin{align*}
        \frac{1}{2\pi} \int_{-\pi}^{\pi} a(p) \, e^{ipx - i\omega(p)t} \, dp =
        \begin{cases}
            t^{-1/3} \Ai(-\theta(x / t) \, t^{2/3}) \, h(x / t) \, a(\hat p(x / t)) + O_{v^+, a, \omega}\left(t^{-3/2}\right), & \text{if $x / t \le \hat v$}, \\
            t^{-1/3} \Ai(-\theta(x / t) \, t^{2/3}) \, h(x / t) \left(a(\hat p(x / t)) + O_{v^+, a, \omega}\left(t^{-1}\right)\right), & \text{if $x / t \ge \hat v$},
        \end{cases}
    \end{align*}
    for each $x, t \in \mathbb{Z}$, $v^+ \in \mathbb{R}$ with $t > 0$ and $0 \le x / t \le v^+ < V^+$, where
    \begin{align}
        \label{eq-airy-fns}
        \nonumber
        \hat v &:= \omega'(0), \quad \hat p(v) := \left(\omega'|_{\partial U}\right)^{-1}(v), &
        \theta(v) &:= \sgn(\hat v - v) \left|\tfrac{3}{2}(v \hat p(v) - \omega(\hat p(v)))\right|^{2/3}, \\
        V^+ &:= \liminf_{p \to \partial D \cap U, \Re p\ne 0} \frac{\Re \omega(p)}{\Re p}, &
        h(v) &:=
        \begin{cases}
            |4\theta(v)|^{1/4} \, / \, |\omega''(\hat p(v))|^{1/2}, & \text{if $v \neq \hat v$}, \\
            |\omega'''(0) / 2|^{-1/3}, & \text{if $v = \hat v$}.
        \end{cases}
    \end{align}
\end{theorem}

Here, $\hat p(v)$ is a critical point of the phase-space Lagrangian $\L(p) := vp - \omega(p)$; it is real for $v < \hat v$ and purely imaginary for $v > \hat v$ by Lemma~\ref{lm-airy-L} below. We shall see that the functions $a(\hat p(v))$, $\theta(v)$, and $h(v)$ are real analytic (by Remark~\ref{rm-airy-analyticity} below), so the main term in the asymptotic formula is real analytic in $x / t$.

The assumption $x / t < V^+$ means that the Lagrangian has a negative real part in the vicinity of $\partial D \cap U$. This is the announced boundary condition, implying the desired interior behaviour of the Lagrangian. Our main results in Sec.~\ref{sec-checkers-results} and computations in Secs.~\ref{sec-airy-checkers}~and~\ref{sec-gamma-checkers} demonstrate that the bound on $x / t$ in the theorem is sharp, in the sense that the assumption $x / t \le v^+ < V^+$ cannot be relaxed to $x / t < V^+$ and the value of $V^+$ cannot be increased in general: the constant in $O_{v^+, a, \omega}(t^{-1})$ blows up as $v^+ \to V^+$.

The point of Theorem~\ref{th-airy} is a rigorous uniform error bound in the asymptotic formula, whereas the expression for the main term is well-known \cite{Wong}. For $x / t < \hat v$, the error estimate order in the theorem is sharp, in the sense that $O(t^{-3/2})$ cannot be replaced with $O(t^{-\alpha})$ for $\alpha > 3/2$, as this would contradict the asymptotic expansions of the Bessel function $J_{vt}(t)$ \cite[Eq.~10.20.4]{NIST-10} and $\Ai'$ \cite[Eq.~9.7.10]{NIST-10}; see Sec.~\ref{sec-airy-bessel}.

The asymptotic formula simplifies, when $v := x / t$ is either close to or far from $\hat v$:

\begin{corollary}
    \label{cor-airy}
    Under the assumptions of Theorem~\ref{th-airy}, if $\delta > 0$, $t\ge1$, $v = \frac{x}{t}$ satisfy $|v - \hat v| \not\in \left(\frac{\delta}{t^{2/3}}; \delta\right)$, then
    \begin{align*}
        \frac{1}{2\pi} \int_{-\pi}^{\pi} a(p) \, e^{ipx - i\omega(p)t} \, dp =
        \begin{cases}
            \sqrt{\frac{2}{\pi t|\omega''(\hat p(v))|}} \cos\left(\frac{2}{3} t\theta(v)^{3/2} - \frac{\pi}{4}\right) a(\hat p(v)) + O_{\delta, v^+, a, \omega}\left(\frac{1}{t^{3/2}}\right), & \text{if } v \le \hat v - \delta, \\
            \sqrt[3]{\frac{2}{t|\omega'''(0)|}}
            \Ai\left((x - \hat v t)\sqrt[3]{\frac{2}{t|\omega'''(0)|}}
            \right) a(0) + O_{\delta, v^+, a, \omega}\left(\frac{1}{t}\right), & \text{if } |v - \hat v| \le \frac{\delta}{t^{2/3}}, \\
            \sqrt{\frac{1}{2\pi t|\omega''(\hat p(v))|}} \exp\left(-\frac{2}{3} t|\theta(v)|^{3/2}\right) \left(a(\hat p(v)) + O_{\delta, v^+, a, \omega}\left(\frac{1}{t}\right)\right), & \text{if } v \ge \hat v + \delta.
        \end{cases}
    \end{align*}
\end{corollary}

\begin{remark}
    Theorem~\ref{th-airy} is a rather general result. One can start with an arbitrary domain $D$ as in the theorem, and take a 1--1 conformal map $\omega'\colon D \cap U\to H^-$ without poles on $\partial U$. It exists by the Riemann mapping theorem, and $\omega'(\partial U) \subset \mathbb{R}$ by the local boundary correspondence principle. Then the antiderivative $\omega$ automatically satisfies the assumptions of the theorem after adding a linear function. Indeed, since $\omega'([0; \pi]) \subset \mathbb{R}$, it follows that $\omega'(\overline p) = \overline{\omega'(p)}$ in $D$ by the reflection principle. Thus, $\omega(\overline p) = \overline{\omega(p)} + \mathrm{const}$. Analogously, $\omega'(-\overline p) = \overline{\omega'(p)} = \omega'(2\pi - \overline p)$, hence $\omega(-p) = -\omega(p) + \mathrm{const}$ and $\omega(p + 2\pi) = \omega(p) + \mathrm{const}$. Then $\omega(p) - \omega(0) - (\omega(2\pi) - \omega(0)) \cdot p / 2\pi$ is an odd real analytic $2\pi$-periodic function in $D$.
\end{remark}

In Secs.~\ref{sec-airy-bessel}--\ref{sec-airy-checkers}, we illustrate the theorem with simple examples $\omega(p) = \sin p$ and $\omega(p) = \arcsin \frac{\sin p}{\sqrt{1 + m^2}}$, which arise in the integral representation of the Bessel function and wave function \eqref{eq-A1-int-osc}, respectively.

\subsection{Conformal change of variable}
\label{sec-airy-substitution}

\begin{figure}
    \centering
    \begin{subfigure}[c]{0.5\textwidth}
        \includegraphics[width=\textwidth]{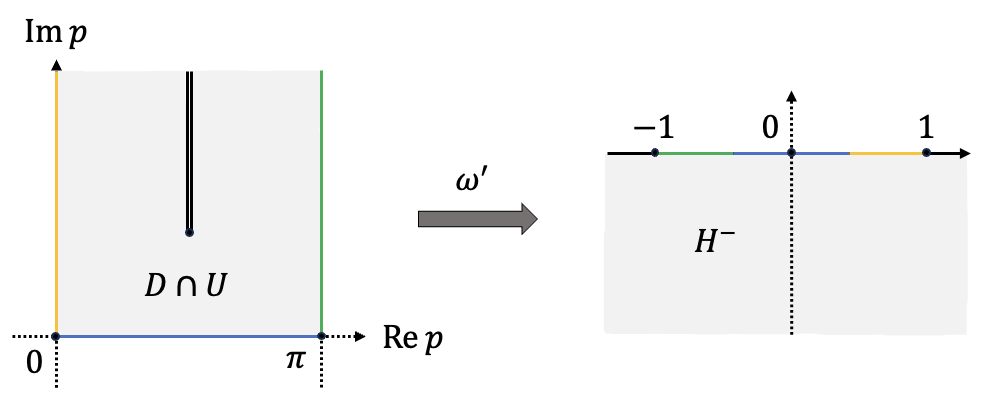}
        \caption{$\omega'(p)$ on $D \cap U$}
        \label{fig-airy-w'-inner}
    \end{subfigure}
    \hspace{0.05\textwidth}
    \begin{subfigure}[c]{0.4\textwidth}
        \includegraphics[width=\textwidth]{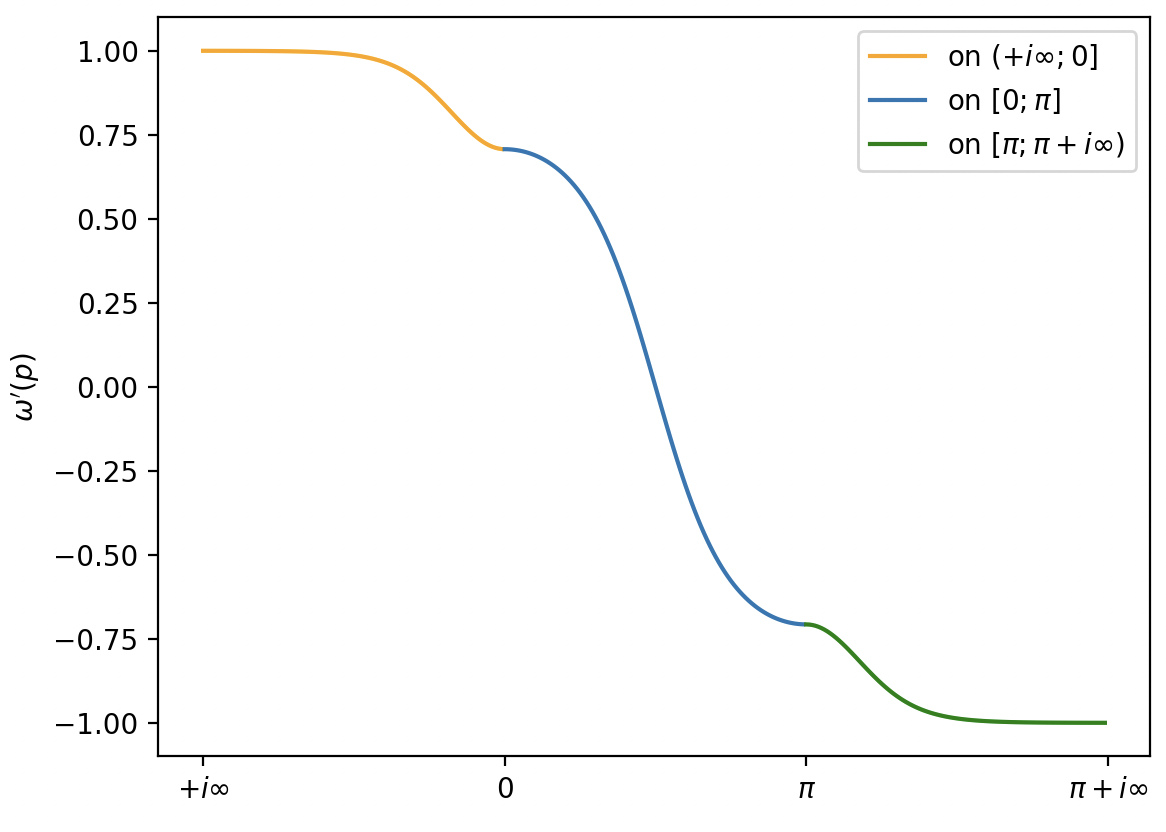}
        \caption{$\omega'(p)$ on $\partial U$}
        \label{fig-airy-w'-boundary}
    \end{subfigure}
    \caption{Illustration of $\omega'(p)$ from Theorem~\ref{th-airy} applied to quantum walks, where $\omega(p)$ is given by \eqref{eq-w}. Note that $\omega'(p) = \frac{\cos p}{\sqrt{m^2 + \cos^2 p}}$ is a bijective conformal mapping of the slit domain $D \cap U$ to the lower half-plane, and it is real and decreasing on $\partial U$. See Lemma~\ref{lm-airy-w}.}
    \label{fig-airy-w'}
\end{figure}

\begin{figure}
    \centering
    \begin{subfigure}[c]{0.47\textwidth}
        \includegraphics[width=\textwidth]{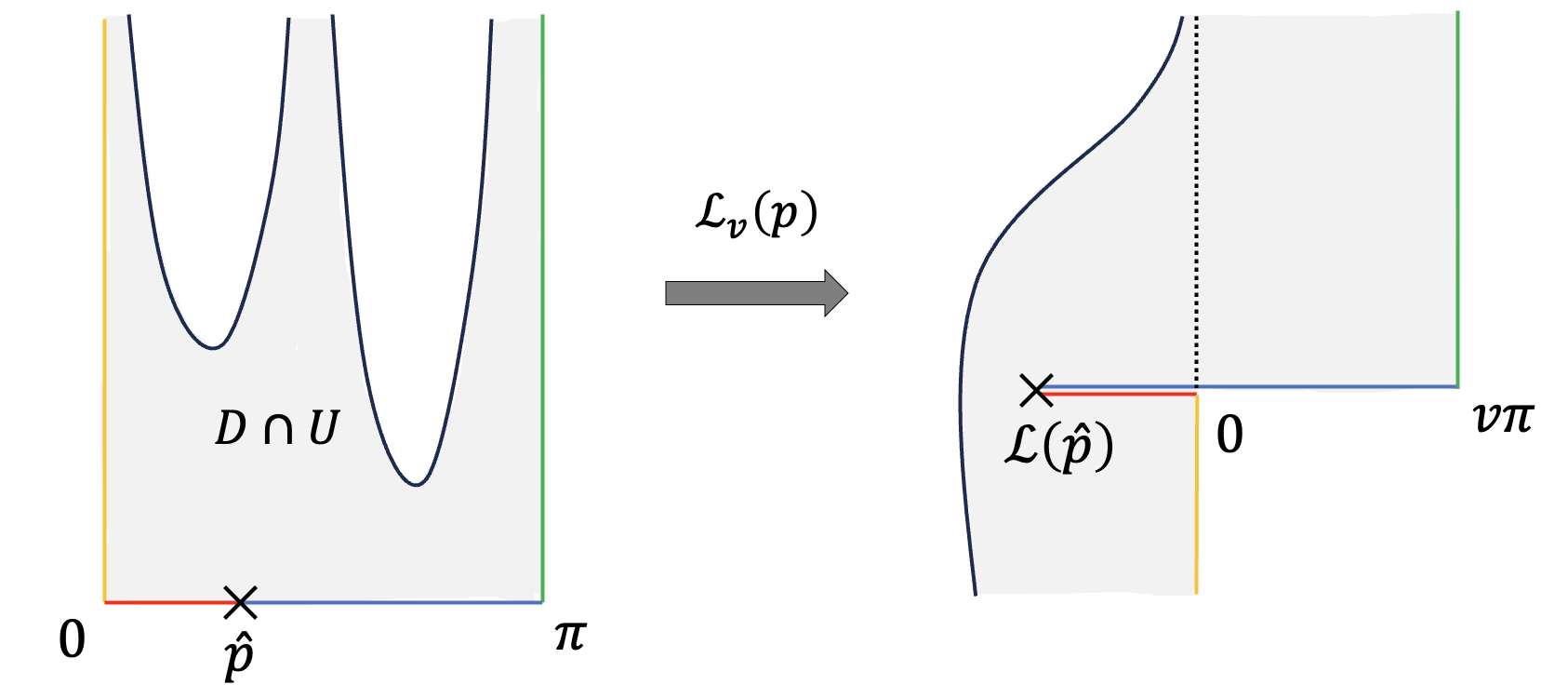}
        \caption{$\L_v(p)$ for $v < \hat v$}
        \label{fig-airy-L-lt}
    \end{subfigure}
    \hspace{0.02\textwidth}
    \begin{subfigure}[c]{0.47\textwidth}
        \includegraphics[width=\textwidth]{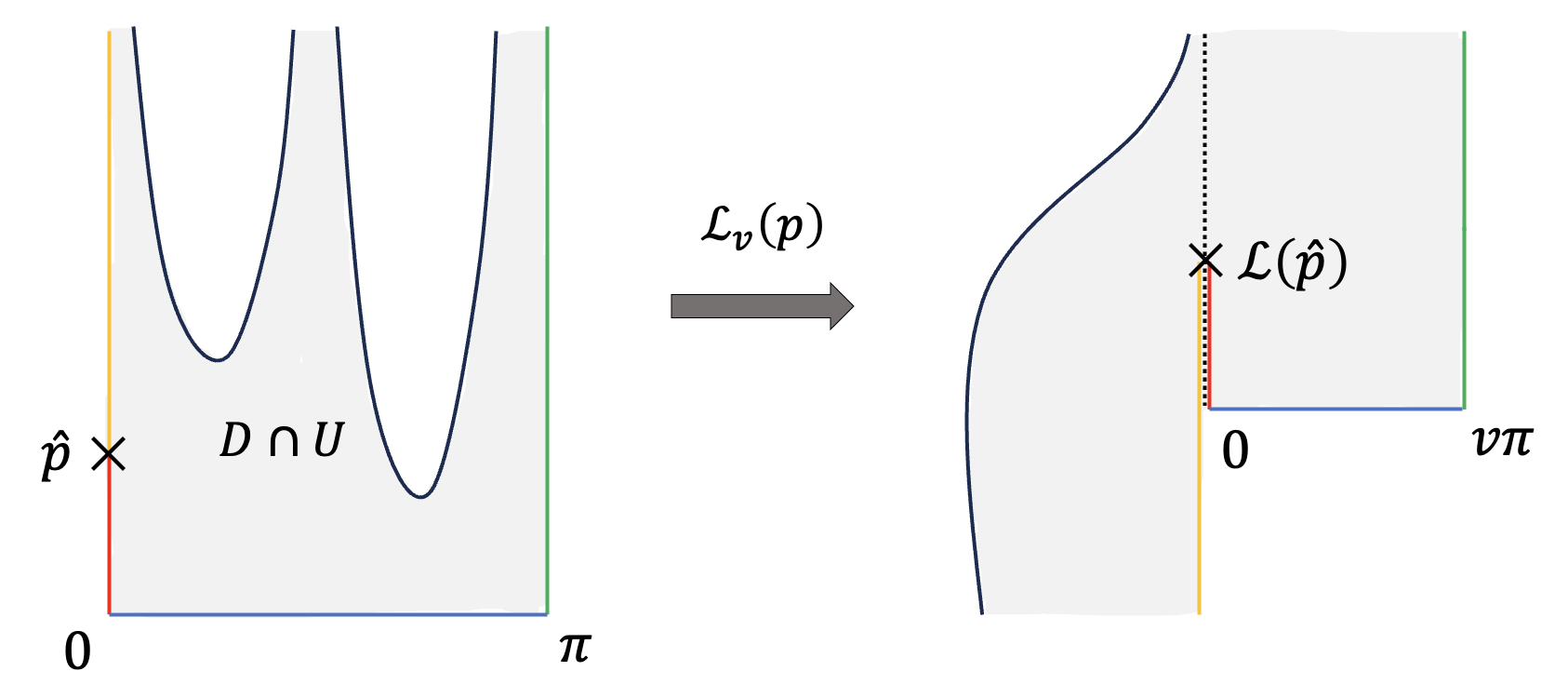}
        \caption{$\L_v(p)$ for $v > \hat v$}
        \label{fig-airy-L-gt}
    \end{subfigure}
    \\[0.4cm]
    \begin{subfigure}[c]{0.47\textwidth}
        \includegraphics[width=\textwidth]{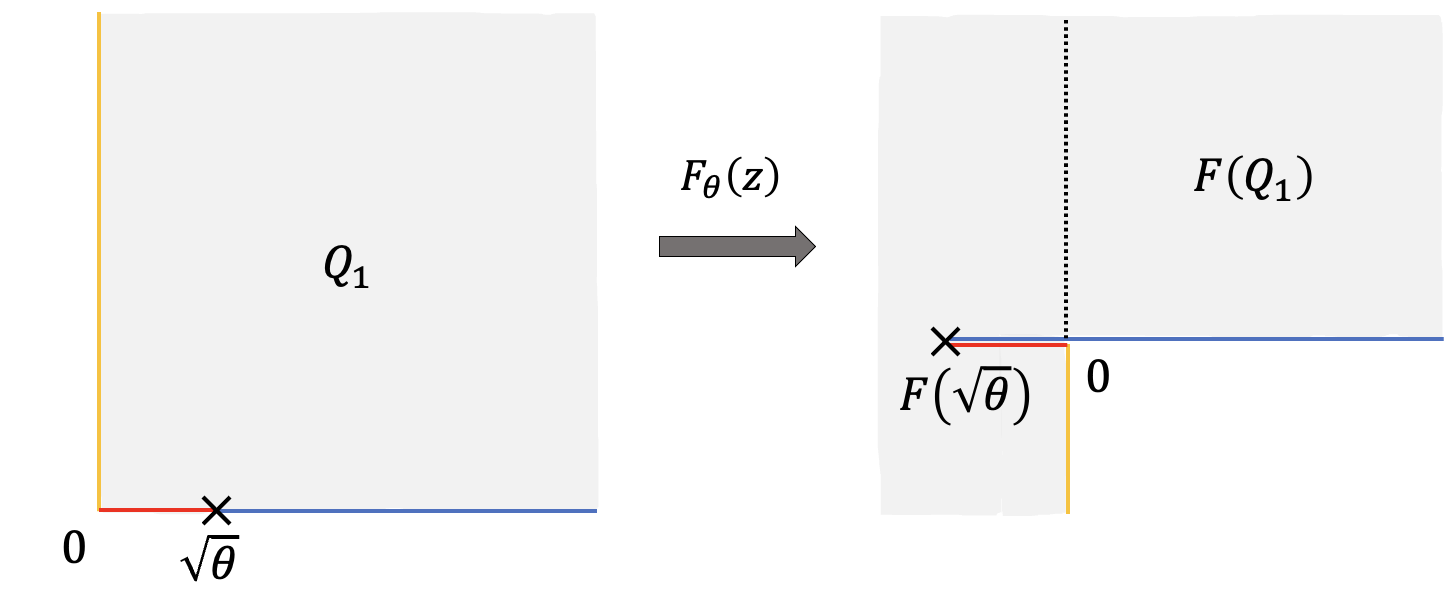}
        \caption{$F_\theta(z)$ for $\theta > 0$}
        \label{fig-airy-F-gt}
    \end{subfigure}
    \hspace{0.02\textwidth}
    \begin{subfigure}[c]{0.47\textwidth}
        \includegraphics[width=\textwidth]{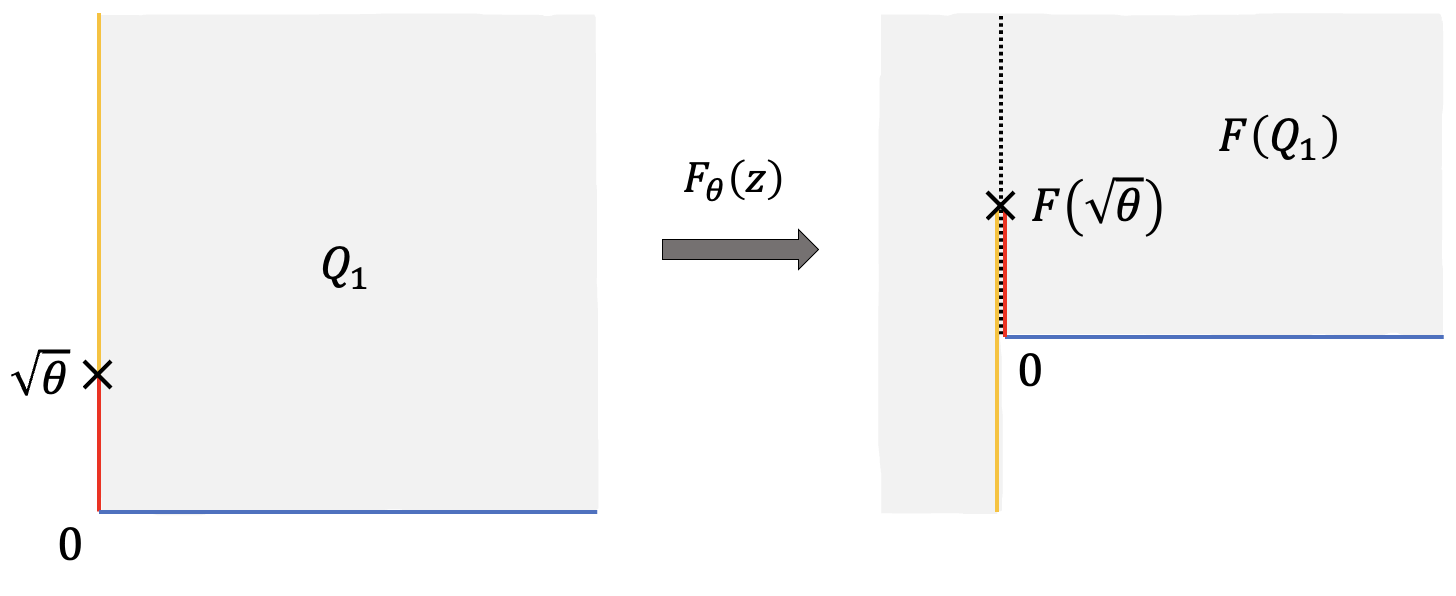}
        \caption{$F_\theta(z)$ for $\theta < 0$}
        \label{fig-airy-F-lt}
    \end{subfigure}
    \\[0.4cm]
    \caption{The Lagrangian $\L_v(p)$ and its normal form $F_\theta(z)$ as conformal mappings. See Lemmas~\ref{lm-airy-L}--\ref{lm-airy-F}.}
    \label{fig-airy-L-F}
\end{figure}

In this subsection, we start the proof of Theorem~\ref{th-airy}. We use the notation and impose the assumptions from the theorem, and assume
\begin{align*}
    v &:= x / t \in V, & V &:= [0; v^+].
\end{align*}
For each $v \in V$, $\theta \in \mathbb{R}$, $p \in D$, and $z \in \mathbb{C}$, define the \emph{Lagrangian} $\L_v(p)$ and its \emph{normal form} $F_\theta(z)$ as
\begin{align*}
    \L(p) = \L_v(p) &:= vp - \omega(p), & F(z) = F_\theta(z) := z^3 / 3 - \theta z.
\end{align*}
The assumption $v^+ < V^+$ implies that there is $\varepsilon = \varepsilon(v^+) > 0$ satisfying
\begin{align}
    \label{eq-airy-eps}
    \Re \L_v(p) < 0 \quad\text{for each $0 \le v \le v^+$ and $p \in D \cap U$ with $\dist(p, \partial D) < \varepsilon$}.
\end{align}
Our goal is to construct a conformal change of variable $z_v(p)$ in a sufficiently large family of domains, such that $\L_v = F_{\theta(v)} \circ z_v$; this is Proposition~\ref{prop-airy-substitution} below. First, Lemma~\ref{lm-airy-w} below derives useful properties of the derivative~$\omega'(p)$ (see Fig.~\ref{fig-airy-w'}). Lemmas~\ref{lm-airy-L}~and~\ref{lm-airy-F} describe the behaviour of $\L_v(p)$ and $F_\theta(z)$ as conformal mappings (see Fig.~\ref{fig-airy-L-F}). Lemma~\ref{lm-airy-univalence} proves the univalence of the Lagrangian in a certain domain and is crucial for showing the injectivity of $z_v(p)$ in Proposition~\ref{prop-airy-substitution}. Finally, Lemma~\ref{lm-airy-contour} is essential for the proof of Theorem~\ref{th-airy} in Sec.~\ref{sec-airy-proof}: it allows us to shift the integration contour into the complex plane and obtain the desired error bound for $v > \hat v$.

\begin{lemma}[Derivative]
    \label{lm-airy-w}
    The derivative $\omega'(p)$ is real on $\partial U$ and decreases as $p$ moves counterclockwise along $\partial U$ (see Fig.~\ref{fig-airy-w'}). We also have $\omega'(0) > 0$, $\omega'(\pi) < 0$, $\omega'(+i\infty) \ge V^+$, $\omega''(0) = 0$, $\omega''(p) \neq 0$ for each $p \in \partial U \setminus \{0, \pi\}$, and $\omega'''(0) < 0$.
\end{lemma}

\begin{proof}
    Recall that $\omega(p)$, hence $\omega'(p)$, is real analytic and $2\pi$-periodic in the domain $D$ containing $\partial U$. Also, $\omega'(p)$ is even because $\omega(p)$ is odd. Thus, for each $p \in \partial U$ we have
    \begin{align*}
        \overline{\omega'(p)} = \omega'(\overline p) =
        \begin{cases}
            \omega'(p), & \text{if $p \in [0; \pi]$}, \\
            \omega'(-p), & \text{if $p \in [0; +i\infty)$}, \\
            \omega'(2\pi - p), & \text{if $p \in [\pi; \pi + i\infty)$},
        \end{cases} = \omega'(p),
    \end{align*}
    hence $\omega'(p) \in \mathbb{R}$. By the principle of orientation, $\omega'$ decreases on $\partial U$, because $\omega'(D \cap U) \subset H^-$.

    Since $\omega'(D \cap U) \subset H^-$, it follows that $\omega''$ never vanishes on $\partial U \setminus \{0, \pi\}$. Also, $\omega''(0) = 0$ and $\omega'''(0) < 0$ because $\omega'$ locally maps a quarter-disk around $0$ in $U$ to a half-disk in $H^-$.

    Since $\omega$ is odd and $2\pi$-periodic, we have $\omega(0) = \omega(\pi) = 0$. By the mean-value theorem, there exists $r \in (0; \pi)$ such that $\omega'(r) = 0$. Since $\omega'$ decreases on $[0; \pi]$, we get $\omega'(0) > 0$ and $\omega'(\pi) < 0$.

    It remains to show that $\omega'(+i\infty) \ge V^+$. First note that $D \cap U$ is simply-connected because $D$ is. Thus, either $D = \mathbb{C}$ (so that $\partial D = \varnothing$), or for each sufficiently large $p \in [0; +i\infty)$, the segment $[p; p + \pi]$ intersects $\partial D$. If $D = \mathbb{C}$, then $\omega'$ is an injective conformal mapping $U \to H^-$ with $\omega'(\partial U) \subset \mathbb{R}$, hence $\omega'(\partial U) = \mathbb{R}$ and $\omega'(+i\infty) = +\infty = V^+$ by the boundary correspondence principle~\cite[Theorem~2.24]{Markushevich-77}. Otherwise, let $p'$ be the leftmost intersection point of $[p; p + \pi]$ with $\partial D$ (which exists since $\partial D$ is closed). Pick $q \in [p; p')$ such that $|q - p'| < 1 / |p|$. Since $\omega$ is odd real analytic, we have $\Re \omega(p) = 0$, thus
    \begin{align}
        \label{eq-airy-hor-seg}
        \frac{\Re \omega(q)}{\Re q} = \frac{\Re \omega(q) - \Re \omega(p)}{\Re q - \Re p} = \Re \omega'(r) \quad\text{for some $r \in [p; q]$}.
    \end{align}
    As $p \to +i\infty$, the point $q$ tends to $\partial D \cap U$, and the right side of \eqref{eq-airy-hor-seg} tends to $\omega'(+i\infty)$ because the arcs $[p; p']$ determine a prime end of $D \cap U$ and the bijective conformal mapping $\omega'$ continuously extends to each prime end (see \cite[Theorems~2.13 and~2.20]{Markushevich-77}). By~\eqref{eq-airy-fns}, the inequality $\omega'(+i\infty) \ge V^+$ follows.
\end{proof}

\begin{lemma}[Lagrangian]
    \label{lm-airy-L}
    In the strip $|\Re p| \le \pi$, the function $\L_v(p)$ has exactly two non-degenerate critical points $\pm \hat p(v) = \pm (\omega'|_{\partial U})^{-1}(v)$ for $v \neq \hat v$: they belong to $(-\pi; \pi)$ for $v < \hat v$, to $(-i\infty; +i\infty)$ for $v > \hat v$, and for $v = \hat v$ there is a unique order-two critical point at the origin. Also, $\L$ takes the intervals comprising $\partial U$ bijectively to the following intervals parallel to the coordinate axes (see Figs.~\ref{fig-airy-L-lt}~and~\ref{fig-airy-L-gt}):
    \begin{align*}
        \begin{cases}
            \L \colon (+i\infty; 0], [0; \hat p], [\hat p; \pi], [\pi; \pi + i\infty) \mapsto (-i\infty; 0], [0; \L(\hat p)], [\L(\hat p); v\pi], [v\pi; v\pi + i\infty), & \text{if $v \le \hat v$}, \\
            \L \colon (+i\infty; \hat p], [\hat p; 0], [0; \pi], [\pi; \pi + i\infty) \mapsto (-i\infty; \L(\hat p)], [\L(\hat p); 0], [0; v\pi], [v\pi; v\pi + i\infty), & \text{if $v \ge \hat v$},
        \end{cases}
    \end{align*}
    where $\L(\hat p) < 0$ for $v < \hat v$, $\L(\hat p) \in (0; +i\infty)$ for $v > \hat v$, and $\L(\hat p) = 0$ for $v = \hat v$.
\end{lemma}

\begin{proof}
    Since $\L'(p) = v - \omega'(p)$, it follows that $p$ is a critical point of $\L$ whenever $\omega'(p) = v$. Clearly, $\L$ has no critical points inside $U$ because $\omega'(D \cap U) \subset H^-$. Since $\omega'$ decreases on $\partial U$, $\omega'(\pi) < 0$, and $\omega'(+i\infty) \ge V^+$ by Lemma~\ref{lm-airy-w}, it follows that $v \in \omega'([0; \pi))$ for $v \le \hat v$ and $v \in \omega'([0; +i\infty))$ for $v \ge \hat v$. Thus, $\L|_{\Cl U}$ has a unique critical point $\hat p = \hat p(v) = (\omega'|_{\partial U})^{-1}(v)$, which belongs to $(0; \pi)$ for $v < \hat v$ and to $(0; +i\infty)$ for $v > \hat v$. When $v \neq \hat v$, the critical point is non-degenerate because $\omega'' \neq 0$ on $(0; \pi) \cup (0; +i\infty)$, and when $v = \hat v$, it has order two because $\omega''(0) = 0$ and $\omega'''(0) \neq 0$. By symmetry, $\L$ has the required critical points.

    Let us find $\L(\partial U)$. Since $\L'$ is real and non-vanishing on $\partial U \setminus \{\hat p\}$, $\L$ maps the intervals between $+i\infty$, $0$, $\hat p$, $\pi$, and $\pi + i\infty$ on $\partial U$ bijectively onto intervals between the limiting values at these endpoints.

    Clearly, $\L(0) = 0$ because $\L$ is odd, and $\L(\pi) = v\pi - \omega(\pi) = v\pi$ because $\omega$ is odd and $2\pi$-periodic. Since $\L'(p) = v - \omega'(p)$ decreases on $[0; +i\infty)$ and increases on $[0;\pi]$, it follows that $\L'(+i\infty) < \L'(\hat p) = 0$, thus $\L(+i\infty) = \int_0^{+i\infty} \L'(p) \, dp = -i\infty$. Similarly, since $\L'(p)$ increases on $[\pi; \pi + i\infty)$, it follows that $\L'(\pi + i\infty) > 0$, thus $\L(\pi + i\infty) = \L(\pi) + \int_\pi^{\pi + i\infty} \L'(p) \, dp = v\pi + i\infty$. As for $\L(\hat p)$, let us consider three cases. For $v < \hat v$, we have $\L(\hat p) = \int_0^{\hat p} \L'(p) \, dp < 0$ since $\L'(p) < 0$ on $[0; \hat p)$. For $v > \hat v$, we have $\L(\hat p) = \int_0^{\hat p} \L'(p) \, dp \in (0; +i\infty)$ since $\L'(p) > 0$ on $[0; \hat p)$. For $v = \hat v$, we have $\L(\hat p) = \L(0) = 0$.
\end{proof}

Recall our convention that $\sqrt{\theta}$ has positive imaginary part for $\theta < 0$.

\begin{lemma}[Normal form]
    \label{lm-airy-F}
    The function $F_\theta(z)$ has exactly two non-degenerate critical points $\pm \sqrt \theta$ for each $\theta \neq 0$ and a unique order-two critical point at the origin for $\theta = 0$. Also, $F|_{Q_1}$ is a bijective conformal mapping onto $F(Q_1) = \mathbb{C} \setminus (\Cl Q_4 \cup [0; F(\sqrt \theta)])$, and $F$ takes the intervals comprising $\partial Q_1$ bijectively to the following intervals parallel to the coordinate axes (see Figs.~\ref{fig-airy-F-gt}~and~\ref{fig-airy-F-lt}):
    \begin{align}
        \label{eq-lm-airy-F}
        \begin{cases}
            F \colon (+i\infty; 0], [0; \sqrt \theta], [\sqrt \theta; +\infty) \mapsto (-i\infty; 0], [0; F(\sqrt \theta)], [F(\sqrt \theta); +\infty), & \text{if $\theta \ge 0$}, \\
            F \colon (+i\infty; \sqrt \theta], [\sqrt \theta; 0], [0; +\infty) \mapsto (-i\infty; F(\sqrt \theta)], [F(\sqrt \theta); 0], [0; +\infty), & \text{if $\theta \le 0$}.
        \end{cases}
    \end{align}
    Additionally, $\theta \mapsto F_\theta(\sqrt \theta)$ is a bijection between $(-\infty; 0]$, $[0; +\infty)$ and $[0; +i\infty)$, $(-\infty; 0]$, respectively.
\end{lemma}

\begin{proof}
    We have $F'(z) = z^2 - \theta = 0$ if and only if $z = \pm \sqrt \theta$. These critical points are non-degenerate for $\theta \neq 0$ because $F''(\sqrt \theta) = 2\sqrt \theta \neq 0$, and for $\theta = 0$ we have an order-two critical point because $F'''(0) = 2 \neq 0$.

    Let us find $F(\partial Q_1)$. Since $F'$ is real and non-vanishing on $\partial Q_1 \setminus \{\sqrt \theta\}$, $F$ maps the intervals between $+i\infty$, $0$, $\sqrt \theta$, $+\infty$ on $\partial Q_1$ bijectively onto intervals between the limiting values at these endpoints. Clearly, $F(0) = 0$, $F(+i\infty) = -i\infty$, and $F(+\infty) = +\infty$ since $F(z) \sim z^3 / 3$ as $z \to \infty$. This implies \eqref{eq-lm-airy-F}. Clearly, the map $\theta \mapsto F_\theta(\sqrt \theta) = -\frac{2}{3} \theta^{3/2}$ is a bijection between $(-\infty; 0]$, $[0; +\infty)$ and $[0; +i\infty)$, $(-\infty; 0]$, respectively.

    The map $F|_{Q_1}$ is conformal and injective by the Noshiro--Warschawski theorem \cite[Theorem~2.16]{Duren-83}, because $\Im F'(z) = \Im(z^2)$ has a constant sign in the convex domain $Q_1$. Since $F(z) \sim z^3 / 3$ as $z \to \infty$, it follows that $F|_{Q_1}$ extends continuously to $\partial Q_1 \cup \{\infty\}$ in the wide sense by $F(\infty):=\infty$. Thus, $\partial F(Q_1) = F(\partial Q_1)$ by a compactness argument. The image $F(Q_1)$ is then determined to be the unique domain with boundary $F(\partial Q_1)$ given by \eqref{eq-lm-airy-F} (up to inessential ambiguity for $\theta = 0$).
\end{proof}

\begin{figure}
    \centering
    \includegraphics[width=0.9\textwidth]{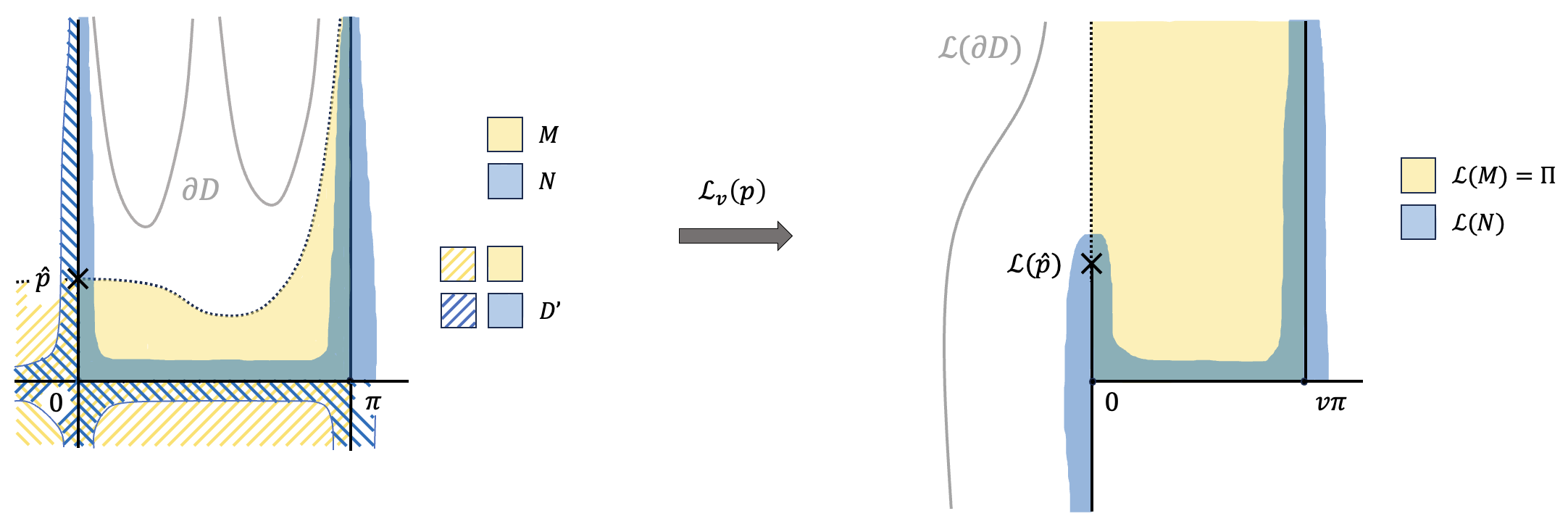}
    \caption{The domains $M$, $N$, $D'$ and their images for $v > \hat v$. See the proof of Proposition~\ref{prop-airy-substitution}.}
    \label{fig-airy-domains}
\end{figure}

\begin{figure}
    \centering
    \begin{subfigure}[c]{0.47\textwidth}
        \includegraphics[width=\textwidth]{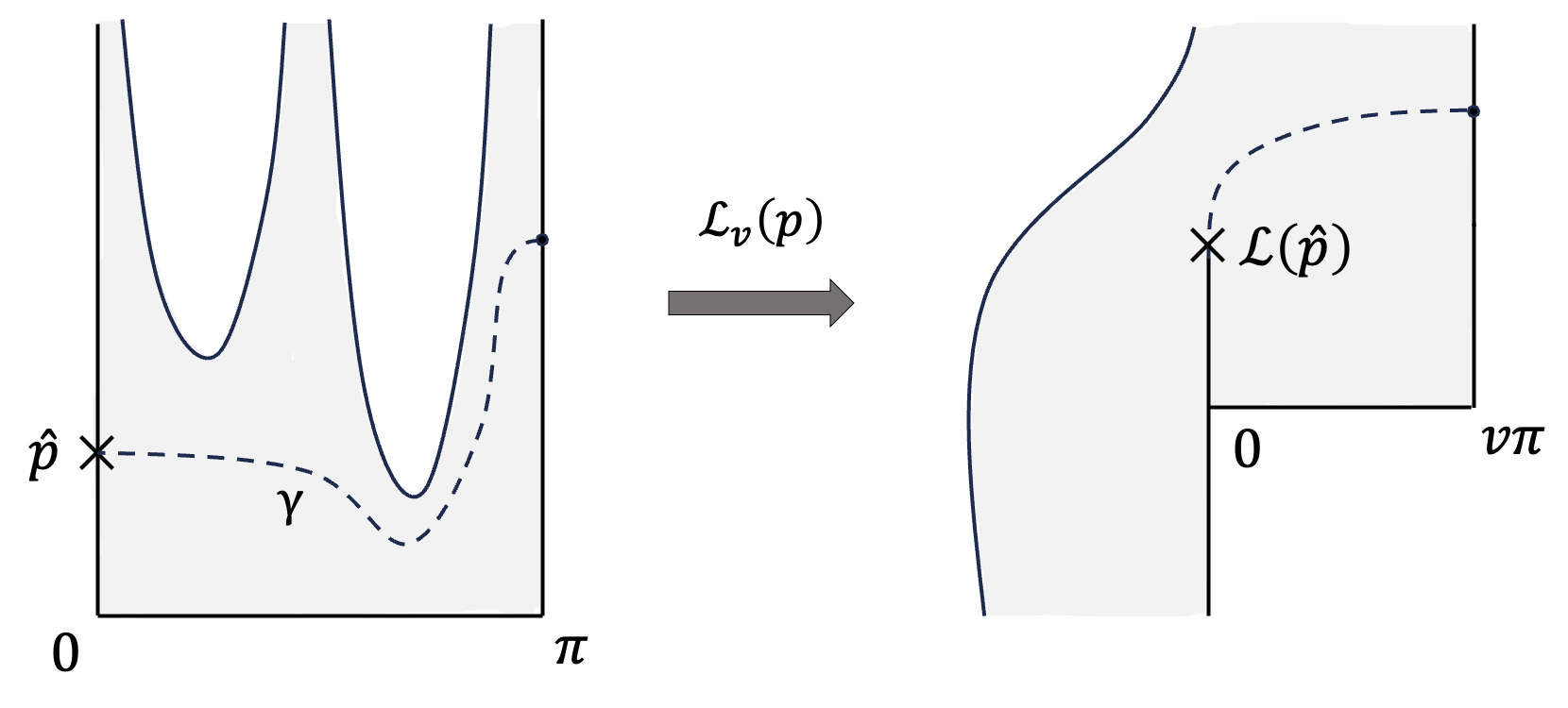}
        \caption{on $p$-plane}
    \end{subfigure}
    \hspace{0.02\textwidth}
    \begin{subfigure}[c]{0.47\textwidth}
        \includegraphics[width=\textwidth]{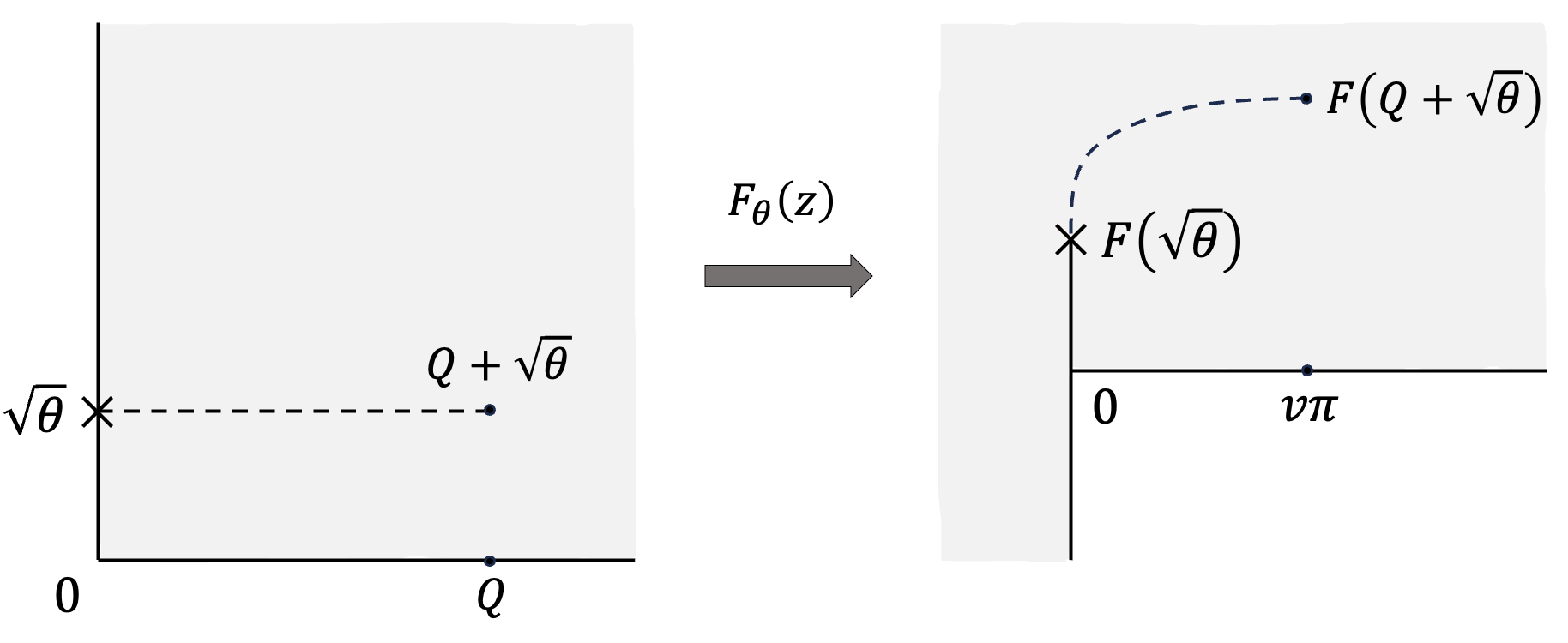}
        \caption{on $z$-plane}
    \end{subfigure}
    \caption{Shifted integration contour on different planes}
    \label{fig-airy-contour}
\end{figure}

We are going to shift the integration contour within the following subsets of $\mathbb{C}$ for each $v > \hat v$ (and therefore need to include them in our change of variable; see Figs.~\ref{fig-airy-domains} and~\ref{fig-airy-contour}):
\begin{align*}
    \Pi = \Pi_v &:= \{z \in \mathbb{C} \colon 0 < \Re z < \pi v, \Im z > 0 \}, \\
    M = M_v &:=
    \begin{cases}
        \text{component of $\L^{-1}(\Pi)\cap U$ whose boundary contains $\pi$}, & \text{if $v > 0$}, \\
        \varnothing, & \text{if $v = 0$}.
    \end{cases}
\end{align*}

\begin{lemma}[Univalence domain]
    \label{lm-airy-univalence}
    For each $v > 0$, the set $\L^{-1}(\Pi)$ contains a neighbourhood of $\pi$ in $U$, so $M$ is well-defined and non-empty, and the restriction $\L|_M$ is a homeomorphism $M \leftrightarrow \Pi$. Furthermore, $M$ is the only component of $\L^{-1}(\Pi) \cap U$ whose boundary has infinitely many common points with $\partial U$.
\end{lemma}

The proof of this lemma is quite technical because we avoid any additional assumptions on the conformal mapping $\omega'$. Most of the alternative approaches would fail in the case when $(\omega')^{-1}(\infty) = \infty$.

\begin{proof}
    For $v > 0$, the set $\L^{-1}(\Pi)$, hence $M$, contains a neighbourhood in $U$ of the following connected set containing $\pi$ (see Fig.~\ref{fig-airy-domains}):
    \begin{align*}
        \Gamma :=
        \begin{cases}
            (\hat p; 0] \cup (0; \pi) \cup [\pi; \pi + i\infty), & \text{if $v \ge \hat v$}, \\
            \left((\L|_{(0; \pi]})^{-1}(0); \pi\right] \cup [\pi; \pi + i\infty), & \text{if $v < \hat v$}.
        \end{cases}
    \end{align*}
    Indeed, by Lemma~\ref{lm-airy-L} and the principle of orientation (Fig.~\ref{fig-airy-L-F}), for each $p \in \Gamma$, the map $\L$ takes a neighbourhood of $p$ in $U$ into $\Pi$. So $M$ is well-defined and non-empty.

    By the principle of orientation, $\partial U \setminus \Cl \Gamma$ is disjoint with $\Cl \, (\L^{-1}(\Pi) \cap U)$. Thus, $M$ is the only component of $\L^{-1}(\Pi) \cap U$ whose boundary has infinitely many common points with $\partial U$.

    Next, let us show that $\L|_M$ is proper. We verify the sequential properness criterion: for each sequence $p_n \in M$ converging to a point $p_0 \in \partial M \cup \{\infty\}$, the sequence of images $\L(p_n)$ has no limit in $\Pi$. Consider the following four cases.

    Case $p_0 \in \partial U$. By continuity, $\L(p_n) \to \L(p_0)$, which does not lie in $\Pi$ by Lemma~\ref{lm-airy-L}.

    Case $p_0 \in D \cap U$. By continuity, $\L(p_n) \to \L(p_0)$. If $p_0 \in \L^{-1}(\Pi)$, then $p_0 \in M$ because $p_0 \in D \cap U$, $\L^{-1}(\Pi) \cap U$ is open, and all $p_n \in M$. This contradicts the assumption $p_0 \in \partial M$, hence $\L(p_0) \not\in \Pi$.

    Case $p_0 \in \partial D \cap U$. This is actually impossible because $\Re \L < 0$ in the $\varepsilon$-neighbourhood of $\partial D$ in ${D \cap U}$ by \eqref{eq-airy-eps}, but $\Re \L(p_n) > 0$ as $p_n \in M$.

    Case $p_0 = \infty$. Passing to a subsequence, we may assume $\Re p_n \to X$ for some $X \in [0; \pi]$ and $\Im p_n \to +\infty$. Let us show that $\lim_{n \to \infty} \L(p_n) = \infty$ by treating $X = 0, \pi$ and $X \in (0; \pi)$ separately.

    For that, we show that $D$ contains a neighbourhood of $[X; X + i\infty)$. Since $\Im \L'(p) = -\Im \omega'(p) > 0$ for each $p \in D \cap U$, it follows that $\Re \L(p)$ increases on each component of $[p_n; \Re p_n] \cap D$ as $\Im p$ decreases. Since $\Re \L < 0$ in the $\varepsilon$-neighbourhood of $\partial D$ in ${D \cap U}$ for some $\varepsilon > 0$ by \eqref{eq-airy-eps} and $\Re \L(p_n) > 0$, it follows that $[p_n; \Re p_n]$ is disjoint with this neighbourhood. Then $\dist([p_n; \Re p_n], \partial D) \ge \varepsilon$. As $n \to +\infty$, we have $\Re p_n \to X$ and $\Im p_n \to +\infty$, and thus $\dist([X; X + i\infty), \partial D) \ge \varepsilon$. Therefore, the half-strip $S := \{p \in H^+ \colon |\Re p - X| < \varepsilon\}$ belongs to $D$ and, without loss of generality, it contains all $p_n$.

    Subcase $X = 0$ or $\pi$. In this case, $\L(X + i\infty) = \infty$ by Lemma~\ref{lm-airy-L}. To show that it equals $\lim_{n \to \infty} \L(p_n)$, let us apply Lindel\"of's theorem for bounded analytic functions \cite[Theorem~1.5.7]{bracci2020continuous}. Consider the analytic function $f(z) := \L(-i \, \widetilde \log \, z)$ of the sector $C := \{e^{ip} \colon p \in S\}$, where $\widetilde \log \, z$ denotes a continuous branch of the logarithm. Let us modify it to a bounded function. The map $f$ is injective in $C \cap H^+$ because $\L$ is injective in $S \cap U$, by $\Im \L'|_{S \cap U} = -\Im \omega'|_{S \cap U} > 0$ and the Noshiro--Warschawski theorem. By symmetry, $f(z)$ is injective in $C \cap H^-$ because $\L(-\overline p) = -\overline{\L(p)}$. Pick a point $q \in f(C)\setminus (\partial f(C\cap H^+)\cup \partial f(C\cap H^-))$. It has at most two preimages, hence a smaller sector $C' := \{e^{ip} \colon |\Re p - X| < \varepsilon', \Im p > 0\}$ for some $0 < \varepsilon' < \varepsilon$ does not intersect neighbourhoods of the preimages. Then the function $(f(z) - q)^{-1}$ is bounded in $C'$. The value $\L(X + i\infty)^{-1} = 0$ is its radial limit at $0$. By Lindel\"of's theorem, the angular limit at $0$ is also $0$. Since $\Re p_n \to X$ and $\Im p_n \to +\infty$, it follows that $\lim_{n \to \infty} (\L(p_n) - q)^{-1}$ is also $0$. Thus $\lim_{n \to \infty} \L(p_n) = \infty$.

    Subcase $X \in (0; \pi)$. Set $\delta := \min(\varepsilon, X, \pi - X)$.
    Let us show finite $\Re \L(X + i\infty)$ and $\L'(X + i\infty)$ exist.

    The limit $\Re \L(X + i\infty) \in \mathbb{R} \cup \{-\infty\}$ exists because $\Re \L$ decreases as $\Im p$ increases. Assume it is $-\infty$. Apply Lindel\"of's theorem to $f(z) := \exp(\L(-i\,\widetilde \log \, z))$, which is bounded in $\{e^{ip} \colon |\Re p - X| < \delta, \Im p > 0\}$ because $\Re\L$ is bounded from above by its maximum on $[X-\delta;X+\delta]$. We conclude that $\lim_{n \to \infty} \exp(\L(p_n)) = \exp \Re \L(X + i\infty) = 0$, which contradicts $\Re \L(p_n) > 0$. Thus, $\Re \L(X + i\infty) \in \mathbb{R}$.

    To show the existence of $\L'(X + i\infty)$, introduce the function $\sigma(p) := \frac{\L'(p) - \Re \L'(X + iy)}{i \Im \L'(X + iy)}$ for each $y > \delta$. It satisfies $\Re \sigma(p) = \frac{\Im \L'(p)}{\Im \L'(X + iy)} > 0$ in the $\delta$-neighbourhood of $X + iy$, and also $\sigma(X + iy) = 1$. By \cite[Corollary~2.3]{Pommerenke-75}, we have $|\sigma'(X + iy)| \le 2 / \delta$, hence
    \begin{align}
        \label{eq-airy-caratheodory}
        |\L''(X + iy)| \le \frac{2}{\delta} \Im \L'(X + iy).
    \end{align}
    Then for $\delta < y_1 < y_2$, we have
    \begin{multline}
        \label{eq-airy-L'-deviation}
        |\L'(X + iy_2) - \L'(X + iy_1)| \le \int_{y_1}^{y_2} |\L''(X + iy)| \, dy \oref{\text{by} \eqref{eq-airy-caratheodory}}{\le} \frac{2}{\delta} \int_{y_1}^{y_2} \Im \L'(X + iy) \, dy = \\
        = \frac{2}{\delta} (\Re \L(X + iy_1) - \Re \L(X + iy_2)) \to 0 \qquad\text{as }y_1, y_2 \to +\infty,
    \end{multline}
    because $\Re \L(X + i\infty) \in \mathbb{R}$. By the Cauchy criterion, there exists a finite limit $\L'(X + i\infty)$.

    We have $\L'(X + i\infty) \ne 0$ by the injectivity of $\L' = v - \omega'$ in $D \cap U$: since $\omega'$ is real on $\partial U$ by Lemma~\ref{lm-airy-w}, the $\L'$-image of a neighbourhood of $\hat p$ in $D \cap U$ already occupies a neighbourhood of $0$ in $H^+$, so the values $\L'(X + iy) \in H^+$ as $y \to +\infty$ eventually collide with the values $\L'(p)$ for some $p \in D \cap U$ near $\hat p$.

    Therefore, $\L'(X + i\infty) \in \mathbb{C} \setminus \{0\}$. Since the 1--1 conformal map $\L' \colon D \cap U \to H^+$ continuously extends to each prime end \cite[Theorem~2.20]{Markushevich-77} and $p_n \in S \subset D$, it follows that $\lim_{n \to \infty} \L'(p_n) = \L'(X + i\infty)$. Then
    $$
    \L(p_n) = \L(p_1) + \int_{p_1}^{p_n} \L'(p) \, dp \to \infty \qquad\text{as $n \to \infty$}
    $$
    because $\Im p_n \to +\infty$. This proves that $\L|_M$ is proper.

    As every proper map between subsets of the plane is closed, $\L(M)$ is closed in $\Pi$. Since $\L$ is non-constant and analytic, $\L(M)$ is open in $\Pi$. By the first paragraph of the proof, $\L(M)$ is non-empty. Hence $\L(M) = \Pi$. Since every proper conformal map is a covering \cite[Lemma~2]{Ho-75}, $M$ is path connected and $\Pi$ is simply-connected and locally path connected, it follows that $\L|_M$ is a homeomorphism $M \leftrightarrow \Pi$.
\end{proof}

\begin{proposition}[Conformal change of variable]
    \label{prop-airy-substitution}
    There exists a family of domains $D'_v \subset D$ symmetric about the coordinate axes and containing $\partial U \cup M$ (Fig.~\ref{fig-airy-domains}), a real analytic family of odd injective functions $z_v \colon D'_v \to \mathbb{C}$, and a real analytic function $\theta \colon V \to \mathbb{R}$, given by \eqref{eq-airy-fns}, such that for each $v \in V$, we have $\L_v|_{D'_v} = F_{\theta(v)} \circ z_v$. Additionally, we have $z_v(\hat p(v)) = \sqrt{\theta(v)}$ and $z'_v(\hat p(v)) = 1 / h(v)$ with notation \eqref{eq-airy-fns}.
\end{proposition}

\begin{proof}
    Pick $\theta(v) \in \mathbb{R}$ so that the critical values $\L_v(\hat p)$ and $F_\theta(\sqrt \theta)$ match (that is, ``align the crosses'' in Fig.~\ref{fig-airy-L-F}). This leads us to the expression for $\theta(v)$ in \eqref{eq-airy-fns}, since $\L_v(\hat p)$ and $F_\theta(\sqrt \theta)$ belong to the appropriate coordinate rays by Lemmas~\ref{lm-airy-L}--\ref{lm-airy-F}.

    The desired domain $D'$ is the union of $\partial U$, $M$, a family of sufficiently small neighbourhoods $N$ of $\partial U$ in $Q_1$, and their symmetric copies about the coordinate axes (see Fig.~\ref{fig-airy-domains}). By choosing $N$ small enough, we may assume that $\L|_N$ is injective, $N \cap \L^{-1}(\Pi) \subset M$, $\L(N) \cap [v\pi; +\infty) = \varnothing$, and also $N$ is simply-connected (the existence of such a family $N$ is proven in Proposition~\ref{prop-airy-neighbourhoods} in Appendix~\ref{sec-boundary-neighbourhoods}).

    The function $z(p) = z_v(p)$ is constructed in several steps. First, we define (see Fig.~\ref{fig-airy-L-F})
    \begin{align*}
        z(p) := (F|_{Q_1})^{-1}(\L(p)) \qquad\text{for each}\quad p \in M \cup N.
    \end{align*}
    Let us show that $z(p)$ is well-defined and injective. By Lemma~\ref{lm-airy-F}, the inverse $(F|_{Q_1})^{-1} \colon F(Q_1) \to Q_1$ exists and $\L(M) \subset \Pi \subset F(Q_1)$. By Lemmas~\ref{lm-airy-L}--\ref{lm-airy-F}, we get $\partial F(Q_1) = \L([0; +i\infty) \cup [0; \pi)) \cup [v\pi; +\infty)$. Then $\L(N) \cap \partial F(Q_1) = \varnothing$ since $\L|_N$ is injective and $\L(N) \cap [v\pi; +\infty) = \varnothing$. Then $\L(N) \subset F(Q_1)$ because $N$ is connected and $\partial \L(N) \ni \L(\hat p) = F(\sqrt \theta)$ (up to inessential ambiguity for $v = \hat v$). So $z(p)$ is well-defined. The maps $z|_M$ and $z|_N$ are injective because $\L|_M$ and $\L|_N$ are, by Lemma~\ref{lm-airy-univalence} and the choice of $N$. Since $\L(M) \subset \Pi$ and $N \cap \L^{-1}(\Pi) \subset M$, it follows that $\L(N \setminus M) \cap \L(M) = \varnothing$, thus $z(N \setminus M) \cap z(M) = \varnothing$. So, $z(p)$ is injective.

    Next, let us extend $z(p)$ to $\partial N \cap \partial Q_1 \supset [0; \pi] \cup [0; +i\infty)$ continuously. Assume, without loss of generality, that $N$ is adjacent to $\partial N \cap \partial Q_1$. The injective conformal mapping $\L|_N$ extends to a homeomorphism between $\partial N \cap \partial Q_1$ and a set of prime ends of $\L(N)$ corresponding to $\L(\partial N \cap \partial Q_1)$ \cite[Theorem~2.20]{Markushevich-77}. By Lemmas~\ref{lm-airy-L}--\ref{lm-airy-F}, the injective conformal mapping $(F|_{Q_1})^{-1}|_{\L(N)}$ extends to a homeomorphism between this set of prime ends and a connected subset of $\partial Q_1$. Hence, the map $z = (F|_{Q_1})^{-1} \circ \L$ extends continuously to $\partial N \cap \partial Q_1$ with $z(\partial N \cap \mathbb{R}) \subset [0; +\infty)$ and $z([0; +i\infty)) \subset [0; +i\infty)$. Furthermore, by Lemmas~\ref{lm-airy-L}--\ref{lm-airy-F}, this extension maps the intervals between $+i\infty$, $0$, $\hat p$, and $\pi$ to the intervals between $+i\infty$, $0$, $\sqrt \theta$, and $(F|_{(0; +\infty)})^{-1}(v\pi)$ (where the order of $0$, $\hat p$ and $0$, $\sqrt \theta$ is flipped when $v > \hat v$). In particular, for each $v$, we have $z(+i\infty) = +i\infty$, $z(0) = 0$, and $z(\hat p) = \sqrt \theta$.

    Now we can extend $z(p)$ analytically to $D'$ using Schwarz reflection principle~\cite[Theorem~9.1]{Markushevich-77}, because $z(\partial N \cap \partial Q_1) \subset \partial Q_1$ and $M \cup N$ is adjacent to $\partial N \cap \partial Q_1$. Since $z(M \cup N) \subset Q_1$, the map $z \colon D' \to \mathbb{C}$ is injective.

    Next, we verify the analyticity of $\theta$ and $z$ in $v$. Note that $\hat p(v) = (\omega')^{-1}(v)$ is analytic for each $v \neq \hat v$ by the inverse function theorem (as $\omega''(\hat p) = -\L''(\hat p) \neq 0$ by Lemma~\ref{lm-airy-L}). Hence, expression~\eqref{eq-airy-fns} for $\theta(v)$ is real analytic for each $v \neq \hat v$, since $\L(\hat p) = v\hat p - \omega(\hat p) \neq 0$ by Lemma~\ref{lm-airy-L}. Let us check its real analyticity at $v = \hat v$. Since $\omega'(p)$ is even, there exists a real analytic function $f_1$ such that
    \begin{align*}
        \omega'(p) = \omega'(0) - f_1(p^2), \qquad\text{where}\quad f_1'(0) = -\frac{\omega'''(0)}{2} > 0 \quad\text{by Lemma~\ref{lm-airy-w}}.
    \end{align*}
    Using the inverse function theorem for sufficiently small $\nu := \hat v - v$, we consequently compute
    \begin{align}
        \nonumber
        \hat p(v)^2 &= f_1^{-1}(\nu) = \nu \left(\frac{-2}{\omega'''(0)} + \nu f_2(\nu)\right), &
        \hat p(v) &= \sqrt \nu \left(\sqrt{\frac{-2}{\omega'''(0)}} + \nu f_3(\nu)\right), \\
        \L(\hat p) &= v \hat p(v) - \omega(\hat p(v)) = \nu^{3/2} \left(-\frac{2}{3} \sqrt{\frac{-2}{\omega'''(0)}} + \nu f_4(\nu)\right), &
        \label{eq-airy-theta-series}
        \theta(v) &= \nu \left(\sqrt[3]{\frac{-2}{\omega'''(0)}} + \nu f_5(\nu)\right)
    \end{align}
    for some real analytic $f_2(\nu), \dots, f_5(\nu)$, because $\hat p(v) \in \partial Q_1$, $F(\sqrt \theta) = \L(\hat p)$, and $\theta(v) \in \mathbb{R}$. This shows that $\theta(v)$ is real analytic.

    Let us show that $z_v(p)$ is a real analytic family. For $p \in D' \cap Q_1$, the function $v \mapsto z_v(p) = F_{\theta(v)}^{-1}(\L_v(p))$ is real analytic because $\theta(v)$ is real analytic, $\L_v|_{D'\cap Q_1}$ and $F_\theta|_{Q_1}$ are analytic and injective, and $(F_\theta|_{Q_1})^{-1}$ forms an analytic family by a version of the inverse function theorem (see Lemma~\ref{lm-inverse-param} in Appendix~\ref{sec-complex-analysis}). Analogously, for $p \in D' \cap \partial Q_1 \setminus \{\hat p\}$, we have $F'(z(p)) \neq 0$, hence $F$ is injective in some family of domains $B \ni z(p)$ (forming a small neighbourhood of $(z(p), v) \in \mathbb{C} \times V$). Then $(F|_B)^{-1} \circ \L$ is analytic in $v$ and coincides with $z_v$ at $p$. For $p \in D' \setminus \{\pm \hat p\}$, the function $v \mapsto z_v(p)$ is analytic by symmetry. Finally, it is analytic at $p = \pm \hat p$ by a version of the removable singularity theorem (see Lemma~\ref{lm-rm-singularity-param} in Appendix~\ref{sec-complex-analysis}).

    It remains to verify $z'(\hat p) = 1 / h(v)$. Differentiating the identity $\L(p) = F(z(p))$, we get
    \begin{align}
        \label{airy-sub-der1}
        \L'(p) &= F'(z(p)) \, z'(p), \\
        \label{airy-sub-der2}
        \L''(p) &= F''(z(p)) \, z'(p)^2 + F'(z(p)) \, z''(p), \\
        \label{airy-sub-der3}
        \L'''(p) &= F'''(z(p)) \, z'(p)^3 + 3F''(z(p)) \, z'(p) \, z''(p) + F'(z(p)) \, z'''(p).
    \end{align}
    If $v \neq \hat v$, then Eq.~\eqref{airy-sub-der2} implies
    \begin{align*}
        \L''(\hat p) &= F''(\sqrt \theta) \, z'(\hat p)^2 + F'(\sqrt \theta) \, z''(\hat p) = 2\sqrt \theta z'(\hat p)^2, \\
        z'(\hat p) &= \sqrt{\frac{\L''(\hat p)}{2\sqrt{\theta}}} = \frac{|\omega''(\hat p)|^{1/2}}{|4\theta|^{1/4}} = 1 / h(v),
    \end{align*}
    where we choose the positive branch of the square root because $z' > 0$ on the coordinate axes by the principle of orientation. If $v = \hat v$, then from \eqref{airy-sub-der3} it follows that
    \begin{align*}
        \L'''(0) &= F'''(0) \, z'(0)^3 + 3F''(0) \, z'(0) \, z''(0) + F'(0) \, z'''(0) = 2z'(0)^3, \\
        z'(0) &= \sqrt[3]{\frac{\L'''(0)}{2}} = (-\omega'''(0) / 2)^{1/3} = 1 / h(v);
    \end{align*}
    the branch is correct because $z'(0) \in \mathbb{R}$.
\end{proof}

\begin{remark}
    \label{rm-airy-analyticity}
    From Proposition~\ref{prop-airy-substitution}, it follows that the functions $\theta(v)$, $\hat p(v)^2$, $a(\hat p(v))$, and $h(v)$ are real analytic. Indeed, $\theta(v)$ is real analytic by the proposition, $\hat p(v)^2 = z_v^{-1}\left(\sqrt{\theta(v)}\right)^2$ is real analytic because $z_v^{-1}$ is odd, $a(\hat p(v))$ is real analytic because $a(p)$ is even, and $h(v) = 1 / z'(\hat p(v))$ is real analytic because $z'(p)$ is even.
\end{remark}

\begin{remark}
    \label{rm-airy-theta}
    We have $\theta'(v) < 0$ for all $v \in V$ and $\theta'(\hat v) = -h(\hat v)$. Indeed, for $v < \hat v$ we have $\L(\hat p) < 0$ by Lemma~\ref{lm-airy-L}, hence $\theta(v) = \left(-\frac{3}{2} \L(\hat p)\right)^{2/3}$ and
    \begin{align*}
        \theta'(v) = -\left(-\frac{3}{2} \L(\hat p)\right)^{-1/3} \left(\frac{\partial \L}{\partial v} (\hat p) + \frac{\partial \L}{\partial p} (\hat p) \cdot \hat p'\right) = -\frac{\hat p}{\sqrt{\theta(v)}} < 0.
    \end{align*}
    This remains true for $v > \hat v$, when $\theta(v) = -\left(-i\frac{3}{2} \L(\hat p)\right)^{2/3}$ by Lemma~\ref{lm-airy-L}. For $v = \hat v$, we have $\theta'(v) = -\sqrt[3]{-2 / \omega'''(0)} = -h(\hat v) < 0$ by \eqref{eq-airy-theta-series} and Lemma~\ref{lm-airy-w}.
\end{remark}

\begin{lemma}[Contour shift]
    \label{lm-airy-contour}
    For $v > \hat v$, there exists a contour $\gamma_v \subset D'$, passing through $\hat p$, starting and ending at $\pm \pi + i\Delta_v$ for some $\Delta_v > 0$, and such that $z(\gamma) = [-\sqrt[3]{3v\pi} + \sqrt \theta; \sqrt[3]{3v\pi} + \sqrt \theta]$ (see Fig.~\ref{fig-airy-contour}).
\end{lemma}

\begin{proof}
    Set $Q := \sqrt[3]{3v\pi}$ and
    \begin{align*}
        \J(q) := F(q + \sqrt \theta) = \frac{(q + \sqrt \theta)^3}{3} - \theta \cdot (q + \sqrt \theta) = \frac{q^3}{3} + \sqrt \theta q^2 - \frac{2}{3} \theta^{3/2}.
    \end{align*}
    Since $\theta < 0$, it follows that for each $q \in (0; Q)$, we have
    \begin{align*}
        \Re \J(q) &= q^3 / 3 \in (0; Q^3 / 3) = (0; v\pi), &
        &\J(Q) = Q^3 / 3 + \sqrt \theta Q^2 - \tfrac{2}{3} \theta^{3/2} \in (v\pi; v\pi + i\infty) = \L((\pi; \pi + i\infty)),\\
        \Im \J(q) &= |\theta|^{1/2} (q^2 - \tfrac{2}{3} \theta) > 0, &
        &\J((0; Q]) \subset \Pi \cup \L((\pi; \pi + i\infty)) \subset \L(D' \cap Q_1),
    \end{align*}
    where the last inclusion holds by Proposition~\ref{prop-airy-substitution}. Applying $(F|_{Q_1})^{-1}$, we get $(\sqrt \theta; Q + \sqrt \theta] \subset z(D')$ and $Q + \sqrt \theta = z(\pi + i\Delta)$ for some $\Delta > 0$. By symmetry, $[-Q + \sqrt \theta; Q + \sqrt \theta] \subset z(D')$ and $\pm Q + \sqrt \theta = z(\pm \pi + i\Delta)$. The desired contour is then $\gamma := z^{-1}([-Q + \sqrt \theta; Q + \sqrt \theta])$. It passes through $\hat p$ because $z^{-1}(\sqrt \theta) = \hat p$ by Proposition~\ref{prop-airy-substitution}.
\end{proof}

\subsection{Proof of the general Airy phenomenon (Theorem~\ref{th-airy})}
\label{sec-airy-proof}

We are ready to apply the change of variable and complete the proof of the general Airy phenomenon by a standard argument; cf.~\cite[Chapter VII Section 4]{Wong}. We use a few simple lemmas from Appendices~\ref{sec-complex-analysis}--\ref{sec-asymptotic-estimates}.

\begin{proof}[Proof of Theorem~\ref{th-airy}]
    Let $v := x / t$. Take $D'$, $\theta(v)$, and $z_v(p)$ from Proposition~\ref{prop-airy-substitution}. Let $p(z)$ be the inverse function of $z(p)$, defined on $z(D')$. We are going to prove a slightly modified asymptotic formula:
    \begin{align}
        \label{eq-airy-I}
        \psi(x, t) := \frac{1}{2\pi} \int_{-\pi}^{\pi} a(p) \, e^{it\L(p)} \, dp =
        \begin{cases}
            t^{-1/3} \Ai\left(-\theta t^{2/3}\right) h(v) \, a(\hat p) + O_{v^+, a, \omega}\left(t^{-3/2}\right), & \text{if $\theta \ge - t^{-2/3}$}, \\
            t^{-1/3} \Ai\left(-\theta t^{2/3}\right) h(v) \left(a(\hat p) + O_{v^+, a, \omega}\left(t^{-1}\right)\right), & \text{if $\theta < -t^{-2/3}$}.
        \end{cases}
    \end{align}
    The theorem follows from \eqref{eq-airy-I} because $\sgn(\theta) = \sgn(\hat v - v)$ by \eqref{eq-airy-fns}, and because for $\theta \in [-t^{-2/3}; 0]$ the second case of the theorem follows from the first case of \eqref{eq-airy-I} since $\min_{\lambda \in [0; 1]} \Ai(\lambda) > 0$, $\min_{v \in [0; v^+]} h(v) > 0$.

    Let us prove the first case of \eqref{eq-airy-I}. We will actually do it for arbitrary $v$, without the restriction $\theta(v) \ge -t^{-2/3}$ (but the resulting error bound is too weak for $\theta < -t^{-2/3}$). Substituting $z = z_v(p)$, we get
    \begin{align*}
        \psi(x, t) = \frac{1}{2\pi} \int_{-\pi}^{\pi} a(p) \, e^{it\L(p)} \, dp = \frac{1}{2\pi} \int_{-z(\pi)}^{z(\pi)} G(z) \, e^{itF(z)} \, dz, \quad\text{where}\quad G(z) = G_v(z) := a(p(z)) \, p'(z).
    \end{align*}
    By the inverse function theorem with a parameter (Lemma~\ref{lm-inverse-param}), $p_v(z)$ is an odd real analytic family, hence $G_v(z)$ is an even real analytic family. Thus, by the equivariant Hadamard lemma (Lemma~\ref{lm-hadamard-equiv}) there exists an even real analytic family of functions $\widetilde G_v(z)$ such that $G_v(z) = G_v(\sqrt{\theta}) + (z^2 - \theta) \, \widetilde G_v(z)$. Substituting this into the integral above, we get
    \begin{align}
        \label{eq-airy-case1-I}
        \psi(x, t) = \frac{1}{2\pi} \int_{-z(\pi)}^{z(\pi)} \left(G(\sqrt{\theta}) + (z^2 - \theta) \, \widetilde G(z)\right) e^{itF(z)} \, dz = \frac{1}{2\pi} \left(G(\sqrt{\theta}) \, I_1 + I_2\right),
    \end{align}
    where
    \begin{align*}
        I_1 &:= \int_{-z(\pi)}^{z(\pi)} e^{itF(z)} \, dz, & I_2 &:= \int_{-z(\pi)}^{z(\pi)} \widetilde G(z)(z^2 - \theta) \, e^{itF(z)} \, dz.
    \end{align*}

    Let us compute $I_1$. Since $F(z(\pi)) = \L(\pi) = v\pi$ is an integer multiple of $\pi / t$, one can estimate the tail using Lemma~\ref{lm-sincos-tail} and substitute $k = t^{1/3} z$ to express $I_1$ in terms of the Airy function as follows:
    \begin{multline}
        \label{eq-airy-case1-I1}
        I_1 = \int_{-z(\pi)}^{z(\pi)} e^{itF(z)} \, dz = \int_{-\infty}^{+\infty} e^{itF(z)} \, dz - 2\int_{z(\pi)}^{+\infty} \cos(tF(z)) \, dz = \\
        = \frac{2}{t^{1/3}} \int_{0}^{+\infty} \cos\left(\frac{k^3}{3} - \theta t^{2/3} k\right) dk + O_{v^+, \omega}\left(\frac{1}{t^2}\right) = \frac{2\pi}{t^{1/3}} \Ai\left(-\theta t^{2/3}\right) + O_{v^+, \omega}\left(\frac{1}{t^2}\right).
    \end{multline}
    Here we used the following lower bound:
    \begin{align}
        \label{eq-airy-theta-lb}
        |z(\pi) - \sqrt{\theta}| = |z(\pi) - z(\hat p)| = \Omega_{v^+, \omega}(1),
    \end{align}
    which follows from Lemma~\ref{lm-airy-L}, the injectivity of $z(p)$, and a compactness argument.

    Next, let us bound $I_2$. Integrating by parts, we get
    \begin{align*}
        I_2 = \frac{1}{it} \int_{-z(\pi)}^{z(\pi)} \widetilde G(z) \, de^{itF(z)} = \frac{\widetilde G(z) \, e^{itF(z)}}{it}\bigg\rvert_{-z(\pi)}^{z(\pi)} - \frac{1}{it} \int_{-z(\pi)}^{z(\pi)} \widetilde G'(z) \, e^{itF(z)} \, dz.
    \end{align*}
    Since $\widetilde G(z)$ is even and the value $F(z(\pi)) = \L(\pi) = v\pi$ is an integer multiple of $\pi / t$, the first term vanishes. Since $\widetilde G'(z)$ is an odd real analytic family of functions, by the equivariant Hadamard lemma (Lemma~\ref{lm-hadamard-equiv}) there exists an odd real analytic family of functions $\widetilde H_v(z)$ and a real analytic function $c = c(v)$, such that $\widetilde G'_v(z) = c(v) \, z + (z^2 - \theta) \, \widetilde H_v(z)$. Thus,
    \begin{align}
        \label{eq-airy-case1-I2}
        I_2 = -\frac{1}{it} \int_{-z(\pi)}^{z(\pi)} \left(c(v) \, z + (z^2 - \theta) \, \widetilde H_v(z)\right) e^{itF(z)} \, dz = -\frac{1}{it} (c \cdot I_3 + I_4),
    \end{align}
    where
    \begin{align*}
        I_3 &:= \int_{-z(\pi)}^{z(\pi)} e^{itF(z)} \, z \, dz, & I_4 &:= \int_{-z(\pi)}^{z(\pi)} \widetilde H(z)(z^2 - \theta) \, e^{itF(z)} \, dz.
    \end{align*}
    Let us bound $I_3$. Estimating the tail using Lemma~\ref{lm-sincos-tail} and substituting $k = t^{1/3} z$, we express $I_3$ in terms of the Airy function derivative as follows:
    \begin{multline}
        \label{eq-airy-case1-I3}
        I_3 = \int_{-z(\pi)}^{z(\pi)} e^{itF(z)} \, z \, dz = \int_{-\infty}^{+\infty} e^{itF(z)} \, z \, dz - 2i\int_{z(\pi)}^{+\infty} \sin(tF(z)) \, z \, dz = \\
        = \frac{2i}{t^{2/3}} \int_{0}^{+\infty} \sin\left(\frac{k^3}{3} - \theta t^{2/3} k\right) k \, dk + O_{v^+, \omega}\left(\frac{1}{t}\right) = -\frac{2\pi i}{t^{2/3}} \Ai'\left(-\theta t^{2/3}\right) + O_{v^+, \omega}\left(\frac{1}{t}\right) = O_{v^+, \omega}\left(\frac{1}{\sqrt{t}}\right),
    \end{multline}
    where we used lower bound \eqref{eq-airy-theta-lb} and the estimate implied by \cite[Eq.~9.7.6,~9.7.10]{NIST-10}:
    \begin{align*}
        \Ai'(u) = O\left(1 + |u|^{1/4}\right) \quad\text{for each $u \in \mathbb{R}$}.
    \end{align*}
    To bound $I_4$, integrate by parts:
    \begin{align}
        \label{eq-airy-case1-I4}
        I_4 = \frac{1}{it} \int_{-z(\pi)}^{z(\pi)} \widetilde H(z) \, de^{itF(z)} = \frac{1}{it} \left(\widetilde H(z) \, e^{itF(z)}\bigg\rvert_{-z(\pi)}^{z(\pi)} - \int_{-z(\pi)}^{z(\pi)} \widetilde H'(z) \, e^{itF(z)} \, dz\right) =  O_{v^+, a, \omega}\left(\frac{1}{t}\right),
    \end{align}
    since the functions $z_v(\pi)$, $\widetilde H_v(z)$, and $\widetilde H'_v(z)$ are uniformly bounded for $z \in [-z_v(\pi); z_v(\pi)]$ by compactness.

    By combining estimates \eqref{eq-airy-case1-I}, \eqref{eq-airy-case1-I1}, \eqref{eq-airy-case1-I2}, \eqref{eq-airy-case1-I3}, \eqref{eq-airy-case1-I4} and simplifying the expressions using Proposition~\ref{prop-airy-substitution}, the first case of asymptotic formula \eqref{eq-airy-I} follows.

    Let us prove the second case of \eqref{eq-airy-I}. We intend to shift the integration contour so that it passes through the critical point, and moreover becomes straight after the change of variable. We choose the contour $\gamma \subset D'$ using Lemma~\ref{lm-airy-contour} so that it starts at $-\pi + i\Delta$ and ends at $\pi + i\Delta$ for some $\Delta > 0$, and $z(\gamma) = [-Q + \sqrt \theta; Q + \sqrt \theta]$, where
    \begin{align}
        \label{eq-airy-Q-lb}
        Q := \sqrt[3]{3v\pi} \ge \sqrt[3]{3\hat v\pi} = \Omega_{\omega}(1).
    \end{align}
    The latter estimate holds because $\hat v > 0$ by Lemma~\ref{lm-airy-w}. Since $D$ is simply-connected and $\partial U \subset D$, we have
    \begin{align*}
        \psi(x, t) = \frac{1}{2\pi} \left(\int_{-\pi}^{-\pi + i\Delta} a(p) \, e^{it\L(p)} \, dp + \int_{\gamma} a(p) \, e^{it\L(p)} \, dp + \int_{\pi + i\Delta}^{\pi} a(p) \, e^{it\L(p)} \, dp\right).
    \end{align*}
    Since $\omega(p)$ is $2\pi$-periodic, the value $\L(\pi + iy) - \L(-\pi + iy)$ is an integer multiple of $2\pi / t$ for each $y \in [0; \Delta]$. Since $a(p)$ is $2\pi$-periodic, $a(\pi + iy) = a(-\pi + iy)$ for each $y \in [0; \Delta]$. Thus, the first and third terms cancel out, and substituting $z = z_v(p)$, we get
    \begin{align*}
        \psi(x, t) = \frac{1}{2\pi} \int_{\gamma} a(p) \, e^{it\L(p)} \, dp = \frac{1}{2\pi} \int_{-Q + \sqrt{\theta}}^{Q + \sqrt{\theta}} G(z) \, e^{itF(z)} \, dz, \quad\text{where}\quad G(z) = G_v(z) := a(p(z)) \, p'(z).
    \end{align*}
    Since $G_v(z)$ is an even real analytic family of functions, by the equivariant Hadamard lemma (Lemma~\ref{lm-hadamard-equiv}) there exists an even real analytic family of functions $\widetilde G_v(z)$ such that $G_v(z) = G_v(\sqrt{\theta}) + (z^2 - \theta) \, \widetilde G_v(z)$. Substituting this into the integral above, we get
    \begin{align}
        \label{eq-airy-case2-I}
        \psi(x, t) = \frac{1}{2\pi} \int_{-Q + \sqrt{\theta}}^{Q + \sqrt{\theta}} \left(G(\sqrt{\theta}) + (z^2 - \theta) \, \widetilde G(z)\right) e^{itF(z)} \, dz = \frac{1}{2\pi} \left(G(\sqrt{\theta}) \, I_1 + I_2\right),
    \end{align}
    where
    \begin{align*}
        I_1 &:= \int_{-Q + \sqrt{\theta}}^{Q + \sqrt{\theta}} e^{itF(z)} \, dz, & I_2 &:= \int_{-Q + \sqrt{\theta}}^{Q + \sqrt{\theta}} \widetilde G(z)(z^2 - \theta) \, e^{itF(z)} \, dz.
    \end{align*}

    Let us compute $I_1$. Estimating the tail using Lemma~\ref{lm-exp-tail} and substituting $k = t^{1/3} z$, we express $I_1$ in terms of the Airy function as follows:
    \begin{multline}
        \label{eq-airy-case2-I1}
        I_1 = \int_{-Q + \sqrt{\theta}}^{Q + \sqrt{\theta}} e^{itF(z)} \, dz = \int_{-\infty + \sqrt{\theta}}^{+\infty + \sqrt{\theta}} e^{itF(z)} \, dz - 2\Re \int_{Q + \sqrt{\theta}}^{+\infty + \sqrt{\theta}} e^{itF(z)} \, dz = \\
        = \frac{2}{t^{1/3}} \int_{0}^{+\infty} \cos\left(\frac{k^3}{3} - \theta t^{2/3} k\right) dk + O\left(\frac{e^{-2/3 \, t|\theta|^{3/2}}}{Q^3 t^{3/2}|\theta|^{1/4}}\right)
        = \frac{2\pi}{t^{1/3}} \Ai\left(-\theta t^{2/3}\right) + O_{\omega}\left(\frac{e^{-2/3 \, t|\theta|^{3/2}}}{t^{3/2}|\theta|^{1/4}}\right),
    \end{multline}
    where we used lower bound \eqref{eq-airy-Q-lb} and the identity
    \begin{align*}
        \int_{-\infty + \sqrt{\theta}}^{+\infty + \sqrt{\theta}} e^{itF(z)} \, dz = \int_{-\infty}^{+\infty} e^{itF(z)} \, dz,
    \end{align*}
    which follows from Lemma~\ref{lm-int-shift} via the substitution $\tilde z = t^{1/3} z$.

    To bound $I_2$, integrate by parts and substitute $q = z - \sqrt{\theta}$:
    \begin{multline}
        \label{eq-airy-case2-I2}
        I_2 = \frac{1}{it} \int_{-Q + \sqrt{\theta}}^{Q + \sqrt{\theta}} \widetilde G(z) \, de^{itF(z)} = \frac{1}{it} \left(\widetilde G(z) \, e^{itF(z)}\bigg\rvert_{-Q + \sqrt{\theta}}^{Q + \sqrt{\theta}}
        - \int_{-Q + \sqrt{\theta}}^{Q + \sqrt{\theta}} \widetilde G'(z) \, e^{itF(z)} \, dz\right) = \\
        = \frac{e^{-2/3 \, t|\theta|^{3/2}}}{it} \left(\widetilde G(q + \sqrt{\theta}) \, e^{itq^3/3-t|\theta|^{1/2}q^2}\bigg\rvert_{-Q}^{Q} - \int_{-Q}^{Q} \widetilde G'(q + \sqrt{\theta}) \, e^{itq^3/3-t|\theta|^{1/2}q^2} \, dq\right) = \\
        = \frac{e^{-2/3 \, t|\theta|^{3/2}}}{it} \, O_{v^+, a, \omega}\left(e^{-t|\theta|^{1/2}Q^2} + \int_{-\infty}^{+\infty} e^{-t|\theta|^{1/2}q^2} \, dq\right) = O_{v^+, a, \omega}\left(\frac{e^{-2/3 \, t|\theta|^{3/2}}}{t^{3/2}|\theta|^{1/4}}\right),
    \end{multline}
    where the second to last equality follows from the boundedness of $\theta(v)$, $\widetilde G_v(q + \sqrt{\theta})$, and $\widetilde G'_v(q + \sqrt{\theta})$ for $v \in [\hat v; v^+]$ and $|q| \le Q \le \sqrt[3]{3v^+\pi}$, and the last equality follows from the Gaussian integral, the bound $e^{-u} = O(1 / \sqrt{u})$ for $u > 0$, and lower bound \eqref{eq-airy-Q-lb}.

    By combining estimates \eqref{eq-airy-case2-I}, \eqref{eq-airy-case2-I1}, \eqref{eq-airy-case2-I2}, simplifying the expressions using Proposition~\ref{prop-airy-substitution}, applying the bound $G(\sqrt{\theta}) = a(\hat p) \, p'(\sqrt \theta) = \Omega_{v^+, \omega}(a(\hat p))$, the assumption $\theta < -t^{-2/3}$, and the lower bound for the Airy function \cite[Eq.~9.7.5~and~\S9.7(iii)]{NIST-10}
    \begin{align*}
        \Ai(u) = \Omega\left(\frac{e^{-2/3 \, u^{3/2}}}{u^{1/4}}\right) \quad\text{for each $u \ge 1$},
    \end{align*}
    the second case of the asymptotic formula \eqref{eq-airy-I} follows.
\end{proof}

\begin{proof}[Proof of Corollary~\ref{cor-airy}]
    The cases $v \le \hat v - \delta$ and $v \ge \hat v + \delta$ follow from Theorem~\ref{th-airy} and the known asymptotic formulae for the Airy function \cite[Eq.~9.7.5,~9.7.9, and~\S9.7(iii)]{NIST-10}. The case $|v - \hat v| \le \delta t^{-2/3}$ follows from Theorem~\ref{th-airy}, Remarks~\ref{rm-airy-analyticity} and~\ref{rm-airy-theta}, and the estimates
    \begin{align*}
        a(\hat p(v)) \, h(v) &= a(\hat p(\hat v)) \, h(\hat v) + O_{v^+, a, \omega}\left(|v - \hat v|\right) = a(0) \cdot \sqrt[3]{\tfrac{-2}{\omega'''(0)}} + O_{v^+, a, \omega}\left(\delta t^{-2/3}\right), \\
        \theta(v) &= \theta(\hat v) + \theta'(\hat v) \, (v - \hat v) + O_{v^+, \omega}\left((v - \hat v)^2\right) = -\sqrt[3]{\tfrac{-2}{\omega'''(0)}} \, (v - \hat v) + O_{v^+, \omega}\left(\delta^2 t^{-4/3}\right), \\
        \Ai\left(-\theta(v) \, t^{2/3}\right) &= \Ai\left(\sqrt[3]{\tfrac{-2}{\omega'''(0)}} \, (v - \hat v) \, t^{2/3} + O_{\delta, v^+, \omega}\left(t^{-2/3}\right)\right) = \Ai\left(\sqrt[3]{\tfrac{-2}{t\omega'''(0)}}\left(x - \hat vt\right)\right) + O_{\delta, v^+, \omega}\left(t^{-2/3}\right).
    \end{align*}
\end{proof}

\subsection{Airy phenomenon for Bessel functions}
\label{sec-airy-bessel}

To illustrate, we apply Theorem~\ref{th-airy} to estimate the integral defining the Bessel function~(see \cite[Sec.~2.9]{Olver-97}):
\begin{equation*}
    J_x(t) = \frac{1}{2\pi} \int_{-\pi}^{\pi} e^{ipx - it\sin p} \, dp.
\end{equation*}
In this case, $\omega(p) = \sin p$, $a(p) = 1$, $D = \mathbb{C}$, $V^+ = +\infty$, and $\omega'(p) = \cos p$ is a 1--1 conformal mapping $U\to H^-$. We compute
$$
\hat p(v) =
\begin{cases}
    \arccos v, & \text{if } v \in [0; 1]; \\
    i\arccosh v, & \text{if } v \in [1; +\infty),
\end{cases} \quad
\omega(\hat p(v)) =
\begin{cases}
    \sqrt{1 - v^2}, & \text{if } v \in [0; 1]; \\
    i\sqrt{v^2 - 1}, & \text{if } v \in [1; +\infty),
\end{cases} \quad
\omega''(\hat p(v)) = -\omega(\hat p(v)),
$$
$$
\small
\theta(v) =
\begin{cases}
    \left|\frac{3}{2}(v\arccos v - \sqrt{1 - v^2})\right|^{2/3}, & v \in [0; 1); \\
    0, & v = 1; \\
    -\left|\frac{3}{2}(v\arccosh v - \sqrt{v^2 - 1})\right|^{2/3}, & v \in (1; +\infty),
\end{cases} \qquad
h(v) =
\begin{cases}
    \frac{\sqrt{2}\left|\frac{3}{2}(v\arccos v - \sqrt{1 - v^2})\right|^{1/6}}{(1 - v^2)^{1/4}}, & v \in [0; 1); \\
    2^{1/3}, & v = 1; \\
    \frac{\sqrt{2}\left|\frac{3}{2}(v\arccosh v -\sqrt{v^2 - 1})\right|^{1/6}}{(v^2 - 1)^{1/4}}, & v \in (1; +\infty).
\end{cases}
$$
Thus Theorem~\ref{th-airy} gives the known asymptotic formula (see~\cite[Sec.~11.10]{Olver-97})
\begin{align*}
    J_x(t) =
    \begin{cases}
        t^{-1/3} \Ai\left(-\theta(x / t) \, t^{2/3}\right) h(x / t) + O\left(t^{-3/2}\right), & \text{if } 0 \le x / t \le 1, \\
        t^{-1/3} \Ai\left(-\theta(x / t) \, t^{2/3}\right) h(x / t) \cdot \left(1 + O_{v^+}\left(t^{-1}\right)\right), & \text{if } 1 \le x / t \le v^+.
    \end{cases}
\end{align*}
In the former case, the constant in $O\left(\ldots\right)$ is independent of $v^+$, since we can apply Theorem~\ref{th-airy} for $v^+ = 1$.

\subsection{Airy phenomenon for quantum walks (proof of Theorem \ref{th-airy-checkers})}
\label{sec-airy-checkers}

\begin{proof}[Proof of Theorem~\ref{th-airy-checkers}]
    To apply Theorem~\ref{th-airy}, we write integral representations~\eqref{eq-A1-int-osc}--\eqref{eq-A2-int-osc} in the form
    \begin{align*}
        \widetilde A_1(x, t, m) &= i^{x - t + 1} \, \frac{1}{2\pi} \int_{-\pi}^{\pi} a_1(p) \, e^{ipx - i\omega(p)t} \, dp, & \widetilde A_2(x, t, m) &= i^{x - t} \, \frac{1}{2\pi} \int_{-\pi}^{\pi} a_2(p) \, e^{ipx - i\omega(p)t} \, dp,
    \end{align*}
    where
    \begin{align*}
        \omega(p) &:= \arcsin \frac{\sin p}{\sqrt{1 + m^2}}, & a_1(p) &:= \frac{m}{\sqrt{m^2 + \cos^2 p}}, & a_2(p) &:= 1 + \frac{\cos p}{\sqrt{m^2 + \cos^2 p}}.
    \end{align*}
    The functions $\omega(p)$, $a_1(p)$, and $a_2(p)$ are defined on the symmetric $2\pi$-periodic domain
    \begin{align*}
        D := \mathbb{C} \setminus \bigcup_{n \in \mathbb{Z}} [\pi / 2 + \pi n \pm i\hat y; \pi / 2 + \pi n \pm i\infty),
    \end{align*}
    where $\hat y := \arcsinh m$. Thus $\partial D \cap U$ is a vertical slit $[\pi / 2 + i\hat y; \pi / 2 + i\infty)$. The derivative
    \begin{align*}
        \omega'(p) = \frac{\cos p}{\sqrt{m^2 + \cos^2 p}}
    \end{align*}
    is a 1--1 conformal mapping of $D \cap U$ to the lower half-plane as the composition of mappings (cf. Fig.~\ref{fig-airy-w'-inner})
    $$
    D \cap U \xrightarrow{p \mapsto \cos p} H^- \setminus [-im; -i\infty) \xrightarrow{z \mapsto z / \sqrt{m^2 + z^2}} H^-.
    $$
    In particular, $\omega'(p)$ is bounded in the complement of any neighbourhood of $\pi / 2 + i \hat y$ in $D \cap U$.

    To compute $V^+$, note that $\Re \omega(p) \to \pi / 2$ as $p \to \pi / 2 + iy$ for each $y \ge \hat y$. Here, the convergence is uniform in $y$ for $\hat y \le y \le 2\hat y$ by compactness. The convergence is uniform for $y \ge 2\hat y$, because
    $$
    |\Re \omega(\pi / 2 + iy + \varepsilon) - \Re \omega(\pi / 2 + iy)| \le \max_{\Im p \ge 2\hat y} |\omega'(p)| \cdot \varepsilon \quad\text{for } \varepsilon > 0.
    $$
    We obtain
    \begin{align*}
        V^+ = \liminf_{p \to \partial D \cap U} \frac{\Re \omega(p)}{\Re p} = \frac{\pi / 2}{\pi / 2} = 1 \qquad\quad\text{and put }\quad v^+ :=
        \begin{cases}
            \hat v,     & \text{if } 0 \le v < \hat v, \\
            1 - \delta, & \text{if } \hat v < v \le 1 - \delta.
        \end{cases}
    \end{align*}
    We compute the remaining quantities in \eqref{eq-airy-fns} for $v \neq \hat v, 0$:
    \begin{gather}
        \nonumber
        \hat p(v) =
        \begin{cases}
            \arccos\frac{mv}{\sqrt{1 - v^2}} = \arctan\frac{\sqrt{1 - (1 + m^2)v^2}}{mv}, & \text{if } v < \hat v, \\
            i\arccosh\frac{mv}{\sqrt{1 - v^2}} = i\arctanh\frac{\sqrt{ (1 + m^2)v^2 - 1}}{mv}, & \text{if } v > \hat v;
        \end{cases}
        \qquad\qquad
        \begin{aligned}
            a_1(\hat p(v)) &= \sqrt{1 - v^2} \ge\sqrt{\delta}, \\
            a_2(\hat p(v)) &= 1 + v \ge 1,
        \end{aligned}
        \\
        \label{eq-FC-computations}
        |\omega''(\hat p(v))| = \tfrac{1}{m} \left|1 - (1 + m^2)v^2\right|^{1/2} \, (1 - v^2), \qquad\qquad
        h(v) = \frac{\sqrt{m}\,|4\theta(v)|^{1/4}}{|1 - (1 + m^2)v^2|^{1/4} \, (1 - v^2)^{1/2}},
    \end{gather}
    and reproduce expression~\eqref{eq-theta-def} for $\theta(v) = \theta(v, m)$ (the expressions in parentheses in~\eqref{eq-theta-def} are nonnegative since their derivatives have the same sign as $v-\hat v$). By Theorem~\ref{th-airy} for $a(p) = a_k(p)$, where $k = 1, 2$, we get
    \begin{align*}
        \widetilde A_k(x, t, m) &= i^{x-t+2-k} \, t^{-1/3} \, a_k(\hat p(v)) \, h(v) \, \Ai\left(-\theta(v) \, t^{2/3}\right)
        \begin{cases}
            +\,O_m\left(t^{-3/2}\right), & \text{if } v < \hat v, \\
            \times \left(1 + O_{m, \delta}\left(t^{-1}\right)\right), & \text{if } v > \hat v.
        \end{cases}
    \end{align*}
    In the case $v < \hat v$, the parameter $v^+$, hence the constant in $O_m\left(\ldots\right)$, does not depend on $\delta$. In the case $v > \hat v$, we have used that $a_k(\hat p(v))$ is bounded away from zero, so $a_k(\hat p(v)) + O_{m, \delta}\left(t^{-1}\right) = a_k(\hat p(v))\left(1 + O_{m, \delta}\left(t^{-1}\right)\right)$. For $v = \hat v, 0$, the theorem follows by the continuity of the main term in Theorem~\ref{th-airy}; see Remarks~\ref{rm-airy-analyticity}--\ref{rm-airy-theta}.
\end{proof}

We conclude this section with a corollary, which is a uniform version of \cite[Propositions~2.1, 3.1, and~4.1]{Sunada-Tate-12} with an improved error bound near $\hat v$ (see row 4 of Tab.~\ref{tab-checkers-asymp}).

\begin{corollary}
    \label{cor-airy-checkers}
    The first five rows of Tab.~\ref{tab-checkers-asymp} provide correct asymptotic formulae for $\widetilde A_1(x, t, m)$ for each $\delta, m > 0$ and integer $x, t \ge 0$ satisfying the corresponding inequality in the first column.
\end{corollary}

\begin{proof}
    Row 3 is Theorem~\ref{th-airy-checkers}. The rest follows from Corollary~\ref{cor-airy}, Eq.~\eqref{eq-FC-computations}, and $\omega'''(0) = -m^2 / (1 + m^2)^{3/2}$.
\end{proof}

\section{Gamma phenomenon}
\label{sec-gamma}

\subsection{General gamma phenomenon}
\label{sec-gamma-general}

Now we turn to estimates of exponentially-damped integrals via conformal mappings (see e.g. Fig.~\ref{fig-gamma-p'-inner}). The general theorem below and its proof showcase a remarkable resemblance to those in the Airy regime.

\begin{theorem}
    \label{th-gamma}
    Let $D \subset \mathbb{C}$ be a simply-connected domain symmetric about the real axis, with a bounded projection on the imaginary axis, and containing $(-\infty; 0]$. Let $a(\omega)$ and $q(\omega)$ be real analytic functions in $D$, and let $p(\omega) := \log \omega + q(\omega)$. Assume $p'$ is a bijective conformal mapping of $D \cap H^+$ to $H^-$, and also there exist finite limits $a(-\infty)$, $a'(-\infty)$, and $p''(-\infty)$. Then
    \begin{align}
        \label{eq-gamma-asymp}
        \frac{1}{2\pi i} \int_{-\infty}^{(0+)} a(\omega) \, e^{\omega t- p(\omega)x} \, d\omega = \frac{(x / e)^x}{|x| \, \Gamma(x)} \cdot \frac{e^{\hat \omega(x / t)t - p(\hat \omega(x / t))x}}{\sqrt{-p''(\hat \omega(x / t))}} \cdot \left(a(\hat \omega(x / t)) + O_{v^-, v^+, a, p}\left(t^{-1}\right)\right)
    \end{align}
    for each $x, t, v^-, v^+ \in \mathbb{R}$ with $t > 0$, $x \neq 0$, and $V^- < v^- \le x / t \le v^+ < V^+$, where
    \begin{equation}
        \label{eq-gamma-fns}
        \begin{aligned}
            \hat \omega(v) &:= (p')^{-1}(1 / v), &
            \Lambda^+ &:= \limsup_{\omega \to \partial D, \, \Im \omega \neq 0} \frac{\Im p(\omega)}{\Im \omega}, \\
            V^- &:=
            \begin{cases}
                1 / p'(-\infty), & \text{if $p'(-\infty) < 0$}, \\
                -\infty, & \text{if $p'(-\infty) \ge 0$},
            \end{cases} &
            V^+ &:=
            \begin{cases}
                1 / \Lambda^+, & \text{if $\Lambda^+ > 0$}, \\
                +\infty, & \text{if $\Lambda^+ \le 0$}.
            \end{cases}
        \end{aligned}
    \end{equation}
\end{theorem}

Here, $\hat \omega(v)$ is a critical point of the phase-space Lagrangian $\L(\omega) := vp(\omega) - \omega$ for $v \neq 0$, which is negative for $v < 0$ and positive for $v > 0$ by Lemma~\ref{lm-gamma-L} below. Hereafter, by a slight abuse of terminology, any point of $(-\infty; 0)$ where the analytic extension of $\L'$ vanishes is called \emph{critical}, although $\L$ itself is discontinuous on $(-\infty; 0)$. The expressions for $\hat \omega(v)$ and $V^-$ in \eqref{eq-gamma-fns} are well-defined by Lemma~\ref{lm-gamma-p}, and the integral in \eqref{eq-gamma-asymp} is well-defined by Lemma~\ref{lm-gamma-int-exist}. By Lemma~\ref{lm-gamma-p}, the limit $p'(-\infty)$ is finite, hence $V^- < 0 \le V^+$. By Remark~\ref{rm-gamma-analyticity}, asymptotic formula~\eqref{eq-gamma-asymp} is also valid for $x = 0$ after continuously extending the main term at $x = 0$ by $0$, which then becomes a real analytic function in $x$.

The assumption $x / t > V^-$ is needed for the convergence of the integral in \eqref{eq-gamma-asymp}. The assumption $x / t < V^+$ means that the Lagrangian has a negative imaginary part in the vicinity of $\partial D$ (in the upper half-plane). Our main results in Sec.~\ref{sec-checkers-results} and computations in Sec.~\ref{sec-gamma-checkers} show that the bound $x / t \le v^+ < V^+$ is sharp; that is, it cannot be relaxed to $x / t < V^+$ and the value of $V^+$ cannot be increased in general.

The first factor on the right side of~\eqref{eq-gamma-asymp} is the ratio of the gamma function and its Stirling approximation (up to a factor of $\sqrt{2\pi |x|}$). The second factor is a typical saddle-point approximation (again, up to $\sqrt{2\pi |x|}$). For large $x$, the first factor becomes $1/\sqrt{2\pi |x|}$, leaving the standard saddle-point approximation only. As a result, the asymptotic formula simplifies, when $v := x / t$ is either close to or far from $0$:

\begin{corollary}
    \label{cor-gamma}
    Under the assumptions of Theorem~\ref{th-gamma}, if $\delta > 0$, $t \ge 1$, and
    $v = \frac{x}{t}$ satisfy $|v| \not\in \left(\frac{\delta}{t}; \delta\right)$, then
    \begin{align*}
        \frac{1}{2\pi i} \int_{-\infty}^{(0+)} a(\omega) \, e^{\omega t- p(\omega)x} \, d\omega =
        \begin{cases}
            \frac{e^{\hat \omega(v)t - p(\hat \omega(v))x}}{i\sqrt{2\pi xp''(\hat \omega(v))}} (e^{2\pi ix} - 1) \left(a(\hat \omega(v)) + O_{\delta, v^-, v^+, a, p}\left(t^{-1}\right)\right), & \text{if } v \le -\delta, \\
            \frac{t^{x-1} e^{-q(0)x}}{\Gamma(x)} (a(0) + O_{\delta, v^-, v^+, a, p}(t^{-1})), & \text{if } |v| \le \delta / t, \\
            \frac{e^{\hat \omega(v)t - p(\hat \omega(v))x}}{\sqrt{-2\pi xp''(\hat \omega(v))}} \left(a(\hat \omega(v)) + O_{\delta, v^-, v^+, a, p}\left(t^{-1}\right)\right), & \text{if } v \ge \delta.
        \end{cases}
    \end{align*}
\end{corollary}

In Sec.~\ref{sec-gamma-beta}--\ref{sec-gamma-checkers}, we illustrate the theorem with examples $p(\omega) = \log(e^\omega - 1)$ and $p(\omega) = \log(e^\omega - 1) - \log(1 + m^2 - e^\omega)$, which arise in the simple random walk and quantum walk, respectively. Further applications include the beta function, hypergeometric functions, L-functions, and residues.

\subsection{Conformal change of variable}
\label{sec-gamma-substitution}

\begin{figure}
    \centering
    \begin{subfigure}[c]{0.5\textwidth}
        \includegraphics[width=\textwidth]{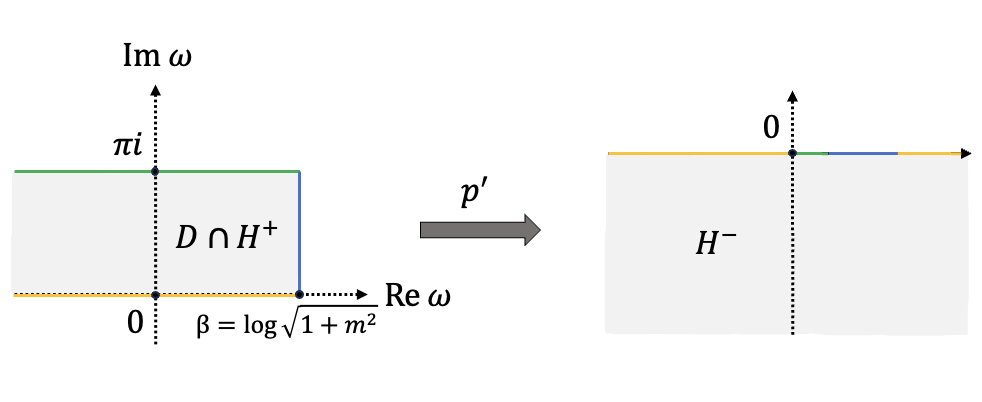}
        \caption{$p'(\omega)$ on $D \cap H^+$}
        \label{fig-gamma-p'-inner}
    \end{subfigure}
    \hspace{0.05\textwidth}
    \begin{subfigure}[c]{0.4\textwidth}
        \includegraphics[width=\textwidth]{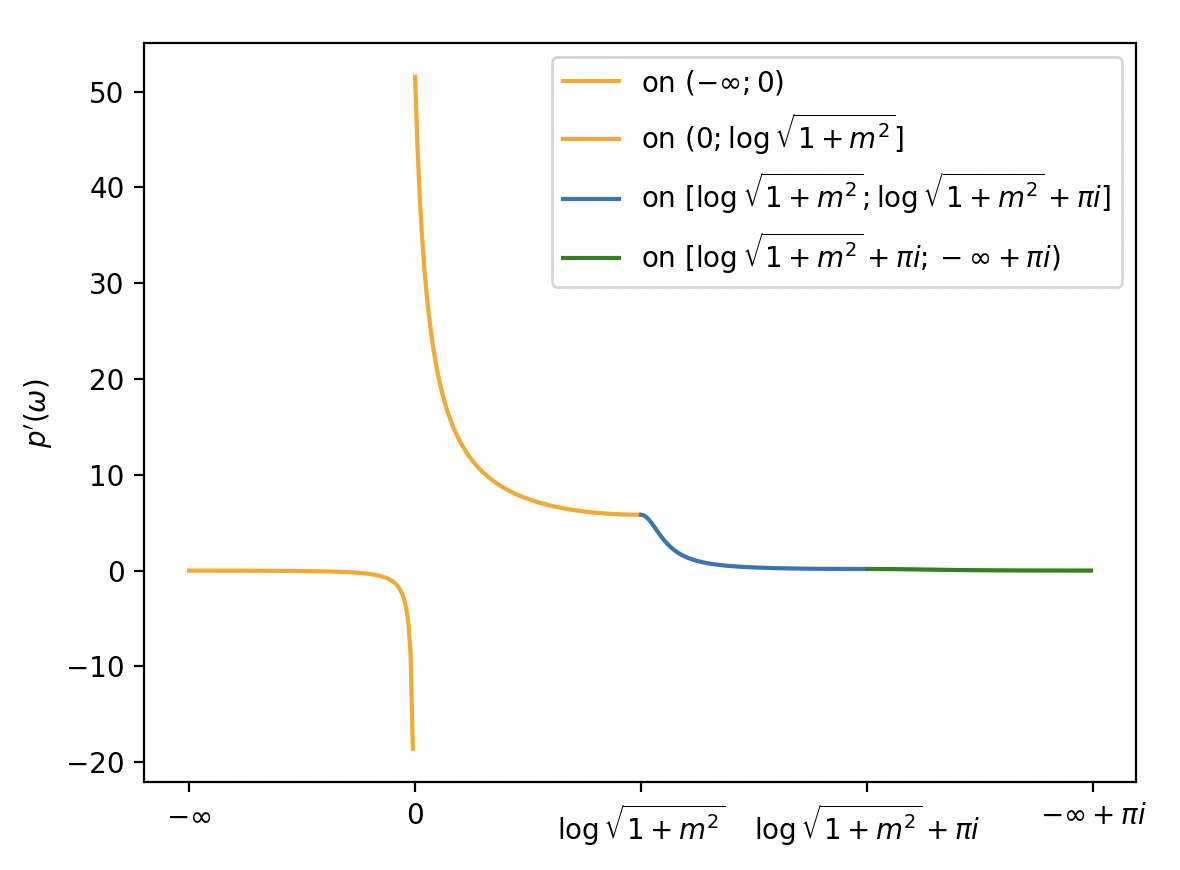}
        \caption{$p'(\omega)$ on $\partial(D \cap H^+)$}
        \label{fig-gamma-p'-boundary}
    \end{subfigure}
    \caption{Illustration of $p'(\omega)$ from Theorem~\ref{th-gamma} applied to quantum walks, where $p(\omega)$ is given by \eqref{eq-p}. Note that $p'(\omega) = \frac{m^2e^\omega}{(e^\omega - 1)(1 + m^2 - e^\omega)}$ is a bijective conformal mapping of $D \cap H^+$ to the lower half-plane, and $p'$ is real and injective on $\partial(D \cap H^+)$. See Lemma~\ref{lm-gamma-p}.}
    \label{fig-gamma-p'}
\end{figure}

\begin{figure}
    \centering
    \begin{subfigure}[c]{0.47\textwidth}
        \includegraphics[width=\textwidth]{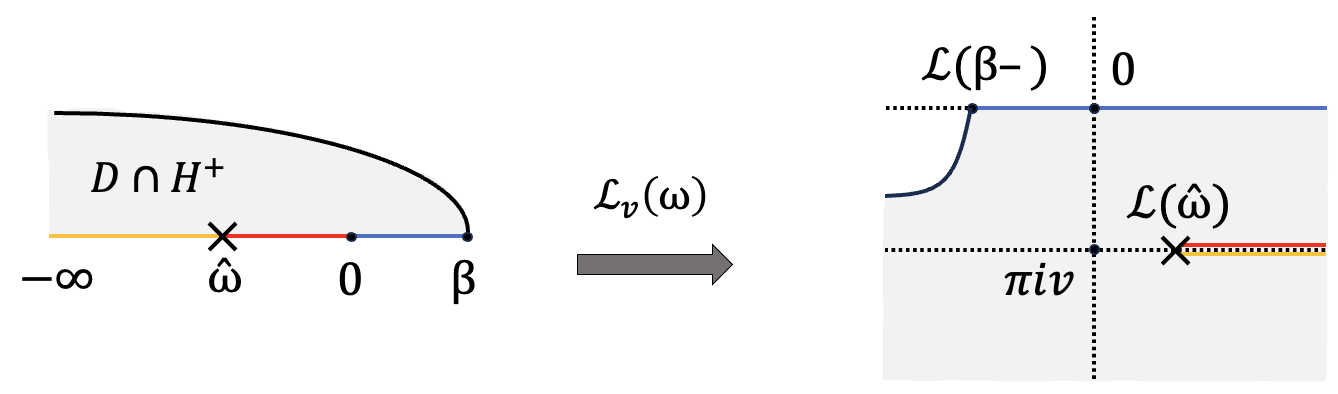}
        \caption{$\L_v(\omega)$ for $v < 0$}
        \label{fig-gamma-L-lt}
    \end{subfigure}
    \hspace{0.02\textwidth}
    \begin{subfigure}[c]{0.47\textwidth}
        \includegraphics[width=\textwidth]{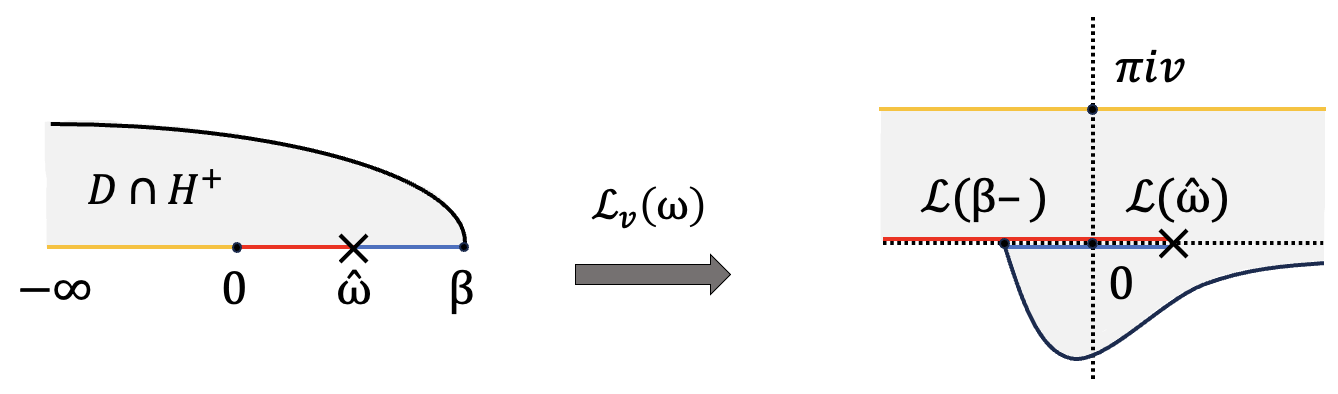}
        \caption{$\L_v(\omega)$ for $v > 0$}
        \label{fig-gamma-L-gt}
    \end{subfigure}
    \\[0.4cm]
    \begin{subfigure}[c]{0.47\textwidth}
        \includegraphics[width=\textwidth]{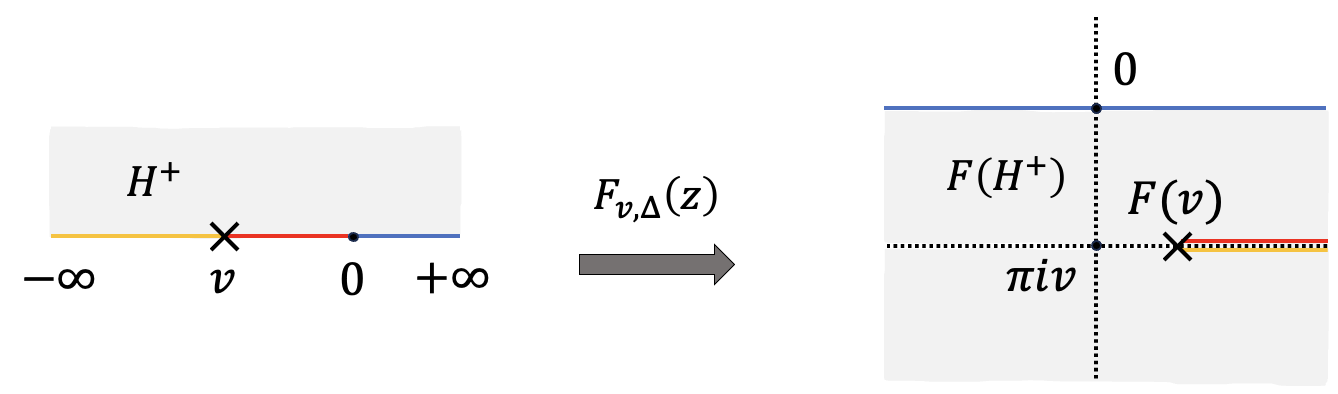}
        \caption{$F_{v, \Delta}(z)$ for $v < 0$}
        \label{fig-gamma-F-lt}
    \end{subfigure}
    \hspace{0.02\textwidth}
    \begin{subfigure}[c]{0.47\textwidth}
        \includegraphics[width=\textwidth]{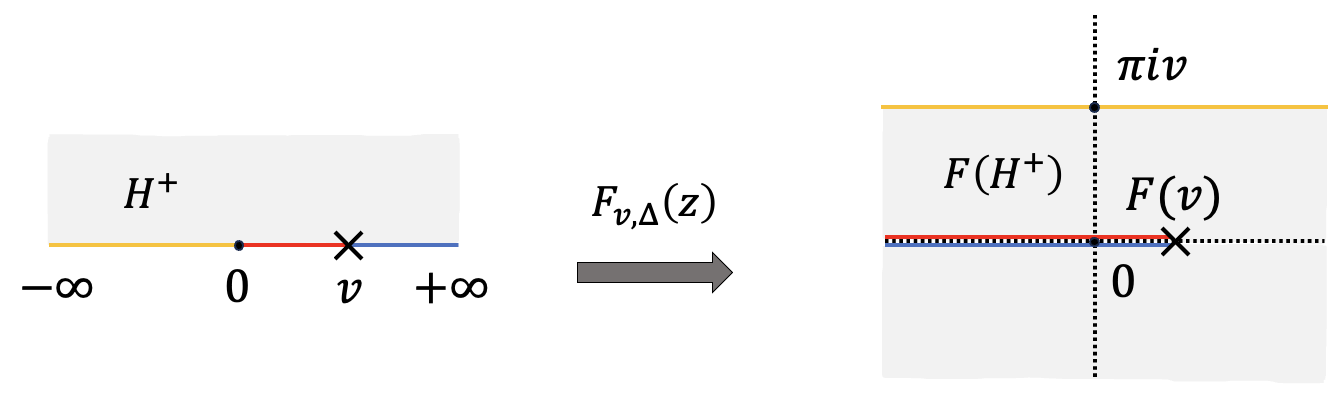}
        \caption{$F_{v, \Delta}(z)$ for $v > 0$}
        \label{fig-gamma-F-gt}
    \end{subfigure}
    \\[0.4cm]
    \caption{The Lagrangian $\L_v(\omega)$ and its normal form $F_{v, \Delta}(z)$ as conformal mappings. See Lemmas~\ref{lm-gamma-L}--\ref{lm-gamma-F}.}
    \label{fig-gamma-L-F}
\end{figure}

In this subsection, which is similar in spirit to Sec.~\ref{sec-airy-substitution}, we start the proof of Theorem~\ref{th-gamma}. We use the notation and impose the assumptions from the theorem, and assume
\begin{align*}
    v &:= x / t \in V, & V &:= [v^-; v^+].
\end{align*}
For each $v \in V$, $\Delta \in \mathbb{R}$, $\omega \in D \setminus \{0\}$, and $z \in \mathbb{C} \setminus \{0\}$, define the \emph{Lagrangian} $\L_v(\omega)$ and its \emph{normal form} $F_{v, \Delta}(z)$ as
\begin{align*}
    \L(\omega) = \L_v(\omega) &:= vp(\omega) - \omega, & F(z) = F_{v, \Delta}(z) &:= v\log z - z + \Delta.
\end{align*}
The assumption $v^+ < V^+$ implies that there is $\varepsilon = \varepsilon(v^+) > 0$ satisfying
\begin{align}
    \label{eq-gamma-eps}
    \Im \L_v(\omega) < 0 \quad\text{for each $0 < v \le v^+$ and $\omega \in D \cap H^+$ with $\dist(\omega, \partial D) < \varepsilon$}.
\end{align}
Our goal is to construct a conformal change of variable $z_v(\omega)$ in a sufficiently large family of domains, such that $\L_v = F_{v, \Delta(v)} \circ z_v$ for suitable $\Delta(v)$; this is Proposition~\ref{prop-gamma-substitution} below. First, Lemma~\ref{lm-gamma-p} below derives useful properties of the derivative~$p'(\omega)$ (see Fig.~\ref{fig-gamma-p'}). Lemmas~\ref{lm-gamma-L}~and~\ref{lm-gamma-F} describe the behaviour of $\L_v(\omega)$ and $F_{v, \Delta}(z)$ as conformal mappings (see Fig.~\ref{fig-gamma-L-F}). Lemma~\ref{lm-gamma-univalence} proves the univalence of the Lagrangian in a certain domain and is crucial for showing the injectivity of $z_v(\omega)$ in Proposition~\ref{prop-gamma-substitution}. Throughout, we denote
$$
(-\infty; \beta) := D \cap \mathbb{R}.
$$

\begin{lemma}[Derivative]
    \label{lm-gamma-p}
    The derivative $p'(\omega)$ is real and decreasing on $(-\infty; 0)$ and $(0; \beta)$, and moreover, it is injective on $(-\infty; 0) \cup (0; \beta)$ (see Fig.~\ref{fig-gamma-p'}). We also have $p'(0\pm) = \pm \infty$, $p'(-\infty) \ne \pm \infty$, $p'(-\infty) \ge 1 / V^-$, $p'(\beta-) \le \Lambda^+$, and $p''(\omega) \neq 0$ for each $\omega \in (-\infty; 0) \cup (0; \beta)$.
\end{lemma}

\begin{proof}
    Recall that $p(\omega) = \log \omega + q(\omega)$, where $q(\omega)$ is real analytic in the domain $D$ containing $(-\infty; 0]$. Thus, $p'(\omega)$ is real analytic in $D \setminus \{0\}$, implying that $p'(\omega) \in \mathbb{R}$ for each $\omega \in (-\infty; 0) \cup (0; \beta)$. By the principle of orientation, $p'$ decreases on $(-\infty; 0)$ and $(0; \beta)$, because $p'(D \cap H^+) \subset H^-$. Moreover, it is injective on $(-\infty; 0) \cup (0; \beta)$: otherwise, the $p'$-images of two disjoint half-disks in $D \cap H^+$ around some $\omega_1 \in (-\infty; 0)$ and $\omega_2 \in (0; \beta)$ would intersect, contradicting the injectivity of $p'$ in $D \cap H^+$.

    Since $p'(D \cap H^+) \subset H^-$, it follows that $p''$ never vanishes on $(-\infty; 0) \cup (0; \beta)$.

    Since $p'(\omega) = 1 / \omega + q'(\omega)$, we have $p'(0\pm) = \pm \infty$. Then $p'(-\infty) \ne \pm \infty$ by the injectivity and continuity on $(-\infty; 0) \cup (0; \beta)$. Note that $p'(-\infty) \ge 1 / V^-$ in both cases $p'(-\infty) < 0$ and $p'(-\infty) \ge 0$ by \eqref{eq-gamma-fns}.

    It remains to show that $p'(\beta-) \le \Lambda^+$. Let us consider two cases:
    \begin{enumerate}
        \item $\beta \in \partial D$. Since $p'(\beta-)$ exists by the monotonicity, there is a function $\varepsilon(x) \searrow 0$ as $x \nearrow \beta$ such that $\lim_{\omega \to \beta} p'(\omega) = p'(\beta-)$ for complex $\omega$ with $\Re \omega < \beta$ and $0 < \Im \omega < \varepsilon(\Re \omega)$. For such $\omega$,
            \begin{align}
                \label{eq-gamma-vert-seg}
                \frac{\Im p(\omega)}{\Im \omega} = \frac{\Im p(\omega) - \Im p(\Re \omega)}{\Im \omega - \Im(\Re \omega)} = \Re p'(\nu) \to p'(\beta-) \quad\text{as $\omega \to \beta$},
            \end{align}
            for some $ \nu = \nu(\omega) \in [\Re \omega; \omega]$. Thus, $p'(\beta-) \le \Lambda^+$ by \eqref{eq-gamma-fns}.
        \item $\beta = +\infty$. Since $D$ has a bounded projection on the imaginary axis, the ray $[x; x + i\infty)$ intersects $\partial D$ for each $x \in (0; +\infty)$. Let $h(x)$ be the lowest intersection point. For $\Re \omega \to +\infty$, $0 < \Im \omega < \Im h(\Re \omega)$, and $\omega - h(\Re \omega) \to 0$, Eq.~\eqref{eq-gamma-vert-seg} still holds because the segments $[\Re \omega; h(\Re \omega)]$ determine a prime end of $D \cap H^+$ and the injective conformal mapping $p'$ continuously extends to each prime end (see \cite[Theorems~2.13 and~2.20]{Markushevich-77}). The inequality $p'(\beta-) \le \Lambda^+$ follows.\\[-1.2cm]
    \end{enumerate}
\end{proof}

\begin{lemma}[Lagrangian]
    \label{lm-gamma-L}
    The function $\L_v(\omega)$ has a unique non-degenerate critical point $\hat \omega(v) = (p')^{-1}(1 / v)$ for $v \neq 0$, which belongs to $(-\infty; 0)$ for $v < 0$, to $(0; \beta)$ for $v > 0$, and for $v = 0$, we have $\L_v(\omega) = -\omega$. Also, $\L$ takes the intervals comprising $(-\infty; \beta) \setminus \{0\}$ bijectively to the following intervals parallel to the real axis (see Figs.~\ref{fig-gamma-L-lt}~and~\ref{fig-gamma-L-gt}):
    \begin{align*}
        \begin{cases}
            \L \colon (-\infty; \hat \omega], [\hat \omega; 0), (0; \beta) \mapsto (\pi iv + \infty; \L(\hat \omega)], [\L(\hat \omega); \pi iv + \infty), (+\infty; \L(\beta-)), & \text{if $v < 0$}, \\
            \L \colon (-\infty; 0), (0; \hat \omega], [\hat \omega; \beta) \mapsto (\pi iv + \infty; \pi iv - \infty), (-\infty; \L(\hat \omega)], [\L(\hat \omega); \L(\beta-)), & \text{if $v > 0$},
        \end{cases}
    \end{align*}
    where $\L(\beta-)$ is either $-\infty$ or a real number, and $\L(\beta-) < \L(\hat \omega)$ for $v > 0$. Additionally, $\L'(-\infty) < 0$.
\end{lemma}

\begin{proof}
    For $v = 0$, we have $\L(\omega) = -\omega$ by definition, and there is nothing to prove. Assume $v \ne 0$.

    Since $\L'(\omega) = vp'(\omega) - 1$, it follows that $\omega$ is a critical point of $\L$ whenever $p'(\omega) = 1 / v$. Clearly, $\L$ has no critical points inside $H^+$ because $p'(D \cap H^+) \subset H^-$. Since $p'$ is continuous and decreases on $(-\infty; 0)$ and $(0; \beta)$, $p'(0\pm) = \pm \infty$, $p'(-\infty) \ge 1 / V^-$, and $p'(\beta-) \le \Lambda^+$ by Lemma~\ref{lm-gamma-p}, it follows that $1 / v \in p'((-\infty; 0))$ for $V^- < v < 0$ and $1 / v \in p'((0; \beta))$ for $0 < v < V^+$. As $p'$ is also injective on $(-\infty; 0) \cup (0; \beta)$, the Lagrangian has a unique critical point $\hat \omega = \hat \omega(v) = (p')^{-1}(1 / v)$ for $v \neq 0$, which belongs to $(-\infty; 0)$ for $v < 0$ and to $(0; \beta)$ for $v > 0$. It is non-degenerate because $p'' \neq 0$ on $(-\infty; 0) \cup (0; \beta)$ by Lemma~\ref{lm-gamma-p}.

    Let us show that $\L'(\omega) < 0$ for $\omega \in \{-\infty\} \cup (-\infty; \min(\hat \omega, 0)) \cup (\max(\hat \omega, 0); \beta)$. For $v < 0$ and $\omega \in \{-\infty\} \cup (-\infty; \hat \omega) \cup (0; \beta)$, we have
    \begin{align}
        \label{eq-gamma-L'-ub}
        \L'(\omega) = vp'(\omega) - 1 < vp'(\hat \omega) - 1 = 0,
    \end{align}
    since $p'(\omega) > p'(\hat \omega)$ by Lemma~\ref{lm-gamma-p}. For $v > 0$ and $\omega \in \{-\infty\}\cup(-\infty; 0) \cup (\hat \omega; \beta)$, the inequality~\eqref{eq-gamma-L'-ub} still holds, because in this case Lemma~\ref{lm-gamma-p} implies $p'(\omega) < p'(\hat \omega)$.

    Let us find $\L((-\infty; \beta) \setminus \{0\})$. Since $\L'$ is real and non-vanishing on $(-\infty; \beta) \setminus \{0, \hat \omega\}$, it follows that $\L$ maps the real intervals between $-\infty$, $0$, $\hat \omega$, and $\beta$ bijectively into horizontal intervals between the limiting values at these endpoints.

    Clearly, $\L(0+) = -\sgn(v) \infty$ and $\L(0-) = -\sgn(v) \infty + \pi iv$ because $\Re \L(\omega) = v(\log|\omega| + \Re q(\omega)) - \Re \omega \sim v\log|\omega|$ as $\omega \to 0$ and $\Im \L(\omega) = v \cdot \arg \omega$ for small real $\omega$. We have $\L(-\infty) = \L(-1) + \int_{-1}^{-\infty} \L'(\omega) \, d\omega = \L(-1) + \infty = \pi iv + \infty$ because $\L'(-\infty) < 0$. As for $\L(\beta-)$, note that $\L(\beta-) = \L(\delta) + \int_\delta^\beta \L'(\omega) \, d\omega \in \mathbb{R} \cup \{-\infty\}$ for any $\delta \in (0; \beta)$. For $v > 0$, setting $\delta = \hat \omega$ and using that $\L'(\omega) < 0$ on $(\hat \omega; \beta)$, we get $\L(\beta-) < \L(\hat \omega)$.
\end{proof}

\begin{remark}
    \label{rem-omega-analyticity}
    The function $\hat \omega(v)$ extended by $0$ at $v = 0$ is real analytic and satisfies $\hat \omega'(0) = 1$ and $\L_v(\hat\omega(v))\to 0$ as $v\to 0$. Indeed, $\hat \omega(v) = (p')^{-1}(1 / v)$ is real analytic for each $v \neq 0$ by the inverse function theorem (as $p''(\hat \omega) = \L''(\hat \omega) / v \neq 0$ by Lemma~\ref{lm-gamma-L}). It is also real analytic at $v = 0$ by the inverse function theorem applied to $1 / p'(\omega)$ at $\omega = 0$: the function $1 / p'(\omega) = \omega / (1 + \omega q'(\omega))$ extends analytically by $0$ at $\omega = 0$ and $(1 / p')'(0) = 1 \neq 0$. In particular, we have $\hat \omega'(0) = 1$, and also $\sgn \hat \omega = \sgn v$ by Lemma~\ref{lm-gamma-L}. Thus $\L_v(\hat \omega(v)) = v\log \frac{\hat \omega(v)}{v} + v\log v + vq(\hat \omega(v)) - \hat \omega(v) \to 0$ as $v \to 0$. In what follows, we set $\hat \omega(0) := 0$.
\end{remark}

\begin{lemma}[Normal form]
    \label{lm-gamma-F}
    The function $F_{v, \Delta}(z)$ has a unique non-degenerate critical point $\hat z = v$ for each $v \neq 0$ and $\Delta \in \mathbb{R}$, and for $v = 0$ we have $F_{v, \Delta}(z) = -z + \Delta$. Also, $F|_{H^+}$ is a bijective conformal mapping onto
    \begin{align*}
        F(H^+) =
        \begin{cases}
            \{z \colon \Im z < \max(\pi v, 0)\} \setminus [F(v); F(v) - \sgn(v) \infty), & \text{if $v \neq 0$}, \\
            H^-, & \text{if $v = 0$},
        \end{cases}
    \end{align*}
    and $F$ takes the intervals comprising $\mathbb{R} \setminus \{0\}$ bijectively to the following intervals parallel to the real axis (see Figs.~\ref{fig-gamma-F-lt}~and~\ref{fig-gamma-F-gt}):
    \begin{align}
        \label{eq-lm-gamma-F}
        \begin{cases}
            F \colon (-\infty; v], [v; 0), (0; +\infty) \mapsto (\pi iv + \infty; F(v)], [F(v); \pi iv + \infty), (+\infty; -\infty), & \text{if $v < 0$}, \\
            F \colon (-\infty; 0), (0; v], [v; +\infty) \mapsto (\pi iv + \infty; \pi iv - \infty), (-\infty; F(v)], [F(v); -\infty), & \text{if $v > 0$}.
        \end{cases}
    \end{align}
\end{lemma}

\begin{proof}
    For $v = 0$, we have $F(z) = -z + \Delta$ by definition, and there is nothing to prove. Assume $v \ne 0$.

    We have $F'(z) = \frac{v}{z} - 1 = 0$ if and only if $z = v$, and clearly $F''(v) = -\frac{1}{v} \neq 0$.

    Let us find $F(\mathbb{R} \setminus \{0\})$. Since $F'$ is real and non-vanishing on $\mathbb{R} \setminus \{0, v\}$, $F$ maps the intervals on $\mathbb{R}$ between $-\infty$, $0$, $v$, $+\infty$ bijectively into horizontal intervals between the limiting values at these endpoints. Clearly, $F(0+) = -\sgn(v) \infty$ and $F(0-) = -\sgn(v) \infty + \pi iv$ because $\Re F(z) \sim v\log|z|$ as $z \to 0$ and $\Im F(z) = v \cdot \arg z$ for all real $z \neq 0$. Also, $F(-\infty) = \pi iv + \infty$ and $F(+\infty) = -\infty$ because $\Re F(z) \sim -z$ as $z \to \pm \infty$. This implies \eqref{eq-lm-gamma-F}.

    The map $F|_{H^+}$ is conformal and injective by the Noshiro--Warschawski theorem \cite[Theorem~2.16]{Duren-83}, because $\Im F'(z) = v \Im(1 / z)$ has a constant sign in the convex domain $H^+$. Since $\Re F(z) \sim v\log|z|$ as $z \to 0$ and $F(z) \sim -z$ as $z \to \infty$, it follows that $F|_{H^+}$ extends continuously to $\partial H^+ \cup \{\infty\}$ in the wide sense by $F(0)$, $F(\infty) := \infty$. Thus, $\partial F(H^+) = F(\partial H^+)$ by compactness. The image $F(H^+)$ is then determined to be the unique domain with boundary $F(\mathbb{R} \setminus \{0\})$ given by \eqref{eq-lm-gamma-F}.
\end{proof}

Similarly to the Airy regime, we need the following subsets of $\mathbb{C}$ for each $v \in V$:
\begin{align*}
    \Pi = \Pi_v &:= \{z \in \mathbb{C} \colon 0 < \Im z < \pi v\}, \\
    M = M_v &:=
    \begin{cases}
        \text{component of $\L^{-1}(\Pi)\cap H^+$ whose boundary contains $0$}, & \text{if $v > 0$}, \\
        \varnothing, & \text{if $v \le 0$}.
    \end{cases}
\end{align*}

\begin{lemma}[Univalence domain]
    \label{lm-gamma-univalence}
    For each $v > 0$, the set $\L^{-1}(\Pi)$ contains a neighbourhood of $0$ in $H^+$, so $M$ is well-defined and non-empty, and the restriction $\L|_M$ is a homeomorphism $M \leftrightarrow \Pi$. Furthermore, $M$ is the only component of $\L^{-1}(\Pi) \cap H^+$ whose boundary has infinitely many common points with $(-\infty; \beta)$.
\end{lemma}

\begin{figure}
    \centering
    \includegraphics[width=0.8\textwidth]{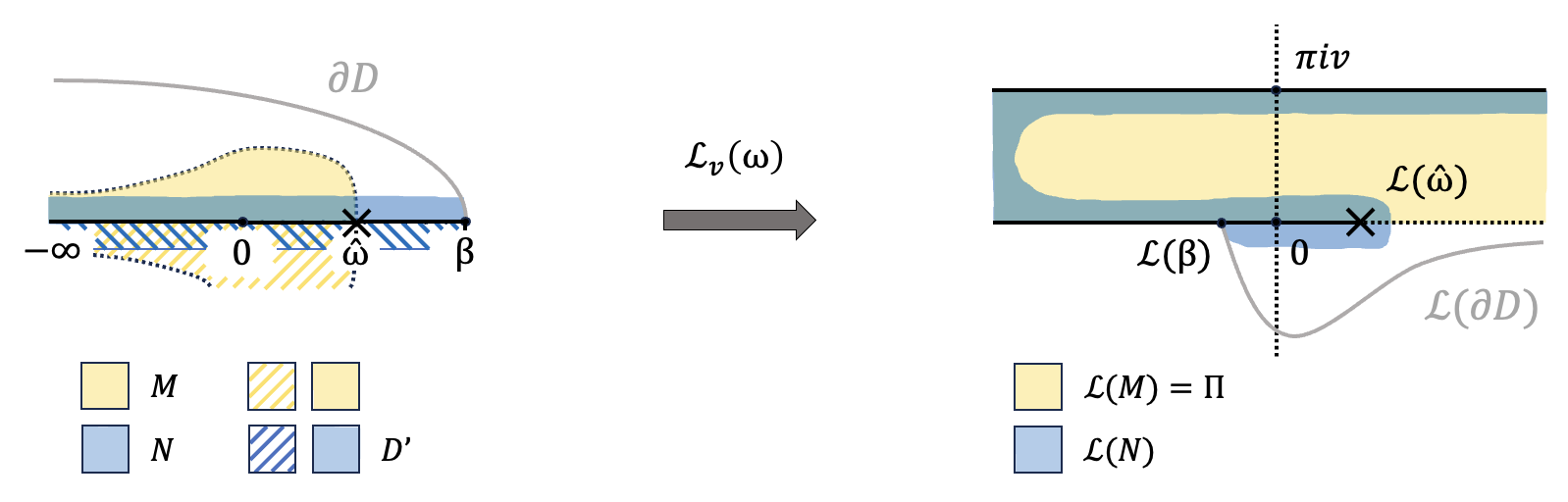}
    \caption{The domains $M$, $N$, $D'$ and their images for $v > 0$. See the proof of Proposition~\ref{prop-gamma-substitution}.}
    \label{fig-gamma-domains}
\end{figure}

\begin{proof}
    For $v > 0$, the set $\L^{-1}(\Pi)$, hence $M$, contains a neighbourhood of $(-\infty; \hat \omega) \ni 0$ in $H^+$ by Lemma~\ref{lm-gamma-L} and the principle of orientation (Figs.~\ref{fig-gamma-L-F}--\ref{fig-gamma-domains}): for each $\omega \in (-\infty; \hat \omega) \setminus \{0\}$, the map $\L$ takes a neighbourhood of $\omega$ in $H^+$ into $\Pi$. It also takes a neighbourhood of $0$ in $H^+$ into $\Pi$ by the principle of orientation applied to the map $\omega \mapsto e^{\L(\omega) / v}$, which has the derivative $e^{q(0)} > 0$ at $\omega = 0$. So $M$ is well-defined and non-empty.

    By the principle of orientation, $(\hat \omega; \beta)$ is disjoint with $\Cl \, (\L^{-1}(\Pi) \cap H^+)$. Thus, $M$ is the only component of $\L^{-1}(\Pi) \cap H^+$ whose boundary has infinitely many common points with $(-\infty; \beta)$.

    Next, let us show that $\L|_M$ is proper. We verify the sequential properness criterion: for each sequence $\omega_n \in M$ converging to a point $\omega_0 \in \partial M \cup \{\infty\}$, the sequence of images $\L(\omega_n)$ has no limit in $\Pi$. Consider the following five cases.

    Case $\omega_0 = 0$. We have $\Re \L(\omega_n) = v(\log|\omega_n| + \Re q(\omega_n)) - \Re \omega_n \to -\infty$ as $\omega_n \to 0$, hence $\L(\omega_n) \to \infty$.

    Case $\omega_0 \in (-\infty; \beta) \setminus \{0\}$. By continuity, $\L(\omega_n) \to \L(\omega_0)$, which is not in $\Pi$ by Lemma~\ref{lm-gamma-L}.

    Case $\omega_0 \in D \cap H^+$. By continuity, $\L(\omega_n) \to \L(\omega_0)$. If $\omega_0 \in \L^{-1}(\Pi)$, then $\omega_0 \in M$ because $\omega_0 \in D \cap H^+$, $\L^{-1}(\Pi)\cap H^+$ is open, and all $\omega_n \in M$. This contradicts the assumption $\omega_0 \in \partial M$, hence $\L(\omega_0) \not\in \Pi$.

    Case $\omega_0 \in \partial D \cap \Cl H^+$. This is actually impossible because $\Im \L < 0$ in the $\varepsilon$-neighbourhood of $\partial D$ in $H^+$ by \eqref{eq-gamma-eps}, but $\Im \L(\omega_n) > 0$ as $\omega_n \in M$.

    Case $\omega_0 = \infty$. Since $D$ has a bounded projection on the imaginary axis, passing to a subsequence, we may assume $\Im \omega_n \to Y$ for some $Y \in [0; +\infty)$ and $\Re \omega_n \to X$ for some $X \in \{\pm \infty\}$. Let us show that $\lim_{n \to \infty} \L(\omega_n) = \infty$.

    For that, we show that $D$ contains a neighbourhood of a horizontal ray approaching $X + iY$. Since $\Im \L'(\omega) = v \cdot \Im p'(\omega) < 0$ for each $\omega \in D \cap H^+$, it follows that $\Im \L$ decreases on each component of $(\omega_n - \infty; \omega_n] \cap D$ as $\Re \omega$ increases. Since $\Im \L < 0$ in the $\varepsilon$-neighbourhood of $\partial D$ in $D \cap H^+$ for some $\varepsilon > 0$ by \eqref{eq-gamma-eps} and $\Im \L(\omega_n) > 0$, it follows that $(\omega_n - \infty; \omega_n]$ is disjoint with this $\varepsilon$-neighbourhood. Then $\dist((\omega_n - \infty; \omega_n], \partial D) \ge \varepsilon$, implying that the half-strips $S_n := \{\omega \in \mathbb{C} \colon \Re \omega < \Re \omega_n, |\Im \omega - \Im \omega_n| < \varepsilon\}$ belong to $D$. We now have two cases:
    \begin{enumerate}
        \item $X = -\infty$. Since $\Im \omega_n \to Y$, it follows that $|\Im \omega_m - Y| < \varepsilon / 2$ for some $m$. Therefore, the half-strip $S := \{\omega \in \mathbb{C} \colon \Re \omega < \Re \omega_m, |\Im \omega - Y| < \varepsilon / 2\}$ belongs to $D$ because it is contained in $S_m$.
        \item $X = +\infty$. In this case, set $S := \{\omega \in \mathbb{C} \colon \Re \omega > 0, |\Im \omega - Y| < \varepsilon\}$. It belongs to $D$ because it is contained in the union $\bigcup_n S_n$.
    \end{enumerate}
    Without loss of generality, we may assume that the half-strip $S$ contains all $\omega_n$ and $\Cl S\subset D$.

    Let us prove that there exist $x_0 \in \mathbb{R}, \delta > 0$ such that $\sgn x_0 = \sgn X$, $x_0 + iY \in S$, and $|\Re \L'(\omega)| \ge \delta$ for each $\omega \in D \cap H^+$, $|\omega| \ge |x_0|$. Assume the contrary. Then there exists a sequence $\omega_j \to \infty$ such that $\Re \L'(\omega_j) \to 0$. Passing to a subsequence, we may assume $\L'(\omega_j) \to z$ for some $z \in [0; -i\infty) \cup \{\infty\}$ because $\L'(D \cap H^+) = H^-$. Since $\L' = vp' - 1$ is a bijective conformal mapping $D \cap H^+ \to H^-$ and $\L'$ is meromorphic at $0$ and $\hat \omega$ with $\L'(0) = \infty$ and $\L'(\hat \omega) = 0$, there is a bounded set $B \subset D \cap H^+$ such that $\L'(B)$ covers a neighbourhood of $z$ in $H^-$. Since $\L'(\omega_j) \to z$, $\omega_j \to \infty$, and $B$ is bounded, we get a contradiction with the injectivity of $\L'$ in $D \cap H^+$.

    Then for large $n$, the value $\Re \L'$ has a constant sign on the horizontal segment $[x_0 + i\Im \omega_n; \omega_n]$, so
    \begin{align*}
        |\Re \L(\omega_n)| \ge \left|\int_{x_0}^{\Re \omega_n} \Re \L'(x + i\Im \omega_n) \, dx\right| - |\Re \L(x_0 + i\Im \omega_n)| \ge \delta |\Re \omega_n - x_0| - O_{v, \varepsilon, x_0, p}(1) \to +\infty
    \end{align*}
    as $n \to \infty$, where $\Re \L(x_0 + iy)$ is bounded by continuity. Thus $\L(\omega_n) \to \infty$, so $\L|_M$ is proper.

    As every proper map between subsets of the plane is closed, $\L(M)$ is closed in $\Pi$. Since $\L$ is non-constant and analytic, $\L(M)$ is open in $\Pi$. By the first paragraph of the proof, $\L(M)$ is non-empty. Hence $\L(M) = \Pi$. Since every proper conformal map is a covering \cite[Lemma~2]{Ho-75}, $M$ is path connected and $\Pi$ is simply-connected and locally path connected, it follows that $\L|_M$ is a homeomorphism $M \leftrightarrow \Pi$.
\end{proof}

\begin{proposition}[Conformal change of variable]
    \label{prop-gamma-substitution}
    There exist a family of domains $D'_v \subset D$ symmetric about the real axis and containing $(-\infty; \beta) \cup M$ (Fig.~\ref{fig-gamma-domains}), a real analytic family of injective functions $z_v \colon D'_v \to \mathbb{C}$, and a real analytic function $\Delta \colon V \to \mathbb{R}$, such that for each $v \in V$, we have $\L_v|_{D'_v} = F_{v, \Delta(v)} \circ z_v$, and moreover, $z(-\infty) = -\infty$, $z(0) = 0$, $z(\hat \omega) = v$,
    \begin{align}
        \label{eq-gamma-substitution}
        z'(\hat \omega) &=
        \begin{cases}
            |v| \sqrt{-p''(\hat \omega)}, & \text{if $v \neq 0$}, \\
            1, & \text{if $v = 0$},
        \end{cases}
        & \text{and} \qquad
        \Delta(v) &=
        \begin{cases}
            \L(\hat \omega) - v\log v + v, & \text{if $v \neq 0$}, \\
            0, & \text{if $v = 0$}.
        \end{cases}
    \end{align}
\end{proposition}

\begin{proof}
    Pick $\Delta(v)$ so that the critical values $\L_v(\hat \omega)$ and $F_{v, \Delta(v)}(v)$ match for each $v \neq 0$, and $\L = F$ for $v = 0$ (that is, ``align the crosses'' in Fig.~\ref{fig-gamma-L-F}). This leads us to the expression for $\Delta(v)$ in \eqref{eq-gamma-substitution}.

    The desired domain $D'$ is the union of $(-\infty; \beta)$, $M$, a family of sufficiently small neighborhoods $N$ of $(-\infty; \beta)$ in $H^+$, and their symmetric copies about the real axis (see Fig.~\ref{fig-gamma-domains}). By choosing $N$ small enough, we may assume that $\L|_N$ is injective, $N \cap \L^{-1}(\Pi) \subset M$, $\L(N) \cap (-\infty; \L(\beta-)) = \varnothing$, and also $N$ is simply-connected (the existence of such a family $N$ is proven in Proposition~\ref{prop-gamma-neighbourhoods} in Appendix~\ref{sec-boundary-neighbourhoods}).

    The function $z(\omega) = z_v(\omega)$ is constructed in several steps. First, we define (see Fig.~\ref{fig-gamma-L-F})
    \begin{align*}
        z(\omega) := (F|_{H^+})^{-1}(\L(\omega)) \qquad\text{for each}\quad \omega \in M \cup N.
    \end{align*}
    Let us show that $z(\omega) $ is well-defined and injective. By Lemma~\ref{lm-gamma-F}, the inverse $(F|_{H^+})^{-1} \colon F(H^+) \to H^+$ exists and $\L(M) \subset \Pi \subset F(H^+)$. By Lemmas~\ref{lm-gamma-L}--\ref{lm-gamma-F}, we get $\partial F(H^+) = \L((-\infty; \beta) \setminus \{0\}) \cup (-\infty; \L(\beta-)]$ (for $\L(\beta-) \ne -\infty$ and $v \ne 0$). Then $\L(N) \cap \partial F(H^+) = \varnothing$ since $\L|_N$ is injective and $\L(N) \cap (-\infty; \L(\beta-)) = \varnothing$. Then $\L(N) \subset F(H^+)$ because $N$ is connected and $\partial \L(N) \ni \L(\hat \omega) = F(v)$. So $z(\omega)$ is well-defined. The maps $z|_M$ and $z|_N$ are injective because $\L|_M$ and $\L|_N$ are, by Lemma~\ref{lm-gamma-univalence} and the choice of $N$. Since $\L(M) \subset \Pi$ and $N \cap \L^{-1}(\Pi) \subset M$, it follows that $\L(N \setminus M) \cap \L(M) = \varnothing$, thus $z(N \setminus M) \cap z(M) = \varnothing$. So, $z(\omega)$ is injective.

    Next, let us extend $z(\omega)$ to $(-\infty; \beta)$ continuously. The injective conformal mapping $\L|_{N}$ extends to a homeomorphism between $(-\infty; \beta)$ and the corresponding set of prime ends of $\L(N)$ \cite[Theorem~2.20]{Markushevich-77}. By Lemmas~\ref{lm-gamma-L}--\ref{lm-gamma-F}, the injective conformal mapping $(F|_{H^+})^{-1}|_{\L(N)}$ extends to a homeomorphism between this set of prime ends and a real interval. Hence, the map $z = (F|_{H^+})^{-1} \circ \L$ extends continuously to $(-\infty; \beta)$ with $z((-\infty; \beta)) \subset \mathbb{R}$. Furthermore, by Lemmas~\ref{lm-gamma-L}--\ref{lm-gamma-F}, this extension maps the intervals between $-\infty$, $\hat \omega$, $0$, and $\beta$ to the intervals between $-\infty$, $v$, $0$, and $(F|_{(\max(v,0); +\infty)})^{-1}(\L(\beta-))$ (where the order of $\hat \omega$, $0$ and $v$, $0$ is flipped when $v > 0$). In particular, for each $v$, we have $z(-\infty) = -\infty$, $z(0) = 0$, and $z(\hat \omega) = v$.

    Now we can extend $z(\omega)$ analytically to $D'$ using Schwarz reflection principle~\cite[Theorem~9.1]{Markushevich-77}, because $z((-\infty; \beta)) \subset \mathbb{R}$ and $M \cup N$ is adjacent to $(-\infty; \beta)$. Since $z(M \cup N) \subset H^+$, the map $z \colon D' \to \mathbb{C}$ is injective.

    Next, we verify the analyticity of $\Delta$ and $z$ in $v$. The function $\Delta(v) = v\log \frac{\hat \omega}{v} + vq(\hat \omega) - \hat \omega + v$ (extended by $0$ at $v = 0$) is real analytic because $\hat \omega(v)$ is (see Remark~\ref{rem-omega-analyticity}), $\hat \omega'(0) = 1$, and $\sgn \hat \omega = \sgn v$ by Lemma~\ref{lm-gamma-L}.

    Let us show that $z_v(\omega)$ is a real analytic family. For $\omega \in D' \cap H^+$, the function $v \mapsto z_v(\omega) = F_{v,\Delta(v)}^{-1}(\L_v(\omega))$ is real analytic because $\Delta(v)$ is real analytic, $\L_v|_{D'\cap H^+}$ and $F_{v,\Delta}|_{H^+}$ are analytic and injective, and $(F_{v,\Delta}|_{H^+})^{-1}$ forms an analytic family by a version of the inverse function theorem (see Lemma~\ref{lm-inverse-param} in Appendix~\ref{sec-complex-analysis}). Analogously, for $\omega \in (0; \beta) \setminus \{\hat \omega\}$, we have $F'(z(\omega)) \ne 0$, hence $F$ is injective in some family of domains $B \ni z(\omega)$ (forming a small neighbourhood of $(z(\omega), v)$ in $\mathbb{C} \times V$). Then $(F|_B)^{-1} \circ \L$ is real analytic in $v$ and coincides with $z_v$ at $\omega$. For $\omega \in (-\infty;0) \setminus \{\hat \omega\}$, the map $v \mapsto z_v(\omega)$ is real analytic by the same argument, only $\L$ and $F$ are replaced with their analytic continuations from $H^+$ to vicinities of $\omega$ and $z(\omega)$, respectively. For $\omega \in D' \cap H^-$, the analyticity holds by symmetry. Finally, $v \mapsto z_v(\omega)$ is analytic at $\omega = \hat \omega, 0$ by a version of the removable singularity theorem (see Lemma~\ref{lm-rm-singularity-param} in Appendix~\ref{sec-complex-analysis}).

    It remains to derive the first equation of~\eqref{eq-gamma-substitution}. Differentiating the identity $\L(\omega) = F(z(\omega))$, we get
    \begin{align*}
        \L'(\omega) &= F'(z(\omega)) \, z'(\omega), & \L''(\omega) &= F''(z(\omega)) \, z'(\omega)^2 + F'(z(\omega)) \, z''(\omega).
    \end{align*}
    Substituting $\omega = \hat \omega$ into the latter formula and using that $z(\hat \omega) = v$, we get
    \begin{align*}
        vp''(\hat \omega) = \L''(\hat \omega) = F''(v) \, z'(\hat \omega)^2 = -\frac{z'(\hat \omega)^2}{v} \quad\text{for $v \neq 0$}.
    \end{align*}
    Thus, $z'(\hat \omega) = +|v| \sqrt{-p''(\hat \omega)}$, with the positive sign in front because $z' > 0$ on $(-\infty; \beta)$ by the principle of orientation. If $v = 0$, then $z(\omega) = \omega$ and $z'(\hat \omega) = 1$.
\end{proof}

\begin{corollary}
    \label{cor-gamma-disk}
    For each $v > 0$, the image $z(D')$ contains the disk $C := \{z \in \mathbb{C} \colon |z| \le v\}$.
\end{corollary}

\begin{proof}
    For each $z \in C \cap H^+$, we have $0 < |z| \sin \arg z < v \cdot \arg z$, thus $\Im F(z) = v \cdot \arg z - |z| \sin \arg z \in (0; v\pi)$, i.e., $F(C \cap H^+) \subset \Pi = \L(M)$ by Lemma~\ref{lm-gamma-univalence}. Hence $C \cap H^+ \subset (F|_{H^+})^{-1}(\L(M)) = z(M)$. Then $z(D')$ contains $C$ by symmetry, as $z(D') \supset (z(-\infty); z(\hat \omega)] \supset [-v; v]$.
\end{proof}

\begin{remark}
    \label{rm-gamma-analyticity}
    The main term of asymptotic formula~\eqref{eq-gamma-asymp} can be expressed in the notation of Proposition~\ref{prop-gamma-substitution} as
    \begin{equation}
        \label{eq-gamma-asymp-anal}
        \frac{t^x e^{-t\Delta(x / t)}}{\Gamma(x) |x|\sqrt{- p''(\hat \omega(x / t))}} \cdot a(\hat \omega(x / t)) = \frac{t^{x-1} e^{-t\Delta(x / t)}}{\Gamma(x) \, z'(\hat \omega(x / t))} \cdot a(\hat \omega(x / t)).
    \end{equation}
    As $x \to 0$, we have $1 / \Gamma(x) \to 0$, $\hat \omega(x / t) \to 0$, and $z'(\hat\omega(x / t)) \to 1$. Therefore, at $x = 0$, function~\eqref{eq-gamma-asymp-anal} extends continuously by zero. Then, asymptotic formula~\eqref{eq-gamma-asymp} becomes valid for $x = 0$: in this case, the integrand is analytic in a neighbourhood of $(-\infty; 0]$ and the integral vanishes. Moreover, function~\eqref{eq-gamma-asymp-anal} is analytic at $x = 0$ because the functions $\Delta(v)$, $\hat \omega(v)$, and $z'(\hat \omega(v))$ are, by Proposition~\ref{prop-gamma-substitution} and Remark~\ref{rem-omega-analyticity}.
\end{remark}

\subsection{Proof of the general gamma phenomenon (Theorem~\ref{th-gamma})}
\label{sec-gamma-proof}

In this subsection, we complete the proof of Theorem~\ref{th-gamma}. We use the notation and impose the assumptions from the theorem and Sec.~\ref{sec-gamma-substitution}. We start with showing that the integral in the theorem is well-defined.

\begin{lemma}
    \label{lm-gamma-int-exist}
    For each real analytic function $A\colon D'\to\mathbb{C}$ that is bounded on $(-\infty; 0]$, and each $x$ and~$t$, the integral $\int_{-\infty}^{(0+)} A(\omega) \, e^{-t\L_{x/t}(\omega)} \, d\omega$ is well-defined, that is, there is a neighbourhood $N$ of $(-\infty; 0]$ in $D'$ such that for each contour $\gamma \colon \mathbb{R} \to N \setminus (-\infty; 0]$ that encircles $(-\infty; 0]$ once counterclockwise and satisfies $\Re \gamma(\pm \infty) = -\infty$, the integral along the contour exists and does not depend on the contour.
\end{lemma}

\begin{proof}
    Since $\L'(-\infty) < 0$ by Lemma~\ref{lm-gamma-L}, it follows that $\Re \L(\omega) > k|\omega| - b$ for some $k, b > 0$ (depending on $x/t$ and $p$) and all $\omega < -1$. Then there is $\delta > 0$ and a continuous function $\varepsilon \colon (-\infty; \delta) \to (0; \delta)$ such that the set $N := \{\omega \colon \Re \omega < \delta, |\Im \omega| < \varepsilon(\Re \omega)\}$ is contained in $D'$, $A(\omega)$ is bounded there, and $\Re \L(\omega) > k |\Re \omega| - b$ for all $\omega$ in this set with $\Re\omega < -1$. The set is the desired neighbourhood of $(-\infty; 0]$. Indeed, for $f(\omega) := A(\omega) \, e^{-t\L(\omega)}$ and all sufficiently large $s_1, s_2$, we have
    \begin{multline*}
        \int\limits_{\gamma([s_1; s_2])} \!\!\!\!f(\omega) \, d\omega = \int\limits_{\gamma(s_1)}^{\Re \gamma(s_1)} \!\!\!\!f(\omega) \, d\omega +
        \int\limits_{\Re \gamma(s_1)}^{\Re \gamma(s_2)} \!\!\!\!f(\omega) \, d\omega + \int\limits_{\Re \gamma(s_2)}^{\gamma(s_2)} \!\!\!\!f(\omega) \, d\omega = O_{k, t, \delta, \varepsilon, A}\left(e^{-kt \min(|\Re \gamma(s_1)|, |\Re \gamma(s_2)|) + bt}\right).
    \end{multline*}
    The same bound holds for all sufficiently small $s_1, s_2 < 0$, only the integrand values on $[\Re \gamma(s_1); \Re \gamma(s_2)]$ should be continuously extended from $H^-$. Thus, the integral along the contour $\gamma$ converges. Passing to the limit $s_2\to \pm\infty$, we get $\int_{\gamma([s_1; \pm \infty))} f(\omega) \, d\omega = \int_{\gamma(s_1)}^{\Re \gamma(s_1)} f(\omega) \, d\omega + \int_{\Re \gamma(s_1)}^{-\infty} f(\omega) \, d\omega$ for sufficiently large $\pm s_1$, which proves the independence of the contour.
\end{proof}

\begin{remark}
    \label{rem-F-int-exist}
    Lemma~\ref{lm-gamma-int-exist} and its proof remain true if $\L(\omega)$ is replaced with $F(\omega)$ because clearly $F'(-\infty) < 0$.
\end{remark}

We are going to perform the change of variable $z_v(\omega)$ given by Proposition~\ref{prop-gamma-substitution}. Let us bound auxiliary functions, arising in the process. Denote by $\omega(z) = \omega_v(z) \colon z(D') \to \mathbb{C}$ the inverse of $z_v(\omega)$, and set
\begin{gather}
    \label{eq-A-def}
    G(z) = G_v(z) := a(\omega_v(z)) \, \omega'_v(z).
\end{gather}
Since $z_v(\omega)$ is a real analytic family by Proposition~\ref{prop-gamma-substitution}, it follows that both $\omega_v(z)$ and $G_v(z)$ are, by Lemma~\ref{lm-inverse-param}.

\begin{lemma}
    \label{lm-gamma-H}
    The family of functions
    \begin{gather}
        \label{eq-H-def}
        H(z) = H_v(z) := \frac{G(z) - G(v)}{F'(z)} = \frac{z}{v - z} (G(z) - G(v)), \qquad v \in V, \quad z \in z_v(D'_v) \setminus \{v\},
    \end{gather}
    extends to a real analytic family of functions on $z_v(D'_v)$. Moreover, $H(z), H'(z) = O_{v^-, v^+, a, p}(1)$ for $z \in (-\infty; 0]$ and $v \in V$.
\end{lemma}

\begin{proof}
    The first assertion follows from Hadamard's Lemma~\ref{lm-hadamard} applied to $G(z)$ and $\hat z(v) = v$. It remains to prove the `moreover' part. Assume without loss of generality that $v^- < 0$.

    Take a small $\delta = \delta_{v^-, v^+, p} > 0$ (depending only on $v^-, v^+, p$) such that $[v^- - \delta; v^+ + \delta] \subset (V^-; V^+)$. Let
    \begin{align*}
        Q = Q_{v^-, v^+, p} := (p')^{-1}\left(\frac{1}{v^- - \delta}\right).
    \end{align*}
    By Lemma~\ref{lm-gamma-p}, $Q$ is well-defined and $Q < \min(\hat \omega, 0)$. Applying $z(\omega)$ (which satisfies $z(-\infty) = -\infty$ and is injective, hence is increasing), we get $z(Q) < \min(v, 0)$. By compactness, $z(Q), z(Q) - v = \Omega_{v^-, v^+, p}(1)$.

    Note that $1 / \L'(\omega) = O_{v^-, v^+, p}(1)$ for $\omega \le Q$, because by Lemma~\ref{lm-gamma-p}, we have
    \begin{align*}
        \L'(\omega) &= vp'(\omega) - 1 \le
        \begin{cases}
            vp'(Q) - 1 = \frac{v}{v^- - \delta} - 1 \le \frac{\delta}{v^- - \delta} < 0, & \text{if } v^- \le v \le 0, \\
            vp'(-\infty) - 1 < \frac{v}{v^+ + \delta} - 1 \le \frac{-\delta}{v^+ + \delta} < 0, & \text{if } 0 < v \le v^+.
        \end{cases}
    \end{align*}

    For $z(Q) \le z \le 0$, the desired bounds on $H(z), H'(z)$ hold because the set $\{(z, v): z_v(Q) \le z \le 0, v^- \le v \le v^+\}$ is compact and $H(z)$ is a real analytic family.

    For $z \le z(Q)$, using the formula $\omega'(z) = F'(z) / \L'(\omega(z))$, we compute
    \begin{align*}
        H(z) &= \frac{1}{F'(z)} \left(a(\omega(z)) \, \omega'(z) - G(v)\right) = \frac{a(\omega(z))}{\L'(\omega(z))} - \frac{zG(v)}{v - z}, \\
        H'(z) &= \left.\left(\frac{a(\omega)}{\L'(\omega)}\right)'\right|_{\omega = \omega(z)} \cdot \omega'(z) - \frac{v G(v)}{(v - z)^2} = \left.\left(\frac{a'(\omega)}{\L'(\omega)^2} - \frac{v a(\omega) \, p''(\omega)}{\L'(\omega)^3}\right)\frac{v - z}{z}\right|_{\omega = \omega(z)} - \frac{v G(v)}{(v - z)^2}.
    \end{align*}
    Since $a(-\infty)$, $a'(-\infty)$, $p''(-\infty)$ are finite and $\omega(z) \le Q$ for $z \le z(Q)$, it follows that $a(\omega(z))$, $a'(\omega(z))$, $p''(\omega(z)) = O_{v^-, v^+, a, p}(1)$. Since $1 / \L'(\omega(z)) = O_{v^-, v^+, p}(1)$, $z(Q) - v = \Omega_{v^-, v^+, p}(1)$, $z(Q) = \Omega_{v^-, v^+, p}(1)$, and $G(v) = a(\omega(v)) \, \omega'(v) = O_{v^-, v^+, a, p}(1)$, it follows that $H(z), H'(z) = O_{v^-, v^+, a, p}(1)$ for $z \le z(Q)$.
\end{proof}

This lemma and Remark~\ref{rem-F-int-exist} imply that the following integrals are well-defined.

\begin{lemma}
    \label{lm-gamma-int}
    We have
    \begin{equation}
        \label{eq-gamma-int}
        \int_{-\infty}^{(0+)} e^{-tF(z)} \, dz = \frac{2\pi i \, e^{-t\Delta} t^{x-1}}{\Gamma(x)} \qquad\text{and}\qquad \int_{-\infty}^{(0+)} e^{-tF(z)} H'(z) \, dz = O_{v^-, v^+, a, p}\left(\frac{e^{-t\Delta} t^{x-1}}{|\Gamma(x)|}\right).
    \end{equation}
\end{lemma}

\begin{proof}
    The first formula is obtained by the change of variable $u = tz$ in Hankel's integral~\cite[\S5.9.2]{NIST-10}
    $$
    \frac{1}{2\pi i} \int_{-\infty}^{(0+)} e^{u} u^{-x} \, du = \frac{1}{\Gamma(x)}.
    $$

    To prove the second formula, consider two cases depending on the sign of $x$.

    Case 1: $x \le 0$. The function $F(z)$ can have a branch point at $0$ and can be discontinuous along the cut $(-\infty; 0]$. By Lemma~\ref{lm-gamma-H} and Remark~\ref{rem-F-int-exist}, the left side can be written as the difference of integrals over $(-\infty; 0]$ with the integrand values continuously extended from the lower and upper half-plane (see Fig.~\ref{fig-gamma-contour-lt}):
    \begin{gather}
        \label{eq-gamma-diff-H}
        \int_{-\infty}^{(0+)} e^{-tF(z)} H'(z) \, dz = (e^{\pi ix} - e^{-\pi ix}) \int_{-\infty}^0 e^{-x \log |z| + tz - t\Delta} H'(z) \, dz.
    \end{gather}
    At the same time,
    \begin{gather}
        \label{eq-gamma-diff}
        \int_{-\infty}^{(0+)} e^{-tF(z)} \, dz = (e^{\pi ix} - e^{-\pi ix})\int_{-\infty}^0 e^{-x \log |z| + tz - t\Delta} \, dz.
    \end{gather}
    Since the integrand is positive, and $H'(z) = O_{v^-, v^+, a, p}(1)$ for $z \in (-\infty; 0]$ by Lemma~\ref{lm-gamma-H}, bound~\eqref{eq-gamma-int} follows.

    \begin{figure}
        \centering
        \begin{subfigure}[c]{0.25\textwidth}
            \includegraphics[width=\textwidth]{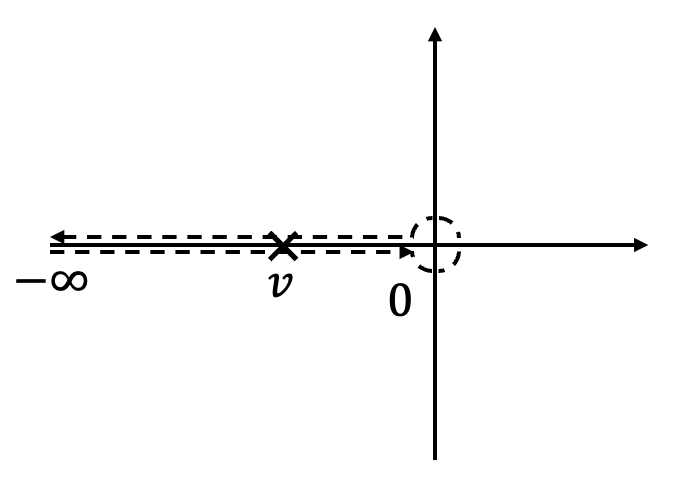}
            \caption{for $v < 0$}
            \label{fig-gamma-contour-lt}
        \end{subfigure}
        \hspace{0.1\textwidth}
        \begin{subfigure}[c]{0.25\textwidth}
            \includegraphics[width=\textwidth]{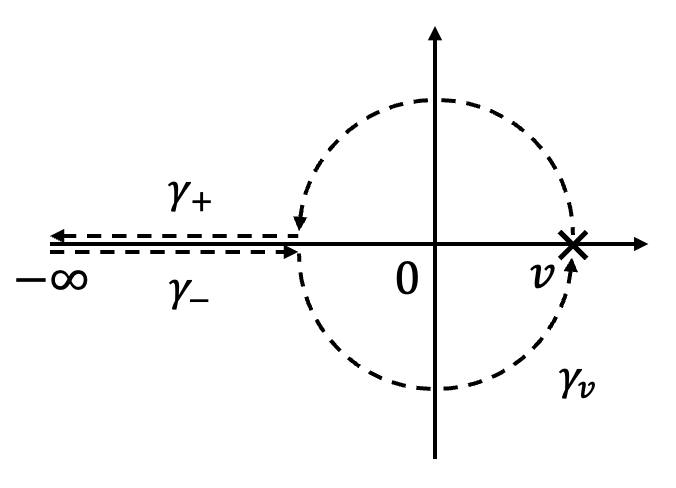}
            \caption{for $v > 0$}
            \label{fig-gamma-contour-gt}
        \end{subfigure}
        \caption{Integration contour from the proof of Lemma~\ref{lm-gamma-int}.}
        \label{fig-gamma-contour}
    \end{figure}

    Case 2: $x > 0$. In this case, $F(z)$ blows up at the origin, and has a unique critical point $v = x / t > 0$. We apply Lemma~\ref{lm-gamma-H} and Remark~\ref{rem-F-int-exist} and choose a special integration contour that bypasses the origin and passes through the critical point; see Fig.~\ref{fig-gamma-contour-gt}. It consists of three parts:
    \begin{enumerate}
        \item $\gamma_- = (-\infty; -v]$, with the integrand values continuously extended from the lower half-plane,
        \item $\gamma_v$, going counterclockwise along the circle of radius $v$ around the origin,
        \item $\gamma_+ = [-v; -\infty)$, with the integrand values continuously extended from the upper half-plane.
    \end{enumerate}
    Corollary~\ref{cor-gamma-disk} guarantees that $\gamma_v$ lies within the domain $z(D')$ of admissible values of $z$.

    To bound the integral along $\gamma_v$, we introduce a new variable $\varphi$ such that $z = ve^{i\varphi}$, $-\pi \le \varphi \le \pi$:
    \begin{multline*}
        \int\limits_{\gamma_v} e^{-tF(z)} H'(z) \, dz = iv^{1-x} \int\limits_{-\pi}^{\pi} e^{x\cos\varphi + ix(\sin\varphi - \varphi) + i\varphi - t\Delta} H'(ve^{i\varphi}) \, d\varphi = O_{v^-, v^+, a, p}\left(e^{-t\Delta} v^{1-x} \int\limits_{-\pi}^{\pi} e^{x\cos\varphi} \, d\varphi\right).
    \end{multline*}
    because $H'(z) = O_{v^-, v^+, a, p}(1)$ for $|z| \le v$ by compactness.

    For $x \ge 1$, using $\cos\varphi = 1 - 2\sin^2(\varphi / 2) \le 1 - 2\varphi^2 / \pi^2$ ($-\pi \le \varphi \le \pi$), we obtain
    \begin{equation*}
        \int_{-\pi}^{\pi} e^{x\cos\varphi} \, d\varphi \le \int_{-\pi}^{\pi} e^{x\left(1 - 2\varphi^2 / \pi^2\right)} \, d\varphi \le \int_{-\infty}^{\infty} e^{x\left(1 - 2\varphi^2 / \pi^2\right)} \, d\varphi = O\left(\frac{e^x}{\sqrt{x}}\right).
    \end{equation*}
    By Stirling's formula, we get
    \begin{gather*}
        \int_{\gamma_v} e^{-tF(z)} H'(z) \, dz = O_{v^-, v^+, a, p}\left(e^{-t\Delta} v^{1-x} \frac{e^x}{\sqrt{x}}\right) = O_{v^-, v^+, a, p}\left(\frac{e^{-t\Delta} t^{x-1}}{\Gamma(x)}\right).
    \end{gather*}

    For $x \in (0; 1)$, we apply the trivial bound $\int_{-\pi}^{\pi} e^{x\cos\varphi} \, d\varphi = O(1)$. Since $x \, \Gamma(x) = \Gamma(x + 1)$ and $x^{-x}$ are bounded on $(0; 1)$, we get the same estimate
    \begin{gather*}
        \int_{\gamma_v} e^{-tF(z)} H'(z) \, dz = O_{v^-, v^+, a, p}\left(e^{-t\Delta} v^{1-x}\right) = O_{v^-, v^+, a, p}\left(\frac{e^{-t\Delta} t^{x-1}}{\Gamma(x)}\right).
    \end{gather*}
    Analogously, we get the bound
    \begin{gather*}
        \int_{\gamma_v} e^{-tF(z)} \, dz  = O_{}\left(\frac{e^{-t\Delta} t^{x-1}}{\Gamma(x)}\right).
    \end{gather*}

    The sum of integrals along the rays $\gamma_-$ and $\gamma_+$ are estimated as in case 1:
    \begin{multline*}
        \int_{\gamma_-} e^{-tF(z)} H'(z) \, dz + \int_{\gamma_+} e^{-tF(z)} H'(z) \, dz = (e^{\pi ix} - e^{-\pi ix})\int_{-\infty}^{-v} e^{-x \log |z| + tz-t\Delta} H'(z) \, dz = \\
        = O_{v^-, v^+, a, p}\left(\left| e^{\pi ix} - e^{-\pi ix}\right| \int_{-\infty}^{-v} e^{-x \log |z| + tz - t\Delta} \, dz\right) = O_{v^-, v^+, a, p}\left(\left|\int_{\gamma_-} e^{-tF(z)}\, dz + \int_{\gamma_+} e^{-tF(z)} \, dz\right|\right) = \\
        = O_{v^-, v^+, a, p}\left(\left|\int_{-\infty}^{(0+)} e^{-tF(z)}\, dz- \int_{\gamma_v} e^{-tF(z)} \, dz\right|\right) = O_{v^-, v^+, a, p}\left(\frac{e^{-t\Delta} t^{x-1}}{\Gamma(x)}\right).
    \end{multline*}

    Combining the resulting bounds for the integral over the three parts, we get the desired bound.
\end{proof}

\begin{proof}[Proof of Theorem~\ref{th-gamma}]
    To compute the given integral
    $$
    I = \frac{1}{2\pi i} \int_{-\infty}^{(0+)} a(\omega) \, e^{\omega t- p(\omega)x} \, d\omega = \frac{1}{2\pi i} \int_{-\infty}^{(0+)} a(\omega) \, e^{-t\L(\omega)} \, d\omega,
    $$
    we perform the conformal change of variable $z(\omega)$ given by Proposition~\ref{prop-gamma-substitution}:
    \begin{equation} \label{eq-gamma-change-of-variable}
        I = \frac{1}{2\pi i} \int_{-\infty}^{(0+)} a(\omega(z)) \, \omega'(z) \, e^{-t\L(\omega(z))} \, dz = \frac{1}{2\pi i} \int_{-\infty}^{(0+)} G(z) \, e^{-tF(z)} \, dz.
    \end{equation}
    Here, $z(\omega)\to -\infty$ as $\omega\to-\infty$ along the real axis, hence as $\omega\to-\infty$ in a neighborhood of $(-\infty;0]$ in $\mathbb{C}$. By Lemma~\ref{lm-gamma-int-exist}, we may assume that the integration contour lies in this neighborhood, and~\eqref{eq-gamma-change-of-variable} follows. Recall that $F(z) = v\log z - z + \Delta(v)$ and $G(z) = a(\omega(z)) \, \omega'(z)$, where $\Delta(v)$ is defined in~\eqref{eq-gamma-substitution}.

    We extract the main term by replacing function $G(z)$ with its value at the critical point $v$ of $F(z)$:
    $$
    I = \frac{1}{2\pi i} G(v) \int_{-\infty}^{(0+)} e^{-tF(z)} \, dz + \frac{1}{2\pi i} \int_{-\infty}^{(0+)} (G(z) - G(v)) \, e^{-tF(z)} \, dz =: \frac{1}{2\pi i} \left(G(v) \, I_1 + I_2\right).
    $$
    Here, $\frac{G(v)}{2\pi i}I_1$ is the anticipated main term, and $I_2$ is the error term. Using the definition of $H(z)$ in \eqref{eq-H-def}, integration by parts, and Lemma~\ref{lm-gamma-int}, we bound the latter as follows:
    \begin{gather*}
        I_2 = \int_{-\infty}^{(0+)} e^{-tF(z)} H(z) \, dF(z) = -\frac{1}{t} \int_{-\infty}^{(0+)} H(z) \, de^{-tF(z)} = \frac{1}{t} \int_{-\infty}^{(0+)} e^{-tF(z)} H'(z) \, dz = O_{v^-, v^+, a, p}\left(\frac{|I_1|}{t}\right).
    \end{gather*}
    Here, $H(z) = O_{v^-, v^+, a, p}(1)$ by Lemma~\ref{lm-gamma-H} and $\Re F(-\infty)=+\infty$ by~\eqref{eq-lm-gamma-F}, so $\lim_{z \to -\infty,z\in\mathbb{R}} H(z) \, e^{-tF(z)} = 0$. By Lemma~\ref{lm-gamma-int-exist}, we may assume that the same holds along the integration contour. Therefore,
    \begin{equation*}
        I = \frac{G(v)}{2\pi i} I_1 + O_{v^-, v^+, a, p}\left(\frac{|I_1|}{t}\right) = \frac{I_1}{2\pi i} \, \omega'(v) \left(a(\hat \omega) + O_{v^-, v^+, a, p}\left(\frac{1}{t}\right)\right)
    \end{equation*}
    because $\omega'(v) = \Omega_{v^-, v^+, p}(1)$ by the injectivity of $\omega(z)$ and compactness. By Lemma~\ref{lm-gamma-int} and~\eqref{eq-gamma-substitution}, we get
    $$
    \frac{I_1}{2\pi i} = \frac{t^{x-1} e^{-t\Delta(v)}}{\Gamma(x)} = \frac{t^{x-1} e^{-t\L(\hat \omega)}}{\Gamma(x)}\left(\frac{x}{t}\right)^x e^{-x} = \frac{e^{\hat \omega t - p(\hat \omega)x}}{t \, \Gamma(x)}\left(\frac{x}{e}\right)^x \qquad\text{and}\qquad \omega'(v) = \frac{1}{z'(\hat \omega)} = \frac{1}{|v|\sqrt{-p''(\hat \omega)}},
    $$
    and the desired asymptotic formula~\eqref{eq-gamma-asymp} follows.
\end{proof}

\begin{proof}[Proof of Corollary~\ref{cor-gamma}]
    For $v \ge \delta$, Stirling's formula yields
    $$
    \frac{(x / e)^x}{|x| \, \Gamma(x)} = \frac{1}{\sqrt{2\pi x}}\left(1 + O_\delta\left(\frac{1}{t}\right)\right),
    $$
    and the third case of the corollary follows. For $v \le -\delta$, using Euler reflection formula
    \begin{equation}
        \label{eq-gamma-euler}
        \Gamma(1 - x) \, \Gamma(x) = \frac{\pi}{\sin(\pi x)},
    \end{equation}
    the identities $\Gamma(1 - x) = -x \, \Gamma(-x)$ and $x^x = (-x)^x e^{\pi ix}$ for $x < 0$, and Stirling's formula, we get
    $$
    \frac{(x / e)^x}{|x| \, \Gamma(x)} = \left(\frac{x}{e}\right)^x \cdot \frac{\sin(\pi x) \, \Gamma(1 - x)}{\pi|x|} = \frac{e^{2\pi ix} - 1}{i\sqrt{-2\pi x}} \left(1 + O_\delta\left(\frac{1}{t}\right)\right),
    $$
    which proves the corollary in the first case.

    Consider the remaining case $|v| \le \delta / t$. Recall that $t \ge 1$. By Proposition~\ref{prop-gamma-substitution}, Lemma~\ref{lm-gamma-L}, and Remark~\ref{rem-omega-analyticity}, we have $\sgn \hat \omega = \sgn v$, $\hat \omega'(0) = 1$, $z'(\hat \omega(0)) = 1$, and $z'(\omega)\ne 0$. Hence, by compactness,
    $$
    \hat \omega(v) = v \exp\left(O_{\delta, v^-, v^+, p}(t^{-1})\right)
    \quad\text{and}\quad \sqrt{-v^2 p''(\hat \omega(v))} = z'(\hat \omega(v)) = \exp\left(O_{\delta, v^-, v^+, p}(t^{-1})\right),
    $$
    with the obvious meaning of $\exp\left(O_{\delta, v^-,v^+, p}(t^{-1})\right)$. This allows us to estimate the other quantities in~\eqref{eq-gamma-asymp}:
    \begin{gather*}
        e^{t \hat \omega(v)} = e^x \exp\left(O_{\delta, v^-, v^+, p}(t^{-1})\right), \qquad p(\hat \omega(v)) =
        \log \hat \omega(v) + q(0) + O_{\delta, v^-, v^+, p}(t^{-1}), \\
        \frac{(x / e)^x}{|x| \, \Gamma(x)} \cdot \frac{e^{\hat \omega(v) t- p(\hat \omega(v))x}}{\sqrt{-p''(\hat \omega(v))}} = \frac{x^x v^{-x} e^{-q(0)x} \exp\left(O_{\delta, v^-, v^+, p}(t^{-1})\right)}{t \, \Gamma(x) \sqrt{-v^2 p''(\hat \omega(v))}} = \frac{t^{x-1} e^{-q(0)x}}{\Gamma(x)} \left(1 + O_{\delta, v^-, v^+, p}(t^{-1})\right). \\[-1.2cm]
    \end{gather*}
\end{proof}

\subsection{Gamma phenomenon for beta function and its relatives}
\label{sec-gamma-beta}

As a toy example, let us apply Theorem~\ref{th-gamma} to the integral representation of the beta function \cite[Sec.~1.6]{Erdelyi-81}:
\begin{equation}
    \label{eq-beta-int}
    B(1 - x, t) = -\frac{1}{2i \sin(\pi (1 - x))} \int_1^{(0+)} (-u)^{-x} (1 - u)^{t-1} du = \frac{1}{2i \sin(\pi x)} \int_{-\infty}^{(0+)} e^{\omega t- \overbrace{\scriptstyle{\log(e^\omega - 1)}}^{p(\omega)}\cdot x} \, d\omega,
\end{equation}
where $x \not\in \mathbb{Z}$, $t > 0$, and $u = 1 - e^\omega$. Set $a(\omega) := 1$, $p(\omega) := \log(e^\omega - 1)$, and $D := \{\omega \in \mathbb{C} \colon -\pi < \Im \omega < \pi\}$.

For each $\omega \in D \cap H^+$, we have $\arg \omega, \arg(e^\omega - 1) \in (0; \pi)$. Thus $\left|\arg \frac{e^\omega - 1}{\omega}\right| < \pi$. The same holds for $\omega \in D \setminus \{0\}$; so the function $q(\omega) := p(\omega) - \log \omega = \log \frac{e^\omega - 1}{\omega}$ extended by $q(0) := 0$ is analytic in $D$. Clearly,
\begin{equation}
    \label{eq-gamma-beta-p'}
    p'(\omega) = 1 + (e^\omega - 1)^{-1}
\end{equation}
bijectively maps the domain $D \cap H^+$ onto $H^-$, and $p''(-\infty) = 0$ is finite.

Let us compute values~\eqref{eq-gamma-fns}. By~\eqref{eq-gamma-beta-p'}, we have $p'(-\infty) = 0$, thus $V^- = -\infty$. To compute $V^+$, note that
$$
\pi > \Im p(\omega) = \arg(e^\omega - 1) > \arg(e^\omega) = \Im \omega
$$
for all $\omega \in D \cap H^+$. Thus, $\limsup_{\omega \to \partial D} \Im p(\omega) / \Im \omega = 1$ and hence $V^+ = 1$. Assume $v := x / t < V^+$.

The value $\hat \omega$ is determined by the equation $p'(\hat\omega) = 1 + (e^{\hat \omega} - 1)^{-1} = 1 / v$, i.e., $\hat \omega = -\log(1 - v)$. Then,
$$
\frac{1}{\sqrt{-p''(\hat \omega)}} = \frac{|v|}{\sqrt{1 - v}} \qquad\text{and}\qquad e^{\hat \omega t- p(\hat \omega)x} = v^{-x} (1 - v)^{x-t}.
$$
For $v \in [v^-; v^+] \subset (-\infty; 1)$, by Theorem~\ref{th-gamma} and Euler reflection formula~\eqref{eq-gamma-euler}, we finally obtain
\begin{align}\label{eq-beta-asymptotics}
    B(1 - x, t) &= \frac{\pi}{\sin(\pi x)} \cdot \frac{(1 - v)^{x-t-1/2}}{e^x \, t^{1-x} \, \Gamma(x)} \left(1 + O_{v^-, v^+}\left(\frac{1}{t}\right)\right) = \frac{\Gamma(1 - x) \, t^{t-1/2}}{e^x (t - x)^{t-x+1/2}} \left(1 + O_{v^-, v^+}\left(\frac{1}{t}\right)\right)
\end{align}
for each $x \not\in \mathbb{Z}$, $t > 0$, and $v^- < v^+ < 1$ such that $x / t \in [v^-; v^+]$.

Note that the same asymptotic formula follows from the identity
$
B(t, 1 - x) = \frac{\Gamma(t) \, \Gamma(1 - x)}{\Gamma(t - x + 1)}
$
and Stirling's formula applied to $\Gamma(t)$ and $\Gamma(t - x + 1)$. This toy example can be interpreted as the gamma phenomenon for the probability of a simple random walk on $\mathbb{Z}$ starting at $0$ to hit $t - 2n$ after $t$ steps:
$$
P(n, t) := \frac{1}{2^t} \binom{t}{n} = \frac{\sqrt{t} \, (t / 2e)^t}{\Gamma(t - n + 1) \, \sqrt{n} \, (n / e)^n} \left(1 + O_{v^-}\left(\frac{1}{t}\right)\right), \qquad n / t \ge v^- > 0.
$$

Remarkably, a slight variation of this argument leads to completely new results.

\begin{remark}
    \label{rm-2F1}
    Gauss hypergeometric function has integral representation similar to \eqref{eq-beta-int} (cf.~\cite[Sec.~15.6.4]{NIST-10}):
    \begin{align}
        \label{eq-hypergeometric}
        {}_2F_1(a, b; c; \xi) &= \frac{i \, \Gamma(1 - b) \, \Gamma(c)}{2\pi \cdot \Gamma(c - b)} \int_1^{(0+)} \frac{(-u)^{b-1} (1 - u)^{c-b-1}}{(1 - \xi u)^a} \, du = \frac{\Gamma(x) \, \Gamma(t - x + 1)}{2\pi i \cdot \Gamma(t)} \int_{-\infty}^{(0+)} a(\omega) \, e^{\omega t- p(\omega)x} \, d\omega,
    \end{align}
    where $a \in \mathbb{R}$, $b < c$, $b \neq 1, 2, \ldots$, $c \neq 0, -1, -2, \ldots$, $\xi \in (-\infty; 1)$, $u = 1 - e^\omega$, $t = c - b$, $x = 1 - b$, $a(\omega) = (1 - \xi(1 - e^\omega))^{-a}$, and $p(\omega) = \log(e^\omega - 1)$ is the same as in \eqref{eq-beta-int}. The new $a(\omega)$ is analytic in the same domain $D$ for $\xi \in [0; 1)$. The limits $a(-\infty)$ and $a'(-\infty)$ are finite. We can again apply Theorem~\ref{th-gamma} because the assumptions on $p(\omega)$ have already been verified. Dropping the factor $\pi/\sin(\pi x)$ and inserting $a(\hat\omega)=\Omega_{a, \xi, v^-, v^+}(1)$ into~\eqref{eq-beta-asymptotics}, we get an asymptotic formula, which is uniform in $b$ and $c$ (but not $a$ and~$\xi$):
    \begin{equation}
        \label{eq-hypergeometric-asymp}
        {}_2F_1(a, b; c; \xi) = \frac{\Gamma(c) (c - b)^{c-b-1/2} (c - 1)^{a-c+1/2} e^{b-1}}{\Gamma(c - b) (c - 1 + \xi(1 - b))^a} \left(1 + O_{a, \xi, v^-, v^+}\left(\frac{1}{c - b}\right)\right),
    \end{equation}
    which holds for $a \in \mathbb{R}$, $b < c$, $b \neq 1, 2, \ldots$, $(1 - b) / (c - b) \in [v^-; v^+] \subset (-\infty; 1)$, and $\xi \in [0; 1)$. This complements the non-uniform formulae~\cite[Sec.~15.12.3, 15.12.7]{NIST-10} involving more complicated special functions.
\end{remark}

\subsection{Gamma phenomenon for quantum walks (proof of Theorem \ref{th-gamma-checkers})}
\label{sec-gamma-checkers}

\begin{proof}[Proof of Proposition~\ref{prop-checkers-int-damp}]
    If $x > t$ and $x - t$ is odd, then $\widetilde A_1(x, t, m) = 0$ by Definition~\ref{def-checkers}. The right side of \eqref{eq-A1-int-damp} also vanishes because the integrand extends analytically to $(-\infty;0]$ and vanishes at $-\infty$. For the same reasons, if $x > t$ and $x - t$ is even, then $\widetilde A_2(x, t, m) = 0$ as well as the right side of \eqref{eq-A2-int-damp}.

    In other cases, we use the relations~\cite[Proposition~3]{SkoUst-24}:
    \begin{align*}
        \widetilde A_1(x, t, m) &= \frac{i(-im)^{t-x}}{(1 + m^2)^{t/2}} \binom{(t + x - 1) / 2}{x} {}_2F_1\left(\frac{x - t + 1}{2}, \frac{x - t + 1}{2}; 1 + x; -\frac{1}{m^2}\right), \\
        \widetilde A_2(x, t, m) &= \frac{(-im)^{t-x}}{(1 + m^2)^{t/2}} \binom{(t + x) / 2}{x} {}_2F_1\left(\frac{x - t}{2}, \frac{x - t + 2}{2}; 1 + x; -\frac{1}{m^2}\right),
    \end{align*}
    which are valid for $x, t \in \mathbb{Z}^2$, $x, t \ge 0$, and $m > 0$. Using integral representation~\eqref{eq-hypergeometric} for the Gauss hypergeometric function, we obtain
    \begin{align*}
        \widetilde A_1(x, t, m) &= -\frac{1}{2\pi} \cdot \frac{(-im)^{t-x}}{(1 + m^2)^{t/2}} \cdot \int_1^{(0+)} \frac{(-u)^{(x-t-1)/2} (1 - u)^{(x+t-1)/2}}{(1 + u / m^2)^{(x-t+1)/2}} \, du, \\
        \widetilde A_2(x, t, m) &= \frac{1}{2\pi i} \cdot \frac{x + t}{x - t} \cdot \frac{(-im)^{t-x}}{(1 + m^2)^{t/2}} \cdot \int_1^{(0+)} \frac{(-u)^{(x-t)/2} (1 - u)^{(x+t-2)/2}}{(1 + u / m^2)^{(x-t)/2}} \, du.
    \end{align*}
    The substitution $u = 1 - e^\omega$ completes the proof.
\end{proof}

\begin{proof}[Proof of Theorem \ref{th-gamma-checkers}]
    Let us first prove the asymptotic formula for $\widetilde A_1(x, t, m)$. Rewrite integral~\eqref{eq-A1-int-damp} as
    \begin{align}
        \label{eq-A1-int-damp-new}
        \widetilde A_1(x, t, m) = \frac{i^{x-t+1} \, m}{(1 + m^2)^{t/2}} \cdot \frac{1}{2\pi i} \int_{-\infty}^{(0+)} a(\omega) \, e^{\omega \tau -  p(\omega)\chi} \, d\omega,
    \end{align}
    where $\tau := (t + x + 1) / 2$, $\chi := (t - x + 1) / 2$, $a(\omega) := \frac{1}{1 + m^2 - e^\omega}$, and $p(\omega) := \log(e^\omega - 1) - \log(1 + m^2 - e^\omega)$.

    Note that for integer $\chi\le 0$, the right side of~\eqref{eq-A1-gamma-asymp} vanishes due to the gamma factor, as does the right side of \eqref{eq-A1-int-damp-new} because the integrand extends analytically to $(-\infty; 0]$ and vanishes at $-\infty$. Thus, we may assume $\chi$ is not a nonpositive integer. We may also assume $\delta < 1$ and $\tau > 4 / \delta + \sqrt{1 + m^2}$, otherwise~\eqref{eq-A1-gamma-asymp} is automatic because for each fixed $\delta$ and $m$, there are only finitely many $x$ and $t$ satisfying $\tau \le 4 / \delta + \sqrt{1 + m^2}$.

    We are going to apply Theorem~\ref{th-gamma} to estimate the integral in~\eqref{eq-A1-int-damp-new}. Consider the domain
    \begin{align*}
        D := \{\omega \in \mathbb{C} \colon \Re \omega < \log \sqrt{1 + m^2}, -\pi < \Im \omega < \pi\}.
    \end{align*}
    Each $\omega \in D \cap H^+$ has the form $\omega = \log \sqrt{1 + m^2} + iu$, where $u \in U$ and $U$ is given by \eqref{def-U}. The derivative
    \begin{align}
        \label{eq-p'}
        p'(\omega) &= \frac{e^\omega}{e^\omega - 1} + \frac{e^\omega}{1 + m^2 - e^\omega}, & p'\left(\log \sqrt{1 + m^2} + iu\right) = \frac{m^2}{2 + m^2 - 2\sqrt{1 + m^2} \cos u}
    \end{align}
    is a 1--1 conformal mapping of $D \cap H^+$ to the lower half-plane, because the cosine function maps $U$ bijectively to $H^-$ (cf. Fig.~\ref{fig-gamma-p'-inner}). As it was explained in Sec.~\ref{sec-gamma-beta}, the function $\log \frac{e^\omega - 1}{\omega}$ extended by $0$ at $0$ is analytic in the strip $|\Im \omega| < \pi$. Also, $\Re \left(1 + m^2 - e^\omega\right) > 0$ for $\omega \in D$. Therefore, the function $q(\omega) := p(\omega) - \log \omega = \log \frac{e^\omega - 1}{\omega} - \log\left(1 + m^2 - e^\omega\right)$ is analytic in $D$. Finally, we have $a(-\infty) = \frac{1}{1 + m^2}$ and $a'(-\infty) = p''(-\infty) = 0$, hence all the assumptions of Theorem~\ref{th-gamma} are satisfied.

    We have $V^- = -\infty$ because $p'(-\infty) = 0$ by \eqref{eq-p'}. To compute $V^+$, let us find two limits: $\omega \to (\pm \pi i-\infty;\pm \pi i+\log \sqrt{1 + m^2}]$ and $\omega \to [\log \sqrt{1 + m^2}-\pi i;\log \sqrt{1 + m^2}+\pi i]$. In the first case, note that
    $$
    |\Im p(s + \pi i) - \Im p(s + \pi i - i\varepsilon)| \le \max_{\substack{\pi - \varepsilon\le \Im \omega \le \pi \\ \Re\omega\le \log \sqrt{1 + m^2}}}
    |p'(\omega)| \cdot \varepsilon \quad\text{for } \varepsilon \in (0; \pi)
    \text{ and }s\le \log \sqrt{1 + m^2},
    $$
    $\Im p(s + \pi i)=\pi$, and $p'(\omega)$ is bounded in the complement of any neighbourhood of $0$ in $D \cap H^+$. Thus,
    \begin{align}
        \label{eq-gamma-limsup-1}
        \limsup_{\omega \to(\pm \pi i-\infty;\pm \pi i+\log \sqrt{1 + m^2}]} \frac{\Im p(\omega)}{\Im \omega} = \frac{\pm \pi}{\pm \pi} = 1.
    \end{align}
    In the second case, since $p'$ is real and decreasing on $[\log \sqrt{1 + m^2}; \log \sqrt{1 + m^2} + \pi i]$ by \eqref{eq-p'} (cf. Fig.~\ref{fig-gamma-p'-boundary}), for each $u \in (0; \pi]$, we have
    $$
    \Im p\left(\log \sqrt{1 + m^2} + iu\right) = \int_{0}^{u} p'\left(\log \sqrt{1 + m^2} + iy\right) dy \le p'\left(\log \sqrt{1 + m^2}\right) \cdot u = \frac{\sqrt{1 + m^2} + 1}{\sqrt{1 + m^2} - 1} \cdot u,
    $$
    and moreover, this estimate becomes sharp in the limit $u \to 0$. Since the convergence of the ratio $\Im p(\omega) / \Im \omega$ (as $\omega \to \log \sqrt{1 + m^2} + iu$) is uniform in $u$ by compactness, we have
    \begin{align}
        \label{eq-gamma-limsup-2}
        \limsup_{\substack{\omega \to [\log \sqrt{1 + m^2}-\pi i; \log \sqrt{1 + m^2}+\pi i], \\ \Im \omega \neq 0}} \frac{\Im p(\omega)}{\Im \omega} = \frac{\sqrt{1 + m^2} + 1}{\sqrt{1 + m^2} - 1}.
    \end{align}
    Combining the equations~\eqref{eq-gamma-limsup-1} and~\eqref{eq-gamma-limsup-2}, we get
    \begin{align*}
        V^+ = \max\left(1, \frac{\sqrt{1 + m^2} + 1}{\sqrt{1 + m^2} - 1}\right)^{-1} = \frac{\sqrt{1 + m^2} - 1}{\sqrt{1 + m^2} + 1}.
    \end{align*}

    Let
    \begin{align}
        \label{eq-gamma-velocity}
        \nu := \frac{\chi}{\tau} = \frac{t - x + 1}{t + x + 1} = \frac{1 - v}{1 + v} + \frac{v}{(1 + v) \tau}, \quad\text{where}\quad v := \frac{x}{t}.
    \end{align}
    The assumption $v\ge 1/\sqrt{1+m^2}+\delta$ implies $\nu \in [\nu^-; \nu^+] := [-1; V^+ (1 - \delta)]$, because $\tau > 1 / \delta + \sqrt{1 + m^2}$.

    Theorem~\ref{th-gamma}, applied to the integral in~\eqref{eq-A1-int-damp-new}, yields
    \begin{align}
        \label{eq-A1-gamma-asymp-new}
        \widetilde A_1(x, t, m) = \frac{i^{x-t+1} \, m}{(1 + m^2)^{t/2}} \cdot \frac{(\chi / e)^\chi}{|\chi| \, \Gamma(\chi)} \cdot \frac{e^{\hat \omega \tau -  p(\hat \omega)\chi}}{\sqrt{-p''(\hat \omega)}} \cdot \left(a(\hat \omega) + O_{m, \delta}\left(\tau^{-1}\right)\right).
    \end{align}
    Let us compute all the values appearing in \eqref{eq-A1-gamma-asymp-new}. The equation $p'(\hat \omega) = 1 / \nu$ defining $\hat \omega$ is quadratic in $e^{\hat \omega}$:
    \begin{align}
        \label{eq-hat-omega}
        e^{2\hat \omega} - e^{\hat \omega} (2 + m^2(1 - \nu)) + 1 + m^2 = 0.
    \end{align}
    Its roots are
    \begin{align}\label{eq-root-hat-omega}
        e^{\hat \omega_\pm} = 1 + \frac{m^2(1 - \nu) \pm m\sqrt \Delta}{2}, \quad\text{where}\quad \Delta = \Delta_\nu := m^2(1 - \nu)^2 - 4\nu.
    \end{align}
    Since $\Delta_{V^+} = \Delta_{1 / V^+} = 0$, it follows that $\Delta > 0$ for $\nu < V^+$. Therefore, there are two possible real values for $\hat \omega$, the smallest of which, $\hat \omega = \hat \omega_-$, lies in the domain $D$. Using the equivalent forms $\frac{e^{\hat \omega} - 1}{\nu} = \frac{m^2 e^{\hat \omega}}{1 + m^2 - e^{\hat \omega}}$ and $me^{\hat \omega} \sqrt{\Delta} = 1 + m^2 - e^{2\hat \omega}$ of~\eqref{eq-hat-omega}--\eqref{eq-root-hat-omega}, we compute
    \begin{align}
        \label{eq-gamma-p''}
        p''(\hat \omega) &= e^{\hat \omega} \left(\frac{1 + m^2}{(1 + m^2 - e^{\hat \omega})^2} - \frac{1}{(e^{\hat \omega} - 1)^2}\right) = -\frac{\sqrt \Delta}{m\nu^2}, \\
        a(\hat \omega) \, e^{\tau \hat \omega - \chi p(\hat \omega)} &= e^{\hat \omega \tau} \left(e^{\hat \omega} - 1\right)^{-\chi} \left(1 + m^2 - e^{\hat \omega}\right)^{\chi-1} =
        \label{eq-gamma-a-exp}
        m^{x-t-1} \nu^{(x-t-1)/2} e^{\hat \omega x} \left(1 + m^2 - e^{\hat \omega}\right)^{t-x}.
    \end{align}
    Let $d = d_v := (1 + m^2)v^2 - 1 > 2\delta > 0$. Using expansion~\eqref{eq-gamma-velocity}, we can rewrite~\eqref{eq-root-hat-omega}--\eqref{eq-gamma-a-exp} as follows:
    \begin{align*}
        \Delta &= \tfrac{4d}{(1 + v)^2} - \tfrac{4(d + v + 1)}{(1 + v)^2 \tau} + O_m\left(\tfrac{1}{\tau^{2}}\right), &
        e^{\hat \omega} &= \tfrac{m + \sqrt d}{mv+ \sqrt d} \left(1 + \tfrac{mv}{2\sqrt d \, \tau} + O_{m, \delta}\left(\tfrac{1}{\tau^{2}}\right)\right), \\
        \sqrt \Delta &= \tfrac{2\sqrt d}{1 + v} \left(1 - \tfrac{d + v + 1}{2d\tau}\right) + O_{m, \delta}\left(\tfrac{1}{\tau^{2}}\right), &
        1 + m^2 - e^{\hat \omega} &= \tfrac{m^2 + m\sqrt d}{1 + v} \left(1 - \tfrac{(v+1)v}{2\sqrt d(\sqrt d+mv)\tau} + O_{m, \delta}\left(\tfrac{1}{\tau^{2}}\right)\right), \\
        p''(\hat \omega) &= -\tfrac{2\sqrt d}{(1 + v)m\nu^2} \left(1 + O_{m, \delta}\left(\tfrac{1}{\tau}\right)\right), &
        a(\hat \omega) \, e^{ \hat \omega \tau -  p(\hat \omega)\chi} &= \tfrac{\nu^{(x-t-1)/2}(\sqrt d + m)^te^{v/(1+v)}\cdot\left(1 + O_{m, \delta}\left({1}/{\tau}\right)\right)}{m(\sqrt d + mv)^x(1 + v)^{t-x}},
    \end{align*}
    because $\Delta, \tfrac{\sqrt d}{1 + v}, e^{\hat \omega}, 1 + m^2 - e^{\hat \omega} = \Omega_{m, \delta}(1)$ for $\nu \in [\nu^-; \nu^+]$ by compactness, $1 + \tfrac{m^2 v - m\sqrt d}{1 + v} = \tfrac{m + \sqrt d}{mv + \sqrt d}$, and
    \begin{align*}
        \frac{(v + 1)v}{2\sqrt d (\sqrt d + mv) \tau} \le \frac{1 + \sqrt{1 + m^2}}{2\sqrt{\delta / 2}(\sqrt{\delta / 2} + m)(4 / \delta + \sqrt{1 + m^2})} \le \frac{1}{2} \qquad\text{because}\qquad d\ge \frac{1}{2} \delta v^2.
    \end{align*}

    Substituting these values into \eqref{eq-A1-gamma-asymp-new} and applying the bounds $a(\hat \omega), p''(\hat \omega) = \Omega_{m, \delta}(1)$ and the estimate
    \begin{align*}
        \tau^{\chi-1} = \frac{1}{\sqrt t} \cdot \sqrt{\frac{2}{1 + v}} \cdot \left(\frac{t + x}{2}\right)^{(t-x)/2} \cdot \exp\left(\frac{1 - v}{2(1 + v)}\right) \cdot \left(1 + O\left(\frac{1}{\tau}\right)\right),
    \end{align*}
    we get an asymptotic formula for $\widetilde A_1(x, t, m)$ equivalent to \eqref{eq-A1-gamma-asymp}:
    \begin{align}
        \label{eq-A1-gamma-asymp-simp}
        \widetilde A_1(x, t, m) = \frac{i^{x-t+1} \, \sqrt m \, (2e(1 + v))^{(x-t)/2} \left(\sqrt{(1 + m^2)v^2 - 1} + m\right)^t \cdot \left(1 + O_{m, \delta}\left((x + t)^{-1}\right)\right)}{(1 + m^2)^{t/2} \, t^{(x-t+1)/2} \, \sqrt[4]{(1 + m^2)v^2 - 1} \left(\sqrt{(1 + m^2)v^2 - 1} + mv\right)^x \Gamma((t - x + 1) / 2)}.
    \end{align}

    The proof of the asymptotic formula for $\widetilde A_2(x, t, m)$ follows the same scheme. Theorem~\ref{th-gamma} should be applied to integral~\eqref{eq-A2-int-damp} with $\tau := (t + x) / 2$, $\chi := (t - x) / 2 \ne 0$, and $a(\omega) := 1$ (for $\chi = 0$, both $\widetilde A_2(t, t, m)$ and its approximation from Theorem~\ref{th-gamma-checkers} evaluate to $(1 + m^2)^{-t/2}$). The domain $D$ and the function $p(\omega)$, hence $V^-$ and $V^+$, remain the same. The remaining calculations are carried out analogously.
\end{proof}

\begin{corollary}
    \label{cor-gamma-checkers}
    The last three rows of Tab.~\ref{tab-checkers-asymp} provide correct asymptotic formulae for $\widetilde A_1(x, t, m)$ for each $\delta, m > 0$ and integer $x, t \ge 0$ satisfying the corresponding inequality in the first column.
\end{corollary}

\begin{proof}
    The formula in row~7 follows from Corollary~\ref{cor-gamma} applied to representation~\eqref{eq-A1-int-damp-new}. In this case, $a(0) = 1 / m^2$, $q(0) = -2\log m$, and $\tau^{\chi - 1} = t^{(t-x-1)/2} (1 + O_\delta(t^{-1}))$.

    The formula in row~8 follows from Theorem~\ref{th-gamma-checkers} and the estimate
    $$
    \frac{1}{\Gamma((t - x + 1) / 2)} \left(\frac{|t - x|}{2e}\right)^{(t-x)/2} = \sqrt{\frac{2}{\pi}} \cdot (-1)^{\frac{x-t}{2}} \left(1 + O\left(\frac{1}{x - t}\right)\right) = \sqrt{\frac{2}{\pi}} \cdot (-1)^{\frac{x-t}{2}} \left(1 + O_\delta\left(\frac{1}{x + t}\right)\right),
    $$
    which is a consequence of the reflection formula~\eqref{eq-gamma-euler}, Stirling's formula, and $(1 + 1 / n)^n = e + O(1 / n)$.

    The formula in row~9 can be proven by directly repeating all the steps of the proof of Theorem~\ref{th-gamma-checkers}. Since for $t = 0$ we have $\nu = -1 + 2 / (x + 1) \in [\nu^-; \nu^+]$, it follows that formula~\eqref{eq-A1-gamma-asymp-new} and computations~\eqref{eq-hat-omega}--\eqref{eq-gamma-a-exp} remain valid. After that, we estimate expressions~\eqref{eq-gamma-p''}--\eqref{eq-gamma-a-exp} as follows:
    \begin{align*}
        p''(\hat \omega) &= -\frac{2\sqrt{1 + m^2}}{m} \left(1 + O_m\left(\frac{1}{x}\right)\right), &
        a(\hat \omega) \, e^{\tau \hat \omega - \chi p(\hat \omega)} &= \frac{i^{x-1}}{m} \left(\sqrt{1 + m^2} - m\right)^x \left(1 + O_m\left(\frac{1}{x}\right)\right).
    \end{align*}
    By the reflection formula~\eqref{eq-gamma-euler} and Stirling's formula, we have
    \begin{align*}
        \frac{(\chi / e)^\chi}{|\chi| \, \Gamma(\chi)} = \frac{2i}{\sqrt{\pi x}} \left(1 + O\left(\frac{1}{x}\right)\right).
    \end{align*}
    Substituting the obtained expressions into \eqref{eq-A1-gamma-asymp-new}, we get the required asymptotic formula.
\end{proof}

\begin{proof}[Proof of Corollary~\ref{cor-A2-A1}]
    To prove \eqref{eq-cor-A2-A1-Airy}, use Theorem~\ref{th-airy-checkers} to express $\widetilde A_2(x, t, m)$ in terms of $\widetilde A_1(x, t, m)$. For $v := x / t \le \hat v$, the estimate follows trivially. For $v > \hat v$, choose $C = C(m)$ such that $\widetilde A_1(x, t, m) \neq 0$ for all $t > C$ and $v \in (\hat v; 1)$; this is possible by Theorems~\ref{th-airy-checkers}~and~\ref{th-gamma-checkers} because $\theta(v,m)<0$ for $v > \hat v$. Then the resulting $(1 + O_{m, \delta}(t^{-1}))^{-1}$ factor is well-defined for all $t > C$ and equals $1 + O_{m, \delta}(t^{-1})$ by expanding the function $z \mapsto 1 / (1 + z)$ around zero, where $|z| < 1 / 2$ except for finitely many pairs $(x, t)$.

    To prove \eqref{eq-cor-A2-A1-Gamma}, fix $\delta = \delta(m) \in (0; 1 - \hat v)$. For $0 \le v \le 1 - \delta$, use Stirling's formula and write
    \begin{equation*}
        \frac{\Gamma((t - x + 1) / 2)}{\Gamma((t - x + 2) / 2)} \sqrt{\frac{t + x}{2}} = \sqrt{\frac{1 + v}{1 - v}} \left(1 + O_\delta\left(\frac{1}{t}\right)\right).
    \end{equation*}
    Eq.~\eqref{eq-cor-A2-A1-Gamma} then follows from~\eqref{eq-cor-A2-A1-Airy} because $\widetilde A_1(x, t, m) = O_{m, \delta}(t^{-1/3})$, which is a consequence of Theorem~\ref{th-airy-checkers} (cf.~Proposition~\ref{prop-maximum} below). For $1 - \delta < v < 1$, the desired estimate is a direct consequence of Theorem~\ref{th-gamma-checkers} and a similar bound $(1 + O_{m, \delta}((x + t)^{-1}))^{-1} = 1 + O_{m, \delta}(t^{-1})$ for $t > C$.
\end{proof}

\section{The peak}
\label{sec-peak}

In this section, we provide an asymptotic formula for the maximum of the wave function of the Hadamard walk for the first time. We also provide an accurate estimate of the position at which the maximum is attained. This improves a rough upper bound for the maximum from \cite[Theorem 2.1]{Maeda-etal-18}, a rough estimate of its position from \cite{Ambainis-etal-01}, and solves an open problem posed in \cite[Problem~1]{SkoUst-22}. For simplicity, we restrict ourselves to the Hadamard walk (case $m = 1$) and to the real part of the wave function ($\widetilde A_1(x, t)$). The same technique is applicable to the imaginary part $\widetilde A_2(x, t)$ and to the probability $|\widetilde A_1(x, t)|^2 + |\widetilde A_2(x - 1, t)|^2$.

\begin{figure}[htbp]
    \centering
    \begin{subfigure}[c]{0.4\textwidth}
        \includegraphics[width=\textwidth]{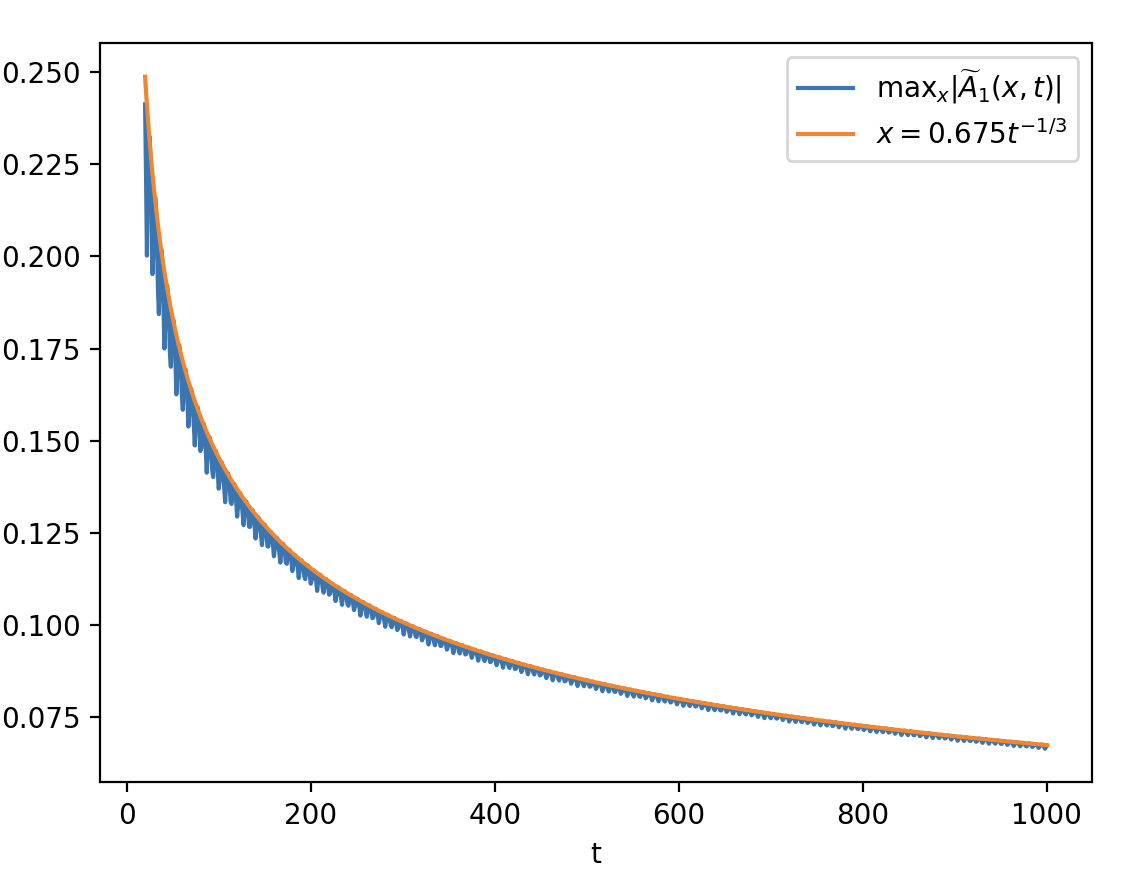}
        \caption{$\max_{x \in [0; t]} |\widetilde A_1(x, t)|$}
        \label{fig-peak-value}
    \end{subfigure}
    \begin{subfigure}[c]{0.4\textwidth}
        \includegraphics[width=\textwidth]{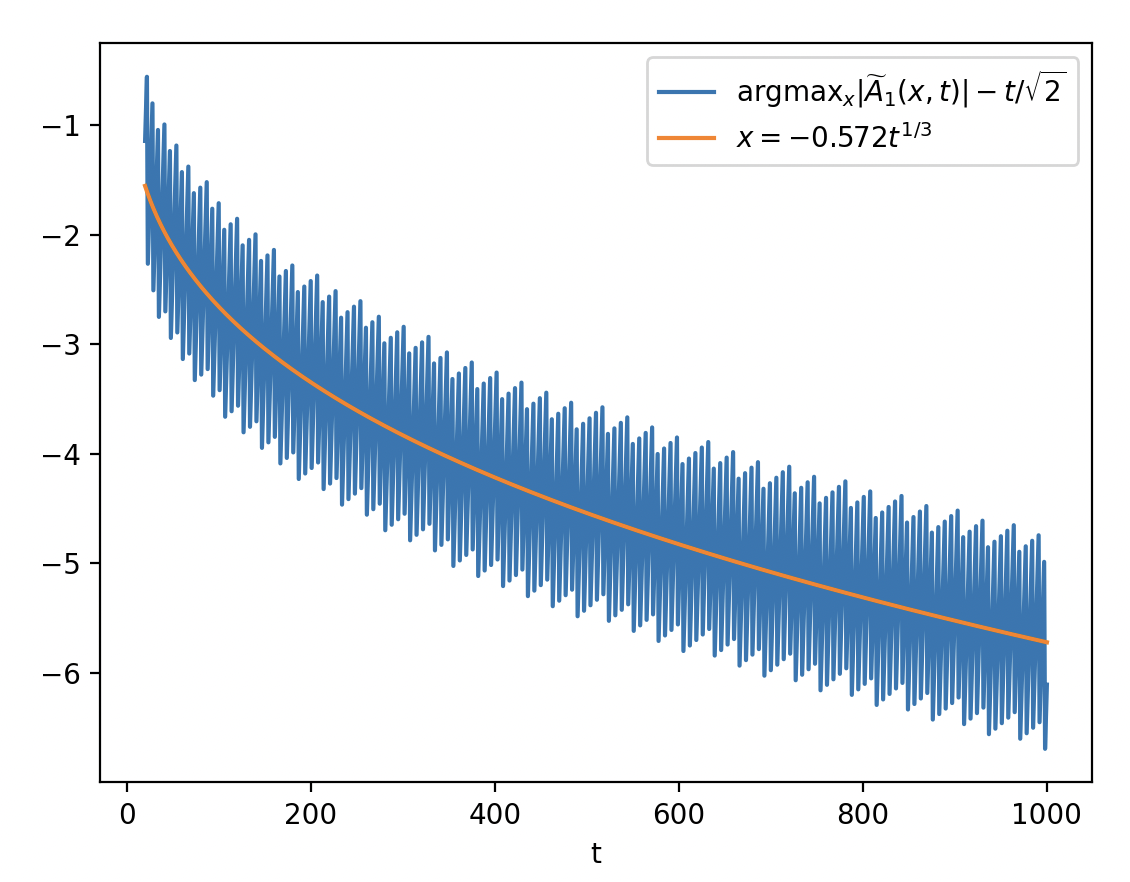}
        \caption{$\argmax_{x \in [0; t]} |\widetilde A_1(x, t)| - t / \sqrt{2}$}
        \label{fig-peak-position}
    \end{subfigure}
    \caption{Peak value and the deviation of its position from the turning point. Orange depicts our approximations from Proposition~\ref{prop-maximum}, where $2^{1/3} \cdot \max_x \Ai(x) \approx 0.675$ and $2^{-5/6} \cdot \argmax_x \Ai(x) \approx -0.572$.}
    \label{fig-peak}
\end{figure}

As seen in Fig.~\ref{fig-checkers-a1-between}, for a fixed $t$, the function $|\widetilde A_1(x, t)|$ has symmetric maxima located near $x = \pm t / \sqrt{2}$. Figs.~\ref{fig-peak-value}--\ref{fig-peak-position} illustrate the value of the right maximum and the deviation of its position from the line $x = t / \sqrt{2}$ as $t$ increases, along with our approximation from Proposition~\ref{prop-maximum} below.

\begin{proposition}
    \label{prop-maximum}
    For each integer $t > 0$, we have
    \begin{align*}
        \max_{\substack{x \in [0; t] \\ x + t \text{ odd}}} |\widetilde A_1(x, t)| &= \max_{x \in \mathbb{R}} \Ai(x) \cdot 2^{1/3} \, t^{-1/3} + O\left(\frac{1}{t}\right), \\
        \argmax_{\substack{x \in [0; t] \\ x + t \text{ odd}}} |\widetilde A_1(x, t)| &= \frac{t}{\sqrt{2}} + \argmax_{x \in \mathbb{R}} \Ai(x) \cdot 2^{-5/6} \, t^{1/3} + O(1).
    \end{align*}
\end{proposition}

Hereafter, to avoid ambiguity, $\argmax f(x)$ is understood as the minimal $\hat x$ such that $f(\hat x)=\max f(x)$.

This is a sample application of the uniform asymptotic formula from Theorem~\ref{th-airy-checkers} (it can also be deduced from a simpler result \cite[Theorem~1]{Zakorko} combined with an elementary monotonicity result \cite[Theorem~3]{KuyanovSlizkov-23}). The asymptotic formulae
known before are helpless for a rigorous proof because the wave function can a priori have a sharp peak in one of
the uncovered white regions in Fig.~\ref{fig-intro-domains-before}. We need the following lemmas.

\begin{lemma}
    \label{lm-maximum-outside}
    There is $\delta > 0$ such that for all integer $t>0$ and $x \in [0; t] \setminus (t / \sqrt{2} - \delta t^{1/3}; t / \sqrt{2} + \delta t^{1/3})$ such that $x + t$ odd, we have
    \begin{align}
        \label{eq-maximum-outside}
        |\widetilde A_1(x, t)| \le t^{-1/3} \Ai(0) + O_\delta\left(\frac{1}{t}\right).
    \end{align}
\end{lemma}

\begin{proof}
    Assume $v := x / t \in [0; 0.9]$. Then, Theorem~\ref{th-airy-checkers} implies the following simplified asymptotic formula:
    \begin{align}
        \label{eq-maximum-A-f}
        |\widetilde A_1(x, t)| = \frac{1}{t^{1/3}}f(v) |\Ai(-\theta(v) \, t^{2/3})| + O\left(\frac{1}{t}\right), \qquad\text{where } \theta(v) := \theta(v, 1), \quad f(v) := \left(\frac{4\theta(v)}{1 - 2v^2}\right)^{1/4}.
    \end{align}
    By Remark~\ref{rm-airy-analyticity}, $f(v)$ and $\theta'(v)$ are continuous on $[0; 0.9]$. By Remark~\ref{rm-airy-theta}, $\theta'(v) < 0$. Thus, there exists $M > 0$ such that $|f(v)| < M$ and $|\theta'(v)| > 1 / M$. Since $\Ai(\pm \infty) = 0$ and $\Ai(0) > 0$, there is $u_0 > 0$ such that $|\Ai(u)| < \Ai(0) / M$ for $|u| \ge u_0$. Take any $\delta \ge u_0 M$. Then for $|x - t / \sqrt{2}| \ge \delta t^{1/3}$, we have
    \begin{align*}
        |\theta(v) \, t^{2/3}| = t^{2/3} \left|\int_{1 / \sqrt{2}}^v \theta'(u) \, du\right| \ge \frac{t^{2/3} |v - 1 / \sqrt{2}|}{M} = \frac{|x - t / \sqrt{2}|}{t^{1/3} M} \ge \frac{\delta}{M} \ge u_0.
    \end{align*}
    By inserting  the resulting estimates into~\eqref{eq-maximum-A-f}, bound~\eqref{eq-maximum-outside} follows for $x / t \in [0; 0.9]$. For $x / t \in (0.9; 1]$, it follows from the estimate $\widetilde A_1(x, t) = O(t^{-1})$, which is a consequence of the first derivative test \cite[Lemma~5.1.2]{Huxley-96} applied to integral~\eqref{eq-A1-int-osc} (see \cite[Remark~9]{SkoUst-22}).
\end{proof}

\begin{lemma}
    \label{lm-maximum-arg}
    There exists $\delta > 0$ such that for each integer $t > 0$, we have
    \begin{align}
        \label{eq-maximum-arg}
        \argmax_{\substack{x \in [0; t] \\ x + t \text{ odd}}} |\widetilde A_1(x, t)| \in [t / \sqrt{2} - \delta t^{1/3}; t / \sqrt{2} + \delta t^{1/3}].
    \end{align}
\end{lemma}

\begin{proof}
    By Lemma~\ref{lm-maximum-outside}, there exists $\delta > 0$ such that for each non-negative $x \not\in (t / \sqrt{2} - \delta t^{1/3}; t / \sqrt{2} + \delta t^{1/3})$, Eq.~\eqref{eq-maximum-outside} holds. Assume without loss of generality $\delta > 2$. Consider the sequence
    \begin{align*}
        x_t := \max \{x \in \mathbb{Z} \mid x / t < 1 / \sqrt{2}, x + t \text{ odd}\} \in (t / \sqrt{2} - \delta t^{1/3}; t / \sqrt{2}).
    \end{align*}
    By Corollary~\ref{cor-airy-checkers} (see row~4 of Tab.~\ref{tab-checkers-asymp}),
    \begin{align}
        \label{eq-maximum-custom-x}
        |\widetilde A_1(x_t, t)| = 2^{1/3} \, t^{-1/3} (\Ai(0) + O(t^{-1/3})) + O_\delta\left(\frac{1}{t}\right) = 2^{1/3} \, t^{-1/3} \Ai(0) + O_\delta\left(t^{-2/3}\right).
    \end{align}
    Comparing \eqref{eq-maximum-outside} and \eqref{eq-maximum-custom-x}, we conclude that the desired maximum point belongs to $[t / \sqrt{2} - \delta t^{1/3}; t / \sqrt{2} + \delta t^{1/3}]$ for sufficiently large $t$. By increasing $\delta$, we may assume this holds for all $t$.
\end{proof}

\begin{proof}[Proof of Proposition \ref{prop-maximum}]
    By Lemma~\ref{lm-maximum-arg}, we can restrict ourselves to $x \in [t / \sqrt{2} - \delta t^{1/3}; t / \sqrt{2} + \delta t^{1/3}]$ and $t > 2^{3/4} \delta^{3/2}$ for some $\delta > 1$. Let us normalize this interval and the wave function by a linear transformation:
    \begin{align*}
        S_t &:= \left\{2^{5/6} \, t^{-1/3} (x - t / \sqrt{2}) \colon x \in [t / \sqrt{2} - \delta t^{1/3}; t / \sqrt{2} + \delta t^{1/3}], x + t \text{ odd}\right\} \subset [-2^{5/6} \delta; 2^{5/6} \delta], \\
        F_t(u) &:= \left|\widetilde A_1\left(t / \sqrt{2} + t^{1/3} \, 2^{-5/6} \, u, \, t\right)\right| \cdot \left(t / 2\right)^{1/3} = |\Ai(u)| + O_\delta(t^{-2/3}), \qquad u \in S_t,
    \end{align*}
    where the latter estimate follows from Corollary~\ref{cor-airy-checkers} (see row~4 of Tab.~\ref{tab-checkers-asymp}). Then
    \begin{align*}
        \max_{\substack{x \in [0; t] \\ x + t \text{ odd}}} |\widetilde A_1(x, t)| &= \max_{u \in S_t} |F_t(u)| \cdot 2^{1/3} \, t^{-1/3} = \max_{u \in [-2^{5/6} \delta; 2^{5/6} \delta]} |\Ai(u)| \cdot 2^{1/3} \, t^{-1/3} + O_\delta\left(\frac{1}{t}\right), \\
        \argmax_{\substack{x \in [0; t] \\ x + t \text{ odd}}} |\widetilde A_1(x, t)| &= \frac{t}{\sqrt{2}} + \argmax_{u \in S_t} |F_t(u)| \cdot t^{1/3} \, 2^{-5/6}
        = \frac{t}{\sqrt{2}}+ \argmax_{u \in [-2^{5/6} \delta; 2^{5/6} \delta]} |\Ai(u)| \cdot t^{1/3} \, 2^{-5/6} + O_\delta(1),
    \end{align*}
    where the last estimate in each line holds by Lemma~\ref{lm-maximum-limit} in Appendix~\ref{sec-asymptotic-estimates} because $\Ai''(\hat u) = \hat u \Ai(\hat u) \ne 0$ at $\hat u := \argmax_{u \in [-2^{5/6} \delta; 2^{5/6} \delta]} |\Ai(u)| = \argmax_{u \in \mathbb{R}} \Ai(u)$ by the assumption $\delta > 1$ and~\cite[\S9.2.1, Tab.~9.9.1]{NIST-10}.
\end{proof}

\subsection{Conclusions and future work}
\label{sec-future}

We have proved a uniform large-time asymptotic formula for the wave function of a quantum walk in the whole plane (Theorems~\ref{th-airy-checkers} and~\ref{th-gamma-checkers}) involving the common Airy phenomenon and a new transition phenomenon governed by the gamma function, which we call the \emph{gamma phenomenon}. We have proved general Theorems~\ref{th-airy} and~\ref{th-gamma} that are similar to each other, producing such uniform asymptotic formulae from conformal mappings associated with the underlying integrals with exponential phase.

A natural next step is to extend these results to hypergeometric functions, quantum walks, and the six-vertex model with more general parameters \cite{Arnault-22,Destri-Vega-87,Ozhegov-24}, as well as related models \cite{Shah-etal-25,Borodin-etal-25}.

The Airy phenomenon is observed in a wide variety of combinatorial problems. It would be interesting to expand the set of examples in which the gamma phenomenon appears. For instance, an asymptotic formula for the number of random surface maps, obtained recently by M.~Wallner et al. (private communication), exactly matches general form~\eqref{eq-gamma-asymp} of the gamma phenomenon. It is interesting how often the phenomenon occurs near the light cone (e.g., in asymptotics of~\eqref{eq-general-solution}), event horizons, or domain boundaries.

Theorems~\ref{th-airy} and~\ref{th-gamma} are only the simplest versions of these phenomena. Collision of two critical points or a critical point and a pole may lead to more complicated expressions with the Airy and gamma functions; see \cite[Theorem~4]{Banderier-random-01},~\cite[Theorem~2.12]{Kuijlaars-Martinez-Finkelshtein-04}, \cite[Eq.~(3.9)]{Bleistein-66}, \cite[Theorem~2.3]{Borodin-Olshansky-03}. It would be interesting to generalize our theorems to those and even more complex collision scenarios, as well as different integration contours. Also, we conjecture that the bijectivity assumption for the derivative can be dropped in Theorems~\ref{th-airy}--\ref{th-gamma}.

\subsubsection*{Acknowledgments}
We thank C.~Banderier, I.~Binder, A.~Borodin, D.~Frolenkov, A.~Gottlieb, and M.~Wallner for inspiring discussions.

The work of M.~Drmota was supported by the Austrian Science Fund (FWF), project [10.55776/F1002].
The work of M.~Skopenkov was supported by KAUST Baseline Funding.
The work of A.~Ustinov was supported by the Theoretical Physics and Mathematics Advancement Foundation ``BASIS''. In 2021--2022, the last three authors were supported by a joint grant of FWF [I5554] and RSF.

The main results of Secs.~\ref{sec-airy},~\ref{sec-gamma}, and~\ref{sec-peak} first appeared in the courseworks by F. Kuyanov in 2024, 2026, and 2023, respectively.

\subsubsection*{AI use statement}
Generative AI tools were only used to find mathematical errors (but not for correcting them), to fix delivery issues in the introduction (each suggestion was implemented manually, revised, and reviewed), and to fix grammatical mistakes. The authors take full credit and responsibility for the entire manuscript's content.

\appendix

\section{Technical construction of boundary neighbourhoods}
\label{sec-boundary-neighbourhoods}

In this appendix, we prove the following results required for Propositions~\ref{prop-airy-substitution} and~\ref{prop-gamma-substitution}.

\begin{proposition}[Near-boundary behaviour, Airy case]
    \label{prop-airy-neighbourhoods}
    With the notation of Sec.~\ref{sec-airy-substitution}, there is a family of simply-connected neighbourhoods $N = N_v$ of $\partial U$ in $Q_1$ such that $\L|_N$ is injective, $N \cap \L^{-1}(\Pi) \subset M$, and $\L(N) \cap [v\pi; +\infty) = \varnothing$ for each $v$.
\end{proposition}

\begin{proposition}[Near-boundary behaviour, Gamma case]
    \label{prop-gamma-neighbourhoods}
    With the notation of Sec.~\ref{sec-gamma-substitution}, there is a family of simply-connected neighbourhoods $N = N_v$ of $(-\infty; \beta)$ in $H^+$ such that $\L|_N$ is injective, $N \cap \L^{-1}(\Pi) \subset M$, and $\L(N) \cap (-\infty; \L(\beta-)) = \varnothing$ for each $v$.
\end{proposition}

In both cases, a family of sufficiently small neighbourhoods will do the job. However, an accurate proof is quite technical, because the existence of such a neighbourhood for each $v$ does not guarantee the existence of a family of neighbourhoods, which is required for our proofs (see Remark~\ref{rem-parameters} below).

Let us introduce a few notions and lemmas. Let $\dot D$ be a domain. A connected subset $\gamma \subset \partial \dot D$ is a \emph{Jordan boundary interval} if each point of $\gamma$ has a neighbourhood $B$ in $\dot D$ such that $\partial B$ is a Jordan curve in $\dot D \cup \gamma$. In particular, $\partial \dot D \setminus \gamma$ is closed. Let $\Gamma \subset \partial \dot D$ be a finite disjoint union of Jordan boundary intervals. An analytic family $\L_v \colon \dot D \to \mathbb{C}$, where $v \in V$, is \emph{locally injective near $\Gamma$}, if each point of $\Gamma$ has a family of neighbourhoods $B_v$ in $\dot D$ such that $\L_v|_{B_v}$ is injective and $\partial \L_v^{-1}(\L_v(B_v)) \cap \Gamma = \partial B_v \cap \Gamma$ for each $v \in V$. For instance, the map $z\mapsto z^2$ of the unit disk is locally injective near $\{e^{i\phi}:0<\phi<\delta\}$ if and only if $\delta\le \pi$.

\begin{lemma}[Local injectivity, Airy case]
    \label{lm-airy-local-injectivity}
    In Sec.~\ref{sec-airy-substitution}, the family $\L|_{D \cap U}$ is locally injective near $\partial U$.
\end{lemma}

\begin{proof}
    For each $p_0 \in \partial U$ and $v_0 \in V$, we shall construct a family of neighbourhoods $B = B_v$ of $p_0$ in $D \cap U$ with $v$ varying in $(v_0 - \delta; v_0 + \delta) \cap V$ for some $\delta > 0$; viewed as subsets of $D \times V$, those families will constitute the desired family of neighbourhoods of $p_0$ in $D\cap U$. To construct $B$, take an open semi-circle $C$ around $p_0$ in $U$ (or a quarter-circle if $p_0 = 0$ or $\pi$) with $\partial C \subset D$; see Fig.~\ref{fig-airy-L-F}. Since $\Im \L' = -\Im \omega' > 0$ in $D \cap U$, it follows that $\L|_C$ is injective by the Noshiro--Warschawski theorem. Then $\L(C)$ is the domain bounded by $\L(\partial C)$ because $\L$ is analytic on $\partial C$. Clearly, $\L(p_0) \not\in \Cl \L(\partial C \setminus \partial U)$. By Lemma~\ref{lm-airy-L}, there are $\delta, r > 0$ such that for each $v \in (v_0 - \delta; v_0 + \delta) \cap V$, the distances from $\L_v(p_0)$ to $\L_v(\partial C \setminus \partial U)$ and $\L_v(C) \cap \L_v(\partial U)\subset\L_v(\partial U)\setminus \L_v(\partial C\cap \partial U)$ are greater than $r$. Let $Q_v$ be the open disk of radius $r$ centered at $\L_v(p_0)$. Then the component of $\L_v^{-1}(Q_v) \cap C$ containing $p_0$ on the boundary is the desired $B_v$. Since $\Cl Q \cap \L(C) \cap \L(\partial U) = \varnothing$, by the principle of orientation it follows that $\partial \L^{-1}(\L(B)) \cap \partial U = \partial B \cap \partial U$.
\end{proof}

\begin{lemma}[Local injectivity, Gamma case]
    \label{lm-gamma-local-injectivity}
    In Sec.~\ref{sec-gamma-substitution}, the family $\L|_{D\cap H^+}$ is locally injective near $(-\infty; \beta) \setminus \{0\}$.
\end{lemma}

\begin{proof}
    This is proved by the same argument using that $\sgn \Im \L_v' = -\sgn v$ in $D \cap H^+$ and $\L_0(\omega) = -\omega$; see Fig.~\ref{fig-gamma-L-F}. For $v_0 = 0$ and $|v| < \delta$, the distance from $\L_v(\omega_0)$ to $\L_v(C) \cap \L_v((-\infty; \beta) \setminus \{0\})$ is bounded away from $0$ by Lemma~\ref{lm-gamma-L} and the principle of orientation, since $\L_v(\hat \omega(v)) \to 0$ as $v \to 0$ by Remark~\ref{rem-omega-analyticity}.
\end{proof}

\begin{lemma}[Local-to-global injectivity with a parameter]
    \label{lm-local-to-global-parameter}
    Assume that an analytic family $\L_v \colon \dot D \to \mathbb{C}$ is locally injective near a finite disjoint union $\Gamma \subset \partial \dot D$ of Jordan boundary intervals and extends continuously to $\Gamma$ (viewed as a function $\dot D \times V \to \mathbb{C}$). Then $\Gamma$ has a family of neighbourhoods in $\dot D$ where $\L_v$ is injective.
\end{lemma}

\begin{proof}
    For each $p_0 \in \Gamma$ and $v_0 \in V$, we construct a family of neighbourhoods $\dot B = \dot B_v$ of $p_0$ in $\dot D$ with $v$ varying in $(v_0 - \delta; v_0 + \delta) \cap V$ for some $\delta > 0$; viewed as subsets of $\dot D \times V$, those families will constitute the desired family $\dot N = \dot N_v$ of neighbourhoods of $\Gamma$ in $\dot D$.

    First, we construct an auxiliary family $B=B_v$. Since $\L$ is locally injective near $\Gamma$, it is injective in a family of neighbourhoods $C = C_v$ of $p_0$ in $\dot D$ such that $\partial \L^{-1}(\L(C)) \cap \Gamma = \partial C \cap \Gamma$. Then $\L(C) \cap \L(\Gamma) = \varnothing$. Since $\Gamma \subset \partial \dot D$ is a union of Jordan boundary intervals, we may assume $\partial C$ is a Jordan curve in $\Gamma \cup \dot D$. Since $\L$ extends to $\Gamma$ continuously, it takes $C$ to the domain bounded by $\L(\partial C)$. We may assume $\L(p_0) \not\in \Cl \L(\partial C \setminus \Gamma)$. Then there are $\delta = \delta(p_0, v_0) > 0$, $r = r(p_0, v_0) > 0$ such that for each $v \in (v_0 - \delta; v_0 + \delta) \cap V$, the distance from $\L_v(p_0)$ to $\L_v(\partial C_v \setminus \Gamma)$ is greater than $r$. Assume also $r$ is less than $1/4$ of the minimal diameter of components of~$\L_v(\Gamma)$ (which is positive by the local boundary correspondence principle since $\Gamma$ is a finite disjoint union of Jordan boundary intervals). Let $Q_v$ be the open disk of radius $r$ centered at $\L_v(p_0)$. Since $\L(C) \cap \L(\Gamma)=\varnothing $ and $\L(\partial C \setminus \Gamma) \cap \Cl Q=\varnothing$, it follows that $\partial(\L_v^{-1}(Q_v) \cap C)\subset\Gamma\cup\L_v^{-1}(\partial Q_v)$. Since $\L$ is continuous, $\L_v^{-1}(Q_v) \cap C$ contains a neighborhood of $p_0$ in $C$. Let $B_v$ be the component of $\L_v^{-1}(Q_v) \cap C$ containing $p_0$ on the boundary. Then $\L|_B$ is injective and $\L(B)$ is a component of $Q\setminus\L(\Gamma)$.

    Let $\dot Q_v$ be the open disk of radius $\dot r := r / 2$ centered at $\L_v(p_0)$. Let $\dot B_v(p_0, v_0)$ be the component of $\L_v^{-1}(\dot Q_v) \cap B$ containing $p_0$ on the boundary.

    To prove that $\L|_{\dot N}$ is injective, assume $\L_v(p) = \L_v(p')$ for some $p \in \dot B := \dot B_v(p_0, v_0)$ and $p' \in \dot B' := \dot B_v(p'_0, v'_0)$, where $|v - v_0| < \delta(p_0, v_0)$ and $|v - v'_0| < \delta(p'_0, v'_0)$. Define $B$, $B'$, $Q$, $Q'$, $r$, $r'$, $\dot r$, and $\dot r'$ analogously.

    Let us show that $Q \cap Q'\cap\L(\Gamma)\ne \varnothing$. Since $\L(p) = \L(p') \in \L(\dot B) \cap \L(\dot B') \subset \dot Q \cap \dot Q'$, it follows that $|\L(p_0) - \L(p_0')| \le |\L(p_0) - \L(p)| + |\L(p') - \L(p'_0)| < \dot r + \dot r' \le \max(r, r')$. So $Q \cap Q'$ contains $\L(p_0)$ or $\L(p_0')$.

    Let us show that $\L(B) \cap \L(B')$ is a component of $Q \cap Q' \setminus \L(\Gamma)$. Assume the converse. Since $Q$ and $Q'$ are convex, $Q \cap Q'$ is connected. Then a pair of points of $\L(B) \cap \L(B')$ is separated by a component of $\L(\Gamma)$ in $Q \cap Q'$. This pair is joined by two paths $\gamma$ and $\gamma'$ in $\L(B)$ and $\L(B')$, because $\L(B)$ and $\L(B')$ are components of $Q \setminus \L(\Gamma)$ and $Q' \setminus \L(\Gamma)$. Then $\gamma \cup \gamma'$ is disjoint with $\L(\Gamma)$ but winds once around a component of $\L(\Gamma) \cap Q \cap Q'$. This is a contradiction, because the radii of $Q$ and $Q'$ are smaller than $1/4$ of the minimal diameter of components of $\L(\Gamma)$. So, $\L(B) \cap \L(B')$ is a component of $Q \cap Q' \setminus \L(\Gamma)$.

    Next, we show that $B \cap B' \ne \varnothing$. Since $\L(B) \cap \L(B')$ is a component of $Q \cap Q' \setminus \L(\Gamma)$ and $Q \cap Q' \cap \L(\Gamma) \ne \varnothing$, it follows that $\partial \L^{-1}(\L(B) \cap \L(B')) \cap \Gamma$ contains some Jordan boundary interval $\gamma \subset \partial \dot D$. The latter is a subset of both $\partial B \cap \Gamma$ and $\partial B' \cap \Gamma$ by local injectivity near $\Gamma$. Then both $B$ and $B'$ are neighbourhoods of $\gamma$ in $\dot D$ because $\Cl (\partial B \setminus \Gamma)\subset \L^{-1}(\partial Q)$ and $\Cl (\partial B' \setminus \Gamma)\subset \L^{-1}(\partial Q')$ are disjoint from $\gamma$. Thus $B \cap B' \ne \varnothing$.

    Consider the set $W$ of points of $\L(B) \cap \L(B')$ having at least two distinct $\L$-preimages in $B \cup B'$. The set $W$ is open by analyticity. It is also closed because $\L|_B$ and $\L|_{B'}$ are homeomorphisms on their images. We have $W \ne \L(B) \cap \L(B')$ because $W$ is disjoint from $\L(B \cap B') \ne \varnothing$. Then $W = \varnothing$, because $\L(B) \cap \L(B')$ is connected. Thus $p$ = $p'$, and $\L|_{\dot N}$ is injective.
\end{proof}

A family of open sets $\Pi_v \subset \mathbb{C}$ is in \emph{special position} with $\L_v(\Gamma)$ if $\Pi_v \cap \L_v(\Gamma) = \varnothing$ and any open disk centered in $\L_v(\Gamma)$ and intersecting $\partial \Pi_v$ contains infinitely many points of $\partial \Pi_v \cap \L_v(\Gamma)$ for each $v \in V$.

\begin{lemma}[Special position]
    \label{lm-special-position}
    In Secs.~\ref{sec-airy-substitution} and~\ref{sec-gamma-substitution}, the families $\Pi_v$ are in special position with $\L_v(\Gamma)$.
\end{lemma}

\begin{proof}
    In Secs.~\ref{sec-airy-substitution} and~\ref{sec-gamma-substitution}, the images $\L_v(\Gamma)$ are given by Lemmas~\ref{lm-airy-L} and~\ref{lm-gamma-L}, respectively; see Figs.~\ref{fig-airy-L-F} and~\ref{fig-gamma-L-F}. Thus, $\Pi_v \cap \L_v(\Gamma) = \varnothing$ and $\partial \Pi_v \cap \L_v(\Gamma)$ does not have isolated points. It remains to show that any open disk centered in $\L_v(\Gamma) \setminus \partial \Pi_v$ and intersecting $\partial \Pi_v \setminus \L_v(\Gamma)$ contains a point of $\partial \Pi_v \cap \L_v(\Gamma)$. In Sec.~\ref{sec-airy-substitution}, any such disk contains $0 \in \partial \Pi_v \cap \L_v(\Gamma)$. In Sec.~\ref{sec-gamma-substitution}, this holds because one of the difference sets is empty.
\end{proof}

\begin{lemma}[Special position with a parameter]
    \label{lm-special-position-param}
    Under the assumptions of Lemma~\ref{lm-local-to-global-parameter}, let $\Pi_v \subset \mathbb{C}$ be a family of convex open sets in special position with $\L_v(\Gamma)$. Assume that for each $v \in V$ such that $\Pi_v \ne \varnothing$, only one component $M_v$ of $\L_v^{-1}(\Pi_v)$ has infinitely many points of $\Gamma$ on the boundary. Then $\Gamma$ has a family of neighbourhoods in $\dot D$ disjoint with $\L^{-1}(\Pi) \setminus M$.
\end{lemma}

\begin{proof}
    Use the notation $B$ and $Q$ from the second paragraph of the proof of Lemma~\ref{lm-local-to-global-parameter}, so that $\L(B)$ is a component of $Q \setminus \L(\Gamma)$. By special position, $\L(\Gamma) \cap \Pi = \varnothing$. Since $Q$ and $\Pi$ are convex, $Q \cap \Pi$ is connected. Hence, $\L(B) \cap \Pi$ and $B \cap \L^{-1}(\Pi)$ are. Assume $\L(B) \cap \Pi \ne \varnothing$ so that $Q\cap\Pi\subset \L(B)$. We have $\L(p_0) \in Q$ but $\L(p_0) \not\in \Pi$ because $\L(p_0) \in \L(\Gamma)$. Thus, $Q$ has common points with both $\Pi$ and $\mathbb{C} \setminus \Pi$, hence with $\partial \Pi$. By special position, it contains infinitely many points of $\partial \Pi \cap \L(\Gamma)$. They all belong to $\partial \Pi \cap \L(\partial B \cap \Gamma)$ because $Q\cap\Pi\subset \L(B)$ and $\L(\partial B\setminus\Gamma)\subset\partial Q$. Thus the boundary of the connected set $B \cap \L^{-1}(\Pi)$ contains infinitely many points of $\Gamma$, hence the set is contained in $M$. So, the families $B$ for all $p_0$ and $v_0$ constitute the desired family of neighbourhoods.
\end{proof}

\begin{lemma}[Regular neighbourhood with a parameter]
    \label{lm-regular-neighborhood}
    If $\dot D$ is simply connected and unbounded, and $\Gamma\subset \partial \dot D$ is a Jordan boundary interval, then any family of neighbourhoods $N_v$ of $\Gamma$ in $\dot D$ contains a family of simply-connected neighbourhoods of $\Gamma$ in $\dot D$.
\end{lemma}

\begin{proof}
    By the assumptions of the lemma and the local boundary correspondence principle, there is a bijective conformal mapping $\phi \colon \dot D \to H^+$ that extends to $\Gamma$ continuously. The desired family of neighbourhoods is $\bigcup_{x \in \phi(\Gamma)} \phi^{-1}((x; x + iy_v(x)))$, where $y_v(x) \in \mathbb{R} \cup \{+\infty\}$ is maximal such that $(x; x + iy_v(x)) \subset \phi(N_v)$.
\end{proof}

\begin{proof}[Proof of Proposition~\ref{prop-airy-neighbourhoods}]
    A family of neighbourhoods $\dot N = \dot N_v$ of $\Gamma := \partial U$ in $D \cap U$, where $\L|_{\dot N}$ is injective and $\dot N \cap \L^{-1}(\Pi) \subset M$, exists by Lemmas~\ref{lm-airy-local-injectivity},~\ref{lm-local-to-global-parameter}--\ref{lm-special-position-param}, and~\ref{lm-airy-univalence}. By injectivity, $\L(\dot N)$ belongs to one component of $\mathbb{C}\setminus \L(\partial U)$, namely, to left one, by Lemma~\ref{lm-airy-L}. By Lemma~\ref{lm-regular-neighborhood}, we may assume that $\dot N$ is simply connected.

    It remains to extend $\dot N$ across $(\pi; \pi + i\infty)$. The component $\ddot N$ of $\dot N \cap\L^{-1}(H^+)$ whose boundary contains $\pi$ forms a family of neighbourhoods of $(\pi; \pi + i\infty)$ in $D \cap U$ because $\L((\pi; \pi + i\infty))\subset H^+$ by Lemma~\ref{lm-airy-L}. Let $\sigma(p) := 2\pi - \overline p$ be the reflection about the line $\Re p = \pi$, and set $N := \dot N \cup (\pi; \pi + i\infty) \cup \sigma(\ddot N)$. Then $N = N_v$ is a family of neighbourhoods of $\partial U$ in $Q_1$.

    By the symmetries and periodicity of $D$ and $\omega$, we have $\sigma(D) = D$ and $\L_v(\sigma(p)) = 2\pi v - \overline{\L_v(p)}$. Thus $\L$ maps $\dot N$, $(\pi; \pi + i\infty)$, and $\sigma(\ddot N)$ into the pairwise disjoint sets
    \begin{align*}
        \{z \colon \Re z < v\pi\}, \qquad \{z \colon \Re z = v\pi, \Im z > 0\}, \qquad \{z \colon \Re z > v\pi, \Im z > 0\},
    \end{align*}
    respectively. Since $\L|_{\dot N}$ is injective, $\L|_{\sigma(\ddot N)}$ is injective by reflection, and $\L|_{(\pi; \pi + i\infty)}$ is injective by Lemma~\ref{lm-airy-L}, it follows that $\L|_N$ is injective. The images of the added sets are disjoint with both $\Pi$ and $[v\pi; +\infty)$, so $N \cap \L^{-1}(\Pi) \subset M$ and $\L(N) \cap [v\pi; +\infty) = \varnothing$.

    Finally, since $\L|_{\dot N}$ is an injective conformal map, it follows that $\L(\dot N)$ is simply connected. Since $\L(\ddot N)$ is a component of $\L(\dot N) \cap H^+$, it follows that $\ddot N$ is simply connected as well. Thus $N$ is simply connected.
\end{proof}

Proposition~\ref{prop-gamma-neighbourhoods} has a more technical proof because $\L_v(\omega)$ is discontinuous at $\omega = 0$.

\begin{proof}[Proof of Proposition~\ref{prop-gamma-neighbourhoods}]
    A desired family of (possibly disconnected) neighbourhoods of $\Gamma := (-\infty; \beta) \setminus \{0\}$ in $D \cap H^+$ exists by Lemmas~\ref{lm-gamma-local-injectivity}--\ref{lm-special-position-param} and~\ref{lm-gamma-univalence}. It remains to add a family of neighbourhoods $B = B_v$ of $0$ in $D \cap H^+$ for each $v_0 \in V$, with $v$ varying in $(v_0 - \delta; v_0 + \delta) \cap V$ for some $\delta > 0$ (and shrink the result).

    The construction of the family $B$ is analogous to the proof of Lemma~\ref{lm-airy-local-injectivity} above, only $Q = Q_v$ is set to be the half-plane $\sgn v_0 \Re z < -r$ with large $r = r(v_0)$ if $v_0 \ne 0$, and the union of an open disk $Q^0$ of small radius $r_0$ centered at $0$ and the half-strip $Q^{\infty} = Q^{\infty}_v = \{z \colon v \Re z < 0, \Im z \in (0; \pi v)\}$ if $v_0 = 0$.

    In the latter case, $Q$ has analogous properties for small $|v|$: since $\L_0(\omega) = -\omega$, the image $\L_v(\Cl(\partial C \setminus \mathbb{R}))$ is outside $Q^0$ for small $r_0$ by compactness. Next, recall that $\L_v((-\infty; \beta) \setminus \{0\})$ is given by Lemma~\ref{lm-gamma-L} and $\hat \omega(v), \L_v(\hat \omega(v)) \to 0$ as $v \to 0$ by Remark~\ref{rem-omega-analyticity}; see Fig.~\ref{fig-gamma-L-F}. Since $\L$ is injective in $C$ and analytic on $\partial C \setminus \{0\}$, it follows that $\L_v(\partial C \setminus \mathbb{R})$ is disjoint with $(-v\infty; \L_v(\hat \omega)]$, hence with $Q^{\infty}_v$ and $Q_v$. Since, in addition, $\L_v(\omega) \to \infty$ as $\omega \to 0$ for $v \ne 0$, it follows that $\L_v(C)$ is the domain bounded by the curve $\L_v(\partial C \setminus \mathbb{R})$ and two rays to $-v\infty$ and $\pi iv - v\infty$ for $v\ne 0$.
    Then $\Cl Q_v \cap \L_v(C) \cap \L_v(\Gamma) = \varnothing$ because $\Cl Q_v \cap \L_v(\Gamma)$ is contained in the union of the two rays. Applying the same argument to the half-circle $\dot C \subset H^+$ of radius $r_0 / 2$ centered at $0$, we conclude that $Q_v \supset \L_v(\dot C)$, because $\L_v(\dot C)$ is the domain bounded by $\L_v(\partial \dot C \setminus \mathbb{R})$ and two similar rays. This implies that $\L_v^{-1}(Q_v) \cap C$ is a family of neighbourhoods of $0$ in $H^+$ for $v_0 = 0$ (this is no longer automatic because $\L_v(\omega)$ is discontinuous at $\omega = 0$). The same holds for $v_0 \ne 0$ and $|v - v_0| < \delta < |v_0|$. So, the component of $\L_v^{-1}(Q_v) \cap C$ containing $0$ on the boundary is the desired family of neighbourhoods $B = B_v(0, v_0)$ of $0$ in $H^+$. For $\omega_0 \ne 0$, take $B = B_v(\omega_0, v_0)$ and $r(\omega_0, v_0)$ from the proof of Lemma~\ref{lm-local-to-global-parameter}.

    Next, let us construct a family of neighbourhoods $\dot N$ of $(-\infty; \beta)$ in $H^+$, where $\L$ is injective. For each $\omega_0 \in (-\infty; \beta)$ and $v_0 \in V$, we construct a family of neighbourhoods $\dot B$.

    For $\omega_0 = v_0 = 0$, let $\dot Q^0$ be the open disk of radius $r_0 / 2$ centered at $0$, $\dot Q^{\infty}_v := Q^{\infty}_v$, and $\dot Q_v := \dot Q^0 \cup \dot Q^{\infty}_v$. For $\omega_0 = 0, v_0 \ne 0$, let $\dot Q$ be the half-plane $\sgn v_0 \Re z < -r(v_0) - r_0$. For $\omega_0 \ne 0$, let $\dot Q_v$ be the open disk of radius $\min(r(\omega_0, v_0), r_0) / 2$ centered at $\L_v(\omega_0)$. Let $\dot B_v(\omega_0, v_0)$ be the component of $\L_v^{-1}(\dot Q_v) \cap B_v$ containing $\omega_0$ on the boundary; it is again a family of neighbourhoods of $\omega_0$ in $H^+$.

    The proof of the injectivity of $\L|_{\dot N}$ is analogous to the proof of Lemma~\ref{lm-local-to-global-parameter}, only the cases $\omega \in \dot B_v(0, 0) \cap \L^{-1}(\dot Q^{\infty})$, $\omega \in \dot B_v(0, 0) \cap \L^{-1}(\dot Q^0)$, $\omega \in \dot B_v(0, v_0)$ with $v_0 \ne 0$, and $\omega \in \dot B_v(\omega_0, v_0)$ with $\omega_0 \ne 0$ are considered separately, and similarly for $\omega'$. In the former case, $\Cl Q \cap \Cl Q'$ still contains an interval $\tilde\gamma \subset \L((-\infty; \beta) \setminus \{0\})$, because the image is given by Lemma~\ref{lm-gamma-L}, and $\L(B)\cap \L(B')$ is adjacent to $\tilde\gamma$; then the same argument implies $B \cap B' \ne \varnothing$ because $B$ and $B'$ are also adjacent to $\gamma\subset\L^{-1}(\tilde\gamma)$ by the local boundary correspondence principle.

    Finally, $B=B_v(0, v_0)$ is disjoint with all the components of $\L^{-1}(\Pi) \cap H^+$ apart from $M$ (cf.~Lemma~\ref{lm-special-position-param}): $\Cl Q$ contains an interval $\tilde\gamma\subset\partial\Pi\cap\L((-\infty; \beta) \setminus \{0\})$, and $\L(B)$ is adjacent to $\tilde\gamma$; then the connected set $\L^{-1}(\Pi)\cap B$ is adjacent to a sub-interval of $(-\infty; \beta)$, hence is contained in $M$. It remains to repeat the proof of Lemma~\ref{lm-regular-neighborhood} for $N_v = \dot N_v \cap \L_v^{-1}(\mathbb{C}\setminus (-\infty;\L(\beta-)])$, where $\L(\beta-)<\L(\hat\omega)$ for $v>0$ by Lemma~\ref{lm-gamma-L}.
\end{proof}

\section{Parameter-dependent versions of lemmas from complex analysis}
\label{sec-complex-analysis}

In this appendix, we prove parameter-dependent versions of standard lemmas, with the notation of Sec.~\ref{sec-notation}.

\begin{lemma}[Inverse function theorem with a parameter]
    \label{lm-inverse-param}
    Let $D_v, E_v \subset \mathbb{C}$, where $v \in V \subset \mathbb{R}$, be families of domains. If $f_v \colon D_v \to E_v$ is an analytic family of bijective functions, then $f_v^{-1} \colon E_v \to D_v$ is.
\end{lemma}

\begin{proof}
    Extending $f_v(z)$ to a domain in $\mathbb{C}\times \mathbb{C}$ and applying the inverse function theorem to the map $(v, z) \mapsto (v, f_v(z))$, we conclude that $f_v^{-1}$ is an analytic family, in particular, depends on $v$ analytically.
\end{proof}

\begin{remark}
    \label{rem-parameters}
    The lemma does not remain true if $D_v$ is not a \emph{family} of domains; e.g., if $D_v=H^+$ for $v\ge 0$, $D_v=H^-$ for $v< 0$, and $f_v(z)=z^2$. This example shows that introducing parameters requires extra care.
\end{remark}

\begin{lemma}[Removable singularity theorem with a parameter]
    \label{lm-rm-singularity-param}
    Let  $D_v \subset \mathbb{C}$, where $v \in V \subset \mathbb{R}$, be a family of domains and $\hat z_1,\dots,\hat z_n \colon V \to \mathbb{C}$ be continuous. If a family of functions $f_v \colon D_v \to \mathbb{C}$ is real analytic outside the curves $z = \hat z_k(v)$ and continuous in $D_v$ for each fixed $v \in V$, then it is real analytic everywhere.
\end{lemma}

\begin{proof}
    Fix some $v \in V$. Then $f_v(z)$ is real analytic in $z$ at $\hat z_k(v)$ by the removable singularity theorem. To establish its real analyticity in $v$ at $z_k := \hat z_k(v) \in D_v$ (possibly coincident with another $\hat z_l(v)$), use Cauchy's formula: $f_v(z_k) = \frac{1}{2\pi i} \int_{\gamma} \frac{f_v(z) \, dz}{z - z_k}$ for a fixed small contour $\gamma$ around $z_k$. Since the integrand depends on $v$ analytically (for the fixed $z_k$), so does $f_v(z_k)$. By Osgood's lemma, $f_v$ is a real analytic family.
\end{proof}

\begin{lemma}[Hadamard's lemma with a parameter]
    \label{lm-hadamard}
    Let $D_v \subset \mathbb{C}$, where $v \in V \subset \mathbb{R}$, be a family of domains symmetric about the real axis. Let $\hat z \colon V \to \mathbb{R}$ be real analytic with $\hat z(v) \in D_v$ for each $v \in V$. For each real analytic family of functions $f_v \colon D_v \to \mathbb{C}$, there is a real analytic family of functions $g_v \colon D_v \to \mathbb{C}$ such that
    \begin{align*}
        f(z) = f(\hat z) + (z - \hat z)\,g(z) \quad\text{for each $v \in V$ and $z \in D_v$}.
    \end{align*}
\end{lemma}

\begin{proof}
    Set
    \begin{align*}
        g(z) :=
        \begin{cases}
            (f(z) - f(\hat z)) / (z - \hat z), & \text{if $z \neq \hat z$}, \\
            f'(\hat z), & \text{if $z = \hat z$}.
        \end{cases}
    \end{align*}
    The family of functions $g(z)=g_v(z)$ is real analytic outside the curve $z = \hat z(v)$ and continuous in $D_v$ for each fixed $v \in V$. By Lemma~\ref{lm-rm-singularity-param}, it is real analytic everywhere.
\end{proof}

\begin{lemma}[Equivariant Hadamard's lemma with a parameter]
    \label{lm-hadamard-equiv}
    Let $D_v \subset \mathbb{C}$, where $v \in V \subset \mathbb{R}$, be a family of domains symmetric about both coordinate axes, and let $\theta\colon V \to \mathbb{R}$ be a real analytic function such that $\sqrt{\theta(v)} \in D_v$ and $0 \in D_v$ for each $v \in V$. For each even real analytic family of functions $G_v\colon D_v \to \mathbb{C}$, there exists an even real analytic family of functions $\widetilde G_v\colon D_v \to \mathbb{C}$ such that
    \begin{align*}
        G(z) = G(\sqrt{\theta}) + (z^2 - {\theta})\,\widetilde G(z) \quad\text{for each $v \in V$ and $z \in D_v$}.
    \end{align*}
    For each odd real analytic family of functions $H_v\colon D_v \to \mathbb{C}$, there is an odd real analytic family of functions $\widetilde H_v\colon D_v \to \mathbb{C}$ and a real analytic function $c = c(v)$ such that
    \begin{align*}
        H(z) = cz + (z^2 - \theta)\,\widetilde H(z) \quad\text{for each $v \in V$ and $z \in D_v$}.
    \end{align*}
\end{lemma}

\begin{proof}
    First we show there is a real analytic family of functions $\widehat G_v(u)$ in $\widehat D = \widehat D_v := \{u \colon \pm \sqrt{u} \in D_v\}$ such that $G(z) = \widehat G(z^2)$ (cf.~Whitney's theorem \cite[Theorem~1]{Whitney-43}). Indeed, set $\widehat G(u) := G(\sqrt{u})$. The resulting function is real analytic in $\widehat D \setminus (-\infty; 0]$. Since $G(u)$ is even, it follows that $\widehat G(u) = G(-i\sqrt{-u})$ is real analytic in $\widehat D \setminus [0;+\infty)$. By the removable singularity theorem, it is also analytic at the origin. Clearly, $\widehat G_v(u) = G_v(\sqrt{u})$ depends analytically on $v$ for each fixed $u$ because $G_v(z)$ does.

    Applying Lemma~\ref{lm-hadamard} to $\widehat G(u)$ at the point $\hat u = \theta$, we get
    \begin{align*}
        \widehat G(u) &= \widehat G(\hat u) + (u - \hat u)\,\widetilde{\widehat G}(u) \qquad\text{and}\qquad G(z) = G(\sqrt \theta) + (z^2 - \theta)\,\widetilde G(z),
    \end{align*}
    where $\widetilde G(z) := \widetilde{\widehat G}(z^2)$ is an even real analytic family of functions.

    Since $0 \in D_v$ and $H(0) = 0$, from Lemma~\ref{lm-hadamard} it follows that $H(z) = zK(z)$ for some even real analytic family of functions $K(z)$. Applying the first part of the lemma to $K(z)$, we get
    \begin{align*}
        H(z) = zK(z) = z(K(\sqrt \theta) + (z^2 - \theta)\,\widetilde K(z)) = K(\sqrt \theta)\,z + (z^2 - \theta)\,z \widetilde K(z),
    \end{align*}
    and setting $\widetilde H(z) := z \widetilde K(z)$ and $c := K(\sqrt \theta)$ proves the second part of the lemma.
\end{proof}

\section{Elementary asymptotic estimates}
\label{sec-asymptotic-estimates}

In this appendix, we prove several elementary estimates required for Secs.~\ref{sec-airy-proof} and~\ref{sec-peak}.

\begin{lemma}
    \label{lm-sincos-tail}
    For each $t, Q > 0$ and $\theta < Q^2$ such that $Q^3 / 3 - \theta Q$ is an integer multiple of $\pi / t$, we have
    \begin{align*}
        \int_Q^{+\infty} \sin\left(t\left(\frac{z^3}{3} - \theta z\right)\right) z\,dz &= O\left(\frac{1}{|Q - \sqrt{\theta}|\,t}\right), \\
        \int_Q^{+\infty} \cos\left(t\left(\frac{z^3}{3} - \theta z\right)\right) dz &= O\left(\frac{1}{|Q - \sqrt{\theta}|^5\,t^2}\right).
    \end{align*}
\end{lemma}

\begin{proof}
    Let us prove the first bound. Integrating by parts, we obtain
    \begin{multline*}
        \int_Q^{+\infty} \sin\left(t\left(\frac{z^3}{3} - \theta z\right)\right) z\,dz = -\int_Q^{+\infty} \frac{z}{t(z^2 - \theta)} \cdot d\cos\left(t\left(\frac{z^3}{3} - \theta z\right)\right) = \\
        = -\frac{1}{t} \left(\frac{z\cos(t(z^3/3 - \theta z))}{z^2 - \theta}\bigg\rvert_Q^{+\infty} + \int_Q^{+\infty} \frac{z^2 + \theta}{(z^2 - \theta)^2} \cdot \cos\left(t\left(\frac{z^3}{3} - \theta z\right)\right) dz\right) = \\
        = -\frac{1}{t}\,O\left(\frac{1}{|Q - \sqrt{\theta}|} \cdot \left|\frac{Q}{Q + \sqrt{\theta}}\right| + \int_Q^{+\infty} \frac{1}{|z - \sqrt{\theta}|^2} \cdot \left|\frac{z^2 + \theta}{(z + \sqrt{\theta})^2}\right| dz\right) = O\left(\frac{1}{|Q - \sqrt{\theta}|\,t}\right),
    \end{multline*}
    where the last equality follows from the bounds
    \begin{align*}
        \left|\frac{Q}{Q + \sqrt{\theta}}\right| \le 1, \quad
        \left|\frac{z^2 + \theta}{(z + \sqrt{\theta})^2}\right| = \left|\frac{z^2 + \theta}{z^2 + \theta + 2z\sqrt{\theta}}\right| \le 1,
        \quad\text{and}\quad \frac{1}{|z - \sqrt{\theta}|^2} \le \frac{2}{(z - Q + |Q - \sqrt{\theta}|)^2} \quad\text{for }z\ge Q.
    \end{align*}
    Let us prove the second bound in the lemma. Integrating by parts, we obtain
    \begin{multline*}
        I := \int_Q^{+\infty} \cos\left(t\left(\frac{z^3}{3} - \theta z\right)\right) dz = \int_Q^{+\infty} \frac{1}{t(z^2 - \theta)} \cdot d\sin\left(t\left(\frac{z^3}{3} - \theta z\right)\right) = \\
        = \frac{\sin(t(z^3/3 - \theta z))}{t(z^2 - \theta)}\bigg\rvert_Q^{+\infty} + \frac{1}{t} \int_Q^{+\infty} \frac{2z}{(z^2 - \theta)^2} \cdot \sin\left(t\left(\frac{z^3}{3} - \theta z\right)\right) dz.
    \end{multline*}
    Since $Q^3/3 - \theta Q$ is an integer multiple of $\pi / t$, the first term vanishes. Integrating by parts again, we get
    \begin{multline*}
        I = -\frac{1}{t} \int_Q^{+\infty} \frac{2z}{t(z^2 - \theta)^3} \cdot d\cos\left(t\left(\frac{z^3}{3} - \theta z\right)\right) = \\
        = -\frac{1}{t^2} \left(\frac{2z\cos(t(z^3/3 - \theta z))}{(z^2 - \theta)^3}\bigg\rvert_Q^{+\infty} + \int_Q^{+\infty} \frac{2(5z^2 + \theta)}{(z^2 - \theta)^4} \cdot \cos\left(t\left(\frac{z^3}{3} - \theta z\right)\right) dz\right) = \\
        = -\frac{1}{t^2}\,O\left(\frac{1}{|Q - \sqrt{\theta}|^3} \cdot \left|\frac{Q}{(Q + \sqrt{\theta})^3}\right| + \int_Q^{+\infty} \frac{1}{|z - \sqrt{\theta}|^4} \cdot \left|\frac{5z^2 + \theta}{(z + \sqrt{\theta})^4}\right| dz\right) = O\left(\frac{1}{|Q - \sqrt{\theta}|^5\,t^2}\right),
    \end{multline*}
    where the last equality follows from the bounds $2\,|z - \sqrt{\theta}| \ge z - Q + |Q - \sqrt{\theta}|$ for $z \ge Q$,
    \begin{align*}
        \left|\frac{Q}{(Q + \sqrt{\theta})^3}\right| \le \frac{1}{|Q + \sqrt{\theta}|^2} \le \frac{1}{|Q - \sqrt{\theta}|^2} \qquad\text{and}\qquad \left|\frac{5z^2 + \theta}{(z + \sqrt{\theta})^4}\right| \le \frac{5}{|z + \sqrt{\theta}|^2} \le \frac{5}{|Q - \sqrt{\theta}|^2}.\\[-1.2cm]
    \end{align*}
\end{proof}

\begin{lemma}
    \label{lm-exp-tail}
    For each $t > 0$, $\theta < 0$, and $Q > 0$, we have
    \begin{align*}
        \int_{Q + \sqrt{\theta}}^{+\infty + \sqrt{\theta}} e^{it(z^3/3 - \theta z)}\,dz = O\left(\frac{e^{-2/3\,t|\theta|^{3/2}}}{Q^3\,t^{3/2}|\theta|^{1/4}}\right).
    \end{align*}
\end{lemma}

\begin{proof}
    Denote $F(z) := z^3 / 3 - \theta z$. Integrating by parts and substituting $q = z - \sqrt{\theta}$, we obtain
    \begin{multline*}
        \int_{Q + \sqrt{\theta}}^{+\infty + \sqrt{\theta}} e^{itF(z)}\,dz = \int_{Q + \sqrt{\theta}}^{+\infty + \sqrt{\theta}} \frac{de^{itF(z)}}{it(z^2 - \theta)} = \frac{e^{itF(z)}}{it(z^2 - \theta)}\bigg\rvert_{Q + \sqrt{\theta}}^{+\infty + \sqrt{\theta}} + \frac{1}{it} \int_{Q + \sqrt{\theta}}^{+\infty + \sqrt{\theta}} \frac{2z}{(z^2 - \theta)^2}\,e^{itF(z)}\,dz = \\
        = \frac{e^{-2/3\,t|\theta|^{3/2}}}{it} \left(\frac{e^{itq^3/3-t|\theta|^{1/2} q^2}}{q(q + 2i|\theta|^{1/2})}\bigg\rvert_{Q}^{\infty} + \int_{Q}^{\infty} \frac{2(q + i|\theta|^{1/2})}{q^2(q + 2i|\theta|^{1/2})^2}\,e^{itq^3/3-t|\theta|^{1/2} q^2}\,dq\right) = \\
        = \frac{e^{-2/3\,t|\theta|^{3/2}}}{it}\,O\left(\frac{e^{-t|\theta|^{1/2} Q^2}}{Q^2} + \frac{1}{Q^3} \int_{-\infty}^{+\infty} e^{-t|\theta|^{1/2} q^2}\,dq\right) = O\left(\frac{e^{-2/3\,t|\theta|^{3/2}}}{Q^3\,t^{3/2}|\theta|^{1/4}}\right),
    \end{multline*}
    where the last equality follows from $\int_{-\infty}^{+\infty} e^{-aq^2}\,dq = \sqrt{\pi / a}$ and $e^{-t|\theta|^{1/2} Q^2} = O\left(\frac{1}{Q\,t^{1/2}|\theta|^{1/4}}\right)$.
\end{proof}

\begin{lemma}
    \label{lm-int-shift}
    For each $\theta \in \mathbb{R}$ and $\Delta > 0$, we have
    \begin{align*}
        \int_{-\infty}^{+\infty} e^{i(z^3 / 3 - \theta z)}\,dz = \int_{-\infty + i\Delta}^{+\infty + i\Delta} e^{i(z^3 / 3 - \theta z)}\,dz.
    \end{align*}
\end{lemma}

\begin{proof}
    For $R, R' > 0$, denote by $I_1$, $I_2$, $I_{\downarrow}$, and $I_{\uparrow}$ the integrals of $e^{i(z^3 / 3 - \theta z)}$ over $[-R'; R]$, $[-R' + i\Delta; R + i\Delta]$, $[-R' + i\Delta; -R']$, and $[R; R + i\Delta]$, respectively. Then $I_2 = I_{\downarrow} + I_1 + I_{\uparrow}$. Substituting $z = R + iy$, we obtain
    \begin{align*}
        |I_{\uparrow}| = \left|\int_{0}^{\Delta} ie^{i((R + iy)^3 / 3 - \theta (R + iy))}\,dy\right| \le \int_{0}^{\Delta} e^{-R^2y + y^3 / 3 + \theta y}\,dy = O_{\theta, \Delta}\left(\int_{0}^{\Delta} e^{-R^2y}\,dy\right) = O_{\theta, \Delta}\left(\frac{1}{R^2}\right),
    \end{align*}
    thus $I_{\uparrow} \to 0$ as $R \to +\infty$. Analogously, $I_{\downarrow} \to 0$ as $R' \to +\infty$. Thus, $\lim_{R, R' \to +\infty} I_1 = \lim_{R, R' \to +\infty} I_2$.
\end{proof}

\begin{lemma}
    \label{lm-maximum-limit}
    Let $F(u)$ be a continuous function on a segment $[l; r]$ with a unique global maximum at some point $\hat u \ne l, r$, satisfying $F''(\hat u) < 0$. Let $S_t \subset [l; r]$ be a sequence of finite non-empty subsets such that the distance between any two consecutive points of $S_t \cup \{l, r\}$ is bounded by $d(t)$. Let $F_t(u)$ be a sequence of functions on $S_t$ such that $|F_t(u) - F(u)|$ is bounded by $g(t)$. Then
    \begin{align}
        \label{eq-maximum-limit-val}
        \max_{u \in S_t} F_t(u) &= F(\hat u) + O_F\left(g(t) + d(t)^2\right), &
        \argmax_{u \in S_t} F_t(u) &= \hat u + O_F\left(\sqrt{g(t)} + d(t)\right).
    \end{align}
\end{lemma}

\begin{proof}
    Denote
    $u_t := \argmax_{u \in S_t} F_t(u)$, $u'_t := \argmin_{u \in S_t} |u - \hat u|$, $G(u) :=
    \begin{cases}
        \frac{F(\hat u) - F(u)}{(u - \hat u)^2}, & \text{if $u \neq \hat u$}, \\
        -F''(\hat u) / 2, & \text{if $u = \hat u$}.
    \end{cases}
    $

    Since $F$ is twice differentiable at the local extremum $\hat u$, we have
    \begin{align*}
        F(u) = F(\hat u) + \frac{F''(\hat u)}{2} (u - \hat u)^2 + o\left((u - \hat u)^2\right).
    \end{align*}
    Then, by the continuity of $F(u)$, the function $G(u)$ is continuous on $[l; r]$. Also, $G(u) > 0$ because $F''(\hat u) < 0$ and $F(\hat u) > F(u)$ for all $u \neq \hat u$. Thus, there exist positive constants $L_F$, $R_F$ such that for each $u \in [l; r]$ we have $L_F < G(u) < R_F$. Substituting $u = u'_t$ and using $G(u) < R_F$, we get
    \begin{align*}
        F(u'_t) = F(\hat u) - G(u'_t)(u'_t - \hat u)^2 > F(\hat u) - R_F \cdot d(t)^2.
    \end{align*}
    Since $u_t = \argmax_{u \in S_t} F_t(u)$ and $|F_t(u) - F(u)| < g(t)$, we get
    \begin{align*}
        F(u_t) \ge F_t(u_t) - g(t) \ge F_t(u'_t) - g(t) \ge F(u'_t) - 2g(t).
    \end{align*}
    Combining the two previous inequalities, we get $F(\hat u) \ge F(u_t) > F(\hat u) - R_F \cdot d(t)^2 - 2g(t).$

    This proves the first equality in \eqref{eq-maximum-limit-val}. To prove the second one, we substitute $u = u_t$ into $G(u) > L_F$:
    \begin{align*}
        (u_t - \hat u)^2 = \frac{F(\hat u) - F(u_t)}{G(u_t)} \le \frac{F(\hat u) - F(u_t)}{L_F} = O_F\left(d(t)^2 + g(t)\right) = O_F\left(\left(d(t) + \sqrt{g(t)}\right)^2\right). \\[-1.6cm]
    \end{align*}
\end{proof}

{\footnotesize
    \bibliographystyle{abbrv}
    \bibliography{asymptotic}
}

\end{document}